\documentclass[12pt]{book}
\usepackage{amsmath}
\usepackage{epsfig}
\def\ex{{\hbox{\rm e}}}

\def\im{{\hbox{\rm Im}}} \def\mod{{\hbox{\rm mod}}} 
\def\tr{{\hbox{\rm Tr}}}

\def\pff{{\hbox{\rm Pf}}}

\def\ie{{\em i.e.}}
\def\bea{\begin{eqnarray}}
\def\eea{\end{eqnarray}}
\def\ret{\nonumber\\{}}
\def\eqs#1{(\ref{#1})}
\def\la#1{\label{#1}}
\def\re{{\hbox{\rm Re}}}
\def\im{{\hbox{\rm Im}}}
\def\ker{{\hbox{\rm ker}}} 
 
\def\mani{\cal{M}}  

\def\ad{\hbox{\rm ad}}
\def\Lie{\hbox{\bf g}}
\def\ind{{\hbox{\rm ind}}}
\def\fr#1#2{\frac{#1}{#2}}
\def\intl{\int\limits}
\def\too{\longrightarrow}
\def\half{\fr{1}{2}} 

\def\to{\rightarrow}

\def\sqr#1#2{{\vcenter{\vbox{\hrule height.#2pt
  \hbox{\vrule width.#2pt height#1pt \kern#1pt
    \vrule width.#2pt}
  \hrule height.#2pt}}}}

\def\deriv{{\cal D}}

\def\raiz{\sqrt{2}}

\def\d#1{{\dot #1}}
\def\dalpha{{\dot\alpha}}

\def\mapright#1{\smash{\mathop{\longrightarrow}\limits^{#1}}}

\font\upright=cmu10 
\font\cmss=cmss10 at 12pt \font\cmsss=cmss8 at 8pt
\font\cmssi=cmss10 at 10pt

\def\IZ{\relax\ifmmode\mathchoice
{\hbox{\cmss Z\kern-.4em Z}}{\hbox{\cmss Z\kern-.4em Z}}
{\lower.4pt\hbox{\cmsss Z\kern-.4em Z}}
{\lower1.2pt\hbox{\cmsss Z\kern-.4em Z}}\else{\cmss Z\kern-.4em
Z}\fi}
\def\iz{\relax\hbox{\cmssi Z\kern -.4em Z}}
\def\mt{\rlap{\cmss T}\kern 3.0pt{\hbox{{\cmss T}}}}
\def\identity{{\upright\rlap{1}\kern 2.0pt 1}}
\def\inbar{\vrule height1.5ex width.4pt depth0pt}
\def\mininbar{\vrule height.75ex width.3pt depth0pt} 
\def\cc{\relax\,\hbox{$\mininbar\kern-.2em{\hbox{\rm\tiny
C}}$}}
\def\IC{\relax\,\hbox{$\inbar\kern-.3em{\rm C}$}}
\def\IR{\relax{\rm I\kern-.18em R}}

\def\IL{\relax{\rm I\kern-.18em L}}
\def\IH{\relax{\rm I\kern-.18em H}}
\def\IB{\relax{\rm I\kern-.18em B}}
\def\ID{\relax{\rm I\kern-.18em D}}
\def\IE{\relax{\rm I\kern-.18em E}}
\def\IF{\relax{\rm I\kern-.18em F}}
\def\IG{\relax\hbox{$\inbar\kern-.3em{\rm G}$}}
\def\IGa{\relax\hbox{${\rm I}\kern-.18em\Gamma$}}
\def\IH{\relax{\rm I\kern-.18em H}}
\def\II{\relax{\rm I\kern-.18em I}}
\def\IK{\relax{\rm I\kern-.18em K}}
\def\IP{\relax{\rm I\kern-.18em P}}
\def\IQ{\relax\hbox{$\inbar\kern-.3em{\rm Q}$}}
\def\cp{\IC\IP}
\def\hat{\widehat}
\def\C#1{{\cal #1}}
\def\cn{{\cal N}}

\begin{document}

\thispagestyle{empty}

\begin{flushright}   US-FT/18-99\\ hep-th/9907123\\ Revised version\\
May, 2000\\
\end{flushright}
\vspace*{20pt}
\bigskip
\begin{center} {\huge Duality in Topological }
\vskip3mm {\huge Quantum Field
Theories\footnote{Doctoral Thesis,  University of
Santiago de Compostela, June 1999.}}
\vskip 0.9truecm

\vspace{3pc}

{\Large{Carlos Lozano}\footnote{Present address: Department of Physics, 
Brandeis University. Waltham, MA 02454. USA.}}

\vspace{1pc}

{\em  Departamento de F\'\i sica de Part\'\i culas,\\
Universidade de Santiago de Compostela,\\ E-15706 Santiago de
Compostela, Spain.\\}
\vspace{.5pc}
{\tt lozano@fpaxp1.usc.es}
\end{center}

\vspace{6pc}
\noindent\hrulefill

\vspace{1pc}

\noindent This thesis presents a thorough analysis of the links between
the
$\cn=4$ supersymmetric gauge theory in four dimensions and its three
topological twisted counterparts. Special emphasis is put in deriving
explicit results in terms of the vacuum structure and low-energy effective
description of four-dimensional supersymmetric gauge theories. A key
ingredient is the realization of the Montonen-Olive duality in the
twisted theories, which is discussed in detail from different viewpoints.

\vspace{1pc}


\noindent\hrulefill

\newpage
\thispagestyle{empty}
\ \ \
\newpage
\thispagestyle{empty}
\begin{center}

{\large {\bf UNIVERSIDADE DE SANTIAGO\\
DE COMPOSTELA}}

\vspace{0.5cm}
{ { Departamento de F\'\i sica de Part\'\i culas}}

\begin{figure}[hbtp]
\begin{center}
\mbox{\psfig{file=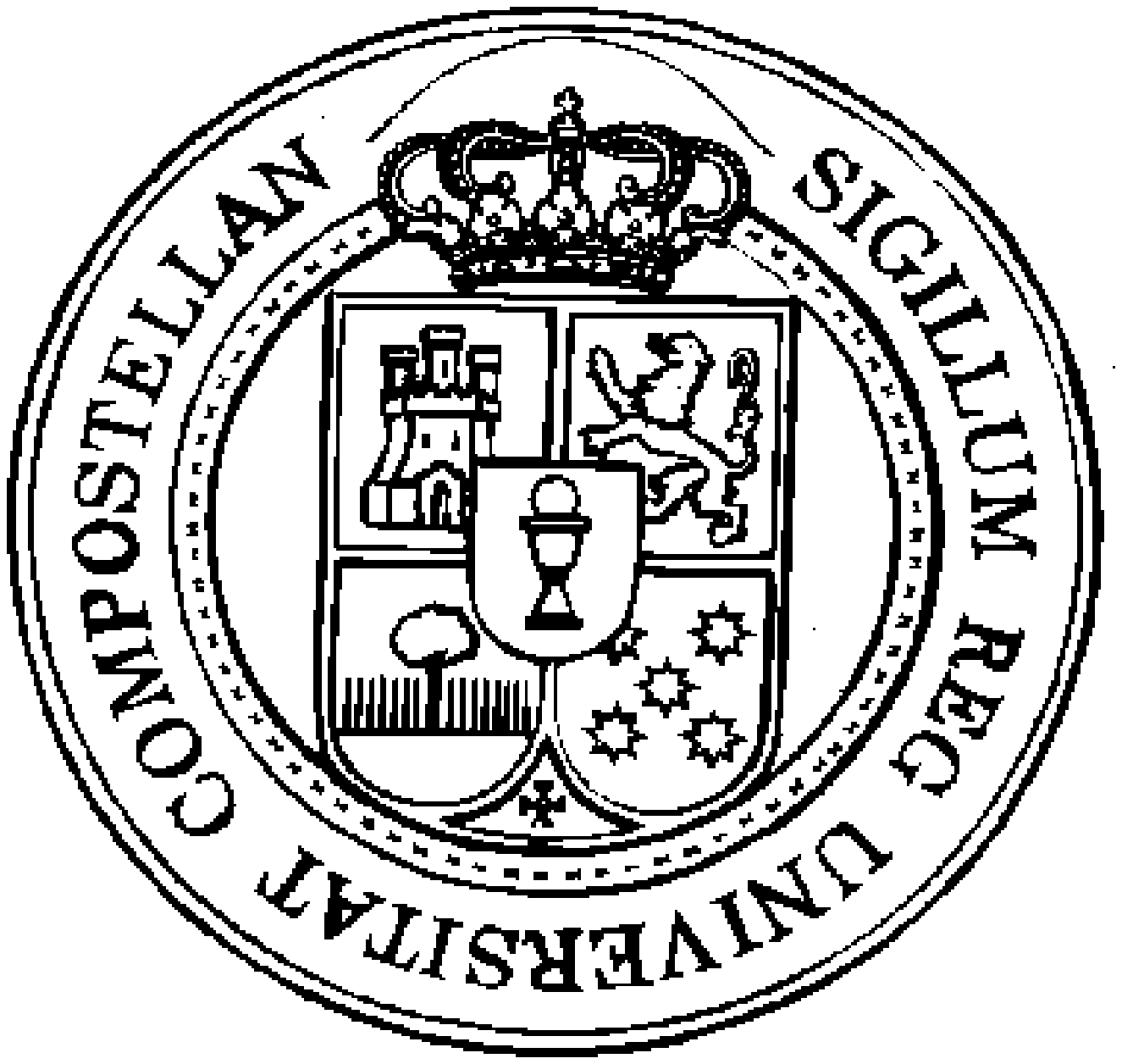,height=3cm}}
\end{center}
\end{figure}


{\Huge {\bf
Duality in Topological\\

Quantum Field Theories\\
}}

\end{center}

\vspace{6cm}

\begin{flushright}
\vspace{0.3cm}
{\em Carlos Lozano Rodr\'\i guez}\\
\vspace{0.3cm}
{\em June 11, 1999}
\end{flushright}

\ \ \ \ \

\newpage
\thispagestyle{empty}
\ \ \ \ \
\newpage
\thispagestyle{empty}
\begin{flushright}
{\it Tiens! Un Donatello parmi des fauves! }\\
L. Vauxcelles
\end{flushright}

\begin{center}
{\large{\bf Acknowledgements}}
\end{center}

I would like to thank my advisor prof. Jos\'e Labastida for 
his patience and understanding over the last four years. I am 
enormously indebted to him for sharing with me his wisdom and his
insight, and for his
permanent advice and encouragement.  

Meeting Marcos Mari\~no has been a most exciting scientific 
and personal experience. He has been a permanent and patient source of 
inspiration, topological advice and weird discussions.

I am most grateful to prof. Alfonso V. Ramallo for many illuminating
and enjoyable discussions. 

I would like to thank the Theory Division at CERN, and specially 
Uli and Fabio, for hospitality  
during the months I spent in the  
``irrenhaus." 

Finally, I would like to thank almost everybody at the Departamento de 
F\'\i sica de Part\'\i culas de la Universidad de Santiago de
Compostela for making these years a hardly forgettable experience.

\newpage
\thispagestyle{empty}
\ \ \ \ \
\newpage

\tableofcontents


\chapter*{Introduction}
\markboth{\footnotesize\bfseries Introduction
}{\footnotesize\bfseries Introduction}
\markright{\textsc {Duality in Topological Quantum Field Theories}}
\addcontentsline{toc}{chapter}{\numberline{}Introduction}

Topological quantum field theory  \cite{tqft} has become a
very fruitful link between physics and mathematics. In four
dimensions, it provides an extremely powerful  framework to
apply and test  different ideas emerged in the context of
duality as a symmetry of extended supersymmetric gauge
theories. Two salient examples are the introduction of the
Seiberg-Witten invariants 
\cite{matilde}\cite{monopole}, and the
strong coupling test of
$S$-duality carried out by Vafa and Witten 
\cite{vw} from the analysis of a twisted version of the
$\cn=4$ supersymmetric gauge theory. Subsequent
generalizations in the framework of the Seiberg-Witten
invariants have enriched the  physics 
\cite{corea}\cite{faro}\cite{lns}\cite{tesis}
\!\!\cite{mmtwo}\cite{mm4}\cite{geog}\cite{moorewitten}
as well as the mathematical literature  
\cite{oko}\cite{pt}\cite{tele}.   

The origin of these connections can be traced back to the
work of Donaldson in the early
eighties. Donaldson addressed the problem of classifying
differentiable structures on four-manifolds by studying 
classical non-Abelian gauge theory. He 
introduced the celebrated Donaldson invariants, which are 
invariants of the diffeomorphism type of smooth
four-manifolds, and which are defined as cohomology
classes on instanton moduli spaces    
\cite{donald}\cite{donaldkr}. 

In 1988, Edward Witten made a decisive contribution to the
field with his work on Donaldson theory from the point of
view  of quantum field theory \cite{tqft}. He showed that in 
a twisted version of the $\cn=2$ supersymmetric gauge theory, 
now known as the Donaldson-Witten theory, one can define
certain correlation functions which are equivalent to the
Donaldson invariants. His formulation was later reinterpreted 
by Atiyah and Jeffrey \cite{jeffrey} in a more geometrical
setting, as the Mathai-Quillen representative \cite{mathai} of
the Thom class of a vector bundle associated to the moduli
space of instantons. 

The realization of the Donaldson
invariants as correlation functions of a physical theory
inmediately raises the question as to whether one can use
physical methods to calculate them, and conversely, whether
topological computations can be used to test assumed facts
about the behaviour of physical models. This has been
repeatedly shown to be the case in a series of works
\cite{geog}\cite{moorewitten}\cite{vw}\cite{wijmp}
\!\!\cite{monopole}.

This thesis is intended to present  a thorough
analysis  of the links between four dimensional ${\cal N}=4$
supersymmetric  Yang-Mills theory and its topologically
twisted counterparts. These theories  appear as effective
worldvolume theories on D3-branes and M5-branes  wrapping
supersymmetric cycles of higher dimensional exotic
compactification  manifolds \cite{BSV}\cite{estrings}, and
provide a promising arena for testing key ideas as
field-theory  dualities, large
$N$ dynamics of supersymmetric gauge theories and,
eventually,  the recent AdS-CFT conjecture
\cite{malda}. 

For the 
${\cal N}=4$ theory, the topological twist can be
performed in three  different ways \cite{yamron}, so one ends
up with three different topological quantum  field theories
which, while living in arbitrary four-dimensional
geometries,  still retain some information
about their  physical origin. The way to unravel and at the
same time take advantage of   this connection between SYM
theories and their twisted counterparts is as  follows. The
twisted theories are topological in the sense that the
partition  function as well as a selected set of correlation
functions are independent  of the metric which defines the
background geometry. In the short distance  regime,
computations in the twisted theory are given exactly by a
saddle-point  calculation around a certain bosonic background
or moduli space, and in fact  the correlation funcions can be
reinterpreted  as describing intersection theory on this
moduli space. This correspondence can  be made more precise
through the Mathai-Quillen construction
\cite{jeffrey}\cite{mathai}, and this topic  was addressed in
full generality for the
${\cal N}=4$ theories  in \cite{ene4}. Unfortunately, it is
not possible to  perform explicit computations from this
viewpoint: the moduli spaces one ends  up with are
generically non-compact, and no precise recipe is known to 
properly compactify them. 

A complementary approach which sheds more light on the
structure of the  twisted theories and allows explicit
computations involves the long-distance  regime, where one
expects that a good description should be provided by the 
degrees of freedom of the vacuum states of the physical
theory on $\IR^4$. If  the physical theory admits an
effective low-energy Seiberg-Witten description
\cite{swi}\cite{swii},  it might appear that the answer for
the twisted, topological theory should come  by integration
over the corresponding moduli space of vacua (the $u$-plane)
of the physical model. This conjecture is indeed true,  and
has been recently put on a sound basis by Moore and Witten
\cite{moorewitten}, who have given  the precise recipe for
integrating over the $u$-plane. In the case of the pure  
${\cal N}=2$ theory with gauge group $SU(2)$, they have been 
able to  reproduce well-known results in Donaldson
theory. This approach is at the heart of the successful
reformulation of the Donaldson theory in terms of the
Seiberg-Witten invariants proposed by Witten in
\cite{monopole}. The question immediately arises as to whether
it is possible to carry out a  similar computation for the
twisted
${\cal N}=4$ theories.

While for the  topological theories related to asymptotically
free ${\cal N}=2$ theories the  interest lies in their
ability to define topological invariants for  four-manifolds,
for the twisted ${\cal N}=4$ theories the topological
character  is used as a tool to perform explicit
computations which may shed light on  the structure of the
physical ${\cal N}=4$ theory. This theory is finite and 
conformally invariant \cite{nseiberg}, and is conjectured to
have a symmetry exchanging strong  and weak coupling and
exchanging electric and magnetic fields \cite{monoli}, which
extends to  a full
$SL(2,\IZ)$ symmetry acting on the microscopic complexified
coupling 
$\tau_0$. It is natural to expect that this property should be
shared by the  twisted theories on arbitrary four-manifolds.
This was checked by Vafa and  Witten for one of the twisted
theories and for gauge group $SU(2)$ \cite{vw}, and it was 
clearly most interesting to extend their computation to
higher rank groups  and to the other twisted theories. In
this context, the
$u$-plane  approach has been applied to the twisted
mass-deformed
${\cal N}=4$ SYM theory with  gauge group
$SU(2)$ \cite{htwist}. This theory is obtained by twisting the
${\cal N}=4$ SYM  theory with bare masses for two of the
chiral multiplets. The physical theory  preserves ${\cal N}=2$
supersymmetry, and its low-energy effective description  was
given by Seiberg and Witten \cite{swii}. The answer for the
twisted theory, which  is a topological field theory with an
arbitrary mass parameter, is given  explicitly and completely
in terms of the periods and the discriminant of the 
Seiberg-Witten solution of the physical model. The partition
function of the twisted  theory transforms as a modular form
and, as in the theory considered by Vafa  and Witten, it is
possible to sharpen the computations by including 
non-Abelian electric and magnetic 't Hooft fluxes
\cite{gthooft} which are exchanged under  duality in the
expected fashion. It is also possible to compute correlation 
functions for a selected set of operators, and these turn out
to transform  covariantly under
$SL(2,\IZ)$, following a pattern which can be reproduced 
with a far more simple topological Abelian model. All
these resuts will be reviewed in chapter \ref{chtwist}.   

It would be very interesting, in the light of the AdS/CFT
correspondence, to  generalize these results to $SU(N)$ gauge
groups, but the generalization involves many non-trivial
constructions which are only available for the pure
${\cal N}=2$ SYM theory \cite{mmtwo}. On the other hand, the
Vafa-Witten twist  seems to be a far more promising example.
In this case, however, the approach based on the
Seiberg-Witten effective description is not useful, but, as
conjectured by Vafa and Witten, one can  nevertheless compute
in terms of the vacuum degrees of freedom of the 
${\cal N}=1$ theory which results from giving bare masses to
all the three  chiral multiplets of the ${\cal N}=4$ theory.
As analyzed in detail in \cite{masas}, the twisted massive
theory is topological on K\"ahler four-manifolds with 
$h^{2,0}\not=0$, and the partition function is invariant
under the perturbation. Under certain general assumptions,
the construction of Vafa and Witten can be  generalized to
$SU(N)$ (at least when $N$ is prime), and the resulting
formula  turns out to satisfy all the required constraints
\cite{sun}. It is still not clear
what the large $N$ limit of this formula  corresponds 
to on the gravity side, but the issue certainly deserves 
further study. These facts will be reviewed in great length in
chapter \ref{chvw}. 

As for the third twisted $\cn=4$ theory, no explicit results 
have been obtained so far, but some general features are
known  which constrain the structure of its topological
observables. The theory is believed to be a certain
deformation of the four-dimensional BF theory, and as such it
describes essentially intersection theory on the moduli
space of complexified flat connections 
\cite{blauthomp}\cite{marcus}. It has also been pointed out
that the theory is {\sl amphicheiral}, which means that it
is invariant to a reversal of the orientation of the
spacetime manifold. The terminology is borrowed from knot
theory,  where an oriented 
knot is said to be amphicheiral if, crudely speaking, it is
equivalent to its mirror image \cite{nudo}. From this
property it can be shown \cite{ene4} that the topological
invariants of the theory are completely independent of
the complexified coupling $\tau_0$. All this will be reviewed
in chapter \ref{chmarcus}.

The organization of the thesis is as follows. In chapter
\ref{tft} we present a general introduction to
topological quantum field theories from a functional
integral perspective. In chapter \ref{chmq} we briefly
review the Mathai-Quillen formalism, with special
emphasis on those aspects which are
relevant to the work presented here. Several excellent
reviews are available
\cite{phyrep}\cite{thompson}\cite{moore}, to
which we refer the reader for details. In chapter
\ref{chene4} we review the structure of the $\cn=4$ theory
and study its  low-energy effective description for gauge
group $SU(2)$. We also comment on several aspects of
electric-magnetic duality on arbitrary four-manifolds. In
chapter \ref{ctwists} we review the twisting of the $\cn=4$
theory and make several remarks about the structure of the
twisted theories. We also discuss the connection
between topological quantum field theories and string
theory. Chapters 
\ref{chvw}, \ref{chtwist} and \ref{chmarcus} deal with each
of the twisted $\cn=4$ theories, both from the viewpoint of
the twist and the Mathai-Quillen approach. They also include  
a thorough description of exact results for these theories
where available. Many technical details are deferred to
apendices at the end of each chapter. Finally, an appendix
contains the conventions used throughout this work.


\chapter{Topological quantum field theory}
\la{tft}
\markboth{\footnotesize\bfseries Topological quantum field theory
}{\footnotesize\bfseries Topological quantum field theory}
\markright{\textsc {Duality in Topological Quantum Field Theories}}

\setcounter{equation}{0}

In this chapter we will study the general structure of topological
quantum field theory (TQFT) from a  functional integral point of view. The 
involved functional integration is not in general well defined but, as in
ordinary quantum field theory, an  axiomatic approach has been 
constructed \cite{axio}, which we shall not attempt to review here. We will
concentrate instead on a naive functional integral approach. Although
not well defined in general, it is the  approach which has shown to be
more successful.

The basic framework will consist of an $n$-dimensional Riemannian 
manifold $X$ endowed with a metric $g_{\mu\nu}$. On this manifold we will
consider  a  set of fields $\{\phi_i\}$, and  a real functional of these
fields $S(\phi_i)$, which will be regarded as the action of the  theory.
We will  consider operators 
${\cal O}_\alpha(\phi_i)$, which will in general be arbitrary  functionals
of the fields. In TQFT these functionals are labelled by  some set of
indices
$\alpha$ carrying topological or group-theoretical data. The vacuum
expectation value (vev) of a product  of these operators is defined as the
following functional integral: 
\begin{equation}
\langle {\cal O}_{\alpha_1} {\cal O}_{\alpha_2} \cdots {\cal O}_{\alpha_p}
\rangle =
\int [D\phi_i]  {\cal O}_{\alpha_1}(\phi_i) {\cal  O}_{\alpha_2}(\phi_i)
 \cdots {\cal O}_{\alpha_p}(\phi_i) \exp\big(-S(\phi_i)\big).
\label{lpsanse}
\end{equation} 
A quantum field theory is considered topological if it
possesses the following property:
\begin{equation}
\frac{\delta}{\delta g^{\mu\nu}} 
\langle {\cal O}_{\alpha_1} {\cal O}_{\alpha_2} \cdots {\cal O}_{\alpha_p}
\rangle = 0,
\label{lpreme}
\end{equation}
i.e. if the vacuum expectation values (vevs) of some set of selected  
operators remain invariant under variations of the metric $g_{\mu\nu}$  on
$X$. If such is the case, these operators are called observables.

There are two ways to guarantee, at least formally, that condition  
(\ref{lpreme}) is satisfied. The first one corresponds to the situation in
which both the action $S$ and the operators ${\cal O}_{\alpha}$,
are metric-independent. These TQFTs are called of the {Schwarz}
type \cite{phyrep}. In the case of Schwarz-type theories one must first 
construct an action which is independent of the metric $g_{\mu\nu}$. The
method is best  illustrated by considering an explicit example. Let us take
into consideration the most  interesting case of this type of theories:
Chern-Simons gauge theory \cite{csgt}.  The data in the Chern-Simons gauge
theory are the following: a  differentiable compact three-manifold $X$, a
gauge group
$G$, which will be taken  simple and compact, and an integer parameter
$k$. The action is the integral of the Chern-Simons form associated to a 
gauge connection $A$ corresponding to the group $G$:
\begin{equation} 
S_{\mbox{\tiny\rm CS}} (A) = \int_X \tr \left(A\wedge d A + \frac{2}{3}
A\wedge A\wedge A\right).
\label{lpvalery}
\end{equation}

Observables are constructed out of operators which do not contain the 
metric
$g_{\mu\nu}$. In gauge-invariant theories, as is the case here, one must 
also demand that these operators be invariant under gauge transformations. 
The basic set of observables in the Chern-Simons gauge theory is provided
by the trace of the holonomy of the gauge connection $A$ in  some
representation
$R$ along a 1-cycle $\gamma$ in $X$, that is,  the Wilson  loop:
\begin{equation}
\tr_R \big( {\hbox{\rm Hol}}_\gamma (A) \big) =
\tr_R {\hbox{\rm P}} \exp \int_\gamma A.
\label{lpsilvie}
\end{equation} 
The vevs are labelled by representations $R_i$ and  
embeddings $\gamma_i$ of $S^1$ into $X$:
\begin{equation}
\langle \tr_{R_1} {\hbox{\rm P}} \ex^{\int_{\gamma_1} A}
  \dots
  \tr_{R_n} {\hbox{\rm P}} \ex^{\int_{\gamma_n} A} \rangle 
   = \int [DA] \tr_{R_1} {\hbox{\rm P}}
\ex^{\int_{\gamma_1} A}
  \dots
  \tr_{R_n} {\hbox{\rm P}} \ex^{\int_{\gamma_n} A}
  \ex^{\frac{i k}{4\pi} S_{\mbox{\tiny CS}} (A) }.
\label{lpencarna}
\end{equation} 
A non-perturbative analysis of the 
theory \cite{csgt}  shows that the invariants associated to the observables
(\ref{lpencarna})  are  knot and link invariants  as 
the
Jones polynomial \cite{jones} and its generalizations. The perturbative
analysis has confirmed this result and has  
shown to provide a very useful framework to study the Vassiliev  
invariants  (see \cite{preport}\cite{farocs} for a brief review.)

Another important set of TQFT of the Schwarz type are the BF
theories \cite{horo}. These can be formulated in any dimension and
are believed to be, as the Chern-Simons gauge theory, exactly solvable
quantum field theories. We will not describe them in this work. They have
recently acquired importance since it has been pointed out that
four-dimensional Yang-Mills theories could be regarded as a deformation of
these theories \cite{cottados}. 

The second way to guarantee (\ref{lpreme}) corresponds to the case in  which
there exists a symmetry, whose infinitesimal form will be denoted by
$\delta$, with the  following properties:
\begin{equation}
\delta {\cal O}_{\alpha}(\phi_i) = 0, 
\;\;\;\;\; T_{\mu\nu}(\phi_i) = \delta G_{\mu\nu}(\phi_i),
\label{lpangela}
\end{equation} where $T_{\mu\nu}(\phi_i)$ is the energy-momentum tensor of
the theory,
\ie
\begin{equation} T_{\mu\nu} (\phi_i) = \frac{\delta}{\delta g^{\mu\nu}}
S(\phi_i),
\label{lprosi}
\end{equation} and $G_{\mu\nu}(\phi_i)$ is some functional of the fields
$\phi_i$.

The fact that $\delta$ in (\ref{lpangela}) is a symmetry of the theory 
means that the transformations $\delta\phi_i$ of the fields are such
that  both 
$\delta S(\phi_i) =0$ and $\delta {\cal O}_{\alpha} (\phi_i) =0 $.  
Conditions (\ref{lpangela}) lead, at least formally, to the following 
relation for vevs:
\begin{eqnarray} & &\frac{\delta}{\delta g^{\mu\nu}} 
\langle {\cal O}_{\alpha_1} {\cal O}_{\alpha_2} \cdots {\cal O}_{\alpha_p}
\rangle  = -\int [D\phi_i]  {\cal O}_{\alpha_1}(\phi_i) {\cal
O}_{\alpha_2}(\phi_i)
 \cdots {\cal O}_{\alpha_p}(\phi_i) T_{\mu\nu}  
\exp\big(-S(\phi_i)\big)
\nonumber \\ & & \,\,\,\,\,\,\, \,\,\,\,\,\,\, = -\int [D\phi_i]
\delta\Big( {\cal O}_{\alpha_1}(\phi_i) {\cal O}_{\alpha_2}(\phi_i)
 \cdots {\cal O}_{\alpha_p}(\phi_i) G_{\mu\nu}  
\exp\big(-S(\phi_i)\big)\Big)
 =  0,
\label{lprosina}
\end{eqnarray} which implies that the quantum field theory can be regarded
as topological. In (\ref{lprosina}) it has been assumed that the action and
the measure $[D\phi_i]$ are invariant under the symmetry  
$\delta$. We have also assumed in (\ref{lprosina}) that the observables are
metric-independent. This is a common situation in this type of  theories,
but it does not have to be so. In fact, in view of
(\ref{lprosina}),  it would be possible to consider a wider class of
operators satisfying:
\begin{equation}
 \frac{\delta}{\delta g_{\mu\nu}}{\cal O}_{\alpha}(\phi_i)=
\delta O^{\mu\nu}_{\alpha}(\phi_i),
\label{lpqueso}
\end{equation} 
with $O^{\mu\nu}_{\alpha}(\phi_i)$ a certain functional of the fields 
of the theory.

This second class of TQFTs is called cohomological or of the Witten
type \cite{phyrep}\cite{coho}. Its most celebrated representatives are 
the four-dimensional  Donaldson-Witten theory \cite{tqft}, which can be
regarded as a certain {\it twisted} version of
the  $\cn=2$ supersymmetric Yang-Mills theory, and the two-dimensional
topological sigma models \cite{tsm}, which are also twisted versions of the
conventional $\cn=2$ sigma model. In fact, all the twisted 
supersymmetric quantum field theories are topological field theories of
this type. 

It is important to remark that the symmetry
$\delta$ must be a scalar symmetry. The reason is that, being a global 
symmetry, the corresponding generator must be covariantly constant and for
arbitrary manifolds this  property, if  satisfied at all, implies
strong restrictions unless  the generator is a singlet under the holonomy
of the manifold.

Most of the TQFTs of cohomological type satisfy the relation:
\begin{equation}
 S(\phi_i)=\delta\Lambda(\phi_i),
\label{lpcaca}
\end{equation}
 for some functional $\Lambda(\phi_i)$. This has far-reaching
consequences, for it means that the topological  observables of the theory
(in particular the partition function itself) are  independent of the
value of the coupling constant. Indeed, let us consider for example  the
vev: 
\begin{equation}
\langle {\cal O}_{\alpha_1} {\cal O}_{\alpha_2} \cdots {\cal O}_{\alpha_p}
\rangle =
\int [D\phi_i]  {\cal O}_{\alpha_1}(\phi_i) {\cal  O}_{\alpha_2}(\phi_i)
 \cdots {\cal O}_{\alpha_p}(\phi_i)  
\exp\left(-\frac{1}{e^2}S(\phi_i)\right).
\label{lpsansed}
\end{equation} Under a change in the coupling constant, $1/e^2\to
1/e^2-\Delta$, one  has (assuming that the observables do not depend on
the coupling), up to first order in $\Delta$:
\begin{align}
\langle {\cal O}_{\alpha_1} {\cal O}_{\alpha_2} \cdots {\cal O}_{\alpha_p}
\rangle&\too 
\langle {\cal O}_{\alpha_1} {\cal O}_{\alpha_2} \cdots {\cal O}_{\alpha_p}
\rangle  \nonumber\\ +\Delta
\int &[D\phi_i] \delta\left[ {\cal O}_{\alpha_1}(\phi_i) {\cal
O}_{\alpha_2}(\phi_i)
 \cdots {\cal O}_{\alpha_p}(\phi_i) 
\Lambda(\phi_i)\exp\left(-\frac{1}{e^2}S(\phi_i)\right)\right]\nonumber\\&= 
\langle {\cal O}_{\alpha_1} {\cal O}_{\alpha_2} \cdots {\cal O}_{\alpha_p}
\rangle.\ret
\label{lpsunset}
\end{align} 
Hence, observables can be computed either in the weak coupling limit, 
$e\to 0$, or in the strong coupling limit, $e\to\infty$. 

So far we have presented a rather general definition of TQFT and made a
series of  elementary remarks. Now we will analyse some aspects of its
structure. We begin by pointing out that given a theory in which
(\ref{lpangela}) holds, one can build correlators which correspond to 
topological invariants (in the sense that they are invariant under
deformations  of the metric  
$g_{\mu\nu}$) just by considering the operators of the theory which are
invariant  under the 
symmetry. We will call these operators observables.
Actually, to be  more precise, we will call observables to certain classes
of those  operators. In virtue of eq. (\ref{lprosina}),  if one of these 
operators can be  written as a symmetry transformation of another
operator, its presence in a  correlation function will make it vanish.
Thus we may identify operators  satisfying (\ref{lpangela}), which differ by
an operator which corresponds to a  symmetry transformation of another
operator. Let us denote the set of the resulting classes by $\{\Phi\}$. 
Actually, in
general, one could identify  bigger sets of operators since two operators
of which one does not  satisfy (\ref{lpangela}) could lead to the
same invariant if they differed by an  operator which is a symmetry
transformation of another operator. For example,  consider
${\cal O}$ such that $\delta{\cal O}=0$ and
${\cal O}+\delta\Gamma$. Certainly, both operators lead to the same
observables. But it may well happen that $\delta^2\Gamma\neq 0$ and  therefore
we have operators, which do not satisfy (\ref{lpangela}), that must be 
identified. The natural way out is to work {\it equivariantly}, which in
this  context means that one must consider only operators which are
invariant under both  
$\delta$ and $\delta^2$. It turns out that in most of the cases (and in  
particular, in all  the cases that we will be considering) $\delta^2$ is a
gauge  transformation,  so in the end all that has to be done is to
restrict the analysis to gauge-invariant operators, a very natural
requirement. Hence, by  restricting the analysis to the  appropriate set
of operators, one has  in  fact,  
\begin{equation}
\delta^2=0. 
\label{lponce}
\end{equation}

Property (\ref{lponce}) has striking consequences on the features of TQFT.
First, the symmetry must be odd, which implies the presence in the theory
of commuting and anticommuting fields. For example, the tensor
$G_{\mu\nu}$ in (\ref{lpangela}) must be  anticommuting. This is the first
appearance of an odd non-spinorial field in TQFT. Those kinds of objects
are standard features of cohomological TQFTs.  Secondly, if  we denote by
$Q$ the operator which implements this symmetry, the  observables of the
theory can be described as the cohomology classes of $Q$: 
\begin{equation}
\{\Phi\} = \frac{\ker \, Q}{ {\hbox{\rm im}}\,Q},
\qquad\qquad\qquad  Q^2=0.
\label{lpdoce}
\end{equation}

Equation (\ref{lpangela}) means that in addition to the Poincar\'e group 
the theory possesses a symmetry generated by an odd version of the
Poincar\'e group. The corresponding odd generators are constructed out of
the tensor
$G_{\mu\nu}$ in much the same way as the ordinary Poincar\'e generators
are built out of $T_{\mu\nu}$. For example, if $P_\mu$ represents the
ordinary momentum operator, there exists a corresponding odd one $G_\mu$
such that 
\begin{equation}
 P_\mu = \{Q, G_\mu\}.
\label{lptrece}
\end{equation}

Let us discuss the structure of the Hilbert space of the theory in virtue
of the symmetries that we have just described. The states of  this space
must correspond to representations of the algebra generated by  the
operators in the Poincar\'e group and $Q$. Furthermore, as follows
from the analysis leading to (\ref{lpdoce}), if one is 
interested only in states
$|\Psi\rangle$ leading to topological  invariants, one must consider states
that satisfy 
\begin{equation} 
Q|\Psi\rangle=0,
\label{lpcatorce}
\end{equation} 
and two states which differ by a $Q$-exact state must be identified.
The odd Poincar\'e group can then be used to generate descendant states
out  of a state satisfying (\ref{lpcatorce}).  The operators $G_\mu$  act
non-trivially on the states and in fact, out of a state satisfying
(\ref{lpcatorce}), we can build additional states using this generator. 
The simplest case consists of
\begin{equation}
\int_{\gamma_1} G_{\mu} |\Psi\rangle, 
\label{lpquince}
\end{equation}  
where $\gamma_1$ is a 1-cycle. One can easily verify using  
(\ref{lpangela}) that this new state satisfies (\ref{lpcatorce}):
\begin{equation} 
Q \int_{\gamma_1} G_{\mu} |\Psi\rangle =
\int_{\gamma_1} \{Q,G_{\mu} \}|\Psi\rangle =
\int_{\gamma_1} P_{\mu} |\Psi\rangle =0.
\label{lpdiezseis}
\end{equation} 
Similarly, one may construct other invariants by tensoring $n$ operators
$G_\mu$ and integrating over $n$-cycles $\gamma_n$:
\begin{equation}
\int_{\gamma_n} G_{\mu_1}G_{\mu_2}...G_{\mu_n} |\Psi\rangle.
\label{lpdiezsiete}
\end{equation}
 Notice that since the operator $G_\mu$ is odd and its algebra
is   Poincar\'e-like the integrand in this expression is an $n$-form. It
is   straightforward to prove that these states also satisfy  condition
(\ref{lpcatorce}). Therefore, starting  from a state  $|\Psi\rangle\in \ker
\, Q$ we have built a set of partners or descendants giving rise to a
topological multiplet. The members of a multiplet have well defined
{\it ghost} number. If one assigns ghost number $-1$ to the operator $
G_{\mu}$, the state in (\ref{lpdiezsiete})  has ghost number $-n$ plus the
ghost number of
$|\Psi\rangle$. Of course, $n$
 is bounded by the dimension of the manifold $X$. Among the states
constructed in this way there may be many which are related via another
state which is
$Q$-exact, \ie which can be written as $Q$  acting on some other state.
Let us try to single out representatives at each level of ghost number in
a given topological multiplet. 

Consider an $(n-1)$-cycle which is the boundary of an $n$-dimensional
surface,
$\gamma_{n-1}=\partial S_n$. If one  builds a state taking such a cycle one
finds ($P_\mu=-i\partial_\mu$)
\begin{equation}
\int_{\gamma_{n-1}} G_{\mu_1}G_{\mu_2}...G_{\mu_{n-1}} |\Psi\rangle=i
\int_{S_n} P_{[\mu_1} G_{\mu_2}G_{\mu_3}...G_{\mu_{n}]}|\Psi\rangle= iQ
\int_{S_n} G_{\mu_1} G_{\mu_2}...G_{\mu_{n}}|\Psi\rangle,
\label{lpdiezocho}
\end{equation}
\ie it is $Q$-exact. The symbols [ ] in (\ref{lpdiezocho}) denote that all
indices between  them must by antisymmetrized. In (\ref{lpdiezocho}) use has
been made of (\ref{lptrece}). This result tells us that  the representatives
we are looking for are built out of the homology cycles of the manifold
$X$. Given a manifold $X$, the homology cycles are equivalence classes
among cycles, the equivalence relation being that two $n$-cycles are
equivalent if they differ by a cycle which is the boundary of an
$(n+1)$-dimensional  surface. Thus, knowledge of the homology of the
manifold on which the TQFT is defined allows us to classify the
representatives among the operators  (\ref{lpdiezsiete}). Let us assume
that
$X$ has dimension $d$ and that its homology cycles are
$\gamma_{i_n}$, $i_n=1,...,d_n$, $n=0,...,d$, $d_n$ being the  dimension
of the 
$n$-homology group, and $d$ the dimension of $X$. Then, the non-trivial
partners or descendants of a given  {\it highest-ghost-number
state} $|\Psi\rangle$ are labelled in the following way:
 \begin{equation}
\int_{\gamma_{i_n}} G_{\mu_1}G_{\mu_2}...G_{\mu_n} |\Psi\rangle,
\,\,\,\,\,\,\,\,\,i_n=1,...,d_n,\,\,\,\,\,\,\, n=0,...,d.
\label{lpdieznueve}
\end{equation}

A  construction similar to the one just described can be made for  fields.
Starting with a field $\phi(x)$ which satisfies 
\begin{equation}
 [Q,\phi(x)]=0,
\label{lpveinte}
\end{equation} 
one can construct other fields using the operators $G_{\mu}$. These
fields, which we will call partners or descendants, are antisymmetric
tensors defined as
\begin{equation}
\phi^{(n)}_{\mu_1\mu_2...\mu_n}(x)=\frac{1}{n!}
[G_{\mu_1},[G_{\mu_2}...[G_{\mu_n},\phi(x)\}...\}\},
\,\,\,\,\,\,\,\, n=1,...,d.
\label{lpveintep}  
\end{equation}  
Using (\ref{lptrece}) and (\ref{lpveinte}) one finds that these fields  
satisfy the so-called {\it topological descent equations}:
\begin{equation}
 d \phi^{(n)} = i [Q,\phi^{(n+1)}\},
\label{lpvseis}
\end{equation} 
where the subindices of the forms have been suppressed for
simplicity,  and the highest-ghost-number field $\phi(x)$ has been denoted
as $\phi^{(0)}(x)$. These equations enclose all the relevant properties of
the observables which are constructed out of  them. They  constitute a
very useful tool to build the observables of the theory. Let us consider
an $n$-cycle and the following quantity:
\begin{equation}
 W^{(\gamma_n)}_\phi = \int_{\gamma_n} \phi^{(n)}.
\label{lpvsiete}
\end{equation} 
The subindex of this quantity denotes the highest-ghost-number field
out of which the form $\phi^{(n)}$ is generated. The superindex  denotes
the order of such a form as well as the cycle used in the
integration. Using the topological descent equations (\ref{lpvseis}) it  is
immediate to prove that
$W^{(\gamma_n)}_\phi$ is indeed an observable:
\begin{equation}
 [Q, W^{(\gamma_n)}_\phi\} =\int_{\gamma_n} [Q,\phi^{(n)}\}
=-i\int_{\gamma_n} d\phi^{(n-1)}=0.
\label{lpvocho}
\end{equation} 
Furthermore, if $\gamma_n$ is a trivial homology cycle,  
$\gamma_n=\partial S_{n+1}$, one obtains that $W^{(\gamma_n)}_\phi$ is
$Q$-exact:
\begin{equation} 
W^{(\gamma_n)}_\phi = \int_{\gamma_n} \phi^{(n)}=
\int_{S_{n+1}} d \phi^{(n)} = i \int_{S_{n+1}} [Q, \phi^{(n+1)}\}
=i\left[Q,\int_{S_{n+1}}\phi^{(n+1)}\right\},
\label{lpvnueve}
\end{equation} 
and therefore its presence in a vev makes it vanish. Thus, similarly 
to the previous analysis leading to (\ref{lpdieznueve}), the observables of 
the theory are operators of the form (\ref{lpvsiete}):
\begin{equation} 
W^{(\gamma_{i_n})}_\phi,\,\,\,\,\,\, i_n=1,...,d_n,\,\,\,\,\,\,  
n=0,...,d,
\label{lptreinta}
\end{equation} 
where, as before, $d_n$ denotes the dimension of the $n$-homology 
group. Of course, these observables are a basis of observables, but one
can  make arbitrary products of them, leading to new ones.

One may wonder at this point how it is possible that there may be
observables which depend on the space-time position $x$ and  nevertheless
lead to topological invariants. For example, an observable containing the
zero form
$\phi^{(0)}(x)$ seems to lead to vevs which depend  on
$x$, since the space-time position $x$ is not integrated over. A closer
analysis, however, shows that this is not the case. As follows from the
topological descent equation (\ref{lpvseis}), the derivative of  
$\phi^{(0)}(x)$ with  respect to $x$ is $Q$-exact and therefore such a
vev is actually independent of the space-time
position. This is completely analogous to the situation in conventional
supersymmetric gauge theories, where it is well-known \cite{venez} that
certain correlation functions (as the gaugino condensate) are independent 
of spatial separation and can be computed in either the short or long
distance limits.


\chapter{The Mathai-Quillen formalism}
\la{chmq}
\markboth{\footnotesize\bfseries The Mathai-Quillen formalism
}{\footnotesize\bfseries The Mathai-Quillen formalism}
\markright{\textsc {Duality in Topological Quantum Field Theories}}

TQFTs of the cohomological type are  conventionally characterized by three
basic data: fields, symmetries, and  equations
\cite{phyrep}\cite{moore}\cite{coho}. The starting point is a configuration
space ${\cal M}$, whose elements  are fields
$\phi_i$ defined on some Riemannian manifold $X$. These fields  are
generally acted on by some group
${\cal G}$ of local transformations  (gauge symmetries, or a diffeomorphism
group, among others), so that one is  naturally led to consider the quotient
space
${\cal M}/{\cal G}$. Within this quotient  space, a certain subset of
special configurations or {\it moduli space}, ${\cal M}_0$, is singled out 
by a set of  equations
$s(\phi_i)=0$:
\begin{equation}
{\cal M}_0=\{\phi_i\in{\cal M}|s(\phi_i)=0\}/{\cal G}.
\label{lpmoduli}
\end{equation} 
This moduli space is generically finite-dimensional. 
The crucial aspect of topological field theories is that the path integral 
localizes to these configurations
\begin{equation}
Z=\int_{\C{M}} \deriv\phi_i\,\ex^{-{\cal S}(\phi_i)}\too \int_{{\cal
M}_0}\cdots,
\la{tplocal}
\end{equation}
and this makes it possible to express partition and correlation
functions as finite dimensional integrals over the moduli space. 

In standard supersymmetric theories, this type of special
configurations typically arise as  supersymmetric or BPS
configurations, which are characterized by the property that they
preserve a certain fraction of the supersymmetries \cite{oli2}. For
example, in $\cn=2$ supersymmetric gauge theories, the supersymmetry
variation of the gaugino reads
\begin{equation}
\delta\psi=F_{\mu\nu}\gamma^{\mu\nu}\epsilon,
\la{bps}
\end{equation}
 so that for $\epsilon$ satisfying $\epsilon=\gamma_5\epsilon$ we find that
instanton configurations with $F^{+}=0$ are invariant under half of the
supersymmetries, while the other half generate fermion zero modes in the
instanton background. In cohomological field theories a similar mechanism
works:  the moduli space is defined as the fixed point set of the
topological symmetry
$\delta$
\cite{yau}. Within this framework, $\delta$ furnishes a representation of
the ${\cal G}$-equivariant cohomology on the field  space. When
${\cal G}$ is the trivial group, $\delta$ is nothing but the de Rham
(or exterior differential) operator on the field space. 

The next step consists of building the topological theory associated to 
this moduli problem. We will do this within the framework of the 
Mathai-Quillen formalism \cite{mathai}. This formalism is the most
geometric of all the approaches leading to the construction of
TQFTs. It can be applied to  any Witten-type theory. It was first
implemented in the context of TQFT by Atiyah and Jeffrey \cite{jeffrey},
and later further developed in two works \cite{blaumathai}\cite{moore}.
The basic idea behind this formalism is the  extension, to the
infinite-dimensional case, of ordinary finite-dimensional geometric
constructions. Soon after the formulation of the first TQFTs it became 
clear that the partition function of these theories was related to the
Euler class of a certain bundle associated to the space of solutions of
the basic equations of the theory. In the finite-dimensional case there
are many different, though equivalent, forms of thinking on the Euler
class, which we will recall bellow. The  Mathai-Quillen formalism
basically consists of generalizing one of these forms to the
infinite-dimensional case. In what follows we will give a brief account of
the fundamentals of the construction. For further details, we refer the 
reader to \cite{thompson}\cite{moore}\cite{laplata}, where comprehensive
reviews on this approach are presented.

\section{Finite-dimensional case}

Let $X$ be an orientable, boundaryless, compact $n$-dimensional  manifold.
Let us consider an orientable vector bundle ${\cal E}\to X$ of rank
${\hbox{\rm rk}}({\cal E})=2m\leq n$ over $X$. For completeness we  recall
that a vector bundle ${\cal E}$, with  a $2m$-dimensional vector space
$F$ as fibre, over a base manifold $X$, is a topological space with a  
continuous projection, $\pi:{\cal E}\to X$, such that 
$\forall x\in X$, $\exists \,U_x\subset X$, open set, $x\in U_x$, ${\cal E}$
is a product space, $U_x\times F$, when restricted to $U_x$. This  means
that there exists a homeomorphism $\varphi:U_x\times F\to \pi^{-1}(U)$
which  preserves the fibres, \ie $\pi(\varphi(x,f))=x$, with $f\in F$.  

There exist two complementary ways of defining the Euler class of  
${\cal E}$,
$e({\cal E})\in H^{2m}(X)$:
\begin{enumerate}
\item In terms of sections. A section $s$ of ${\cal E}$ is a map
$s:X\to {\cal E}$ such that $\pi(s(x))=x$. A {\it generic} section is  one
which  is transverse to the zero section, and which therefore vanishes on
a set of dimension $n-2m$. In this context 
$e({\cal E})$ shows up as the Poincar\'e dual (in $X$) of the homology  class
defined by the zero locus of a generic section of ${\cal E}$.  
\item  In terms of characteristic classes. The approach makes use of  the 
Chern-Weil theory, and gives a representative 
$e_{\nabla}({\cal E})$ of $e({\cal E})$ associated to a connection  
$\nabla$ in
${\cal E}$: 
\begin{equation} 
e_{\nabla}({\cal E})=(2\pi)^{-m}\pff (\Omega_{\nabla}),
\label{lptreintauno}
\end{equation} 
where $\pff(\Omega_{\nabla})$ stands for the Pfaffian of the
curvature  
$\Omega_{\nabla}$, which is an antisymmetric matrix of two-forms. The
representative
$e_{\nabla}({\cal E})$ can be written in ``field-theoretical" form:
\begin{equation} 
e_{\nabla}({\cal E})=(2\pi)^{-m}\int  
d\chi\ex^{\half\chi_a\Omega^{ab}_{\nabla}\chi_b},
\label{lptreintados}
\end{equation} 
by means of a set of real Grassmann-odd variables $\chi_a$,
$a=1,\ldots ,2m$, satisfying the Berezin rules of integration:
\begin{equation}
\int d\chi_a \chi_b = \delta_{ab}.
\label{lptreintatres}
\end{equation}
\end{enumerate}

If ${\hbox{\rm rk}}({\cal E})=2m=n={\hbox{\rm dim}}(X)$, one can  evaluate
$e({\cal E})$ on $X$ to obtain the Euler number of ${\cal E}$ in two 
different ways:  
\bea
\chi({\cal E})&=&\sum_{x_k:s(x_k)=0} (\pm1),\nonumber \\
\chi({\cal E})&=& \intl_X e_{\nabla}({\cal E}).\nonumber \\ 
\label{lptreintacuatro}
\eea 
In the first case, one counts signs at the zeroes of a generic
section (notice that in this case a generic section has 
a zero locus $s(x)=0$  consisting of a finite set of isolated points.) In the
second case, one integrates the differential form (\ref{lptreintauno}) over
$X$. Of course, both results coincide, and do not depend either on the
section
$s$ (as long as it is generic) or on the connection $\nabla$.  When
$2m<n$ one can evaluate $e({\cal E})$ on $2m$-cycles or  equivalently take
the product with elements of $H^{n-2m}(X)$ and evaluate it on $X$.  

In the particular case that  ${\cal E} \equiv TX$ the expression
$\chi({\cal E})=\sum_{x_k:s(x_k)=0} (\pm1)$, which gives the Euler  number
of the base manifold $X$,  can be generalized to a non-generic vector 
field $V$ (which is a section of the tangent bundle)
\begin{equation}
\chi(X)=\chi(X_V),
\label{lptreintacinco}
\end{equation} 
where $X_V$ is the zero locus of $V$, which is not necessarily  
zero-dimensional.

In this framework the Mathai-Quillen formalism gives a representative of 
the Euler class $e_{s,\nabla}({\cal E})$, which interpolates between the 
two approaches described above. It depends explicitly on both a section
$s$ and a connection $\nabla$ on ${\cal  E}$:
\bea e_{s,\nabla}({\cal E})&\in&[e({\cal E})],\nonumber\\
\chi({\cal E})&=&\intl_X e_{s,\nabla}({\cal E}),\qquad  ({\rm if}\,\, 2m=n).
\label{lptreintaseis}
\eea The construction of $e_{s,\nabla}$ is given by the formalism. First,
it provides an explicit representative of the Thom class \cite{tu},
$\Phi({\cal E})$, of ${\cal E}$. Let ${\cal E}\to X$ be a vector bundle of
rank
$2m$ with  fibre $F$, and let us consider the cohomology of forms with
Gaussian decay along  the fibre. By integrating the form along the fibre, 
one has an explicit isomorphism (the Thom isomorphism) between $k$ forms
over ${\cal E}$  and $k-2m$ forms over $X$. This isomorphism can be  made
explicit with the  aid of the Thom class, whose representative
$\Phi({\cal E})$ is a closed 
$2m$-form over ${\cal E}$ with Gaussian decay along the fibre such that 
its integral over the fibre is unity. In terms of this form, and given any
arbitrary $p$-form $\omega$ over $X$, its image under the Thom isomorphism
is the $p+2m$ form $\pi^*(\omega)\wedge\Phi({\cal E})$, which by
construction has Gaussian decay along the fibre. $\pi^*(\omega)$ is the
pull-back of
$\omega$ by  the projection $\pi:{\cal E}\to X$. If $s$ is any section of
${\cal E}$, the pull-back of the Thom form under $s$, $s^*\Phi({\cal E})$,
is a  closed form in the same cohomology class as the Euler class $e({\cal
E})$. If
$s$ is a  generic section, then $s^{*}\Phi({\cal E})$ is the Poincar\'e
dual  of the zero locus of $s$. Mathai and Quillen constructed an explicit
representative, $\Phi_{\nabla}({\cal E})$, of the Thom form in terms of  a
connection $\nabla$ in ${\cal E}$. Its pull-back by a section $s$,
$e_{s,\nabla}({\cal E})=s^{*}\Phi_{\nabla}({\cal E})$, is  represented as a
Grassmann integral:
\begin{equation} 
e_{s,\nabla}({\cal E})=(2\pi)^{-m}\int d\chi\ex^{-\half \vert
s\vert^2+\half\chi_a\Omega^{ab}_{\nabla}
\chi_b+i\nabla s^a\chi_a}.
\label{lptreintasiete}
\end{equation} 
As a consistency check, note that, as follows from (\ref{lptreintados}),  
$e_{s=0,\nabla}({\cal E})=e_{\nabla}({\cal E})$,
{\ie} the pull-back of the Mathai-Quillen representative by the zero 
section gives back the Euler class of ${\cal E}$. $e_{s,\nabla}({\cal E})$
is a closed
$2m$-form. This can be verified after integrating over the Grassmann-odd
variables  
$\chi_a$. It is closed because the exponent is invariant under the
transformations
\begin{equation}
\delta s=\nabla s,\qquad \delta \chi_a=is_a(x).
\label{lptreintaocho}
\end{equation}
These transformations square ``on-shell" (that is, using the $\chi$
equation of motion $i\nabla s^a=\Omega_\nabla^{ab}\chi_b$) to
rotations by the curvature matrix $\Omega_\nabla$, 
\begin{equation}
\delta^2 s^a=\Omega_\nabla^{ab} s_b,\qquad \delta^2
\chi^a=\Omega_\nabla^{ab}\chi_b.
\label{lptreintaocho2}
\end{equation}

 The Mathai-Quillen
representative
\eqs{lptreintasiete} interpolates between the two different
approaches to the Euler class of a vector bundle described above. This
statement can be made more precise as follows. The construction of
$e_{s,\nabla}({\cal E})$  is such that it is cohomologous to
$e_{\nabla}({\cal E})$ for any choice of a generic section $s$. Take for  
example the case $n=2m$, and rescale $s\to\gamma s$. Nothing should
change, so in  particular:
\begin{equation} 
\chi({\cal E})=\intl_X e_{\gamma s,\nabla}({\cal E}).
\label{lpcuarentaseis}
\end{equation} 
We can now study (\ref{lpcuarentaseis}) in two different limits:
\begin{enumerate}

\item Limit $\gamma\to 0$: after using (\ref{lptreintados}),
$\chi({\cal E})=(2\pi)^{-m}\int\pff (\Omega_{\nabla})$.
\item Limit $\gamma\to\infty$: the curvature term in  (\ref{lptreintasiete})
can be neglected, and the integration in  (\ref{lptreintasiete})
localizes to a finite sum over the ``moduli space" $s(x)=0$, leading to
$\chi({\cal E})=\sum_{x_k:s(x_k)=0}(\pm 1)$. These signs are generated by
the ratio of the determinants of $\nabla s$ and its modulus, which result
from the Gaussian integrations after expanding around each zero
$x_k$.
\end{enumerate}

This localization property can be restated in terms of the Grassman odd
scalar symmetry $\delta$. Notice that the fixed point
of the symmetry transformations \eqs{lptreintaocho} is the set 
$s_a(x)=0=\nabla s$, which is precisely the moduli space to which the 
integration \eqs{lpcuarentaseis} localizes. This is a simple 
realization of the general result due to Witten \cite{yau} that we mentioned
above. 

\section{Infinite-dimensional case}

We now turn into the study of the infinite-dimensional case. The main
complication that one finds in this case is that $e({\cal E})$ is not
defined. By taking advantage of what we have learned so far, we could try
to  generalize the Mathai-Quillen formalism to define something analogous to
an Euler  class for
${\cal E}$. It turns out that this is actually possible, and the outcome of 
the construction is what is called a regularized Euler number for the  bundle
${\cal E}$. Unfortunately, it depends  explicitly on the  section  chosen
for the construction, so it is important to make good selections.

The outline of the construction is as follows.  First  recall that, as
stated in (\ref{lptreintacinco}), in the finite-dimensional case
$\chi(X)=\chi(X_V)$ when $V$ is non-generic, \ie when its zero locus,
$X_V$,  has dimension ${\hbox{\rm dim}}(X_V)<2m$. For $X$ 
infinite-dimensional the  idea is to introduce a vector field $V$ with
finite-dimensional zero locus.  The regularized Euler number of ${\cal E}$
would then be defined as:
\begin{equation}
\chi_V(X)=\chi(X_V),
\label{lpochentaseis}
\end{equation} 
which explicitly depends on $V$.  By analogy with the
finite-dimensional case one expects that:
\begin{equation}
\chi_V(X)=\intl_X e_{V,\nabla}(TX),
\label{lpochentasiete}
\end{equation} 
as a functional integral, where $e_{V,\nabla}(TX)$ is meant
to be the Mathai-Quillen representative for the corresponding Euler class. 

In  general, the regularized Euler number $\chi_s({\cal E})$ of an
infinite-dimensional vector bundle ${\cal E}$ is given by:
\begin{equation}
\chi_s({\cal E})=\intl_X e_{s,\nabla}({\cal E}),
\label{lpochentaocho}
\end{equation}  
where $e_{s,\nabla}({\cal E})$ is given by the
Mathai-Quillen formalism.  The construction follows the pattern of
the finite-dimensional case. But it is important to remark that  
(\ref{lpochentaocho}) makes sense only when the zero locus of  $s$,
$X_s$, is finite-dimensional.
$\chi_s({\cal E})$ is then the Euler number of  some
finite-dimensional vector bundle over $X_s$, and it corresponds to
the regularized Euler number of the infinite-dimensional bundle
${\cal E}$. Of course,
$\chi_s({\cal E})$ depends on
$s$, but if $s$ is naturally associated to ${{\cal E}}$ one expects
to obtain interesting topological information. For topological
quantum field theories,
$s$ is given by the fixed points of the fermionic symmetry $\delta$,
which in turn comes from the underlying supersymmetry of the theory.
The expression in the exponential in
\eqs{lptreintasiete} is the action of the topological model, and  the
first term, $-\half \vert s\vert^2$, gives the purely  bosonic terms
in the action, while
$\nabla s^a\chi_a$ gives the kinetic energy  of the fermions (notice
that
$\nabla s$ is a one-form on $X$, so one can see it  as a fermion). 
The situation where there are local symmetries in the problem  (such
as conventional gauge symmetries) is more involved, but one can
easily  extend the formalism to cover these cases as explained in
great detail in 
\cite{moore}. We will review  the construction when we describe the
twisted 
$\cn=4$ theories below.


\chapter{The $\cn=$ 4 supersymmetric gauge theory 
in four dimensions}
\la{chene4}
\markboth{\footnotesize\bfseries The $\cn=4$ supersymmetric gauge theory 
in four dimensions
}{\footnotesize\bfseries The $\cn=4$ supersymmetric gauge theory 
in four dimensions}
\markright{\textsc {Duality in Topological Quantum Field Theories}}

In this chapter we will review some aspects of the four-dimensional $\cn=4$ 
supersymmetric gauge theory. We will describe the Seiberg-Witten solution 
for the low-energy effective description of the  $\cn=4$ theory
with gauge group $SU(2)$, and we will make several remarks concerning the
Montonen-Olive strong/weak-coupling duality symmetry of the theory on 
arbitrary four-manifolds.

\section{$\cn=4$ supersymmetric gauge theory}

We begin with the standard $\cn=4$ supersymmetric gauge theory on flat 
${\IR}^4$. Our conventions regarding spinor notation are almost  as in Wess and
Bagger \cite{wessba}, with some differences that we
conveniently  compile in the final appendix. 

The $\cn=4$ supersymmetric Yang-Mills
theory is unique once the gauge group
$G$ and the microscopic coupling
$\tau_0={\frac{\theta_0}{2\pi}}+{\frac{4\pi i}{e^2_0}}$ are
fixed. The Lagrangian and supersymmetry transformations can be constructed
from the $\cn=1$ supersymmetric Yang-Mills theory in $10$ dimensions by 
 dimensional reduction on a six-torus \cite{scherk}. The ten-dimensional
Lorentz  group $SO(1,9)$ splits as $SO(1,3)\times SO(6)_I$, where $SO(1,3)$ is
the four-dimensional  Lorentz group, while $SO(6)_I\simeq SU(4)_I$, which
corresponds to the rotation group in the extra dimensions, appears in the
four-dimensional theory as a global symmetry group (it is in fact the
automorphism
 or $\C{R}$ symmetry group of the $\cn=4$ supersymmetry algebra).  
  The field content of the model is the following: a gauge field
$A_{\alpha\dalpha}$, four Weyl gauginos $\lambda_u{}^\alpha$ and their complex
conjugates $\bar\lambda^u{}_{
\dalpha}$ (which come from the $10$-dimensional Majorana-Weyl
gaugino) transforming respectively in the representations
${\bf 4}$ and 
${\bf\bar 4}$ of  
$SU(4)_I$ ($u,v,w,\ldots=1,2,3,4$  are $SU(4)_I$ indices), and six scalars
$\phi_{uv}$ (which come from the $10$-dimensional gauge field) in the
${\bf 6}$ of
$SU(4)_I$. All the fields above take values in the adjoint representation of 
some
compact  Lie group $G$. Being in the  representation ${\bf 6}$, the scalars
$\phi_{uv}$ satisfy antisymmetry and self-conjugacy constraints:
\bea
&&\phi_{uv}=-\phi_{vu},\ret 
&&\phi^{uv}=(\phi_{uv})^{\dag}=\phi^{*}_{vu}=
-\half\epsilon^{uvwz}\phi_{wz};\quad \epsilon_{1234}=\epsilon^{1234}=+1.\ret
\label{valle4}
\eea

The action for the model in Euclidean space is:
\bea
{\cal S}&=& \frac{1}{e_0^2}\int d^4 x\, \tr\, \bigl\{\,-\frac{1}
{8}\nabla_{\!\alpha
\dalpha}\phi_{uv}\nabla^{\dalpha\alpha}\phi^{uv} 
-i\lambda_v{}^\alpha\nabla_{\!\alpha\dot\alpha}
\bar\lambda ^{v\dot\alpha} -\frac{1}{4} F_{mn} F^{mn}
\ret 
&-& \frac{i}{\raiz}\,\lambda_u{}^\alpha [\lambda_{v\alpha},\phi^{uv}]
+\frac{i}{\raiz}\,\bar\lambda^u{}_{\dot\alpha}[\bar\lambda^
{v\dot\alpha},\phi_{uv}]+\frac{1}{16}[\phi_{uv},\phi_{wz}]
[\phi^{uv},\phi^{wz}]
\,\bigr\}\ret 
&-&\frac{i\theta_0}{32\pi^2}\int d^4 x\,\tr\,\bigl\{\, *  F_{mn}F^{mn}
\,\bigr\}.
\ret
\label{Boris}
\eea 
 We have introduced the covariant derivative $\nabla_{\!\alpha\dalpha}
=\sigma^m{}_{\alpha\dalpha}(\partial_m+i[A_m,~])$ (together with its 
corresponding field strength $F_{mn}=
\partial_m A_n -\partial_n A_m +i[A_m ,A_n]$) and the trace $\tr$ in the
fundamental representation. The action \eqs{Boris} is invariant under the
following four supersymmetries (in $SU(4)_I$ covariant notation):
\bea
 \delta A_{\alpha\dalpha} &=& -2i\bar\xi^u{}_\dalpha\lambda_{u
\alpha}+2i\bar\lambda^u{}_\dalpha\xi_{u\alpha },\ret 
\delta\lambda_{u\alpha} &=&  -iF^{+}{}_{\!\alpha}{}^{\!\beta}\xi_{u\beta}+
i{\raiz}\bar\xi^{v\dot\alpha}\nabla_{\alpha\dot\alpha}
\phi_{vu}-i\xi_{w\alpha}[\phi_{uv},\phi^{vw}],\ret
\delta\phi_{uv}&=&{\raiz}\bigl\{\xi_u{}^\alpha\lambda_{v\alpha}
-\xi_v{}^\alpha\lambda_{u\alpha} +
\epsilon_{uvwz}\bar\xi^w{}_{\dalpha}\bar\lambda^{z\dalpha}\bigr\},\ret
\label{Vian4}
\eea 
where $F^{+}{}_{\!\alpha}{}^{\!\beta}=\sigma^{mn}{}_{\!\alpha}{}^{\!\beta}
F_{mn}$. Notice that there are no auxiliary fields in the action \eqs{Boris}.
Correspondingly, the transformations
\eqs{Vian4} close the supersymmetry algebra on-shell.   

In $\IR^4$, the global symmetry group of $\cn=4$ supersymmetric theories is
${\cal H} =SU(2)_L\otimes SU(2)_R\otimes SU(4)_I$, where  ${\cal K}=
SU(2)_L\otimes SU(2)_R$ is the rotation group $SO(4)$. The supersymmetry
generators responsible for the transformations \eqs{Vian4} are
$Q^u{}_\alpha$ and $\bar Q_{u\dalpha}$ They  transform  as $({\bf 2},{\bf
1},{\bf
\bar 4})\oplus({\bf 1},{\bf 2},{\bf 4})$ under ${\cal H}$.

From the point of view of $\cn=1$ superspace, the theory contains one $\cn=1$
vector  multiplet and three $\cn=1$ chiral multiplets. These supermultiplets
are represented  in $\cn=1$ superspace  by superfields
$V$ and $\Phi_s$ ($s=1,2,3$), which  satisfy the  constraints $V=V^{\dag}$ and
$\bar D_\dalpha \Phi_s=0$, being 
$\bar D_\dalpha$ a superspace covariant derivative. The physical component
fields of these superfields are:
\bea
V &\longrightarrow &\;   A_{\alpha\dalpha},\; 
\lambda_{4\alpha},\; \bar\lambda^4{}_{\dalpha},\ret 
\Phi_s, \Phi^{\dag s} &\longrightarrow &\; B_s,\;
\lambda_{s\alpha},\; B^{\dag s},\;
\bar\lambda^s{}_{\dalpha}.\ret
\label{ctres}
\eea  
In terms of these fields, the $SU(4)_I$ tensors that we introduced above are
defined as follows: 
\bea
&&\{{\bf 4}\}  \to \lambda_u =
\{\lambda_1,\lambda_2,\lambda_3,\lambda_4\},\ret
&&\{{\bf 6}\} \to
\phi_{uv} \sim \{B_s,B^{\dag s}\},\ret  
&&\{{\bf \bar 4}\} \to
\bar\lambda^u = 
\{\bar\lambda^1,\bar\lambda^2,\bar\lambda^3,\bar\lambda^4\},\ret
\label{cnt}
\eea  
where by $\sim$ we mean precisely:
\begin{equation}
\phi_{uv}=\!\begin{pmatrix} 
                     \!0&-\!B^{\dag 3}&\!B^{\dag 2}&-\!B_1\\
                     \!B^{\dag 3}&0&\!-B^{\dag 1}&\!-B_2\\
                     \!-B^{\dag 2}&\!B^{\dag 1}&\!0&\!-B_3\\
                     \!B_1&        \!B_2&       \!B_3&\! 0\\
\end{pmatrix},\quad
\phi^{uv}=\!\begin{pmatrix}
                   \!0&\!B_3&\!-B_2&\!B^{\dag 1}\\
                   \!-B_3&\!0&\!B_1&\!B^{\dag 2}\\
                   \!B_2&\!-B_1&\!0&\!B^{\dag 3}\\
                   \!-B^{\dag 1}&\!-B^{\dag 2}&\!-B^{\dag 3}&\!0\\
\end{pmatrix}.
\label{Inclan}
\end{equation}

The action \eqs{Boris} takes the following form in $\cn=1$
superspace:
\bea
{\cal S}\!\!\! &=&\!\!\! -{\frac{i}{4\pi}}\tau_0\int d^4 xd^2
\theta\, \tr (W^2) + {\frac{i}{4\pi}}\bar\tau_0\int d^4 x d^2
\bar\theta\, \tr (W^{\dag 2}) \ret &+& \!\!\!\! \frac{1}{e^2_0}
\sum_{s=1}^3 \int d^4 xd^2 \theta d^2
\bar\theta\, \tr(\Phi^{\dag s} \ex^V \Phi_s\ex^{-V}) \ret   
&+&\!\!\!\!\frac{i\raiz}{e^2_0}
\int d^4x d^2\theta \, \tr\bigl\{\Phi_1[\Phi_2 ,\Phi_3]\bigr\} +
\frac{i\raiz}{e^2_0}\int d^4 xd^2\bar\theta\,\tr\bigl\{\Phi^{\dag
1}  [\Phi^{\dag 2},\Phi^{\dag 3}]\bigr\}, \ret
\label{cuno}
\eea
where $W_\alpha =-\frac{1}{16}\bar{D}^2 \ex^{-V}D_\alpha
\ex^{V}$ is the supersymmetric field strength.
 
\section{Electromagnetic duality in the $\cn=4$ theory}
\label{n_4}

Within the last five years, electromagnetic duality has become a powerful tool
to unravel the structure of strongly coupled quantum gauge theories and
string theory. In many theories with $\cn=1$ \cite{ene1} and $\cn=2$
\cite{swi}\cite{swii} supersymmetry, duality shows up as a symmetry of the
effective low-energy description, and it plays a prominent role in
disentangling the infrared dynamics -- see \cite{luis} for a review. On the
other hand, in the $\cn=4$ theory and some special $\cn=2$ theories, the
duality is exact, in the sense that it is conjectured to hold valid at all
energy scales. 

The massless
$\cn=4$ supersymmetric theory has zero beta function, and it is  believed 
to be exactly finite and conformally invariant, even non-perturbatively
\cite{nseiberg}.  The
full duality group is actually $SL(2,\IZ)$, which includes a $\IZ_2$
corresponding to the interchange of electric and magnetic charges along with
the interchange of strong and weak coupling as originally proposed some
twenty years  ago by Montonen and Olive \cite{monoli}\cite{oli1}. 

The purpose of this section is to discuss some aspects of this duality.  We
will start by briefly reviewing the definition of the magnetic group
$H^{v}$ dual to a compact Lie group $H$ \cite{gno}. We follow the discussions
in \cite{hms} and \cite{suansi}. In a gauge theory 
with unbroken gauge group $H$ one can consider two kinds of quantum numbers. 
{\bf Electric} quantum numbers are fixed by the
representations under which the fields transform, and take
values in the weight lattice $\Lambda_{\mbox{\rm\tiny
weight}}^{H}$ of $H$. {\bf Magnetic} quantum numbers are 
topological in nature,
and are related to gauge configurations over two-spheres. 
They are in fact the winding numbers of the equatorial
transition functions $f(\phi):S^2\too H$, and the Dirac
quantization condition, as generalized by Goddard, Nuyts and Olive \cite{gno}, 
forces them to take values in the coweight lattice $\Lambda_{\mbox{\rm\tiny
coweight}}^{H}$ of $H$, which is dual to 
$\Lambda_{\mbox{\rm\tiny weight}}^{H}$.    

The following step is to realize  
that the magnetic quantum numbers of $H$ are the electric
quantum numbers of another {\sl dual} group $H^{v}$, whose
weight lattice is dual to that of $H$, and whose root
lattice $\Lambda_{\mbox{\rm\tiny root}}^{H^{v}}$ is the
coroot lattice $\Lambda_{\mbox{\rm\tiny coroot}}^{H}$ of $H$
spanned by the simple coroots
$\vec\alpha{}^{v}_{i}=2\frac{\vec\alpha_i}{\vec\alpha^2}$
which is dual to $\Lambda_{\mbox{\rm\tiny root}}^{H}$.
Therefore, the true symmetry group of the theory is
actually $H\otimes H^{v}$. Now, the Montonen-Olive duality 
conjecture \cite{monoli} follows simply as the statement that 
the electric and magnetic factors are exchanged under an 
inversion of the coupling constant $e_0\to 1/e_0$.  

Let us consider the example $H=SU(N)$ in detail. Since
$SU(N)$ is simply laced, $SU(N)$ and $SU(N)^v$ have the same
Lie algebra $su(N)$. Also, we can identify the coweight
lattice of $SU(N)$ with the root lattice of $su(N)$.
Thus, $\Lambda_{\mbox{\rm\tiny weight}}^{SU(N)^v}=
\Lambda_{\mbox{\rm\tiny
coweight}}^{SU(N)}=\Lambda_{\mbox{\rm\tiny
root}}^{SU(N)}=\Lambda_{\mbox{\rm\tiny
root}}^{SU(N)^v}$. But since $\Lambda_{\mbox{\rm\tiny
weight}}^{SU(N)^v}=\Lambda_{\mbox{\rm\tiny
root}}^{SU(N)^v}$, $SU(N)^v$ has no center
($\Lambda_{\mbox{\rm\tiny
weight}}^{H}/\Lambda_{\mbox{\rm\tiny root}}^{H}\simeq
Center(H))$, and therefore $SU(N)^v=SU(N)/\IZ_N$.

 Now we can wonder as to how we expect to see Montonen-Olive 
duality in the $\cn=4$ theory. First of all, in presence of a non-zero
$\theta_0$ angle, the original $\IZ_2$ transformation $e_0\to 1/e_0$ is 
extended to a full $SL(2,\IZ)$ symmetry acting on $\tau_0$\footnote{
On arbitary four-manifolds $X$, the symmetry group may be reduced to a
subgroup of $SL(2,\iz)$ as shown in \cite{vw}\cite{verlinde}\cite{sdual}.} 
\begin{equation}
\tau_0\too\frac{a\tau_0+b}{c\tau_0+d},\qquad a,b,c,d\in\IZ,\quad ad-bc=1.
\la{sl2z}
\end{equation}
Then, under $\tau_0\to -1/\tau_0$ we expect $G$ to be exchanged with its
dual $G^v$. This $SL(2,\IZ)$ duality imposes strong constraints on the
spectrum of the theory, and those have led to a series of successful 
tests on the BPS part of spectrum
\cite{suansi}\cite{harvey}\cite{oli2}\cite{sen}.

In addition to this, in the $\cn=4$ theory all the fields take values in the
adjoint representation of $G$. Hence, as pointed out by 't Hooft
\cite{gthooft},  if
$H^2(X,\pi_1(G))\not=0$ it is possible to consider non-trivial $G/Center(G)$
gauge configurations  with discrete magnetic 't Hooft flux 
through the two-cycles of $X$. In fact, 
$G/Center(G)$ bundles on $X$ are classified by the instanton number and a
characteristic class 
$v\in H^2(X,\pi_1(G))$. For example, if $G=SU(2)$, we have $SU(2)/\IZ_2=SO(3)$
and $v$ is the  second Stiefel-Whitney  class $w_2(E)$ of the gauge bundle 
$E$. This 
Stiefel-Whitney class can be represented in De Rham cohomology by a cohomology 
class (a class of differential two-forms under the equivalence relation 
$a_2\sim b_2\Leftrightarrow a_2= b_2+dc_1$, where $a_2,b_2$ are two-forms and 
$c_1$ is a one-form) in $H^2(X,\IZ)$ defined modulo $2$, \ie, $w_2(E)$ and 
$w_2(E)+2\omega$, with $\omega\in H^2(X,\IZ)$, represent the same 't Hooft flux, 
so if  $w_2(E)=2\lambda$, for some 
$\lambda\in H^2(X,\IZ)$, then the gauge configuration is trivial in 
$SO(3)$ (it has no 't Hooft flux.)

Similarly, for $G=SU(N)$ one can fix fluxes in $H^2(X,\IZ_N)$ (the 
corresponding 
Stiefel-Whitney class is defined modulo $N$.) One has therefore a family
of partition functions $Z_v(\tau_0)$, one for each magnetic flux $v$. The
$SU(N)$ partition function is obtained by considering the zero flux
partition function (up to a factor whose origin is explained in \cite{vw}), 
while the $SU(N)/\IZ_N$ partition function is obtained by summing over all
$v$, and both are to be exchanged under $\tau_0\to-1/\tau_0$. The action of
$SL(2,\IZ)$ on the $Z_v$ should be compatible with this exchange, and
according to \cite{vw}\cite{wiads2}, the $\tau_0\to-1/\tau_0$ operation 
mixes the $Z_v$ by a discrete Fourier transform
\begin{equation}
Z_v(-1/\tau_0)=N^{-b_2/2}\left(\frac{\tau_0}{i}\right)^{w/2}
\left(\frac{\bar\tau_0}{-i}\right)^{\bar w/2}\sum_{u\in H^2(X,\IZ)}
\ex^{2\pi i u\cdot v/N}\,Z_u(\tau_0).
\la{zvtransf}
\end{equation}
The modular weights $w$ and $\bar w$ are linear combinations of the Euler
characteristic $\chi$ and the signature $\sigma$ of $X$. 
This pattern has been checked for the physical $\cn=4$
theory on $T^4$ in \cite{italia}. 

As for the  twisted $\cn=4$ theories to be described below, it would be 
natural to expect that they should behave in the same way under duality. 
In fact, for the twisted theories, and in generic circumstances, the partition
function  depends holomorphically on $\tau_0$, so one would actually expect
that they should obey \eqs{zvtransf} with $\bar w=0$. This was checked by Vafa
and  Witten for one of the twisted theories and for gauge group $SU(2)$ 
\cite{vw}, and their result has been generalized to $G=SU(N)$ (with $N$
prime) in \cite{sun}.  Similar results
have been recently derived for another twisted version of the
$\cn$=$4$ theory within the $u$-plane approach \cite{htwist}. 
We will review these results in great detail in chapters \ref{chvw} and 
\ref{chtwist} below.

\section{The Seiberg-Witten solution}

The massless $\cn=4$ theory has a moduli space of vacua  in the Coulomb phase
consisting of  several equivalent copies which are interchanged by the
$SU(4)_I$ symmetry.  Each of these copies corresponds to one of the scalar
fields $\phi_{uv}$   developing a non-zero vacuum expectation value. There is
a classical singularity at the origin of the moduli space, which is very
likely to  survive even in the quantum regime.  

Life gets more interesting if one deforms the 
$\cn=4$ supersymmetric theory by giving bare   
masses, $m\int d^4 x d^2 \theta\tr{(\Phi_1\Phi_2)}+{\hbox{\rm h.c.}}\;$,  
to two of the chiral multiplets. This mass-deformed theory still 
retains $\cn=2$ supersymmetry: the massive superfields build up an $\cn=2$ 
hypermultiplet, while the remaining chiral superfield, together with 
the vector superfield, build up an $\cn=2$ vector multiplet. The low-energy 
effective description of this theory was worked out, for the $SU(2)$ gauge 
group, by Seiberg and Witten in 
\cite{swii}. Their results were subsequently extended to $SU(N)$ by Donagi
and Witten in
\cite{donagi}, where a  link to the $SU(N)$ Hitchin integrable system was
established. D'Hoker and Phong have analyzed the $SU(N)$ theory in terms of
the elliptic Calogero-Moser integrable system \cite{dhoker1}, and have
extended the computations to arbitrary gauge groups \cite{dhoker2}.

 Some quantitative discrepancies  between the proposed solution
and  explicit instanton calculations have been pointed out in 
\cite{instanton}. 
The explicit structure of the effective theory for gauge group $SU(2)$ has 
been much clarified by Ferrari \cite{ferrari}, who has 
also given a detailed account of the BPS spectrum.  

For gauge group $SU(2)$ and for generic values of the mass parameter, the moduli 
space of physically inequivalent vacua forms a one complex-dimensional compact 
manifold (the $u$-plane). This manifold parametrizes a family of elliptic curves, 
which encodes all the relevant information about the low-energy effective 
description of the theory. The explicit solution is given by the 
curve:

\begin{equation}
Y^{2}=\prod_{j=1}^{3}\Bigl(X-e_{j}(\tau_0 )z-\frac{1}{4}e_{j}^{2}(\tau_0
)  m^{2}\Bigr),
\label{elliptic}
\end{equation}
where
\begin{equation}e_1(\tau_0 )=\frac{1}{3}(\vartheta _4^4 + \vartheta
_3^4),\quad e_2(\tau_0 )=-\frac{1}{3}(\vartheta _2^4 + \vartheta _3^4),\quad
e_3(\tau_0 )=\frac{1}{3}(\vartheta _2^4 -\vartheta _4^4),
\label{spin}
\end{equation}
and 
$\vartheta _2$, $\vartheta _3$ and $\vartheta _4$ 
are the Jacobi theta functions -- see the appendix to chapter \ref{chtwist}
below for more details. Notice that the curve \eqs{elliptic} depends
explicitly on $\tau_0$ through the modular forms $e_j$, so $SL(2,\IZ)$ duality
is actually built in from the start. 

The parameter $z$ in \eqs{elliptic} 
is a global gauge-invariant coordinate on the moduli space and it 
is a modular form of weight $2$ under the microscopic duality group. 
It differs from the physical order parameter 
$\langle \tr \, \phi ^2\rangle$ by instanton corrections \cite{instanton},  
which are not predicted by the Seiberg-Witten solution. The precise 
relation is given by: 
\begin{equation}
z=\langle \tr\, \phi ^2\rangle - \frac{1}{8} m^2 e_1(\tau_0 ) +
m^2 \sum_{n=1}^{\infty}c_{n}q_0^{n},\qquad\qquad q_0=\ex^{2i\pi\tau_0}.
\label{instant}
\end{equation}
Notice that the instanton corrections $c_{n}$ are invisible in the 
double-scaling limit $q_0\to 0$, $m\to\infty$, with $4m^4
q_0=\Lambda_{0}{}^4$,  under which the mass-deformed theory flows towards the
pure $\cn=2$ gauge  theory and $z\to u= \langle \tr\, \phi ^2\rangle$. Here
$\Lambda_0$ is the  dynamically generated scale of the $\cn=2$, $N_f=0$
theory. 

The low-energy description breaks down at certain points $z_i$ where the elliptic 
curve degenerates. This happens whenever any two of the roots of the cubic 
polynomial $\prod^{3}_{j=1}\left(X-e_{j}z-(1/4)e_{j}^{2} 
m^{2}\right)$ coincide. These singularities, which from the physical point of 
view are interpreted as due 
to BPS-saturated multiplets becoming massless, are located at the points 
\cite{swii}:
\begin{equation}z_i= \frac{m^2}{4} e_i
\label{singularity}
\end{equation}
These BPS states are generically dyons carrying non-zero magnetic charge, and
can be constructed in the semiclassical regime by quantizing the
zero modes of the elementary fields of the theory in the background of the
dyon \cite{harvey}. For the massless $\cn=4$ theory, these zero modes build up
a short $\cn=4$ vector multiplet with spins up to $1$ \cite{bertin}, and this
explains why the monopoles (with unit magnetic charge and no electric charge) 
can be dual,  in the  Montonen-Olive sense, to the elementary vector bosons. 
A non-zero value of the mass $m$  lifts half of the zero modes, and the
remaining ones build up the $\cn=2$ hypermultiplets which produce the
singularities.  

Following Ferrari \cite{ferrari}, we choose $\vert q_0\vert$
small, 
$m$ large, with 
$m^4 q_0\sim \Lambda_{0}{}^4$.  Under these circumstances, at 
strong (effective) coupling, there are two singularities at $z_2$, $z_3$, 
with $\vert z_2 - z_3\vert \sim \Lambda_0{}^2$, which flow to the 
singularities 
of the pure gauge theory in the double-scaling limit. At weak (effective) 
coupling, there is a third singularity, located at $z_1$, due to an 
electrically charged (adjoint) 
quark becoming massless. For this choice of parameters, 
we have the explicit formulas:   
\begin{equation}
k^2= \frac{\vartheta _2(\tau)^4}{\vartheta _3 (\tau)^4}
=\frac{\vartheta _2(\tau_0) ^4}{\vartheta _3 (\tau_0)^4}\,
\frac{z-z_1}{z-z_3}\raise 2pt\hbox{,}\quad
k^{'2}= 1-k^2=\frac{\vartheta _4 (\tau) ^4}{\vartheta _3 (\tau)^4}
=\frac{\vartheta _4 (\tau_0) ^4}{\vartheta _3 (\tau_0)^4}\,
\frac{z-z_2}{z-z_3},
\label{modulus}
\end{equation}
relating the coordinate $z$ 
to the modulus $k$ of the associated elliptic 
curve (\ref{elliptic}). Here $\tau=\tau_0-\frac{3im^4}{\pi
a^4}\ex^{2i\pi\tau_0}+\cdots$ \cite{instanton} is the complexified effective
coupling of the  low-energy theory, and enters the formalism as the ratio of
the two basic  periods of the elliptic curve.  The first period of the curve
is given by  the formula: 
\begin{equation}
\frac{da}{ dz}= 
\frac{\sqrt{2}}{\pi} \frac{1}{\vartheta_{3}(\tau_0)^{2}\sqrt{z-z_{3}}}\,
K(k),
\label{period}
\end{equation}
where 
\begin{equation}
K(k)= \frac{\pi}{2}\vartheta_3(\tau)^2
\label{otroperiod}
\end{equation}
is the complete elliptic integral of the first kind, and $a$ is
related to the vacuum expectation value of the Higgs field, 
$\langle\phi\rangle=a\sigma_3/2$. The second period
can be  computed from (\ref{period}) as $\frac{d a_D}{ d z}=\tau 
\frac{d a}{d z}$. Owing to  the cuts and non-trivial monodromies present on
the 
$u$-plane\footnote{``$z$-plane" would be more 
accurate here, but the former terminology is by now so widespread that we prefer 
to stick to it.}, $\frac{d a_D}{ d z}$ is not globally defined, and the actual 
formulas are somewhat more involved \cite{ferrari}. 
In any case, the final expression for the $u$-plane integral will be invariant 
under monodromy transformations, so the above naive expression is sufficient 
for our purposes.

Around each of the singularities we have the following series expansion:
\begin{equation}
z=z_j +\kappa_j\, q_j{}^{\half}+\cdots
\label{kapas}
\end{equation}
where $q_j=\ex^{2\pi i\tau_j}$ is the good local coordinate at each 
singularity: $\tau_1=\tau$ for the semiclassical singularity at $z_1$, 
$\tau_2=\tau_D=-{1}/{\tau}$ for the monopole singularity at $z_2$, and 
$\tau_3=\tau_d=-{1}/{(\tau-1)}$ for the dyon singularity at $z_3$. 

Using (\ref{modulus}), one can readily compute:
\begin{equation}
\kappa_1(\tau_0)=4m^2\left(\frac{\vartheta_3\vartheta_4}{\vartheta_2}
\right)^4, \quad
\kappa_2(\tau_0)=-4m^2\left(\frac{\vartheta_2\vartheta_3}{\vartheta_4}
\right)^4,\quad
\kappa_3(\tau_0)=4m^2\left(\frac{\vartheta_2\vartheta_4}{\vartheta_3}
\right)^4.
\label{yespadas}
\end{equation}
At the singularities, each of the periods has a finite limit when 
expressed in terms of the appropriate local coordinate:
\bea
\left(\frac{d a}{ d z}\right)^{\,\,2}_1 &=& \frac{2}{ m^2}\frac{1}{ ( 
\vartheta_3(\tau_0)\vartheta_4(\tau_0))^4}, \ret
\left(\frac{d a_D }{ d z}\right)^{\,\,2}_2 &=& \frac{2}{ m^2}\frac{1}{ ( 
\vartheta_2(\tau_0)\vartheta_3(\tau_0))^4}, \ret 
\left(\frac{d (a_D -a)}{ d z}\right)^{\,\,2}_3 &=& -\frac{2}{ m^2}
\frac{1}{(\vartheta_2(\tau_0)\vartheta_4(\tau_0))^4}.
\ret
\label{zhivago}
\eea

\vfil
\newpage


\chapter{Twisting the $\cn=$ 4 supersymmetric gauge theory}  
\la{ctwists}
\markboth{\footnotesize\bfseries Twisting the $\cn=4$ supersymmetric gauge 
theory}{\footnotesize\bfseries Twisting the $\cn=4$ supersymmetric gauge theory}
\markright{\textsc {Duality in Topological Quantum Field Theories}}

The twist in the context of supersymmetric four-dimensional gauge theories 
was introduced by Witten in \cite{tqft}, where he showed that a twisted
version of the $\cn=2$ supersymmetric gauge theory with gauge group 
$SU(2)$ is a relativistic field-theory realization of the Donaldson theory
of four-manifolds. Soon after Witten's breakthrough, Yamron
\cite{yamron} generalized the construction to the $\cn=4$ supersymmetric
gauge theory and described the
structure of two of the possible non-equivalent twists of these theories and
pointed out the existence of a third one. 

The purpose of this chapter is to describe the possible twists of the 
$\cn=4$ supersymmetric gauge theory. It is intended to provide a general
introduction to the next three chapters. We assume that the reader is
familiar with the analogous (yet simpler) procedure in
$\cn=2$ theories \cite{kungfu}\cite{phyrep}\cite{tqft}\cite{wijmp}. 

In four dimensions, 
the global symmetry group of the extended
supersymmetric gauge theories is of the form ${\cal H}=SU(2)_L\otimes
SU(2)_R\otimes{\cal I}$, where ${\cal K}= SU(2)_L\otimes SU(2)_R$ is the
rotation group, and ${\cal I}$ is the chiral ${\cal R}$-symmetry group. 
The supercharges $Q^i_\alpha$, $\bar Q_{\dalpha j}$ transform under 
${\cal H}$ as $({\bf 2},{\bf 1},{\bf N})$ and $({\bf 1},{\bf 2},
{\bf\bar N})$, where ${\bf N}$ is generically an $\cn$-dimensional
representation of the ${\cal R}$-symmetry group -- $\cn$ is the number of
independent supersymmetries. The aim of the twist is to extract from these
supercharges one (or several)  global scalar
fermionic symmetries -- as described in chapter \ref{tft} -- which always
exist regardless of the space-time topology. Now to create a scalar 
supercharge out of spinor supercharges one has to modify somehow the action
of the rotation group on the supercharges. The idea is to gauge a subgroup
of the ${\cal R}$-symmetry group with the spin connection in such a way 
that at least a linear combination of the original supercharges be a singlet
under a combined Lorentz plus chiral rotation -- see \cite{baryon} for a
review. Depending on how we choose this subgroup, we will obtain different
theories after the twisting\footnote{Note than on a general four-manifold
the holonomy group is $SU(2)_L\otimes SU(2)_R$, so the twist will be an
observable effect, that is, the twisted theory and the physical theory
will be inequivalent. However, on a hyper-K\"ahler manifold the holonomy
is simply $SU(2)_R$, so if one gauges the chiral current  with
the $SU(2)_L$ part of the holonomy only, the twisting is trivial, and the
twisted and physical theories are equivalent.}.  
While in
$\cn=2$ supersymmetric gauge theories  the ${\cal R}$-symmetry group is at
most
$U(2)$ and thus the twist is essentially unique, in the
$\cn=4$  supersymmetric gauge theory the  
${\cal R}$-symmetry group is $SU(4)$ and there are three different
possibilities, depending on how we embedd the rotation group into the
${\cal R}$-symmetry group \cite{ene4}\cite{vw}\cite{yamron}.  

The possible choices are found just by analyzing
how the ${\bf 4}$  of $SU(4)_I$ splits in terms of representations of the
rotation group ${\cal K}$. There are just three possibilities which will
give a topological symmetry for a given choice of the $SU(2)$ component of
${\cal K}$: (i)
${\bf 4}\to({\bf 2},{\bf 1})\oplus ({\bf 2},{\bf 1})$, (ii)
${\bf 4}\to({\bf 2},{\bf 1})\oplus ({\bf 1},{\bf 1})\oplus ({\bf 1},{\bf
1})$ and (iii) ${\bf 4}\to ({\bf 2},{\bf 1})\oplus ({\bf 1},{\bf 2})$, each
of which leads to a different topological quantum field theory. Choosing
the other $SU(2)$ component of ${\cal K}$ one would obtain the equivalent 
twists:
${\bf 4}\to({\bf 1},{\bf 2})\oplus ({\bf 1},{\bf 2})$, 
${\bf 4}\to({\bf 1},{\bf 2})\oplus ({\bf 1},{\bf 1})\oplus ({\bf 1},{\bf
1})$ and
${\bf 4}\to ({\bf 1},{\bf 2})\oplus ({\bf 2},{\bf 1})$. As described below 
these alternative twists are related to the previous ones by a reversal of
the orientation of the four-manifold
$X$.

Cases (i) and (iii) lead to topological field theories with
two supercharges. One of these (i) was considered by Vafa and Witten
\cite{vw} in order to carry out an explicit test of $S$-duality on several
four-manifolds, and will be analyzed in full detail in chapter \ref{chvw}. 
It has the unusual feature that the virtual dimension of its moduli space is
exactly zero. This feature was analysed from the  perspective of balanced
topological field theories in  \cite{balanced}, while the underlying
structure had already been anticipated within the framework of
supersymmetric quantum mechanics in 
\cite{cofield}. 

Case (iii) was first discussed in  
\cite{marcus},  where it was shown to correspond to a topological theory 
of complexified flat gauge connections.  This idea was pursued  further in
\cite{blauthomp},  where a link to supersymmetric BF theories in four
dimensions was established. We will study this theory in chapter
\ref{chmarcus}, where it will be shown that the theory is amphicheiral, this
meaning  that the twist with either $SU(2)_L$ or
$SU(2)_R$ leads essentially  to the same theory. 

The remaining possibility (ii) leads to the ``half-twisted theory",  a
topological theory  with only one BRST supercharge \cite{yamron}. This
feature is reminiscent of the situation in twisted $\cn=2$ supersymmetric
gauge theories,  and in fact \cite{ene4}, the theory is a close  relative of
the non-Abelian monopole theory 
\cite{corea}\cite{nabm}\cite{polynom}\cite{tesis}, 
the non-Abelian
generalization of Witten's monopole theory 
\cite{monopole}, for the special case in which the matter fields are in the
adjoint representation of the gauge group -- see chapter \ref{chtwist}
below. 

\section{Twisted $\cn=4$ supersymmetry and the Mathai-Quillen approach}

In the forthcoming chapters we will analyze to exhaustion the three cases 
described above, both from the viewpoint of the twisting of $\cn=4$
supersymmetry and the Mathai-Quillen approach. It is well known that
topological quantum field theories obtained after twisting
$\cn=2$ supersymmetric gauge theories can be formulated in the framework of
the Mathai-Quillen formalism \cite{jeffrey}\cite{moore}\cite{tesis}. 
One would expect that a similar formulation should exist for the $\cn=4$
case.  Though it turns out that this is so, there is an important issue that
has to be addressed to clarify  what  is meant by a Mathai-Quillen
formulation in the latter case. Twisted
$\cn=2$ supersymmetric gauge theories have an off-shell formulation such
that the topological quantum field theory action can be expressed as a
$Q$-exact expression, where $Q$ is the generator of the topological
symmetry. Actually, this is true only up to a $\theta$-term. However,
the ${\cal R}$-symmetry group of the $\cn=2$ supersymmetric gauge theories
contains a $U(1)$ factor which becomes the ghost-number symmetry of the
topological theory. But this chiral $U(1)$  is anomalous, so one can
actually get rid of the $\theta$-term with a chiral rotation. As a result
of this, the observables in the topological theory are unsensitive to
$\theta$-terms up to a rescaling. What remains is just the 
$Q$-exact part of the action which is precisely the one obtained in the
Mathai-Quillen formalism.

On the other hand, in $\cn=4$ supersymmetric gauge theories 
$\theta$-terms are observable. There is no chiral anomaly and these 
terms can not be shifted away as in the
$\cn=2$ case. This means that in the twisted theories one might have a
dependence on the coupling constants (in fact, this was one of the key
observations in
\cite{vw} to make a strong coupling test of $S$-duality.) This being so we
first have to clarify what one expects to be the form of the twisted
theories in the framework of the Mathai-Quillen formalism. To do this let 
us focus on the part of the action  of a twisted theory
(coming from any gauge theory with extended supersymmetry) involving 
the gauge field strength,
\begin{equation}  
{\cal S}_{X} = -\frac{1}{ 4 e^2} \int_X \sqrt{g} d^4x \,
\tr(F^{\mu\nu}F_{\mu\nu}) -\frac{i\theta}{ 16\pi^2} \int_X \tr(F\wedge F) +
\dots,
\label{runrun}
\end{equation}  
where $X$ is an oriented Riemannian four-manifold endowed with a metric 
$g_{\mu\nu}$. We are using conventions such that,
\begin{align}  
k=\frac{1}{ 16\pi^2} \int_X \tr(F\wedge
F)&=\frac{1}{32\pi^2}\int_X
\sqrt{g}\tr( *  F_{\mu\nu}F^{\mu\nu})\ret &= \frac{1}{32\pi^2}\int_X
\sqrt{g}\tr\bigl\{\,( F^{+})^2-(F^{-})^2\bigr\},
\la{knumber}
\end{align}    
is the instanton number of the gauge configuration. Decomposing the
field strength
$F$ into its self-dual and anti-selfdual parts,
\begin{equation}  F^\pm_{\mu\nu} = \frac{1}{ 2}(F_{\mu\nu}\pm\frac{1}{
2}\epsilon_{\mu\nu\rho\sigma} F^{\rho\sigma}),
\la{coque}
\end{equation}
we see that \eqs{runrun} can be written in the following two forms:
\begin{align} 
{\cal S}_X  &=-\frac{1}{ 2 e^2} \int_X \sqrt{g} d^4x \,
\tr(F^{+\mu\nu}F^+_{\mu\nu})
 -2\pi i\tau \frac{1}{ 16\pi^2} \int_X \tr(F\wedge F) + \dots \ret
     &= -\frac{1}{ 2 e^2} \int_X \sqrt{g} d^4x \,
\tr(F^{-\mu\nu}F^-_{\mu\nu})-
      2\pi i\bar\tau \frac{1}{ 16\pi^2} \int_X \tr(F\wedge F) + \dots, \ret
\label{tsi}
\end{align}  
where
\begin{equation}
\tau = \frac{\theta }{ 2\pi} + \frac{4\pi i}{ e^2}.
\la{latau}
\end{equation}

 What we
intend to discuss here is the difference between the two possible choices
which are present when picking up one of the components of the rotation
group. It turns out that choosing one of them, say, the left or twist $T$,
one must consider the first form of the action in \eqs{tsi} since then,
after working out its off-shell  formulation, it can be written as 
\begin{equation}  
{\cal S}^T_X =  \frac{1}{ 2 e^2} \int_X \sqrt{g} d^4x \,
\{Q,\Lambda\} -
 2\pi i\tau \frac{1}{ 16\pi^2} \int_X \tr(F\wedge F),
\la{terouno}
\end{equation}  
for some $\Lambda$, while it one chooses the other one, the
right or twist
$\tilde T$, one finds, 
\begin{equation} 
{\cal S}^{\tilde T}_X =  \frac{1}{ 2 e^2} \int_X \sqrt{g}
d^4x \, \{\tilde Q,\tilde\Lambda\} -
2\pi i\bar\tau \frac{1}{16\pi^2} \int_X \tr(F\wedge F),
\la{terodos}
\end{equation}  
for some $\tilde\Lambda$ and some $\tilde Q$. These actions correspond to
an orientable four-manifold $X$ with a given orientation. The actions of
the two twists are related in the following way:
\begin{equation} {\cal S}^{T}_X  = {\cal S}^{\tilde T}_{\tilde X}
\Big|_{\tau\rightarrow -\bar\tau}, 
\label{titin4}
\end{equation}  
where the four-manifolds $X$ and $\tilde X$ are related
by a reversal of orientation. 

For twisted theories coming from (asymptotically free) $\cn=2$
supersymmetric gauge theories, observables do not depend on the coupling
constant $e$ because it appears only in a term which is
$Q$-exact. They do not depend on $\tau$ either, up to a rescaling, due to
the chiral anomaly.  In the case of twisted theories $\cn=4$ theories,
however, the partition function and the observables do depend on $e$ and
$\theta$  through $\tau$. In both cases one needs to consider only one of
the types of twist, say $T$, since, according to \eqs{titin4}, the other
just leads to the observables that one would obtain by working on $\tilde
X$ instead of $X$. In the first case this statement is exact and in the
second case one must supplement it with the replacement $\tau\rightarrow
-\bar\tau$. Therefore one can say that up to a reversal of orientation
there is only one possible twist from $\cn=2$ supersymmetric gauge
theories and three, as stated in
\cite{yamron} and described in detail above, from the $\cn=4$ theory.

After these remarks on the twisting procedure we will state what is meant
by a Mathai-Quillen formulation of topological quantum field theories
resulting after twisting  $\cn=4$ supersymmetric gauge theories. The
Mathai-Quillen formulation builds out of a moduli problem a
representative of the Thom class associated to a certain  vector
bundle. This representative can always be written as an integral of the
exponential of a $Q$-exact expression. The three twists of $\cn=4$, after
working out their off-shell formulation, can be written as in
\eqs{terouno}. We will present for each case the moduli problem which in
the context of the Mathai-Quillen approach leads to the $Q$-exact part of
the action. In other words, we will find out the geometrical content
which is behind each of the three twists.

\section{String theory and twisted $\cn=4$ topological field theories}
This section is intended to give and overview of the known relations between 
string theory and topological quantum field theories. Almost everything of 
what we will be saying is well-known and nothing is new. 

Several isolated links between string theory and topological quantum field 
theories have been known for some time. In 1991, Harvey and Strominger 
pointed out that  $S$-matrix elements describing zero momentum scattering 
of spacetime axions off fivebranes in heterotic string theory are proportional 
to the Donaldson polynomials \cite{hs}. Indeed, the set of ground states of 
$N$ static fivebranes is precisely the moduli space of $N$-instanton 
configurations on the four-dimensional manifold transverse to the fivebranes 
\cite{chs}, and the transitions between different ground states are 
governed by effective vertices which are formally identical to the topological 
observables which represent the Donaldson invariants in the twisted $\cn=2$ 
supersymmetric gauge theory \cite{tqft}. The effective theory describing the
dynamics and  interactions of the fivebranes is a six-dimensional sigma
model with target  the moduli space of instantons, but one could be tempted
to conjecture that  the twisted $\cn=2$ theory could provide a dual effective
description of the  dynamics. 

A second connection in the same direction stems from a result of 
Taubes \cite{taubes} pointing out a relation between the Seiberg-Witten 
invariants \cite{monopole}, which capture the essential topological 
information of many twisted four-dimensional supersymmetric theories, 
and the Gromov invariants, which arise in the context of two-dimensional 
topological sigma models \cite{tsm}. This relation has been discussed in 
\cite{coreatres} in connection with the Vafa-Witten theory for $G=SU(2)$. 

New, recent developments in string theory have unveiled more concrete and 
truly unexpected relations. Bershadsky, Sadov and Vafa \cite{BSV}, and
Minahan,  Nemschansky, Vafa and Warner \cite{estrings}, have pointed out that
the three twisted $\cn=4$  theories appear naturally in compactifications of
string/M theory as  world-volume effective theories of IIB D-threebranes or
M-fivebranes wrapping  supersymmetric cycles of the compactification space.
The idea is as  follows -- see \cite{blauthomp}\cite{vafita} for a review.
Consider a certain  compactification of type IIB theory on a $d$-dimensional
compact manifold $W_d$,  and a D3-brane with worldvolume $X_4$ wrapped
around a supersymmetric $4$-cycle  of $W_d$ (this is a $4$-dimensional
submanifold of $W_d$ defined by the property that a D-brane wrapping around
it preserves some  supersymmetries \cite{becker}.) The effective theory on
the threebrane is generically an
$\cn=4$  supersymmetric gauge theory, with the six real scalars describing
the position  of the brane in the ambient space. Or more precisely, the
scalars are sections  of the normal bundle of $X_4$ in $M_{10}$ (the
$10$-dimensional space where the  full string theory lives), which is
$6$-dimensional. But since $X_4$ is embedded  in $W_d$, $d-4$ of the scalars
will be twisted as they are actually sections of  the normal bundle of $X_4$
relative to $W_d$. Now by supersymmetry, the rest of  the fields in the
$\cn=4$ multiplet will be twisted as well, so the general  result is that
the effective theory is a certain twisted version of the $\cn=4$ theory.
Which twisted version one ends up with depends on the details of the 
compactification. Consider for example the Vafa-Witten theory. As we will
see  in the next chapter, the theory contains three real scalar fields,
while the  remaining three are twisted to a self-dual two-form. Thus the
compact space 
$W_d$ must be $7$-dimensional. Likewise, as there are two independent
twisted  supersymmetries, and the D-brane generically preserves $1/2$ of the 
supersymmetries left unbroken by the compactification, we need a $7$
manifold  preserving $1/8$-th of the supersymmetry. This is a $7$-manifold
of $G_2$  holonomy, which indeed has supersymmetric $4$-cycles whose normal
bundle is  precisely $\Omega^{2,+}(X)$ (the bundle of self-dual two-forms on
$X$),  as expected. A similar analysis \cite{BSV} shows that the second
twist arises  in compactifications on an $8$-dimensional $Spin(7)$-holonomy
manifold, while  the amphicheiral theory is realized on supersymmetric
cycles of $SU(4)$-holonomy  Calabi-Yau $4$-folds. In all these cases the
structure of the normal bundle  and the number of unbroken supersymmetries
agree with what is expected for each  of the twisted theories.

As for the M5-brane, it is  known that \cite{dijkfvbr} the effective theory 
on a worldvolume of the form $\IR^4\times T^2$ is again the $\cn=4$ gauge 
theory (the gauge coupling constant is just the modular parameter of the 
torus, and the $SL(2,\IZ)$-duality of the $\cn=4$ theory
follows simply from the $SL(2,\IZ)$ action on the modular parameter of
$T^2$.) If we now consider more general worldvolumes of the form $X\times
T^2$,  where $X$ is a supersymmetric cycle of a given compactification
manifold,  the effective theory on $X$ should be again a twisted $\cn=4$
theory. This fact has been exploited in \cite{estrings} to study certain
six-dimensional tensionless strings. 
Actually, in all these cases the twisted $\cn=4$ theories should correspond 
to the 
long-wavelength limit of the effective theory on the curved branes, whose 
correct description should involve a twisted version of 
the appropriate Born-Infeld theory on the branes 
\cite{chicagoyale}\cite{bspence}.

All these connections seem to suggest a deep relation between string theory 
and topological field theories which would be worth to study in the future.  
The most promising avenue to explore and exploit these 
connections seems to be the recent AdS/CFT conjecture
\cite{klebanov1}\cite{malda}\cite{wiads1} -- see \cite{klebanov2} for a
review. Gopakumar and Vafa  \cite{gopakui}\cite{gopakuii} have recently
shown that in the large $N$ ('t Hooft) limit, the Chern-Simons theory is dual
to  a certain topological closed string theory. 
As for the twisted $\cn=4$ theories, no real progress has been made so far 
(but see \cite{hull}\cite{wiads2}, where some interestings issues in this
direction are addressed.)


\chapter{The Vafa-Witten theory}
\la{chvw}
\markboth{\footnotesize\bfseries The Vafa-Witten theory
}{\footnotesize\bfseries The Vafa-Witten theory}
\markright{\textsc {Duality in Topological Quantum Field Theories}}

In this chapter we will concentrate on the first of the topological
theories that can be constructed by twisting the $\cn=4$ theory. The
twisted theory, for gauge group $SU(2)$, was considered  by Vafa and
Witten
\cite{vw} to carry out an explicit test of $S$-duality on several
four-manifolds. The partition
function of this theory computes, under certain circumstances, the Euler
characteristic of instanton moduli spaces, making  it possible to fix, by
comparing with known mathematical results, several unknown modular
functions which can not be determined otherwise. The final  computation
requires the introduction of a clever mass perturbation which, while
breaking down the $\cn=4$ supersymmetry of the physical theory down to 
$\cn=1$, still preserves one of the topological symmetries of the theory. 
A similar approach was  introduced by Witten in
\cite{wijmp} to obtain the first explicit results for  the
Donaldson-Witten theory just before the far more powerful
Seiberg-Witten approach was available. However, this 
 approach is restricted to K\"ahler manifolds with $b^{+}_2>1$.  
Vafa and Witten conjectured that, in the case of the theory at hand, the
perturbation does not affect the final result for the  partition function. 
We will carefully analyze and confirm their conjecture and extend their 
computation to gauge group $SU(N)$, with $N$ prime. 

\section{The twisted theory}

 After the twisting,  the  symmetry group of the theory becomes 
${\cal H'} =SU(2)'_L\otimes SU(2)_R\otimes SU(2)_F$, where $SU(2)_F$ is a
subgroup of $SU(4)_I$ that commutes with the branching ${\bf
4}\to({\bf 2},{\bf 1})\oplus ({\bf 2},{\bf 1})$ and remains in the theory as
a residual isospin group. Under ${\cal H'}$, the supercharges split up as,
\begin{equation} 
Q^v{}_\alpha\to Q^i,\;Q^i{}_{\alpha\beta},\qquad \bar Q_{v\dalpha}\to
\bar Q^i_{\alpha\dalpha},
\la{elven4}
\end{equation} 
where the index $i$  labels the fundamental representation of
$SU(2)_F$. The twist has produced two scalar supercharges, the $SU(2)_F$
doublet
$Q^i$. These scalar charges are defined in terms of the original supercharges
as follows:
\bea
&&Q^{i=1}\equiv Q^{v=1}_{\alpha =1} +Q^{v=2}_{\alpha=2},
\ret  
&&Q^{i=2}\equiv Q^{v=3}_{\alpha =1} +Q^{v=4}_{\alpha=2}.
\ret
\la{osiete}
\eea

The fields of the $\cn=4$ multiplet decompose under ${\cal H'}$ as follows:
\bea
A_{\alpha\dalpha} &\too& A_{\alpha\dalpha},\ret 
\lambda_{v\alpha}&\too&
\chi_{i\beta\alpha},~
\eta_{i},\ret 
\bar\lambda^v{}_{\dalpha} &\too&
\psi^{i\alpha}{}_{\dalpha},\ret 
\phi_{uv}&\too& \varphi_{ij},~G_{\alpha\beta}.\ret
\label{poocho4}
\eea 
Notice that the fields $\chi^i_{\alpha\beta}$ and $G_{\alpha\beta}$ are
symmetric in their spinor indices and therefore can be regarded as components of
self-dual two-forms. $\varphi_{ij}$ is also symmetric in its isospin indices and
thus transforms in the representation ${\bf 3}$ of
$SU(2)_F$. Some of the definitions in
\eqs{poocho4} need clarification. Our choices for the anticommuting fields are, 
\bea
\chi_{i=1 (\alpha\beta)}\!\!&=&\!\!
\begin{cases}
\chi_{i=1 (11)}=\lambda_{v=1,\alpha=1},\\
\chi_{i=1(12)}=\half(\lambda_{v=1,\alpha=2}+\lambda_{v=2,\alpha=1}),
\\ \vdots
\end{cases}
\ret 
\eta_{i=1}\!\!&=&\!\!\lambda_{v=1,\alpha=2}-\lambda_{v=2,\alpha=1},\ret
\psi^{i=1,\alpha=1,2}_\dalpha \!\!&=&\!\!\bar\lambda^{v=1,2}_\dalpha,
\ret
\la{onueve}
\eea 
while for the scalars $\phi_{uv}$:
\bea
\varphi_{ij}\!\!&=&\!\!\begin{pmatrix}
\phi_{12}&\half(\phi_{14}-\phi_{23})\\
\half(\phi_{14}-\phi_{23})&\phi_{34}
\end{pmatrix},
\ret 
G_{\alpha\beta}\!\!&=&\!\!\begin{pmatrix}
\phi_{13}&\half(\phi_{14}+\phi_{23})\\
\half(\phi_{14}+\phi_{23})&\phi_{24}
\end{pmatrix}.\ret
\label{noventa}
\eea

In terms of the twisted fields, the $\cn=4$ action \eqs{Boris} takes the form 
(remember that we are still on flat $\IR^4$):
\begin{align}
{\cal S}^{(0)} &= \frac{1}{ e^2_0}\int d^4 x\, \tr\, \bigl\{\,\frac{1}{
4}\nabla_{\alpha
\dalpha}\varphi_{ij}\nabla^{\dalpha\alpha}\varphi^{ij}
-\frac{1}{4}\nabla_
{\alpha\dalpha}G_{\beta\gamma}\nabla^{\dalpha\alpha}G^{\beta\gamma} 
-i\psi^{j\beta}{}_{\dot\alpha}\nabla^{\dot\alpha\alpha}
\chi_{j\alpha\beta}
\ret 
&-\frac{i}{2}\psi^j_{\alpha\dot\alpha}\nabla^{\dot\alpha\alpha}
\eta_j -\frac{1}{4} F_{mn} F^{mn}
-\frac{i}{\raiz}\,\chi_i{}^{\alpha\beta}[\chi_{j\alpha\beta},\varphi^{ij}]
+\frac{i}{\raiz}\,\chi^{i\alpha}{}_\beta[\chi_{i\alpha\gamma},G^{\beta\gamma}]
\displaybreak\ret 
&-\frac{i}{\raiz}\,\chi^i{}_{\alpha\beta}[\eta_i,G^{\alpha\beta}]-
\frac{i}{{2\raiz}}\,\eta_i[\eta_j,\varphi^{ij}]
+\frac{i}{\raiz}\,\psi^{i\alpha}{}_{\dot\alpha}[\psi_{i\beta}{}^
{\dot\alpha},G_\alpha{}^\beta]
\ret
&-\frac{i}{\raiz}\,\psi^{i\alpha}{}_{\dot\alpha}[\psi^j{}_{\alpha}{}^
{\dot\alpha},\varphi_{ij}]+\frac{1}{ 4}[\varphi_{ij},\varphi_{kl}]
[\varphi^{ij},\varphi^{kl}] 
-\half[\varphi_{ij},G_{\alpha\beta}][\varphi^{ij},G^{\alpha\beta}] 
\ret
&+\frac{1}{ 4}[G_{\alpha\beta},G_{\gamma\delta}]
[G^{\alpha\beta},G^{\gamma\delta}]\,\bigr\} 
-\frac{i\theta_0}{32\pi^2}\int d^4
x\,\tr\,\bigl\{\, *  F_{mn}F^{mn}
\,\bigr\}.
\ret
\la{nuno4}
\end{align}
 
The $Q^i$-transformations of the twisted theory can be readily obtained from the
corresponding $\cn=4$ supersymmetry transformations. These last transformations are
generated by $\xi_v{}^\alpha Q^v{}_\alpha + 
\bar\xi^v{}_\dalpha \bar Q_v{}^\dalpha$. According to our conventions, to obtain
the $Q^i$-transformations one must set $\bar\xi^v{}_\dalpha=0$ and make the
replacement:  
\begin{equation}
\xi_{v\alpha}=\begin{cases}
\xi_{v=(1,2)\alpha}\to \epsilon_{i=
1}C_{\beta=(1,2)\alpha},\\
\xi_{v=(3,4)\alpha}\to \epsilon_{i= 2}C_{\beta=(1,2)\alpha},
\end{cases}
\la{ndos}
\end{equation} 
where $C_{\beta\alpha}$ (or $C_{\dot\beta\dalpha}$, $C_{ij}$) is the antisymmetric
(invariant) tensor of
$SU(2)$ with the convention $C_{21}=C^{12}=+1$. The resulting transformations are:
\bea
\delta A_{\alpha\dalpha} &=& 2i\epsilon_j\psi^j_{\alpha\dalpha},\ret
\delta F^{+}_{\alpha\beta}&=&2\epsilon_i\nabla_{(\alpha}{}^{\dot\alpha}
\psi^i{}_{\beta)\dot\alpha},\ret 
\delta\psi^{i\alpha}{}_{\dot\alpha} &=& 
-i{\raiz}\epsilon_j\nabla^{\alpha}{}_{\dot\alpha}
\varphi^{ji} + i{\raiz}\epsilon^i\nabla_{\beta\dot\alpha} G^{\beta\alpha},\ret
\delta\chi_{i\alpha\beta} &=&  -i\epsilon_i F^{+}_{\alpha\beta}
-i\epsilon_i[G_{\gamma\alpha},G^{\gamma}{}_{\beta}]-
2i\epsilon_j[G_{\alpha\beta},\varphi^j{}_i],\ret 
\delta\eta_i &=&2i\epsilon_k[\varphi_{ij},\varphi^{jk}],\ret
\delta\varphi_{ij}&=&\raiz\epsilon_{(i}\eta_{j)},\ret 
\delta G_{\alpha\beta}&=&\raiz\epsilon^i \chi_{i\alpha\beta},\ret
\la{ntres}
\eea 
where, for example,
$\epsilon_{(i}\eta_{j)}=\half (\epsilon_{i}\eta_{j}+\epsilon_{j}\eta_{i})$. The
transformations \eqs{ntres} satisfy the on-shell algebra 
$[\delta_1,\delta_2]=0$ modulo a non-Abelian gauge transformation generated by the
scalars
$\varphi_{ij}$. For example, 
$ [\delta_1,\delta_2]G_{\alpha\beta}= -4\raiz i\epsilon_1{}^{\!i}
\epsilon_2{}^{j}[\varphi_{ij}, G_{\alpha\beta}]$. The algebra closes
on-shell, and one has to impose the equations of motion for the
anticommuting fields
$\psi^i_{\alpha\dalpha}$ and
$\chi^i_{\alpha\beta}$. In terms of the generators
$Q^i$, the algebra takes the form:
\bea
\{Q^1,Q^1\}&=& \delta_g (\varphi_{22}),\ret
\{Q^1,Q^2\}&=& \delta_g (\varphi_{12}),\ret
\{Q^2,Q^2\}&=& \delta_g (\varphi_{11}),\ret
\la{damned}
\eea 
where by $\delta_g (\varphi_{22})$ we denote the non-Abelian gauge 
transformation
generated by $\varphi_{22}$. As explained in \cite{yamron}, it is possible to
realize the algebra off-shell by inserting the auxiliary fields $N_{\alpha\beta}$
(symmetric in its spinor indices) and
$M_{\alpha\dalpha}$ in the transformations of 
$\psi^i_{\alpha\dalpha}$ and $\chi^i_{\alpha\beta}$. This is the opposite to the
situation one encounters in the associated physical $\cn=4$ theory, where an
off-shell formulation in terms of unconstrained fields is not possible. After some
suitable manipulations,  the off-shell formulation of the twisted theory takes the
form:  
\begin{align}
{\cal S}^{(1)}&=\frac{1}{e^2_0}\int d^4 x\, \tr
\bigl\{\,\frac{1}{4}\nabla_{\!\alpha\dalpha}
\varphi_{ij}\nabla^{\dalpha\alpha}\varphi^{ij} +\frac{i}{\raiz}
M^{\dalpha\alpha}\nabla_{\beta\dot\alpha}G^{\beta}{}_\alpha 
-i\psi^{j\beta}{}_{\dot\alpha}\nabla^{\dot\alpha\alpha}
\chi_{j\alpha\beta} 
\ret
&-\frac{i}{ 2}\psi^j_{\alpha\dot\alpha}\nabla^{\dalpha\alpha}
\eta_j +\frac{i}{ 2} N^{\alpha\beta} F^{+}_{\alpha\beta}
-\frac{i}{\raiz}\,\chi_i{}^{\alpha\beta}[\chi_{j\alpha\beta},\varphi^{ij}]
+\frac{i}{\raiz}\,\chi^{j\alpha}{}_\beta[\chi_{j\alpha\gamma},G^{\beta\gamma}]
\ret
&-\frac{i}{\raiz}\,\chi^j{}_{\alpha\beta}[\eta_j,G^{\alpha\beta}]-
\frac{i}{{2\raiz}}\,\eta_i[\eta_j,\varphi^{ij}]
+\frac{i}{\raiz}\,\psi^{j\alpha}{}_{\dot\alpha}[\psi_{j\beta}{}^
{\dot\alpha},G_\alpha{}^\beta]-\frac{1}{4}M_{\alpha\dot\alpha}M^{\dalpha\alpha}
\ret
&-\frac{i}{\raiz}\,\psi^{i\alpha}{}_{\dot\alpha}[\psi^j{}_{\alpha}{}^
{\dot\alpha},\varphi_{ij}]+ \frac{1}{ 4}[\varphi_{ij},\varphi_{kl}]
[\varphi^{ij},\varphi^{kl}] 
-\half[\varphi_{ij},G_{\alpha\beta}][\varphi^{ij},G^{\alpha\beta}] 
\ret 
&+ \frac{1}{4}N_{\alpha\beta}N^{\alpha\beta}+ \frac{i}{
2}N_{\alpha\beta} [G_\gamma{}^{\alpha},G^{\gamma\beta}]\,\bigr\} 
-2\pi i\tau_0\frac{1}{32\pi^2}\int d^4
x\,\tr\,\bigl\{\, *  F_{mn}F^{mn}
\,\bigr\}.
\ret
\la{nocho}
\end{align}

The corresponding off-shell transformations are:
\bea
\delta A_{\alpha\dalpha} &=& 2i\epsilon_j
\psi^j_{\alpha\dalpha},\ret
\delta F^{+}_{\alpha\beta}&=& 2\epsilon_i\nabla_{(\alpha}{}^{\dot\alpha}
\psi^i{}_{\!\beta)\dot\alpha},\ret 
\delta\psi^i{}_{\!\alpha\dot\alpha}&=& 
-i{\raiz}\epsilon_j\nabla_{\alpha\dot\alpha}
\varphi^{ji} +
\epsilon^i M'_{\alpha\dot\alpha},\ret 
\delta\chi_{i\alpha\beta} &=& 
-2i\epsilon_j[G_{\alpha\beta},\varphi^j{}_{\!i}]+\epsilon_i N'_{\alpha\beta},\ret
\delta\eta_i &=&2i\epsilon_j[\varphi_{ik},\varphi^{jk}],\ret
\delta\varphi_{ij}&=&\raiz\epsilon_{(i}\eta_{j)},\ret 
\delta G_{\alpha\beta}&=&\raiz\epsilon^j \chi_{j\alpha\beta},\ret
\delta M'_{\alpha\dot\alpha}&=&\epsilon^i\bigl\{\,-i\nabla_{\alpha\dot\alpha}
\eta_i+2\raiz i[\psi^j_{\alpha\dot\alpha},\varphi_{ij}]\,\bigr\},\ret 
\delta N'_{\alpha\beta}&=&\epsilon^i\bigl\{\,
\raiz i[\eta_i ,G_{\alpha\beta}]-2\raiz i[\chi_{j\alpha\beta},\varphi^j{}_{\!i}]\,
\bigr\}.\ret
\la{pnseis}
\eea  
With the aid of the transformations \eqs{pnseis} it is easy (but rather lengthy)
to show that the action \eqs{nocho} can be written as a double
$Q$-commutator plus a $\tau_0$-dependent term, that is,
\begin{equation}
\epsilon^2{\cal S}^{(1)}=\delta^2 \C{F} -\epsilon^2 2\pi i k \tau_0 =
-\half\epsilon^2\{Q^i,[Q_i,\C{F}]\}-\epsilon^2 2\pi i k \tau_0,
\la{jackie4}
\end{equation}  
(here $\delta \equiv \epsilon^i [Q_i,\}$ and $k$ is the instanton number),
with 
\bea
\C{F} &=&\fr{1}{ e^2_0}\int d^4
x\,\bigl\{\,\fr{i}{{2\raiz}}\,F^{+}_{\alpha\beta}G
^{\alpha\beta}+\fr{1}{{4\raiz}}\,N_{\alpha\beta}G^{\alpha\beta}+ 
\fr{1}{8}\,\psi_{j\alpha\dalpha}\psi^{j\dalpha\alpha}\ret 
&+&\fr{i}{{6\raiz}}G_{\alpha\beta}[G^\beta{}_\gamma ,G^{\gamma\alpha}]-
\fr{i}{{12\raiz}}\,\varphi_{ij}[\varphi^j{}_k
,\varphi^{ik}]\,\bigr\}.\ret
\la{nnueve4}
\eea

The next step is to couple the theory to an arbitrary background metric 
$g_{\mu\nu}$ of Euclidean signature. This can be done as follows: {\sl first},
covariantize the expression \eqs{nnueve4} and the transformations \eqs{pnseis}, 
and {\sl second}, define the new action to be $\delta^2 \C{F}_{cov}$. The
resulting action is:
\begin{align}
{\cal S}^{(1)}_{c}&=\fr{1}{ e^2_0}\int_X  d^4 x\,\sqrt{g} \tr
\bigl\{\,\fr{1}{4}\nabla_{\!\alpha\dalpha}
\varphi_{ij}\nabla^{\dalpha\alpha}\varphi^{ij} +\fr{i}{\raiz}
M^{\dalpha\alpha}\deriv_{\beta\dot\alpha}G^{\beta}{}_\alpha 
-i\psi^{j\beta}{}_{\dot\alpha}\deriv^{\dot\alpha\alpha}
\chi_{j\alpha\beta}  
\ret &-\fr{i}{2}\psi^j_{\alpha\dot\alpha}\nabla^{\dalpha\alpha}
\eta_j +\fr{i}{2} N^{\alpha\beta} F^{+}_{\alpha\beta}
-\fr{i}{\raiz}\,\chi_i{}^{\alpha\beta}[\chi_{j\alpha\beta},\varphi^{ij}]
+\fr{i}{\raiz}\,\chi^{j\alpha}{}_\beta[\chi_{j\alpha\gamma},G^{\beta\gamma}]
\ret
&-\fr{i}{\raiz}\,\chi^j{}_{\alpha\beta}[\eta_j,G^{\alpha\beta}]-
\fr{i}{{2\raiz}}\,\eta_i[\eta_j,\varphi^{ij}]
+\fr{i}{\raiz}\,\psi^{j\alpha}{}_{\dot\alpha}[\psi_{j\beta}{}^
{\dot\alpha},G_\alpha{}^\beta]-\fr{1}{4}M_{\alpha\dalpha}M^{\dalpha\alpha}
\ret
&-\fr{i}{\raiz}\,\psi^{i\alpha}{}_{\dot\alpha}[\psi^j{}_{\alpha}{}^
{\dot\alpha},\varphi_{ij}] +\fr{1}{ 4}[\varphi_{ij},\varphi_{kl}]
[\varphi^{ij},\varphi^{kl}] 
-\half[\varphi_{ij},G_{\alpha\beta}][\varphi^{ij},G^{\alpha\beta}] 
\ret 
&+ \fr{1}{4}N_{\alpha\beta}N^{\alpha\beta}+ 
\fr{i}{2}N_{\alpha\beta}[G_\gamma{}^{\alpha},G^{\gamma\beta}]\,\bigr\}
-2\pi  i\tau_0\fr{1}{32\pi^2}\int_X  d^4 x\,\sqrt{g}\tr\,\bigl\{\, * 
F_{\mu\nu}F^{\mu\nu}
\,\bigr\},
\ret
\la{nnocho}
\end{align} 
where we have introduced the full covariant derivative
$\deriv_{\alpha\dalpha}$. The action \eqs{nnocho} is, by construction, 
invariant under the appropriate covariantized version of the
transformations
\eqs{pnseis}. However, it is not manifestly real because it
contains fields in the fundamental representation of $SU(2)_F$, which
are complex, and it is not possible to  assign a non-trivial
ghost number to the fields in \eqs{nnocho}. Now, the action of a
topological cohomological field theory has to be real, since we will
eventually  interpret it as a real differential form defined on a
certain moduli space.  Likewise, it has to posses a non-trivial
ghost-number symmetry which, from the  geometrical viewpoint,
corresponds to the de Rham grading on the moduli  space. 

To overcome these problems we break the $SU(2)_F$ internal symmetry
group  of the theory down to its Cartan subgroup. This allows to
introduce a non-anomalous ghost number in the theory (basically twice
the corresponding charge under the diagonal  generator $T_3$). With
respect to this ghost number, the field content of the theory can be
reorganized as follows (we follow the notation in \cite{vw}): with
ghost number
$+2$, we have the scalar field
$\phi\equiv \varphi_{11}$; with ghost number $+1$, the anticommuting
fields
$\psi_ {\alpha\dalpha}\equiv i\psi_{1\alpha\dalpha}$,
$\tilde\psi_{\alpha\beta}\equiv \chi_{1\alpha\beta}$ and $\zeta\equiv 
i\eta_1$; with ghost number $0$, the gauge connection
$A_{\alpha\dalpha}$, the scalar field $C\equiv i\varphi_{12}$, the
self-dual two-form  $B_{\alpha\beta}\equiv G_{\alpha\beta}$ and the
auxiliary fields
$H_{\alpha\beta}\equiv iN_{\alpha\beta}$ and 
$\tilde H_{\alpha\dalpha}\equiv M_{\alpha\dalpha}$; with ghost number
$-1$,
 the anticommuting fields $\chi_ {\alpha\beta}\equiv
i\chi_{2\alpha\beta}$,
$\tilde\chi_{\alpha\dalpha}\equiv \psi_{2\alpha\dalpha}$ and
$\eta\equiv 
\eta_2$; and finally, with ghost number $-2$,  the scalar field
$\bar\phi\equiv
\varphi_{22}$. These fields are related to those  in the  $\cn=4$ 
theory as follows:
\begin{equation}
\begin{align}
 \lambda_{{\tilde  1}1}&= \tilde\psi^{+}_{11},&
\lambda_{{\tilde  3}1}&= -i\chi^{+}_{11},&
\bar\lambda^{\tilde 1}_{\dalpha}&= \tilde\chi_{2\dalpha},\ret
\lambda_{{\tilde  1}2}&= \tilde\psi^{+}_{12}-\fr{i}{2}\zeta,&
\lambda_{{\tilde  3}2}&= \half\eta-i\chi^{+}_{12},&
\bar\lambda^{\tilde 2}_{\dalpha}&= -\tilde\chi_{1\dalpha},\ret
\lambda_{{\tilde  2}1}&= \tilde\psi^{+}_{12}+\fr{i}{2}\zeta,&
\lambda_{{\tilde  4}1}&= -\half\eta-i\chi^{+}_{12},&
\bar\lambda^{\tilde 3}_{\dalpha}&= i\psi_{2\dot\alpha},\ret
\lambda_{{\tilde  2}2}&= \tilde\psi^{+}_{22},&
\lambda_{{\tilde  4}2}&= -i\chi^{+}_{22},&
\bar\lambda^{\tilde 4}_{\dalpha}&= -i\psi_{1\dot\alpha},\ret
B_1 &=  -B^{+}_{12}+iC,&
B_2&= -B^{+}_{22},&
B_3&= -\bar\phi,\ret
B^{\dag}_1&=  -B^{+}_{12}-iC, &
B^{\dag}_2&=  B^{+}_{11}, &
B^{\dag}_3&= -\phi.\ret
\end{align}
\label{che}
\end{equation}
($\tilde 1$, $\tilde 2$, etc., are $SU(4)_I$ indices).

Notice that now we can consistently assume that all the
fields above are real,  in order to guarantee the reality of the topological
action.

In terms of these new fields, and after making the shifts:
\bea
\tilde H'_{\alpha\dalpha}&=& \tilde
H_{\alpha\dalpha}+\raiz\,\nabla_{\!\alpha\dalpha}C,\ret 
H'_{\alpha\beta}&=&
H_{\alpha\beta}+2i[B_{\alpha\beta},C],\ret
\la{greisen}
\eea
the action \eqs{nnocho} takes the form: 
\begin{align}
{\cal S}^{(2)}_c&=\fr{1}{ e^2_0}\int_X  d^4 x\,\sqrt{g}\, \tr
\Bigl\{\ret &\half\deriv_{\alpha\dalpha}
\bar\phi\deriv^{\dalpha\alpha}\phi -\fr{1}{4}
\tilde {H'}{}^{\dalpha\alpha}\bigl 
(\,\tilde H'_{\alpha\dalpha}-2\raiz \deriv
_{\alpha\dalpha}C-2\raiz i\,
\deriv_{\beta\dot\alpha}B^{\beta}{}_{\!\alpha}\,\bigr )
\ret  
&-\fr{1}{4}
H^{'\alpha\beta}\bigl (\, H'_{\alpha\beta}-2\,
F^{+}_{\alpha\beta}-2\,[B_{\gamma\alpha},B_\beta{}^{\!\gamma}]-4i\,
[B_{\alpha\beta},C]\,\bigr )-i\psi^{\beta}{}_{\dot\alpha}\deriv^{\dot\alpha\alpha}
\chi_{\alpha\beta} 
\ret 
&-i\tilde\chi^{\beta}{}_{\dot\alpha}\deriv^{\dot\alpha\alpha}
\tilde\psi_{\alpha\beta} 
-\fr{1}{2}\tilde\chi_{\alpha\dot\alpha}\deriv^{\dalpha\alpha}
\zeta+\fr{1}{2}\psi_{\alpha\dot\alpha}\deriv^{\dalpha\alpha}
\eta-\fr{i}{\raiz}\,\tilde\psi^{\alpha\beta}[\tilde\psi_{\alpha\beta},\bar
\phi]
\ret 
&+\fr{i}{\raiz}\,\chi^{\alpha\beta}[\chi_{\alpha\beta},\phi]
-i\raiz\,\tilde\psi^{\alpha\beta}[\chi_{\alpha\beta},C]
-\raiz\,\tilde\psi^{\alpha}{}_\beta\,[\chi_{\alpha\gamma},B^{\beta\gamma}]
\ret
&+\fr{i}{\raiz}\,\chi_{\alpha\beta}[\,\zeta,B^{\alpha\beta}]
+\fr{i}{\raiz}\,\tilde\psi_{\alpha\beta}[\,\eta,B^{\alpha\beta}]+
\fr{i}{{2\raiz}}\,\zeta\,[\zeta,\bar\phi]-
\fr{i}{{2\raiz}}\,\eta\,[\eta,\phi]
\ret
&-\fr{i}{{\raiz}}\,\zeta\,[\eta,C]
+\raiz\,\psi_{\alpha\dot\alpha}[\tilde\chi_{\beta}{}^
{\dot\alpha},B^{\alpha\beta}]-
\fr{i}{\raiz}\,\tilde\chi^{\alpha}{}_{\dot\alpha}[\tilde\chi_{\alpha}{}^
{\dot\alpha},\phi]
\ret
&+\fr{i}{\raiz}\,\psi^{\alpha}{}_{\dot\alpha}[\psi_{\alpha}{}^
{\dot\alpha},\bar\phi] - i\raiz\,\psi^{\alpha}{}_{\dot\alpha}[\tilde\chi_{\alpha}{}^
{\dot\alpha},C] -\half[\phi,\bar\phi]^2 +2[\phi,C][\bar\phi,C]
\ret
&-[\phi,B_{\alpha\beta}][\bar\phi,B^{\alpha\beta}] 
 \,\Bigr\} -2\pi i\tau_0\fr{1}{32\pi^2}\int_X  d^4 x\,\sqrt{g}\tr\,
\bigl\{\,*F_{\mu\nu}F^{\mu\nu}
\,\bigr\}.
\ret
\la{Phoebe4}
\end{align} 

After integrating out the auxiliary fields in \eqs{Phoebe4} we find for the
bosonic part of the action not involving the scalars $\phi$ and $\bar\phi$
the following expression:
\begin{align}
\int_X  d^4 x\,\sqrt{g}\, \tr
\bigl\{\,\half
\bigl (\,\deriv_{\alpha\dalpha}C&+i\deriv_{\beta\dot\alpha}
B^{\beta}{}_{\!\alpha}\,\bigr )^2
\ret  
&+\fr{1}{4} \bigl (\, F^{+}_{\alpha\beta}+
[B_{\gamma\alpha},B_\beta{}^{\!\gamma}]+2i\,[B_{\alpha\beta},C]\,\bigr
)^2\,\bigr\}.\ret
\la{one4}
\end{align}
Expanding the squares in this expression  one obtains,
\begin{align}
\int_X  &d^4 x\,\sqrt{g}\, \tr
\bigl\{\,-\fr{1}{2}
\deriv_\mu C\deriv^\mu C
-\half\,(\,\deriv_{\beta\dot\alpha}B^{\beta}{}_{\!\alpha}
\deriv_{\gamma}{}^{\!\dalpha}B^{\gamma\alpha}-
F^{+\alpha\beta}[B_{\gamma\alpha},B_\beta{}^{\!\gamma}]\,)\ret 
 &-\fr{1}{2}F^{+}_{\mu\nu}F^{+\mu\nu}+
[B^{+}_{\mu\nu},B^{+}_{\tau\lambda}][B^{+\mu\nu},B^{+\tau\lambda}]+
2[B^{+}_{\mu\nu},C][B^{+\mu\nu},C]\,\bigr\}.\ret
\la{two4}
\end{align}
where we have used
$\deriv_{\alpha\dalpha}=\sigma^m{}_{\alpha\dalpha}\deriv_m$ and
$F^{+}_{\alpha\beta}\equiv\sigma^{\mu\nu}_{\alpha\beta}F^{+}_{\mu\nu}$, 
$B_{\alpha\beta}\equiv\sigma^{\mu\nu}_{\alpha\beta}B^{+}_{\mu\nu}$.  Let us
now focus on the expression within the parenthesis. Further expansion leads
to the identity:
\begin{align}
\int_X & d^4 x\,\sqrt{g}\,\tr\,\bigl\{\,-\half
\deriv_{\beta\dalpha}B^\beta{}_{\!\alpha}\deriv_
{\gamma}{}^{\!\dalpha}B^{\gamma\alpha} + \half
F^{+\alpha\beta}[B_{\gamma\alpha},B_\beta{}^{\!\gamma}\,]\,\bigr\}
\ret   &=\int_X  d^4 x\,\sqrt{g}\,\tr\,\bigl\{\,-\deriv_\mu
B^{+}_{\nu\lambda}\deriv^\mu B^{+\nu\lambda}-{\half}R
B^{+}_{\mu\nu}B^{+\mu\nu}+R_{\mu\nu\tau\lambda}
B^{+\mu\nu}B^{+\tau\lambda}\,\bigr\},\ret
\la{Beatrice}
\end{align} 
(using again  
$B_{\alpha\beta}\equiv\sigma^{\mu\nu}_{\alpha\beta}B^{+}_{\mu\nu}$).  If we
now express the Riemann tensor in \eqs{Beatrice} in terms of its irreducible
components 
\begin{align} 
R_{\mu\nu\tau\lambda}&=\half(g_{\mu\tau}R_{\nu\lambda}-
g_{\mu\lambda}R_{\nu\tau}-g_{\nu\tau}R_{\mu\lambda}+
g_{\nu\lambda}R_{\mu\tau})\ret
&-\fr{R}{6}(g_{\mu\tau}g_{\nu\lambda}-g_{\nu\tau} g_{\mu\lambda})+
C_{\mu\nu\tau\lambda},\ret
\la{Riemann}
\end{align}   
with $C_{\mu\nu\tau\lambda}$ the Weyl tensor, we finally obtain,
\begin{align}
\int_X & d^4 x\,\sqrt{g}\,\tr\,\bigl\{\,-\half
\deriv_{\beta\dalpha}B^\beta{}_{\!\alpha}\deriv_
{\gamma}{}^{\!\dalpha}B^{\gamma\alpha} + \half
F^{+\alpha\beta}[B_{\gamma\alpha},B_\beta{}^{\!\gamma}\,]\,\bigr\}
\ret   &=\int_X  d^4 x\,\sqrt{g}\,\tr\,\bigl\{\,-\deriv_\mu
B^{+}_{\nu\lambda}\deriv^\mu B^{+\nu\lambda} -B^{+\mu\nu}\bigl
(\,\fr{1}{6}R\,(g_{\mu\tau}g_{\nu\lambda}-g_{\nu\tau}
g_{\mu\lambda})-C_{\mu\nu\tau\lambda}\,\bigr) B^{+\tau\lambda}\,\bigr\}.
\ret
\la{Turandot4}
\end{align}   
Thus we see that when we put the  twisted  theory on 
general curved backgrounds we must include  the non-minimal 
couplings in \eqs{Turandot4} to ensure that the naive action \eqs{nuno4} be
supersymmetric. 

The associated fermionic symmetry splits up as well into BRST ($Q^{+}\equiv
Q^1$) and  anti-BRST ($Q^{-}\equiv iQ^2$) parts. The explicit formulas are:
\begin{align}
[Q^{+}, A_{\alpha\dalpha}] &= -2\psi_{\alpha\dalpha},& 
[Q^{-}, A_{\alpha\dalpha}] &= -2\tilde\chi_{\alpha\dalpha},\ret 
\{Q^{+},\psi_{\alpha\dot\alpha}\} &=  -{\raiz}\deriv_{\alpha\dot\alpha}\phi, &
\{Q^{-},\tilde\chi_{\alpha\dalpha}\} &=  {\raiz}\deriv_{\alpha\dot\alpha}\bar\phi,
\ret 
\big[Q^{+},\phi\big]&=0,&
[Q^{-},\bar\phi\,]&=0,\ret
\big[Q^{+}, B_{\alpha\beta}\big]&=\raiz\tilde\psi_{\alpha\beta},&
\big[Q^{-}, B_{\alpha\beta}\big]&=-\raiz\,\chi_{\alpha\beta},\ret
\{Q^{+},\tilde\psi_{\alpha\beta}\}&=2i\,[B_{\alpha\beta},\phi],&
\{Q^{-},\chi_{\alpha\beta}\}&=2i\,[B_{\alpha\beta},\bar\phi],\ret
\big[Q^{+},C\big]&=\fr{1}{\raiz}\zeta,&
[Q^{-},C]&=-\fr{1}{\raiz}\eta,\ret
\{Q^{+},\zeta\,\} &=4i\,[C,\phi],&
\{Q^{-},\eta\,\} &=4i\,[C,\bar\phi],\ret
\big[Q^{+},\bar\phi\big]&=\raiz\,\eta,&
[Q^{-},\phi]&=\raiz\,\zeta,\ret
\{Q^{+},\eta\,\} &=2i\,[\bar\phi,\phi],&
\{Q^{-},\zeta\,\} &=-2i\,[\phi,\bar\phi],\ret
\{Q^{+},\tilde\chi_{\alpha\dalpha}\} &= \tilde H'_{\alpha\dalpha},&
\{Q^{-},\psi_{\alpha\dalpha}\} &=  -\tilde
H'_{\alpha\dalpha}+2\raiz\,\deriv_{\alpha\dalpha}C,\ret
\big[Q^{+},\tilde H'_{\alpha\dalpha}\big]&= 2\raiz
i\,[\tilde\chi_{\alpha\dalpha},\phi],&
[Q^{-},\tilde
H'_{\alpha\dalpha}]&=-2\,\deriv_{\alpha\dalpha}\eta + 2\raiz
i\,[\psi_{\alpha\dalpha},\bar\phi\,]\ret
&{} &
&-4\raiz i\,[\,\tilde\chi_{\alpha\dalpha},C],\ret
\{Q^{+},\chi_{\alpha\beta}\} &= H'_{\alpha\beta},&
\{Q^{-},\tilde\psi_{\alpha\beta}\} &= H'_{\alpha\beta}
-4i\,[B_{\alpha\beta},C\,],\ret 
\big[Q^{+},H'_{\alpha\beta}\big]&=2\raiz i\,[\chi_{\alpha\beta},\phi],&
[Q^{-},H'_{\alpha\beta}]&=-2\raiz
i\,[\tilde\psi_{\alpha\beta},
\bar\phi]-4\raiz i\,[\chi_{\alpha\beta},C]\ret 
&{} & &-2\raiz i\,[B_{\alpha\beta},\eta],\ret
\la{papagayo4}
\end{align}   
satisfying the algebra,
\bea
&&\{Q^{+},Q^{+}\}=\delta_g(\phi),\ret
&&\{Q^{-},Q^{-}\}=\delta_g(-\bar\phi),\ret
&&\{Q^{+},Q^{-}\}=\delta_g(C) .\ret
\la{leoon4}
\eea

The $\tau_0$-independent part of the action \eqs{Phoebe4} can be written
either as a BRST  ($Q^{+}$) commutator or as an anti-BRST ($Q^{-}$)
commutator. Let us 
 focus on the former possibility. The appropriate ``gauge" fermion turns out
to be:
\bea
\Psi&=& \fr{1}{ e^2_0}\int_X  d^4 x\,\sqrt{g}\, \tr
\bigl\{\,-\fr{1}{4}
\tilde \chi^{\dalpha\alpha}\bigl (\,\tilde H'_{\alpha\dalpha}-2\raiz\deriv
_{\alpha\dalpha}C-2\raiz i\,
\deriv_{\beta\dot\alpha}B^{\beta}{}_{\!\alpha}\,\bigr )
\ret  
&-&\fr{1}{ 4}
\chi^{\alpha\beta}\bigl (\, H'_{\alpha\beta}-2\,
F^{+}_{\alpha\beta}-2\,[B_{\gamma\alpha},B_\beta{}^{\!\gamma}]-4i\,
[B_{\alpha\beta},C]\,\bigr )\,\bigr\}
\ret  
&+&\fr{1}{ e^2_0}\int_X  d^4 x\,\sqrt{g}\,
\tr
\bigl\{\, \fr{1}{{2\raiz}}\bar\phi\,\bigl
(\,\deriv_{\alpha\dalpha}\psi^{\dot\alpha\alpha}+i\raiz
\,[\tilde\psi_{\alpha\beta},B^{\alpha\beta}] -i\raiz\,[\zeta,C]\,\bigr
)\,\bigr\}
\ret
&-&\fr{1}{e^2_0}\int_X  d^4 x\,\sqrt{g}\, \tr
\bigl\{\,\fr{i}{4}\eta[\phi,\bar\phi]\,\bigr\}.\ret
\la{Mazinger4}
\eea

For reasons of future convenience we will rewrite \eqs{Mazinger4} in vector 
indices.
With the definitions, $X_{\alpha\dalpha}
\buildrel{\hbox{\rm\tiny def}}\over =
\sigma^\mu_{\alpha\dalpha}X_\mu$, and, 
$Y_{\alpha\beta}\buildrel {\hbox{\rm\tiny def}}\over =
\sigma^{\mu\nu}_{\alpha\beta}Y_{\mu\nu}$, for any two given fields $X$ and
$Y$,
\eqs{Mazinger4} takes the form:
\bea
\Psi&=& \fr{1}{ e^2_0}\int_X  d^4 x\,\sqrt{g}\, \tr
\bigl\{\,\half
\tilde \chi^\mu\bigl (\,\tilde H'_\mu -2\raiz\deriv_\mu  C+4\raiz\,
\deriv^\nu B^{+}_{\nu\mu}\,\bigr )
\ret 
&+&\half\chi^{+\mu\nu}\bigl (\, H^{'+}_{\mu\nu}-2\,
F^{+}_{\mu\nu}-4i\,[B^{+}_{\mu\tau},B^{+\tau}{}_{\!\nu}]-4i
[B^{+}_{\mu\nu},C]\,\bigr )\,\bigr\}
\ret 
&-&\fr{1}{ e^2_0}\int_X  d^4 x\,\sqrt{g}\, \tr
\bigl\{\, \fr{1}{{2\raiz}}\bar\phi\,\bigl
(\,2\deriv_\mu\psi^\mu+2\raiz i
\,[\tilde\psi^+_{\mu\nu},B^{+\mu\nu}] +\raiz i\,[\zeta,C]\,\bigr )\,\bigr\}
\ret
&-&\fr{1}{ e^2_0}\int_X  d^4 x\,\sqrt{g}\, \tr
\bigl\{\,\fr{i}{4}\eta[\phi,\bar\phi]\,\bigr\}.\ret
\la{mermelada}
\eea

The gauge fermion, in turn, can itself be written as an anti-BRST commutator
\eqs{jackie4}:
\begin{align}
\Psi&=\Big\{Q^{-},\fr{1}{ e^2_0}\int_X  d^4 x\,\sqrt{g}\, \tr\bigl (\, 
-\fr{1}{{2\raiz}}B^{\alpha\beta}\,\bigl
(\,F^{+}_{\alpha\beta}-\fr{1}{2}\,
H'_{\alpha\beta}+\fr{1}{3}[\,B_{\alpha\gamma},B_{\beta}{}^{\!\gamma}\,]\,
\bigr )
\ret 
&+\fr{i}{{2\raiz}}C\,[\phi,\bar\phi\,]+\fr{1}{4}\,
\psi_{\alpha\dalpha}\tilde\chi^{\dalpha\alpha}\,\bigr )\,\Big\}.\ret
\la{Misha4}
\end{align}

\section{The topological action in the Mathai-Quillen approach}

We have described so far the  
Vafa-Witten theory as a twisted version of the $\cn=4$ theory.  
The twisting  procedure has been repeatedly shown to be a very powerful
technique  for the construction of topological quantum field theories.
However, it suffers from serious drawbacks, the main one being that 
it is not possible to identify from the very beginning the underlying 
geometrical structure that is involved. Rather, in most of the cases the
underlying geometrical scenario is unveiled only after a careful 
 analysis with techniques borrowed from conventional quantum field  
theory is carried out \cite{tqft}. In what follows, we will change our 
scope and try to concentrate on the geometrical formulation of the 
theory. We will make use of the Mathai-Quillen formalism (see chapter
\ref{chmq}), which is very well suited  for our purposes. Let us recall
briefly the fundamentals of this  approach. In the framework of topological
quantum field theories of  cohomological type \cite{coho}, one deals with a
certain set of fields  (the field space, ${\cal M}$), on which the action
of a  symmetry group,
${\cal G}$, which is usually a  local symmetry group, is defined. An
appropriate  set of  basic equations
 imposed on  the fields singles out a certain subset (the moduli space) of
${\cal M}/{\cal G}$. The topological quantum field theory associated to  
this moduli problem studies intersection theory on the corresponding  moduli
space. In this context, the Mathai-Quillen formalism involves  the following
steps. Given the field space ${\cal M}$, the basic  equations of the problem
are introduced as sections of an appropriate vector bundle ${\cal V}\to
{\cal M}$, in such a way that the zero locus   of these sections, modded out
by the gauge symmetry, is precisely  the  relevant moduli space. The   
Mathai-Quillen  
 formalism  allows the computation of a certain representative  of the Thom
class of the vector  bundle $\C{E}={\cal V}/\C{G}$, which turns out to 
be the
exponential of the action of the  topological  field  theory under
consideration. The integration on
${\cal M}$ of the pullback  under the sections of  the Thom class of ${\cal
V}$ gives the Euler characteristic of the bundle,  which is the basic
topological invariant associated to the moduli problem.

\subsection{The Vafa-Witten problem}

The analysis starts from two  basic equations involving the self-dual part
of the gauge connection $F^{+}$, a certain scalar field $C$ and a bosonic
self-dual two-form $B^{+}$, all taking values in the adjoint representation
of some compact finite-dimensional  Lie group $G$. These equations
are:   
\begin{equation}
\begin{cases}
\deriv_\mu C+\raiz \deriv^\nu B^{+}_{\nu\mu}=0,\\
F^{+}_{\mu\nu}-\fr{i}{2}[B^{+}_{\mu\tau},B^{+\tau}{}_{\!\nu}]-
\fr{i}{\raiz}
[B^{+}_{\mu\nu},C]=0.
\end{cases}
\la{equations4}
\end{equation}   
One can consider  the equations above as defining a certain moduli problem,
and our aim is to construct the topological quantum field theory which
corresponds to it  within the framework of the Mathai-Quillen formalism. Our
analysis will follow closely that in \cite{abmono}\cite{nabm}\cite{tesis}. 
It
should be pointed out that the Mathai-Quillen construction for this twist
was  already contained, yet not explicitly constructed, in \cite{vw}, and 
was also studied in the context of ``balanced" topological field  theories
by Dijkgraaf and Moore in 
\cite{balanced}, while the basic  structure had already been discussed from
the viewpoint of supersymmetric quantum  mechanics by Blau and Thompson in
\cite{cofield}.

Recently, the Mathai-Quillen formalism has been applied to the twist  under
consideration  in \cite{wang}. The
construction presented in  that work differs from ours in the role assigned
to the field 
$C$.

\subsection{The topological framework}

The geometrical setting is a certain oriented, compact Riemannian 
four-manifold
$X$, and the  field space is ${\cal M}={\cal A}\times\Omega^0(X,\ad
P)\times\Omega^{2,+} (X,\ad P)$, where ${\cal A}$ is the space of
connections on a principal 
$G$-bundle $P\to X$, and the second and third factors denote, respectively,
the
$0$-forms and self-dual differential forms of degree two on $X$ taking 
values in the Lie algebra of $G$. $\ad P$ denotes the adjoint bundle of 
$P$, $P\times_{\ad}\Lie$. The space of sections of this bundle,
$\Omega^0(X,\ad P)$, is the Lie algebra of the group 
${\cal G}$ of gauge transformations (vertical automorphisms)  of the bundle
$P$, whose action on the field space is given locally by:
\begin{align}
g^{*}(A)&=i(dg)g^{-1}+gAg^{-1},\ret
g^{*}(C)&=gCg^{-1},\ret
g^{*}(B^{+})&=gB^{+}g^{-1},
\la{jauje}
\end{align}

\noindent where $C\in \Omega^0(X,\ad P)$ and $B^{+}\in \Omega^{2,+}(X,\ad
P)$. In terms of the covariant derivative, $d_A =d+i[A,~]$, the infinitesimal form
of the transformations \eqs{jauje}, with $g={\hbox{\rm
exp}}(-i\phi)$ and
$\phi\in \Omega^0(X,\ad P)$, takes the form:
\begin{align}
\delta_g(\phi)A&=d_A\phi,& &{}\ret
\delta_g(\phi)C&=i[C,\phi],& &{}\ret
\delta_g(\phi)B^{+}&=i[B^{+},\phi].& &{}
\la{jauja}
\end{align}   

The tangent space to the field space at the point $(A,C,B^{+})$ is the vector
space $T_{(A,C,B^{+})}{\cal M}=\Omega^1(X,\ad P)\oplus\Omega^0(X,\ad
P)\oplus\Omega^{2,+} (X,\ad P)$. On
$T_{(A,C,B^{+})}{\cal M}$ we can define a gauge-invariant Riemannian metric
(inherited from that on $X$) as follows:
\begin{equation}
\langle (\psi,\zeta,\tilde\psi^{+}),(\theta,\xi,\tilde\omega^{+})\rangle =\int_X
\tr(\psi\wedge *\theta)+\int_X \tr(\zeta\wedge *\xi)+
\int_X \tr(\tilde\psi^{+}\wedge *\tilde\omega^{+})
\la{metrica4}
\end{equation}   
where $\psi,\theta\in\Omega^1(X,\ad P)$,
$\zeta,\xi\in\Omega^0(X,\ad P)$ and
$\tilde\psi^{+},\tilde\omega^{+}\in \Omega^{2,+} (X,\ad P)$. 

To introduce the basic equations \eqs{equations4} in this 
framework we proceed as
follows. On the field space
$\mani$ we build a trivial vector bundle ${\cal V}=
\mani\times{\cal F}$, where the fibre is in this case 
 ${\cal F}= \Omega^1(X,\ad P)\oplus\Omega^{2,+} (X,\ad P)$. The basic 
equations
\eqs{equations4} can then  be identified to be a section 
$s:{\cal M}\to {\cal V}$ of the
vector bundle ${\cal V}$. In our case, the section reads, with a 
certain choice  of
normalization that makes easier the comparison with the results in  
the first part of the chapter:
\begin{equation}  
s(A,C,B^{+}) =\bigl (\raiz(\deriv_\mu C+\raiz \deriv^\nu B^{+}_{\nu\mu}),\,-2(
F^{+}_{\mu\nu}-\fr{i}{2}[B^{+}_{\mu\tau},B^{+\tau}{}_{\!\nu}]-
\fr{i}{\raiz}
[B^{+}_{\mu\nu},C])\bigr ).
\la{seccion}
\end{equation}   
Notice that this section is gauge ad-equivariant, and the zero
locus of the associated section  
${\tilde s}:{\cal M}/{\cal G}\to {\cal V}/{\cal G}=\C{E}$ gives precisely
the  moduli space of the topological theory. It would be desirable to
compute  the dimension of this moduli space. The best we can do is to build
the  corresponding deformation complex whose index is known to compute, under
certain assumptions, the dimension of the tangent space to the moduli space.
This index provides what is called the virtual dimension of the moduli
space. The deformation complex that corresponds to our moduli space is the
following:
\bea
0\too\Omega^0(X,\ad P)&&\!\!\!\!\!\!\!\!\!\mapright{{\cal C}}\Omega^1(X,\ad
P)\oplus\Omega^0(X,\ad P)\oplus\Omega^{2,+} (X,\ad P)\ret
&&\mapright{ds}\Omega^1(X,\ad P)\oplus\Omega^{2,+}(X,\ad P)\too 0.\ret
\la{complejo}
\eea  
The map ${\cal C}:\Omega^0(X,\ad P)\too T{\cal M}$, given by
(recall
\eqs{jauja}):
\begin{equation}  
{\cal C}(\phi)=(d_A\phi,i[C,\phi],i[B^{+},\phi]),\quad
\phi\in\Omega^0(X,\ad P),
\la{butraguenho4}
\end{equation} 
defines the vertical tangent space (gauge orbits) to the principal 
${\cal G}$-bundle. The map $ds:T_{(A,C,B^{+})}{\cal M}\too {\cal F}$ is  
given by
the linearization of the basic equations \eqs{equations4}
\bea
ds(\psi,\zeta,\tilde\psi^{+})=\biggl (\raiz(\deriv_\mu
\zeta+i[\psi_\mu,C]+\raiz
\deriv^\nu \tilde\psi^{+}_{\nu\mu}+ i\raiz[\psi^\nu,B^{+}_{\nu\mu}]),\ret
-2\bigl(\,2(\deriv_{[\mu}\psi_{\nu]})^{+}+i[\tilde\psi^{+}_{\tau[\mu},
B^{+\tau}{}_{\!\nu]}]-\fr{i}{\raiz}[\tilde\psi^{+}_{\mu\nu},C]
-\fr{i}{\raiz}[B^{+}_{\mu\nu},\zeta]\,\bigr)\,\biggr ).\ret
\la{lordaeron}
\eea   
Under suitable conditions (which happen to be the same vanishing theorems  
discussed in \cite{vw}), the index of the complex \eqs{complejo} computes de 
dimension of ${\hbox{\rm Ker}}(ds)/{\hbox{\rm Im}}({\cal C})$, that is, the
dimension of the moduli space under consideration. To calculate its index, the
complex \eqs{complejo} can be split up into the Atiyah-Hitchin-Singer instanton
deformation complex  
\cite{ahs} for anti-self-dual (ASD) connections,
\begin{align}   
&(1)~0\too\Omega^0(X,\ad P)\mapright{d_{\!A}}\Omega^1(X,\ad P)
\mapright{p^{+}\!d_{\!A}}\Omega^{2,+}(X,\ad P)\too 0,\la{ahs4}\\
\intertext{and the complex associated to the operator,}
&(2)~D=p^{+}\!d^{*}_A+d_A:\Omega^0(X,\ad P)\oplus\Omega^{2,+} (X,\ad P)
\too\Omega^1(X,\ad P),\ret
\la{fanny}
\end{align}  
 which is also the ASD instanton deformation complex. They
contribute with opposite signs and therefore the net contribution 
to the index is
zero, leaving us with the result that the virtual dimension of 
the moduli space is zero. 

\subsection{The topological action}

We now proceed to construct the topological action associated to 
this moduli
problem, and we will do it within the Mathai-Quillen formalism.  
The Mathai-Quillen
form gives a representative of the Thom class of the  bundle
${\cal E}={\cal M}\times_{\cal G} {\cal F}$, and the integration 
over
${\cal M}/{\cal G}$ of the pullback of this Thom class under the 
section
$\tilde s:{\cal M}/{\cal G}\to {\cal E}={\cal M}\times_{\cal G} 
{\cal F}$ gives the
(generalized) Euler characteristic of ${\cal E}$, which is to be 
identified, from
the field-theory point of view, with the partition function of 
the associated
topological quantum field theory. 

As a first step to construct the topological theory which corresponds to
the moduli problem defined by the basic equations \eqs{equations4}, we
have to give explicitly the field content and the BRST symmetry of the
theory. This will help to clarify the structure of the topological
multiplet we introduced in sect. $2$. In the field space ${\cal M}= {\cal
A}\times\Omega^0(X,\ad P)\times\Omega^{2,+} (X,\ad P)$ we have the gauge
connection $A_\mu$, the scalar field $C$ and the self-dual two-form
$B^{+}_{\mu\nu}$, all with ghost number $0$. In the (co)tangent space
$T_{(A,C,B^{+})}{\cal M}=\Omega^1(X,\ad P)\oplus\Omega^0(X,\ad
P)\oplus\Omega^{2,+} (X,\ad P)$ we have the anticommuting fields
$\psi_\mu$, $\zeta$ and $\tilde\psi^{+}_{\mu\nu}$, all with ghost number
$1$ and which are to be interpreted as differential forms on the moduli
space. In the fibre ${\cal F}= \Omega^1(X,\ad P)\oplus\Omega^{2,+} (X,\ad
P)$ we have anticommuting fields with the quantum numbers of the equations,
namely a one-form 
$\tilde\chi_\mu$ and a self-dual two-form $\chi^{+}_{\mu\nu}$, both with
ghost number $-1$, and their superpartners, a commuting one-form $\tilde 
H_\nu$ and a commuting self-dual two-form $H^{+}_{\mu\nu}$, both with ghost
number $0$ and which appear as auxiliary fields in the associated  field
theory. And finally, associated to the gauge symmetry, we have  a commuting
scalar field $\phi\in\Omega^{0}(X,\ad P)$ with ghost number
$+2$ \cite{coho}, and a multiplet of scalar fields $\bar\phi$ (commuting 
and
with ghost number $-2$) and $\eta$ (anticommuting and with ghost number
$-1$), both also in $\Omega^{0}(X,\ad P)$ and which enforce the horizontal
projection ${\cal M}\to {\cal M}/{\cal G}$ \cite{moore}. The BRST symmetry 
of the
model is given by:
\begin{align}
[Q, A_\mu] &=\psi_\mu,&
\{Q,\psi_{\mu}\} &= \deriv_{\mu}\phi, \ret
[Q,C]&=\zeta,&
\{Q,\zeta\,\} &=i\,[C,\phi],\ret
[Q,B^{+}_{\mu\nu}]&=\tilde\psi^{+}_{\mu\nu},& 
\{Q,\tilde\psi^{+}_{\mu\nu}\}&=i\,[B^{+}_{\mu\nu},\phi],\ret
[Q,\phi]&=0, & &{}\ret
\{Q,\tilde\chi_{\mu}\} &= \tilde H_{\mu},&
[Q,\tilde H_{\mu}]&=i\,[\tilde\chi_{\mu},\phi],\ret 
\{Q,\chi^{+}_{\mu\nu}\} &= H^{+}_{\mu\nu},& 
[Q,H^{+}_{\mu\nu}]&=i\,[\chi^{+}_{\mu\nu},\phi],\ret 
[Q,\bar\phi]&=\eta,&
\{Q,\eta\,\} &=i\,[\bar\phi,\phi].
\ret &{}& &{}
\la{macarena4}
\end{align} 
The BRST generator $Q$ satisfies the algebra 
$\{Q,Q\}=\delta_g(\phi)$,  and can be seen to correspond to 
the Cartan model for
the ${\cal G }$-equivariant cohomology of ${\cal M}$ .

We are now ready to write the action for the topological field theory under
consideration. Instead of writing the full expression for the Mathai-Quillen form,
we define the action to be $\{Q,\Psi\}$ for some appropriate gauge invariant gauge
fermion $\Psi$ \cite{moore}. The use of gauge fermions was introduced in the
context of topological quantum field theory in
\cite{perni} (see \cite{phyrep} for a review). As  is explained in
detail in
\cite{moore}, the gauge fermion consists of two basic pieces, a localization 
gauge fermion, which essentially involves the equations defining the moduli
problem and which in our case takes the form:

\begin{align}
\Psi_{\text{loc}}&=\langle(\tilde\chi,\chi^{+}),s(A,C,B^{+})
\rangle +\langle (\tilde\chi,\chi^{+}),(\tilde H,H^{+})\rangle
\ret 
&=\int_X\sqrt{g}\,\tr\,\bigl\{\,\,\half\chi^{+}_{\mu\nu}\bigl 
(\,H^{+\mu\nu} -2(
F^{+\mu\nu}-\fr{i}{2}[B^{+\mu\tau},B^{+}_\tau{}^{\!\nu}]-\fr{i}{\raiz}
[B^{+\mu\nu},C])\bigr )
\ret 
&+\tilde\chi_\mu\bigl (\,\tilde H^\mu+\raiz
\,(\deriv^\mu C+\raiz\deriv _\nu B^{+\nu\mu})\,\bigr )\,\bigr\},
\ret
\la{localiz}
\end{align}   
and a projection gauge fermion, which enforces the horizontal
projection,
 and which can be written as:
\begin{equation}
\Psi_{\text{proj}}=\langle\bar\phi,{\cal C}^{\dag}(\psi,
\zeta,\tilde\psi)\rangle_{\hbox{\bf g}},
\la{project}
\end{equation}    
where $\langle,\rangle_{\hbox{\bf g}}$ denotes the gauge
invariant  metric in
$\Omega^0(X,\ad P)$, and the map 
${\cal C}^{\dag}:T{\cal M}\to \Omega^0(X,\ad P)$ is the adjoint of 
the map
${\cal C}$ \eqs{butraguenho4} with respect to the Riemannian 
metrics
\eqs{metrica4} in $T{\cal M}$ and $ \Omega^0(X,\ad P)$. The adjoint
of 
${\cal C}(\phi)$ is readily computed to be:
\begin{equation} 
 {\cal C}^{+}(\psi,\zeta,\tilde\psi)=-\deriv_\mu \psi^\mu 
+\fr{i}{2}[\tilde\psi^{+}_{\mu\nu},B^{+\mu\nu}]+i[\zeta,C],
\la{adjoint}
\end{equation} 
 where $(\psi,\zeta,\tilde\psi)\in T_{(A,C,B^{+})}{\cal M}$. This  
leaves for the
projection fermion \eqs{project} the expression:
\begin{equation}
\Psi_{\text{proj}}=\int_X \sqrt{g}\,\tr\,\bigl\{\,
\bar\phi\bigl (\,-\deriv_\mu\psi^\mu
+\fr{i}{2}[\tilde\psi^{+}_{\mu\nu}, B^{+\mu\nu}]+i[\zeta,C]\,\bigr
)\,\bigr\}.
\la{projectn}
\end{equation}

In the Mathai-Quillen formalism the action is built out of the terms
\eqs{localiz} and \eqs{projectn}.  However, as in the case of the Mathai-Quillen
formulation of Donaldson-Witten  theory  \cite{jeffrey}, one must add another
piece to the gauge fermion to make full contact with the corresponding twisted
supersymmetric theory. In  our case, this extra term is:
\begin{equation}
\Psi_{\text{extra}}=-\int_X \sqrt{g}\,\tr\,\bigl\{\,
\fr{i}{2}\eta[\phi,\bar\phi]\,\bigr\}.
\la{extra}
\end{equation}

It is now straightforward to see that after the rescalings
\begin{align}
 A'&=A,&  C'&=-\fr{1}{\raiz}C,& \tilde\chi'&=\raiz\tilde\chi,\ret
\psi'&=-\fr{1}{2}\psi,& \zeta'&=-\zeta,& 
\tilde H'&=\raiz H,\ret
\phi'&=\fr{1}{{2\raiz}}\phi,&
 B^{+'}&=\half B^{+},& \chi^{+'}&=\chi^{+},\ret 
\bar\phi'&=-2\raiz \bar\phi,&
\eta'&=-2\eta,&
\tilde\psi^{+'}&=\fr{1}{{2\raiz}}\tilde\psi^{+},\ret   
H^{+'}&=H^{+},& &{} & &{}\ret
\la{redef}
\end{align}  
and with the identification $Q\equiv Q^{+}$ we recover, in terms of the
primed fields,  the twisted model  we analyzed in the first part of the
chapter and that is encoded in  
\eqs{papagayo4} and \eqs{mermelada}. 

And what about $Q^{-}$? What is its role in this game? In fact,  the
theory admits two   Mathai-Quillen descriptions, related to each other by 
the Weyl group of $SU(2)_F$,  in such a way that the roles of $Q^{+}$ and
$Q^{-}$ are interchanged,  as are the roles of $\psi$ and $\tilde\chi$,
$\chi^{+}$ and $\tilde\psi^{+}$, 
$\zeta$ and $\eta$, and $\phi$ and $\bar\phi$. The corresponding moduli
space  is defined by eqs. (\ref{equations4}) with the  substitution $C\to
-C$, and the  theory  localizes -- as was proved in \cite{vw} -- actually
on the intersection of  both  moduli spaces, which is defined by the
equations 
\begin{equation}
\begin{cases} 
\deriv_{\mu}C=0,\qquad\qquad 
\deriv^\nu B^{+}_{\nu\mu}=0,\\
F^{+}_{\mu\nu}-\fr{i}{2}[B^{+}_{\mu\tau},B^{+\tau}{}_{\!\nu}]=0,\quad
[B^{+}_{\mu\nu},C]=0.
\end{cases}
\label{ecuacions}
\end{equation}

Alternatively, one can consider both $Q^{+}$ and $Q^{-}$ at once within the 
framework of what Dijkgraaf and Moore \cite{balanced} have called the
equivariant cohomology of iterated superspaces. The resulting topological
field theories have two BRST operators $d_{\pm}$ and the virtual dimension
of the corresponding moduli spaces is zero. The
simplifying property of this type of theories is that the topological action
can be derived (up to theta terms) from an {\sl action potential} $\C{F}$
\begin{equation}
\C{S}=d_+d_-\C{F}, 
\end{equation}
-- this is just the content of eq. \eqs{jackie4}. $\C{F}$ is a kind of Morse
function on field space, and under certain assumptions the path
integral can be seen to localize to the critical points of $\C{F}$
\cite{balanced}.

\section{Observables}

In this section we will analyze the structure of the possible
topological observables of
the theory. Observables are operators which are $Q$-invariant but are
not $Q$-exact. A quick look at either the $Q^{+}$ or $Q^{-}$ transformations 
shows that the observables are basically the same as in the ordinary
Donaldson-Witten theory. Indeed, from \eqs{papagayo4} or \eqs{macarena4} one
finds that the trio $A_\mu$, $\psi_\mu$ and $\phi$ transforms such as to
ensure that the operators
\begin{align}
W_0 =& \tr(\phi^2), &
W_2 =& \tr\,\left(\half \psi \wedge \psi + \fr{1}{\raiz}\phi\wedge 
F\right),\ret 
W_1 =& -\raiz\tr(\phi \wedge\psi), &
W_3 =& -\fr{1}{2}\tr(\psi \wedge  F), \ret
\la{gutavo}
\end{align}
satisfy the descent equations,
\begin{equation}
[Q^{+}, W_i\} = d W_{i-1},
\la{descendido}
\end{equation}
which as we know imply that,
\begin{equation}
{\cal O}^{(\gamma_j)} = \int_{\gamma_j} W_j,
\la{noguta2}
\end{equation}
being $\gamma_j$ homology cycles of $X$, are observables. Of course, 
as usual, this set can be enlarged for gauge groups possessing other 
independent Casimirs besides the quadratic one. We can also build a 
similar set of observables $\tilde{\cal O}^{(\gamma_j)}$ in the 
cohomology of $Q^{-}$ by exchanging $\phi$ by $\bar\phi$ and $\psi$ 
by $\tilde\chi$. While the observables ${\cal O}^{(\gamma_j)}$ 
have positive ghost numbers given by $4-j$, the observables 
$\tilde{\cal O}^{(\gamma_j)}$ have negative ghost numbers $j-4$.    
The unusual feature to the first twist is that the ghost number symmetry 
is non-anomalous. This means essentially that the only vacuum expectation values
which are  non-vanishing are those containing sets of observables with zero
overall  ghost number. Although very appealing, the idea of combining
observables  from the $Q^{+}$ and $Q^{-}$ cohomologies with opposite ghost
numbers to  get non-zero vacuum expectation values is not useful. The
reason is that, for these combined operators to be truly  topological
invariants, they must lie  simultaneously in the cohomologies of both
$Q^{+}$ and
$Q^{-}$. The only possible candidate appears to be the  theta-term, which
does not lead to any new invariant. Hence, the only non-trivial observable
is the partition function itself. Notice that the topological action 
splits as 
\begin{equation}
{\cal S}=
\{Q^+,\Psi\}- 2\pi i k \tau_0,
\la{jackie5}
\end{equation}  

 Now, owing to (\ref{jackie5}), the partition
function  depends on  the microscopic couplings $e_0$  and $\theta_0$
only through the combination $2\pi ik \tau_0$,  and in particular this
dependence is a priori holomorphic (were the orientation of the manifold
$X$ reversed, the partition function would depend  anti-holomorphically
on $\tau_0$). However there could be
situations in which,  because of some sort of holomorphic anomaly,
the partition function would  acquire an explicit anomalous
dependence on $\bar\tau_0$. This seems to  be the case, for
example, for the theory defined on 
$\cp^{2}$ \cite{vw} and, more generally, on manifolds with
$b^{+}_2=1$ \cite{estrings}. Somewhat related results have been derived
for the Donaldson-Witten theory in the context of the $u$-plane 
formalism \cite{moorewitten}. 

But \eqs{jackie5} has also far-reaching consequences on the structure of
the partition function, which can be written as
\begin{equation}
Z(\tau_0)=\sum_k \C{Z}_k \,\ex^{2\pi i k \tau_0},
\la{laracroft}
\end{equation}  
where $\C{Z}_k$ is the partition function computed with action
$\{Q^+,\Psi\}$  at fixed instanton number $k$. Now since the $\C{Z}_k$ do
not depend on $e_0$, one can take the weak-coupling limit
$e_0\to 0$ where the computations are given exactly by a saddle-point
calculation around the  bosonic background defined by the Vafa-Witten
equations \eqs{equations4}. Unfortunately, it is
not possible to  perform explicit computations from this
viewpoint: the moduli space is non-compact, and no precise recipe is 
known to  properly compactify it. The way out is to exploit the metric
independence of the partition function to go to the long-distance limit and
compute in terms of the low-energy degrees of freedom of the physical
theory. For the Vafa-Witten theory this is the $\cn=1$ theory which results
from giving bare masses to the chiral multiplets of the $\cn=4$ theory. In
what follows we will review the construction thoroughly and apply it to the
computation of the partition function for gauge group $SU(N)$.  

\section{Mass perturbations and reduction to $\cn=1$}

It is a well-known fact that on complex manifolds the exterior
differential $d$ splits into the Dolbeaut operators $\partial$ 
and $\bar\partial$, and that this splitting is completely effective 
on K\"ahler manifolds, where the de Rham cohomology is equivalent
to the cohomology of  $\partial$. In a similar way, as pointed out
in
\cite{wijmp}, on a  K\"ahler manifold the number of BRST charges of
a twisted supersymmetric theory is doubled, in such a way that, for
example,  the  Donaldson-Witten theory has an enhanced
$\cn_T=2$ topological symmetry, while the
Vafa-Witten theory has a $\cn_T=4$ topological symmetry. In each
case, one of the BRST charges comes from the underlying $\cn=1$ 
subalgebra which corresponds to the formulation of the physical
theory in 
$\cn=1$ superspace. By suitably adding mass terms for some of the 
chiral superfields in the theory, one can break the extended
($\cn=2$ or
$\cn=4$)  supersymmetry of the physical theory down to $\cn=1$.
For the reason sketched  above, the corresponding twisted massive
theory on K\"ahler manifolds should still retain at least  one
topological symmetry. One now exploits the metric independence of
the  topological theory. By scaling up the metric in the
topological theory, 
$g_{\mu\nu}\to tg_{\mu\nu}$, one can take the limit $t\to\infty$.
In this  limit, the metric on $X$ becomes nearly flat, and it is 
reasonable that the computations in the topological field  theory
can be performed in terms of the  vacuum structure of the $\cn=1$
theory. 

One could wonder as to what the effect of the perturbation may
be. The introduction of a mass perturbation may  (and in general
will) distort the original topological field theory.  This poses
no problem in the case of the  Donaldson-Witten theory, as Witten
was able to prove that the perturbation is topologically trivial,
in the sense that it affects the theory in an important but
controllable way \cite{wijmp}.  As for the Vafa-Witten theory
\cite{coreatres}\cite{masas}\cite{vw},  we will see below that 
the twisted massive theory is topological on K\"ahler four-manifolds  
with $h^{2,0}\not=0$, and the partition function is actually invariant
under the perturbation.

In what follows we  will make use of the transformations 
generated by $Q^{+}$ only, which we rewrite as follows
\begin{align}
[Q^{+}, A_{\alpha\dalpha}] &= -2\psi_{\alpha\dalpha},& 
[Q^{+},\bar\phi]& =\raiz\,\eta,\ret
\{Q^{+},\psi_{\alpha\dot\alpha}\}& = 
-{\raiz}{\cal D}_{\alpha\dot\alpha}\phi, &
\{Q^{+},\eta\,\} &=2i\,[\bar\phi,\phi],\ret      
\big[ Q^{+},\phi\big]&=0 &
\{Q^{+},\tilde\chi_{\alpha\dalpha}\}& = 
\tilde H_{\alpha\dalpha}+\raiz s_{\alpha\dalpha},\ret
\big[Q^{+}, B^{+}_{\alpha\beta}\big]&=\raiz\tilde\psi^{+}_{\alpha\beta},&
[Q^{+},\tilde H_{\alpha\dalpha}]&=
2\raiz i\,[\tilde\chi_{\alpha\dalpha},\phi]-\raiz 
[Q^{+},s_{\alpha\dalpha}]
,\ret
\{Q^{+},\tilde\psi^{+}_{\alpha\beta}\}&=2i\,[B^{+}_{\alpha\beta}, \phi],
&
\{Q^{+},\chi^{+}_{\alpha\beta}\} &= H^{+}_{\alpha\beta}+
s_{\alpha\beta},\ret
\big[Q^{+},C\big]&=\fr{1}{\raiz}\zeta,&
[Q^{+},H^{+}_{\alpha\beta}]&=2\raiz i\,
[\chi^{+}_{\alpha\beta},\phi]-[Q^{+},s_{\alpha\beta}],\ret
\{Q^{+},\zeta\,\}& = 4i\,[C,\phi],\ret
\label{papagayo}
\end{align}  
where 
\bea
s_{\alpha\dalpha}&=&\deriv_{\alpha\dalpha}C+
i\deriv_{\beta\dalpha}B^{+\beta}{}_\alpha,\ret
s_{\alpha\beta}&=& F^{+}_{\alpha\beta}+[B^{+}_{\gamma\alpha},
B^{+}_\beta{}^{\!\gamma}]+2i\,[B^{+}_{\alpha\beta},C].\ret
\label{ecuaciones}
\eea

Written in this form (compare with \eqs{papagayo4}), the on-shell
transformations are simply obtained by setting 
$H^{+}_{\alpha\beta}=0=\tilde H_{\alpha\dalpha}$ in (\ref{papagayo}). 

Notice that $s_{\alpha\dalpha}$ and $s_{\alpha\beta}$ are precisely the
components of the section \eqs{seccion} which defines the Vafa-Witten theory
in the Mathai-Quillen approach. Now it is easy to see why the theory
localizes to field configurations satisfying \eqs{equations4}.  According to
Witten's fixed-point theorem, the contributions  to the partition
function of the theory, which is the only non-trivial  observable owing to
the vanishing of the ghost number anomaly,  come from the fixed points of
the BRST symmetry. But according to (\ref{papagayo}), the fixed points 
are precisely (with the auxiliary fields set to zero)
$s_{\alpha\dalpha}=0=s_{\alpha\beta}$. 

One of the main 
ingredients in the analysis in \cite{vw} is the existence, on certain 
four-manifolds (basically of the K\"ahler type), of a suitable 
vanishing theorem which guarantees that all the solutions to  
 eqs. (\ref{equations4}) are of the form: 

\begin{equation}
F^{+}_{\alpha\beta}=0,\quad 
B^{+}_{\alpha\beta}=0, \quad C=0,
\label{vanishing}
\end{equation}
that is, that the moduli space reduces to the moduli space of ASD
connections.  In fact, under these circumstances, the partition function of
the theory  computes, for each value of the instanton number, the Euler
characteristic  of the corresponding instanton moduli space. Observe that
the vanishing  theorem allows only positive instanton numbers to contribute
to the  partition function; the presence of negative instanton number
contributions  would signal a failure of the vanishing theorem.

\subsection{The twist on K\"ahler manifolds}

On a four-dimensional K\"ahler manifold the holonomy group is contained in
$SU(2)_R\otimes U(1)_L$, where $U(1)_L$ is a certain subgroup of $SU(2)_L$.
Under this reduction of the holonomy, left-handed spinors $\psi_\alpha$
decompose into pieces $\psi_1$ and $\psi_2$ of opposite  $U(1)_L$ charges, in
such a way that if the manifold is also spin, the spinor bundle $S^+$ has a
decomposition $S^+\simeq K^\half \oplus  K^{-\half}$, where  $K^\half$ is
some square root of the canonical bundle of $X$, $K=\bigwedge^2_{\bf
{C}}T^{*}X$.  Following \cite{wijmp}, we can pick a complex structure on $X$ 
by taking the $1$-forms
$(\sigma_\mu)_{1\dalpha}dx^\mu$ to be of type $(1,0)$, and the $1$-forms    
$(\sigma_\mu)_{2\dalpha}dx^\mu$ of type $(0,1)$. With this choice, the
self-dual $2$-form $(\sigma_{\mu\nu})_{\alpha\beta}dx^\mu\wedge dx^\nu$ can
be regarded as a $(2,0)$-form for $\alpha=\beta=1$, as a $(0,2)$-form for
$\alpha=\beta=2$, and as a $(1,1)$-form for $\alpha=1, \beta=2$. This
decomposition corresponds to the splitting 
$\Omega^{2,+}(X)=\Omega^{2,0}(X)\oplus\Omega^{0,2}(X)\oplus
\varpi\Omega^{0}(X)$, valid on any K\"ahler surface ($\varpi$ stands for the
K\"ahler form).  

With respect to the complex structure of the manifold, the fields of the
theory
 naturally split into objects that can be thought of as components of forms
of type $(p,q)$. For example, the connection $1$-form $A_{\alpha\dalpha}
(\sigma_\mu)^{\dalpha\alpha}dx^\mu$  splits up into a $(1,0)$-form 
$A_{2\dalpha}(\sigma_\mu)^{\dalpha}_1 dx^\mu$ and a $(0,1)$-form 
$A_{1\dalpha}(\sigma_\mu)^{\dalpha}_2 dx^\mu$. Likewise, the self-dual
$2$-form 
$B^{+}_{\alpha\beta}(\sigma_{\mu\nu})^{\alpha\beta}dx^\mu\wedge dx^\nu$ gives
rise to a $(2,0)$-form  
$B^{+}_{22}(\sigma_{\mu\nu})_{11}dx^\mu\wedge dx^\nu$ a $(0,2)$-form
$B^{+}_{11}(\sigma_{\mu\nu})_{22}dx^\mu\wedge dx^\nu$
and  a $(1,1)$-form for 
$B^{+}_{12}(\sigma_{\mu\nu})_{12}dx^\mu\wedge dx^\nu=B^{+}_{12}\varpi$. 
Notice that in our conventions the field $B^{+}_{11}$ would correspond 
to the $(0,2)$-form $\bar\beta$, $B^{+}_{22}$  
to the $(2,0)$-form $\beta$ and  $B^{+}_{12}$ to the $0$-form $b$ in 
\cite{vw}. Note
that the field $B^{+}_{12}$ can be thought of as a scalar field on $X$. In
fact, we shall see in a moment that it naturally combines with the scalar
field $C$ into two complex scalars $B^{+}_{12}\pm iC$. Something similar
happens with the other self-dual $2$-forms $\chi^{+}$ and $\tilde\psi^{+}$. 

Let us recall that in our conventions the BRST operators $Q^{\pm}$ are 
obtained 
from the $\cn=4$ supercharges $Q^v{}_\alpha$, with the recipe
\begin{equation}
Q^{+}=Q^{\tilde 3}{}_1+Q^{\tilde 4}{}_2,\qquad 
Q^{-}=i(Q^{\tilde 1}{}_1+Q^{\tilde 2}{}_2).
\label{rabindranah}
\end{equation}
In the K\"ahler case, each of the individual components $Q^{\tilde 1}{}_1$, 
$Q^{\tilde 2}{}_2$, $Q^{\tilde 3}{}_1$ and $Q^{\tilde 4}{}_2$ is 
well-defined under the holonomy $SU(2)_R\otimes U(1)_L$. It is therefore
possible to define four charges, of which only   $Q^{\tilde 4}{}_2$ is
related to the underlying construction  in $\cn=1$ superspace. Hence, it is
the only topological symmetry that should be expected to survive after the
mass terms are plugged in. 

In what
follows, we will be interested only in $Q^{\tilde 3}{}_1$ and $Q^{\tilde
4}{}_2$. The corresponding transformation laws (with parameters $\rho_2$ and  
$\rho_1$ respectively) can be extracted from the $\cn=4$ supersymmetry 
transformations \eqs{Vian4} by setting: 
\begin{equation}
\bar\xi^{v\dalpha}=0,\quad \xi_{\tilde 1\alpha}= 
\xi_{\tilde 2\alpha}=0,\quad
\xi_{\tilde 3\alpha}=\rho_2 C_{2\alpha}, \quad 
\xi_{\tilde 4\alpha}=\rho_1 C_{1\alpha},
\label{tagore}      
\end{equation}
The corresponding BRST charges will be denoted by $Q_1=Q^{\tilde 4}{}_2$
and $Q_2=Q^{\tilde 3}{}_1$. The on-shell transformations turn out to be:
\begin{align}
[Q_1,A_{1\dot\alpha}] &= -2\psi_{1\dot\alpha},&
[Q_1,A_{2\dot\alpha}] &=0,\ret 
\big[Q_1,F^{+}_{11}\big]&=-2iD_{1\dalpha}\psi_1{}^\dalpha,&
\big[Q_1,F^{+}_{22}\big]&=0,\ret
\big[Q_1,F^{+}_{12}\big]&=-iD_{2\dalpha}\psi_1{}^\dalpha,&
\left\{Q_1, \half\eta-i\chi^{+}_{12}\right\}&= 0,\ret 
\{Q_1,\psi_{1\dot\alpha}\}&=0,&
\{Q_1,\psi_{2\dot\alpha}\}&=-\raiz D_{2\dot\alpha}\phi,\ret
\big[Q_1,\phi\big]&=0, &
\big[Q_1, \bar\phi\big]&= \raiz\left(\half\eta-i\chi^{+}_{12}\right),\ret
\big[Q_1, B^{+}_{11}\big] &= 0,&
\big[Q_1, B^{+}_{22}\big] &= \raiz\tilde\psi^{+}_{22},\ret
\big[Q_1, B^{+}_{12}+iC\big]& =0 ,&
\big[Q_1, B^{+}_{12}-iC\big] &=\raiz\left(\tilde\psi^{+}_{12}-\fr{i}{2}
\zeta\right) ,\ret
\{Q_1, \tilde\psi^{+}_{11}\}&=2i[ B^{+}_{11},\phi],&
\{Q_1, \tilde\psi^{+}_{22}\}&=0,\ret
\left\{Q_1,\tilde\psi^{+}_{12}+\fr{i}{2}\zeta\right\}
&=-2i[\phi,B^{+}_{12}+iC],&
\left\{Q_1, \tilde\psi^{+}_{12}-\fr{i}{2}\zeta\right\}&=0,\ret
\{Q_1,\chi^{+}_{11} \}&= -2[B^{+}_{12}+iC,B^{+}_{11}],&
\{Q_1,\chi^{+}_{22} \}&= F^{+}_{22},\ret
\{Q_1, \tilde\chi_{1\dalpha}\}&=-\raiz iD_{2\dalpha} B^{+}_{11},&
\{Q_1, \tilde\chi_{2\dalpha}\}&=-\raiz iD_{2\dalpha}(B^{+}_{12}+iC),\ret
\left\{Q_1, \half\eta +i\chi^{+}_{12}\right\}&= 
 -i[\phi,\bar\phi]+iF^{+}_{12}\ret
+{i}[B^{+}_{12}-iC,&B^{+}_{12}+iC]+{i}[B^{+}_{11},
B^{+}_{22}],\ret
\label{sandia}
\end{align}
for $Q_1$. The $Q_2$ transformations are easily computed from 
(\ref{papagayo}) and 
\eqs{sandia}  after using $Q^{+}=Q_1+Q_2$ and read:
\begin{align}
[Q_2,A_{1\dot\alpha}] &=0,&
[Q_2,A_{2\dot\alpha}] &= -2\psi_{2\dot\alpha},\ret
\big[Q_2,F^{+}_{11}\big]&=0,&
\big[Q_2,F^{+}_{22}\big]&=-2iD_{2\dalpha}\psi_2{}^\dalpha,\ret
\big[Q_2,F^{+}_{12}\big]&=-iD_{1\dalpha}\psi_2{}^\dalpha,&
\left\{Q_2, \half\eta+i\chi^{+}_{12}\right\}&= 0,\ret
\{Q_2,\psi_{1\dot\alpha}\}&=-\raiz D_{1\dot\alpha}\phi,&
\{Q_2,\psi_{2\dot\alpha}\}&=0,\ret 
\big[Q_2,\phi\big]&=0, &
\big[Q_2, \bar\phi\big] &= \raiz\left(\half\eta+i\chi^{+}_{12}\right),\ret 
\big[Q_2, B^{+}_{11}\big] &= \raiz\tilde\psi^{+}_{11},&
\big[Q_2, B^{+}_{22}\big]& = 0,\ret 
\big[Q_2, B^{+}_{12}-iC\big]& =0 ,&
\big[Q_2, B^{+}_{12}+iC\big] &=\raiz\left(\tilde\psi^{+}_{12}+
\fr{i}{2}\zeta\right) ,\ret
\{Q_2, \tilde\psi^{+}_{11}\}&=0,&
\{Q_2, \tilde\psi^{+}_{22}\}&=2i[B^{+}_{22},\phi],\ret
\left\{Q_2, \tilde\psi^{+}_{12}-\fr{i}{2}\zeta\right\}&=
-2i[\phi,B^{+}_{12}-iC],&
\{Q_2,\tilde\psi^{+}_{12}+\fr{i}{2}\zeta\}&=0,\ret
\{Q_2,\chi^{+}_{11} \}&= F^{+}_{11},&
\{Q_2,\chi^{+}_{22} \}&= 2[B^{+}_{12}-iC,B^{+}_{22}],\ret
\{Q_2, \tilde\chi_{1\dalpha}\}&=\raiz iD_{1\dalpha}(B^{+}_{12}-iC),&
\{Q_2, \tilde\chi_{2\dalpha}\}&=\raiz iD_{1\dalpha} B^{+}_{22},\ret
\left\{Q_2, \half\eta-i\chi^{+}_{12}\right\}& =
-i\big[\phi,\bar\phi\big]-iF^{+}_{12}\ret
-{i}[B^{+}_{12}-iC,&B^{+}_{12}+iC]-{i}[B^{+}_{11}, B^{+}_{22}].\ret
\label{pavia}
\end{align}
It is straightforward to verify that $(Q_1)^2=(Q_2)^2=0$ on-shell, while 
$\{Q_1,Q_2\}$ gives a gauge transformation generated by $\phi$.  

Notice that these equations are compatible with the 
$U(1)$ symmetry (which will be further exploited below) 
\begin{align}
B^{+}_{11}&\to \ex^{i\alpha}B^{+}_{11},& B^{+}_{22}&\to
\ex^{-i\alpha}B^{+}_{22}, & B^{+}_{12}\pm iC &\to  \ex^{\mp
i\alpha}(B^{+}_{12}
\pm iC),\ret 
\tilde\psi^{+}_{11}&\to \ex^{i\alpha}\tilde\psi^{+}_{11},&
\tilde\psi^{+}_{22}&\to
\ex^{-i\alpha}\tilde\psi^{+}_{22}, 
&\tilde\psi^{+}_{12}\pm\fr{i}{2}\zeta&\to 
\ex^{\mp i\alpha}(\tilde\psi^{+}_{12}\pm\fr{i}{2}\zeta),\ret
\la{uunosim}
\end{align}
which does not act on the rest of the fields. 

\subsection{Mass perturbations}

We now turn to the discussion of the possible ways of (softly) breaking
$\cn=4$ supersymmetry by suitably adding mass terms for the chiral
multiplets. Let us
 consider first the situation that arises on a flat ${\IR}^4$. By adding a
bare mass term for just one of the chiral multiplets, say $\Phi_1$, 
\begin{equation}
\Delta L^{(1)} = m\int d^4 xd^2\theta\tr{(\Phi_1)^2}+{\hbox{\rm h.c.}}, 
\label{pertu}
\end{equation}
$\cn=4$ supersymmetry is broken down to $\cn=1$. The corresponding low-energy
effective theory, at scales below $m$, is $\cn=1$ supersymmetric  QCD, with 
$SU(2)$ as gauge group, coupled to two  massless chiral superfields in the
adjoint representation with a (tree-level)  quartic superpotential induced
by integrating out the massive superfield. As  shown in 
\cite{phases},  this theory has a moduli space  of vacua where both a Coulomb
and a Higgs phase coexist.  On  the other hand, equal bare mass terms for
two of the chiral multiplets, 
\begin{equation}
\Delta L^{(2)} = m\int d^4 xd^2\theta\tr{(\Phi_1\Phi_2)}+{\hbox{\rm h.c.}}, 
\label{pertur}
\end{equation} 
preserve $\cn=2$ supersymmetry, whereas if the mass terms are different: 
\begin{equation}
{\Delta}' L^{(2)} = m_1\int d^4 xd^2\theta\tr{(\Phi_1)^2}+
m_2\int d^4 xd^2\theta\tr{(\Phi_2)^2}+{\hbox{\rm h.c.}}, 
\label{perturb}
\end{equation}
$\cn=4$ supersymmetry is again broken down to $\cn=1$. However, both theories 
flow in the
infrared to a pure $\cn=2$ supersymmetric gauge theory, which has a moduli 
space of vacua  
in the Coulomb phase. 
Finally, mass terms for 
the three 
chiral multiplets, no matter whether the mass parameters are equal or not,
preserve only $\cn=1$ supersymmetry. Of the three inequivalent ways of
breaking
$\cn=4$ supersymmetry down to $\cn=1$, we must choose the one in terms of
which  the analysis of the vacuum structure of the resultant $\cn=1$ theory
is simplest.  The appropriate choice is \cite{vw} 
\begin{equation}
\Delta L^{(3)} = m\int d^4
xd^2\theta\tr{\bigl((\Phi_1)^2+(\Phi_2)^2+(\Phi_3)^2
\bigr)} +{\hbox{\rm h.c.}}, 
\label{perturba}
\end{equation}
in terms of which the classical vacua of the resulting $\cn=1$ theory can be
classified by the complex conjugacy classes of homomorphisms of the $SU(2)$
Lie algebra to that of $G$. In the case that $G=SU(2)$, for example, there
are three discrete vacua, corresponding to the three singularities of the
mass-deformed $\cn=4$  
 supersymmetric gauge theory with gauge group $SU(2)$ \cite{swii}.     

On general curved manifolds the naive construction sketched
above simply does not work. As explained in \cite{vw}\cite{wijmp},
superpotentials of a twisted theory on K\"ahler manifolds must
transform as $(2,0)$-forms. This  comes about as follows. Let us
consider the situation for $\cn=2$ theories, as  it turns out that
all the other cases can be reduced to it. A general  superpotential
can be written in $\cn=1$ superspace as 
\begin{equation}
\int d^4 x d^2\bar\theta\, \overline{\C{W}} + {\text h.c.},
\la{tortela}
\end{equation}
where $\overline{\C{W}}$ is an anti-holomorphic function of the chiral 
superfields. Now, 
viewed as part of the $\cn=2$ superspace (with coordinates 
$\theta^{i}_{\alpha}$ and $\bar\theta^{\dalpha}_{i}$, $i=1,2$), 
the measure $d^2\bar\theta$ carries an $SU(2)_I$ index, 
$d^2\bar\theta=d\bar\theta_{2\dalpha}d\bar\theta_{2}{}^{\dalpha}$,
and therefore  after the twist on a K\"ahler four-manifold it
carries non-trivial charge under 
$U(1)_L$. In fact, after the twist $d^2\bar\theta$ becomes a $(2,0)$-form 
according to the above conventions, so $\overline{\C{W}}$ must be a 
$(0,2)$-form on $X$, and therefore $\C{W}$ must be a $(2,0)$-form as stated. 

Now, according to our conventions, two of the
chiral superfields,
$\Phi_1$ and
$\Phi_3$ (whose scalar components are $B^{+}_{12}\pm iC$ and $\phi$,
$\bar\phi$ resp.) are scalars in the twisted model, while the third one,
$\Phi_2$ (whose scalar components are $B^{+}_{11}$ and  $B^{+}_{22}$), is a
$(2,0)$-form. A suitable  mass term for $\Phi_2$ and one of the other scalar
superfields, say $\Phi_1$, can be readily written down  and
reads:

\begin{equation}
\Delta L(m) = m\int_X d^2\theta\tr{(\Phi_1\Phi_2)}+{\hbox{\rm h.c.}}
\label{perturm}
\end{equation}
In \eqs{perturm} $m$ is just a (constant) mass parameter. A mass term for the
remaining superfield $\Phi_3$ requires the introduction of the 
$(2,0)$-form\footnote{Of course, this sets on the manifold $X$ the
constraint 
$h^{(2,0)}(X)\not=0$, which for K\"ahler manifolds is equivalent to 
demanding $b^{+}_2>1$. This excludes, for example, the case of $\cp^2$.} 
$\omega$ \cite{wijmp}:
\begin{equation}
\Delta L(\omega) = \int_X
\omega\wedge d^2{\bar z}d^2\theta\tr{(\Phi_3)^2}+{\hbox{\rm h.c.}}
\label{perturbation}
\end{equation}

Therefore we now turn to studying the effect of the following mass terms for 
the chiral multiplets $\Phi_1$, $\Phi_2$ and  $\Phi_3$:
\bea
\Delta L (m,\omega)&=& m\int_X d^2\theta\tr{(\Phi_1\Phi_2)}+
m\int_X d^2\bar\theta\tr{(\Phi^{\dag}_1\Phi^{\dag}_2)}\ret
&+&\int_X d^2\theta\omega\tr{(\Phi_3)^2}+
\int_X d^2\bar\theta\bar\omega\tr{(\Phi^{\dag}_3)^2},
\ret
\label{pera}
\eea
where, for simplicity, $\omega=\omega_{11}=(\sigma_{\mu\nu})_{11}
\omega_{\tau\lambda}\epsilon^{\mu\nu
\tau\lambda}$ 
stands for the only non-vanishing component of the $(2,0)$-form 
$\omega$, while $\bar\omega=\bar\omega_{22}=\omega_{11}^{*}$ stands for the
only non-vanishing component of the $(0,2)$-form $\bar\omega$ conjugate to 
$\omega$. 

After expanding the fields and integrating out the auxiliary fields one gets 
the contributions
\bea
&&-2\raiz i \omega B_3[B^{\dag}_1, B^{\dag}_2]-2\raiz i \bar\omega 
B^{\dag}_3[B_1, B_2]-4\vert\omega\vert^2 B_3 B^{\dag}_3\ret
&&-\omega\lambda_{\tilde 3}{}^{\alpha} \lambda_{\tilde 3\alpha}-\bar\omega
\bar\lambda^{\tilde 3}{}_{\dalpha} \bar\lambda^{\tilde 3\dalpha}\ret
&&-2\raiz i m B_2[B^{\dag}_2, B^{\dag}_3]-2\raiz i m 
B^{\dag}_2[B_2, B_3]-m^2 B_2 B^{\dag}_2\ret
&&-m\lambda_{\tilde 1}{}^{\alpha} \lambda_{\tilde 2\alpha}-m
\bar\lambda^{\tilde 1}{}_{\dalpha} \bar\lambda^{\tilde 2\dalpha}\ret
&&-2\raiz i m B_1[B^{\dag}_3, B^{\dag}_1]-2\raiz i m 
B^{\dag}_1[B_3, B_1]-m^2 B_1 B^{\dag}_1.\ret
\label{manzana}
\eea
The $\cn=1$ transformations for the fermions get modified as follows:
\bea
&&\delta \lambda_{\tilde 1\alpha}=\ldots -\raiz\,\xi_{\tilde 4\alpha}m 
B^{\dag}_2,\ret
&&\delta \lambda_{\tilde 2\alpha}=\ldots -\raiz\,\xi_{\tilde 4\alpha}m 
B^{\dag}_1,\ret
&&\delta \lambda_{\tilde 3\alpha}=\ldots -2\raiz\,\xi_{\tilde 4\alpha}
\bar\omega B^{\dag}_3\ret
\label{peach}
\eea
(and their corresponding complex conjugates). 
In terms of the twisted fields the mass contributions are -- see \eqs{che}:

\bea
\tr\!\!\!\!\!\!\!\!\! &&\Bigl\{-2\raiz
i\omega\bar\phi[B^{+}_{12}+iC,B^{+}_{11}]+ 2\raiz  i\bar\omega
\phi[B^{+}_{12}-iC,B^{+}_{22}]-4\vert\omega\vert^2\phi\bar\phi\ret
&&-2i\omega\chi^{+}_{11}\left(\half\eta-i\chi^{+}_{12}\right)+
\bar\omega\psi_{2\dalpha}
\psi_2{}^{\dalpha}-2\raiz imB^{+}_{22}[B^{+}_{11},\phi]\ret
&&-2\raiz imB^{+}_{11}[B^{+}_{22},\bar\phi]+m^2 B^{+}_{11}B^{+}_{22}+ m
\left(\tilde\psi^{+}_{12}+\fr{i}{2}\zeta\right)\left(\tilde\psi^{+}_{12}-
\fr{i}{2}\zeta\right)
\ret &&+m\tilde\psi^{+}_{11}\tilde\psi^{+}_{22}
+m\tilde\chi_{2\dalpha}\tilde\chi_1{}^\dalpha +
\raiz im \phi[B^{+}_{12}+iC,B^{+}_{12}-iC]\ret &&-\raiz im \bar\phi
[B^{+}_{12}+iC,B^{+}_{12}-iC]- m^2 \vert B^{+}_{12}+iC\vert^2\Bigr\}.\ret
\label{uva}
\eea
Notice that the mass terms \eqs{uva} explicitly break the ghost number
symmetry, but they preserve the $U(1)$ symmetry \eqs{uunosim}. 

The $Q_1$ transformations \eqs{sandia}, which are the only ones to survive
the  perturbation a priori, also get modified in a way that is dictated
by the underlying $\cn=1$ structure, so that in view of  \eqs{peach} they
become:
 
\bea
\{Q^{(m,\omega)}_1, \tilde\psi^{+}_{11}\}&=&2i[ B^{+}_{11},\phi]-
\raiz m B^{+}_{11},\ret
\left\{Q^{(m,\omega)}_1,\tilde\psi^{+}_{12}+\fr{i}{2}\zeta\right\}
&=&-2i[\phi,B^{+}_{12}+iC]
+\raiz m(B^{+}_{12}+iC),\ret
\{Q^{(m,\omega)}_1,\chi^{+}_{11} \}&=& -2[B^{+}_{12}+iC,B^{+}_{11}]
+2\raiz i\bar\omega\phi.\ret                                             
\label{sandy}
\eea
(The rest of the transformations remain the same.) 
Notice that the fixed-point equations which stem from \eqs{sandy} are
precisely    the $F$-flatness conditions as derived from the superpotential 
\begin{equation}
i\raiz \tr\left(\Phi_1[\Phi_2,\Phi_3]\right)+m\tr\left(\Phi_1\Phi_2\right)
+\omega\tr(\Phi_3)^2.
\label{sakharov}
\end{equation}

We can analyse  eqs. \eqs{sandy} following \cite{vw}. They admit a
trivial  solution $B^{+}_{11}=B^{+}_{12}=C=\phi=0$, where  the
gauge group is unbroken and which reduces at low energies to  the
$\cn\,$=$\,1$ pure
$SU(N)$ gauge theory (which has $N$ discrete vacua).  In
addition to this trivial vacuum, eqs. \eqs{sandy} admit a non-trivial
fixed point (the irreducible embedding in \cite{vw}) in which $\phi$, and
therefore $B^{+}_{11}$, $B^{+}_{12}$ and
$C$, are not zero.  On flat space-time this solution corresponds to a Higgs
vacuum in which the gauge group is completely broken.  All these vacua 
have a mass gap: the irreducible
embedding is a Higgs vacuum, while the presence of a mass gap in
the trivial vacua is a well-known feature of the $\cn=1$ SYM theory.
When $N$ is prime, these are the only relevant vacua of the
$\cn=1$ theory. There are other, more general, solutions to 
\eqs{sandy} which leave different subgroups of $G$ unbroken.
However, in all these solutions the unbroken gauge group contains
$U(1)$ factors and one expects on general grounds that they
should not contribute to the partition function \cite{vw}. 
On the other hand, when $N$ is not prime, there are additional
contributions coming from embeddings for which the unbroken gauge 
group is $SU(d)$, where $d$ is a positive divisor of $d$. The
low-energy theory is again an $\cn=1$ $SU(d)$ gauge theory wihout
matter with $d$ massive discrete vacua. 
Now in the long-distance limit, the partition function
is given as a finite sum over the contributions of the discrete
massive vacua of  the resulting
$\cn=1$ theory.  For
$G=SU(N)$  the number of such vacua is given by the sum of the
positive divisors of $N$ 
\cite{donagi}. The contribution of each vacuum is universal
(because of the  mass gap), and can be fixed by comparing to
known mathematical results \cite{vw}.  

This analysis is  valid only on hyper-K\"ahler 
manifolds, as in this situation one can pick a section of $H^{(2,0)}$ which
vanishes nowhere. On more general K\"ahler manifolds, this picture must be 
corrected near the zeroes of the two-form $\omega$ along the lines proposed
in \cite{vw}\cite{wijmp}. 

\vskip1cm

With the mass terms added, the action $ S+\Delta L(m,\omega)$ is only 
invariant under $Q^{(m,\omega)}_1$.  To deal with the mass terms
proportional to $m$, we shall proceed as follows. We will modify the
$Q_2$ transformations by appropriately introducing mass terms (proportional
to
$m$), in such a way that $Q^{+}(m)=Q^{(m)}_1+Q^{(m)}_2$ (with mass $m$, and 
$\omega=0$ at this stage)  be a symmetry of the original action plus mass
perturbations. We will show this by proving that $L+\Delta L(m,\omega=0)$ is
actually $Q^{+}(m)$-exact. To this end we make the replacements:
\bea
\{Q_2, \tilde\psi^{+}_{22}\}=2i[ B^{+}_{22},\phi]&\too& 
\{Q^{(m)}_2, \tilde\psi^{+}_{22}\}=2i[ B^{+}_{22},\phi]+\raiz mB^{+}_{22}
,\ret
\left\{Q_2, \tilde\psi^{+}_{12}-\fr{i}{2}\zeta\right\}=
-2i[\phi,B^{+}_{12}-iC]&\too&
\left\{Q^{(m)}_2,\tilde\psi^{+}_{12}-\fr{i}{2}\zeta\right\}\ret 
&=&-2i[\phi,B^{+}_{12}-iC] -2\raiz m(B^{+}_{12}-iC)
\ret
\label{mesa}
\eea
(the rest of the transformations remain the same). 
Notice that still $(Q^{(m)}_2)^2=0$. Next we spell out the
$Q^{+}(m)=Q^{(m)}_1+Q^{(m)}_2$-transformations: 

\bea 
&&[Q^{+}(m), B^{+}_{11}\,] = \raiz\,\tilde\psi^{+}_{11},\ret
&&[Q^{+}(m), B^{+}_{22}\,] = \raiz\,\tilde\psi^{+}_{22},\ret  
&&[Q^{+}(m), B^{+}_{12}\pm iC] =\raiz\left(\tilde\psi^{+}_{12}\pm 
\fr{i}{2}\zeta\right),\ret
&&\{Q^{+}(m), \tilde\psi^{+}_{11}\}=2i[ B^{+}_{11},\phi]-\raiz\, m
B^{+}_{11},\ret
&&\{Q^{+}(m), \tilde\psi^{+}_{22}\}=2i[ B^{+}_{22},\phi]+\raiz\, m
B^{+}_{22},\ret
&&\left\{Q^{+}(m), \tilde\psi^{+}_{12}\pm \fr{i}{2}\zeta\right\}=
2i[B^{+}_{12}\pm iC,\phi\,]
\pm\raiz\, m (B^{+}_{12}\pm iC).\ret
\label{silla}
\eea
On any of these fields (which we denote generically by $X$) 
the charge $Q^{+}(m)$ satisfies the algebra:
\begin{equation}
(Q^{+}(m))^2 X=2\raiz\, i\,[X,\phi\,]+2mqX, 
\label{equivariant}
\end{equation}
where  $q =-1$  for $B^{+}_{11}$, $\tilde\psi^{+}_{11}$,  
$B^{+}_{12}-iC$ and $\tilde\psi^{+}_{12}-\fr{i}{2}\zeta$, and  $q =+1$
for $B^{+}_{22}$, $\tilde\psi^{+}_{22}$,  
$B^{+}_{12}+iC$ and $\tilde\psi^{+}_{12}+\fr{i}{2}\zeta$. 
Notice that these charge assingments  are compatible with the $U(1)$ symmetry
that we discussed above, and in fact one can see the ``central charge" 
$\delta_q X=2mqX$ arising in the algebra \eqs{equivariant} as an
infinitesimal 
$U(1)$ transformation with parameter $m$. 

We also extend the $Q^{+}(m)$ transformation off-shell by declaring  
its action on $\tilde H_{\alpha\dalpha}$ to be:
\bea
&&[Q^{+}(m),\tilde H_{1\dalpha}]=\ldots -2m\tilde\chi_{1\dalpha},\ret
&&[Q^{+}(m),\tilde H_{2\dalpha}]=\ldots +2m\tilde\chi_{2\dalpha}.\ret
\label{papaya}
\eea
In this way, $Q^{+}(m)$ closes on $\tilde
H_{1\dalpha},\tilde\chi_{1\dalpha}$  with $q=-1$, and on $\tilde
H_{2\dalpha},\tilde\chi_{2\dalpha}$ with $q=+1$.
 
Let us now prove that the above modifications suffice to render the $m$ mass
terms $Q^{+}(m)$-exact:   
\begin{equation}
\fr{1}{{2\raiz}}\,m(\tilde\psi^{+}_{22}B^{+}_{11}-
\tilde\psi^{+}_{11}B^{+}_{22})
\buildrel Q^{+}(m)\over \longrightarrow m^2 B^{+}_{11}B^{+}_{22}+
m\tilde\psi^{+}_{11}\tilde\psi^{+}_{22}-\raiz\,
imB^{+}_{22}[B^{+}_{11},\phi],
\label{Rimbaud}
\end{equation}
and 
\bea
&&-\fr{1}{{2\raiz}}m\left\{(B^{+}_{12}-iC)\left(\tilde\psi^{+}_{12}+
\fr{i}{2}\zeta\right)-
(B^{+}_{12}+iC)\left(\tilde\psi^{+}_{12}-\fr{i}{2}\zeta\right)\right\}
\buildrel Q^{+}(m)\over \longrightarrow \ret
&&-m^2 \vert B^{+}_{12}+iC\vert^2+\raiz im\phi[B^{+}_{12}+iC,B^{+}_{12}-iC]
+m\left(\tilde\psi^{+}_{12}+\fr{i}{2}\zeta\right)\left(\tilde\psi^{+}_{12}-
\fr{i}{2}\zeta\right). 
\ret
\label{Apollinaire}
\eea
Notice, moreover, that these terms are likewise   
$Q^{(m,\omega)}_{1}$-exact:
\begin{equation}
-\fr{1}{{\raiz}}m\tilde\psi^{+}_{11}B^{+}_{22}
\,\buildrel Q^{(m,\omega)}_1\over \longrightarrow m^2 B^{+}_{11}B^{+}_{22}+
m\tilde\psi^{+}_{11}\tilde\psi^{+}_{22}-\raiz imB^{+}_{22}[B^{+}_{11},\phi],
\label{isa}
\end{equation}
and
\bea
&&-\raiz m\left\{(B^{+}_{12}-iC)\tilde\psi^{+}_{12}\right\}\,
\buildrel Q^{(m,\omega)}_1\over \longrightarrow 
-m^2 \vert B^{+}_{12}+iC\vert^2+\raiz im\phi[B^{+}_{12}+iC,B^{+}_{12}-iC]
\ret&&+m\left(\tilde\psi^{+}_{12}+\fr{i}{2}\zeta\right)\left(\tilde
\psi^{+}_{12}-\fr{i}{2}\zeta\right). 
\ret
\label{ennes}
\eea
But we have not yet reproduced the terms -- see \eqs{uva}:
$-\raiz im\bar\phi[B^{+}_{12}+iC,B^{+}_{12}-iC]$, 
$-2\raiz imB^{+}_{11}[B^{+}_{22},\bar\phi]$
and $m\tilde\chi_{2\dalpha}\tilde\chi_1{}^\dalpha$. These come from pieces 
already present in the gauge fermion. Explicitly,  
\begin{equation}
 \tr
\left\{-\fr{1}{4}
\tilde \chi^{\dalpha\alpha}\tilde H_{\alpha\dalpha}\right\} 
\buildrel Q^{+}(m)\over \longrightarrow 
m\tilde\chi_{2\dalpha}\tilde\chi_1{}^\dalpha ,
\label{sumathi}
\end{equation}
and 
\bea
&&\tr \left\{
\fr{i}{{2}}\bar\phi[\tilde\psi_{\alpha\beta},B^{\alpha\beta}]
+\fr{i}{{2}}[\zeta,C]\right\}\buildrel Q^{+}(m) \over \longrightarrow
\ret &&-\raiz im\bar\phi[B^{+}_{12}+iC,B^{+}_{12}-iC] 
-2\raiz imB^{+}_{11}[B^{+}_{22},\bar\phi].\ret
\label{fujita}
\eea

The analysis of the terms containing the $(2,0)$-form $\omega$ can be carried
out essentialy as in the Donaldson-Witten theory. The perturbation breaks up 
into a
$ Q^{(m,\omega)}_1$-exact piece:

\begin{align}
\{ Q^{(m,\omega)}_1,&\tr(\raiz i\omega\bar\phi\chi^{+}_{11})\}=\ret
&\tr\Biggl\{-2\raiz i\omega\bar\phi[B^{+}_{12}+iC,B^{+}_{11}] 
-4\vert\omega\vert^2\phi\bar\phi
-2i\omega\chi^{+}_{11}\left(\half\eta-i\chi^{+}_{12}\right)\Biggr\},\ret
\label{pinha}
\end{align}
and an operator of ghost number $+2$:
\begin{equation}
J(\bar\omega)=\int_X \tr\left(2\raiz i\bar\omega
\phi[B^{+}_{12}-iC,B^{+}_{22}]+\bar\omega\psi_{2\dalpha}
\psi_2{}^{\dalpha}\right).
\label{cereza}
\end{equation}
Equation \eqs{cereza} is not very useful as it stands. To rewrite it in a 
more convenient
form we note that from (\ref{papagayo}) it follows that:
\begin{equation}
2\raiz i\bar\omega\tr\left\{\phi[B^{+}_{12}-iC,B^{+}_{22}]\right\}=
\raiz i\tr\left(\{Q^+,\bar\omega\phi\chi^+_{22}\}\right)-\raiz i\bar\omega
\tr\left(\phi F^+_{22}\right).
\label{ricardos}
\end{equation}
Hence, 
\begin{equation}
J(\bar\omega)=\{Q^+,\cdots\}+\underbrace{\int_X \bar\omega\tr\left(
\psi_{2\dalpha}\psi_2{}^{\dalpha}-\raiz i\phi
F^+_{22}\right)}_{I(\bar\omega)} ,\qquad [Q^+,I(\bar\omega)]=0.
\label{canon}
\end{equation}
Moreover, as the $m$ mass term does not enter in any of the above
calculations, the results also hold for $Q^+(m)$.

The preceding analysis implies that if we denote vacuum expectation values
in the twisted theory (which has topological symmetry $Q^{+}$and action $L$) 
by $\langle \ldots\rangle$, in the completely perturbed theory (with action 
$L+\Delta L(m,\omega)$ and symmetry $Q^{(m,\omega)}_1$) by 
$\langle \ldots\rangle_{m,\omega}$, and in the equivariantly extended theory 
(with action $L+\Delta L(m)$ and symmetry $Q^{+}(m)$) by  
$\langle \ldots\rangle_{m}$, the situation for the partition function is 
the following:
\begin{equation}
\langle 1\rangle_{m,\omega}=
\left\langle \ex^{-J(\bar\omega)}\ex^{-\Delta L(m)}\right\rangle =
\left\langle \ex^{-J(\bar\omega)}\right\rangle_{m}. 
\label{merlin}
\end{equation}
 
In the first equality we have discarded the $Q^{(m,\omega)}_1$-exact term 
\eqs{pinha}. Notice that it is also possible, for the same reason, to
discard the  terms in \eqs{isa} and \eqs{ennes}. This leaves the
$Q_1^{(m,\omega)}$-closed action
$L+\Delta^{(1)}+J(\bar\omega)$, where $\Delta^{(1)}$ are the mass terms
\eqs{sumathi} and \eqs{fujita}, i.e.
\begin{equation}
\Delta^{(1)}= m\int_X\tr\left(\tilde\chi_{2\dalpha}\tilde\chi_1^\dalpha-
\raiz i\bar\phi[B^{+}_{12}+iC,B^{+}_{12}-iC]-2\raiz i
B^{+}_{11}[B^{+}_{22},\bar\phi]\right).
\label{petete}
\end{equation}
Notice that $\Delta^{(1)}$ has ghost number $-2$, while $J(\bar\omega)$ has
ghost number $+2$. Also $L+\Delta^{(1)}$ is $Q^{+}(m)$-closed
(in fact it is $Q^{+}(m)$-exact up to a $\theta$-term). 
Hence, we can trade
$J(\bar\omega)$ for 
$\{Q^{+}(m),\cdots\}+I(\bar\omega)$ and discard the $Q^+(m)$-exact piece in
\eqs{ricardos}. We are left with the action 
\begin{equation}
\underbrace{L+\Delta^{(1)}}_{Q^+(m)-{\rm exact}}+\underbrace{I(\bar\omega)}_{
Q^+(m)-{\rm closed}}.
\label{menelao}
\end{equation}
Now, as noted in \cite{wijmp} in a closely related context,
$I(\bar\omega)$ (or rather $J( \bar\omega)$) is the $F$-term of the chiral
superfield $\Phi_3$;  therefore, it cannot develop a vacuum expectation
value if supersymmetry is to remain unbroken. Strictly speaking, this
applies to
$\langle
\psi_2\psi_2\rangle$. As for the remaining term
$\phi[B^{+}_{12}-iC,B^{+}_{22}]$, one can readily check that  it vanishes on
the moduli space. Hence,                     
\begin{equation}
\left\langle \ex^{I(\bar\omega)}\right\rangle_m=\langle 1\rangle_m=
\left\langle
\ex^{-\Delta^{(1)}}\right\rangle.
\label{unbelievable}
\end{equation}

Finally, since $\Delta^{(1)}$ has ghost number $-2$, its vev in the original
theory must vanish as well, if the ghost number symmetry is to remain
unbroken.  Hence, under these assumptions, the partition function is indeed 
invariant under the perturbation as stated in \cite{vw}.

\subsection{Equivariant extension of the Thom form}

On K\"ahler manifolds there is a $U(1)$ symmetry \eqs{uunosim} 
acting on the moduli space. 
This symmetry was already noted in \cite{vw} within the discussion of the 
vanishing theorem, which guarantees localization on the moduli space of ASD 
connections, but not  further use of it was made. 
We have discussed it in the previous section in connection with the mass
perturbations. Its action on the different fields is the following: 

\begin{gather}
\begin{cases}
B^{+}_{11}\to \ex^{-it}B^{+}_{11},\\
B^{+}_{12}-iC\to  \ex^{-it} (B^{+}_{12}-iC),\\ 
\tilde\psi^{+}_{11}\to  \ex^{-it}\tilde\psi^{+}_{11},\\
\tilde\psi^{+}_{12}-\fr{i}{2}\zeta
\to \ex^{-it}(\tilde\psi^{+}_{12}-\fr{i}{2}\zeta),\\ 
\tilde\chi_{1\dalpha }\to \ex^{-it}\tilde\chi_{1\dalpha},\\
\tilde H_{1\dalpha}\to \ex^{-it}\tilde H_{1\dalpha},
\end{cases}
\qquad
\begin{cases}
B^{+}_{22}\to  \ex^{it}B^{+}_{22},\\ 
B^{+}_{12}+iC\to \ex^{it} (B^{+}_{12}+iC),\\
\tilde\psi^{+}_{22}\to \ex^{it}\tilde\psi^{+}_{22},\\
\tilde\psi^{+}_{12}+\fr{i}{2}\zeta\to 
\ex^{it}(\tilde\psi^{+}_{12}+\fr{i}{2}\zeta),\\
\tilde\chi_{2\dalpha}\to \ex^{it}\tilde\chi_{2\dalpha},\\
\tilde H_{2\dalpha}\to \ex^{it}\tilde H_{2\dalpha}.
\end{cases}
\label{circulo}
\end{gather}

The gauge field $A$, the antighosts $\chi^{+}_{\alpha\beta}$ and $\eta$,
and the scalar fields $\phi$ and $\bar\phi$, carry no charge under this
$U(1)$. These transformations can be thought of as defining the
one-parameter flow associated to the action on the field space ${\cal M}$ of
the following vector field $X_{\cal M}\in T_{(A,B^+,C)}{\cal M}$:
\begin{equation}
X_{\cal M}=\left(0,-iB^{+}_{11}, iB^{+}_{22}, -i(B^{+}_{12}-iC),
i(B^{+}_{12}+iC)\right).
\label{vectorcillo}
\end{equation}

From the viewpoint of the Mathai-Quillen formalism, the unperturbed twisted
theory provides a representation of the
${\cal G}$-equivariant de Rham cohomology (in the Cartan model) on the moduli
space. However, the formulation is not equivariant with respect to the $U(1)$
action. In other words, the perturbed action is not invariant (i.e. it is not
equivariantly closed) under the unperturbed twisted supercharge. On the other
hand, it is invariant under the perturbed twisted supercharge. In fact,  the
twisted supercharge $Q^{+}(m)$ of the perturbed theory can be interpreted as
the generator of the $ U(1)$-equivariant extension of the ${\cal
G}$-equivariant de Rham cohomology on the moduli space. This connection
between  massive extensions of twisted  supersymmetric
theories and equivariant cohomology was exploited in \cite{eq}, where the
explicit construction leading to the idea of the  equivariant extension was
carried out in detail. In what follows, we will try  to adapt the
construction in \cite{eq} to our problem. We intend to be as sketchy as
possible, and therefore refer the reader  to the work cited above for the
minute details of the construction. 

The idea underlying the construction is the following. Prior to the 
perturbation, we have a topological field theory which admits a 
Mathai-Quillen description with BRST charge $Q^{+}$. This means, among other
things, that the corresponding Lagrangian is a $Q^{+}$-commutator.  After
adding the mass terms proportional to $m$, it is possible to modify the
$Q^{+}$ transformation laws so that the perturbed Lagrangian can be written 
as a $Q^{+}(m)$-commutator as well, where $Q^{+}(m)$ are the modified
topological transformations. In view of this, it would be tempting to assume 
that there has to be a standard Mathai-Quillen construction associated to the
new topological theory. However, the perturbation has not changed the
geometrical setting of the problem, so there is a priori no reason why the 
Mathai-Quillen formulation should change at all. In fact, it does not, and
it turns out that the perturbed theory admits no standard Mathai-Quillen
formulation. However, as pointed out in \cite{eq}, the  formalism allows a
natural generalization in those situations  in which there is an additional
symmetry group acting on the moduli space. The geometrical construction
involved is an equivariant extension \cite{atibott} of the Thom form of
${\cal E}$ within the framework of the Mathai-Quillen formalism.    
 
The Mathai-Quillen formalism provides an explicit representative of the Thom
form of the oriented vector bundle ${\cal E}={\cal M}\times_{\cal G}{\cal
F}$.  The bundle ${\cal E}$ is awkward to work with, and it is preferable to
work equivariantly, i.e. to regard ${\cal E}$ explicitly as an associated
vector  bundle to the ${\cal G}$-principal vector bundle ${\cal M}\times
{\cal F}\to {\cal E}$. The Mathai-Quillen representative of the Thom form of
${\cal E}$ is
${\cal G}$-equivariantly closed and basic on ${\cal M}\times {\cal F}$ (and
hence descends naturally to ${\cal E}$). In the Weil model for the ${\cal
G}$-equivariant cohomology of ${\cal E}$, the Mathai-Quillen form is an
element in ${\cal W}({\bf g})\otimes\Omega^{*}(F)$ (${\cal W}({\bf g})$  is
the Weil algebra of $G$) given by \cite{moore}:
\begin{equation}
U=\ex^{-\vert x\vert^2}\int D\chi\,{\hbox{\rm exp}}\Biggl
(\fr{1}{4}\chi_i  K_{ij}\chi_j +i\chi_i(dx_i+\theta_{ij}x_j)\Biggr).
\label{thom}
\end{equation}
In (\ref{thom}) $x_i$ are orthonormal coordinates on the fibre ${\cal F}$,
and 
$dx_i$ are their corresponding differentials. The $\chi_i$ are Grassmann
orthonormal coordinates for the fibre, while $K$ and $\theta$ are the
generators of ${\cal W}({\bf g})$. The Chern-Weil homomorphism, which
essentially substitutes the universal realizations $K$ and $\theta$ by the 
actual curvature and connection in ${\cal M}\times {\cal F}$, gives the link
between the Universal representative $U$ and the Thom form $\Phi({\cal
E})$.   The important point is that while $U$ is ${\cal G}$-equivariantly
closed by  construction, it is not equivariantly closed with respect to the
$U(1)$ action. It seems natural to look for a redefinition of the
representative (\ref{thom}), which is  $U(1)$-equivariantly closed.
The equivariant extension of $U$ with respect to the $U(1)$ action
simply amounts to finding a suitable form $p$ such that $U+p$ is 
$U(1)$-equivariantly closed \cite{atibott}. Within  the framework of the
Mathai-Quillen formalism this amounts to replacing the curvature $K$ with a
new equivariant curvature 
$K_{U(1)}$ \cite{eq}, which is just the original curvature $2$-form $K$ plus
an operator $L_\Lambda$ involving the infinitesimal $U(1)$ action and the
connection $1$-form $\theta$. In the Cartan model, which is the best suited
to topological field theories, the connection form is set to zero, and hence
the equivariant extension of the curvature is just the original one plus an
operator implementing the infinitesimal $U(1)$ action. This may sound rather 
abstract, so we now proceed to the
actual construction. The main ingredients are a $U(1)$ action defined on the
moduli space and the fibre ${\cal F}$, under which the metrics on both the
moduli space and the fibre must be invariant, while the section $s:{\cal
M}\to {\cal V}$ has to transform equivariantly; that is, if $\phi^{\cal
M}_t$ and 
$\phi^{\cal F}_t$ denote the action of $U(1)$ on ${\cal M}$ and 
${\cal F}$ respectively, then 
\begin{equation}
s\cdot\phi^{\cal M}_t=\phi^{\cal F}_t\cdot s 
\label{hola}
\end{equation}
This can be easily verified in the present problem in view of the form of $s$
\eqs{ecuaciones} and the $U(1)$ actions \eqs{circulo}. As for the metrics,
it suffices to show that for two vector fields $(0,X^{+},x)$ and
$(0,Y^{+},y)$, their scalar product is invariant under the $U(1)$ action
\eqs{circulo}. According to our conventions, $\tr
X^{+}_{\alpha\beta}Y^{+\alpha\beta}=-4\tr X\wedge *Y$, so a natural
definition for the metric on the field space would be as follows ($\langle |
\rangle$ denotes the scalar product on $T{\cal M}$):
\bea
&&\left\langle (0,X^{+},x)|(0,Y^{+},y)\right\rangle = 
-\int_X \tr \left( X^{+}_{\alpha\beta}
Y^{+\alpha\beta}\right)+2\int_X \tr *(xy)=\ret &&-\int_X\tr\,\big(X^{+}_{11}
Y^{+}_{22}+X^{+}_{22}Y^{+}_{11}\big)+\int_X \tr\, \big[ (X^{+}_{12}+ix)
(Y^{+}_{12}-iy)+(X^{+}_{12}-ix)(Y^{+}_{12}+iy)\big],
\ret
\label{cernch}
\eea
which is indeed invariant under the $U(1)$ action. 

To incorporate the $U(1)$
action to the Mathai-Quillen construction for the theory, we
modify the $Q^{+}$ transformations of the ghosts and the
auxiliary fields charged under 
$U(1)$ by replacing the curvature $\phi$ with its equivariant extension 
$\phi(t)=\phi+{\cal L}_t$, where  ${\cal L}_t$ generates on the fields an 
infinitesimal $U(1)$ transformation. 
According to \eqs{circulo}, this affects only   
$\tilde\psi^{+}_{\alpha\beta}$, $\zeta$ and $\tilde H_{\alpha\dalpha}$. In
view of (\ref{papagayo}), the new transformations read:
\bea 
\{Q^{+}(t),\tilde\psi^{+}_{11}\}&=&2i([ B^{+}_{11},\phi]-itB^{+}_{11}),\ret
\{Q^{+}(t), \tilde\psi^{+}_{22}\}&=&2i([B^{+}_{22},\phi]+
itB^{+}_{22}),\ret
\left\{Q^{+}(t), \tilde\psi^{+}_{12}\pm \fr{i}{2}\zeta\right\}&=&
2i\bigl([B^{+}_{12}\pm iC,\phi]\pm it(B^{+}_{12}\pm iC)\bigr),\ret
\big[Q^{+}(t),\tilde H_{1\dalpha}\big]&=&
2\raiz i([\tilde\chi_{1\dalpha},\phi]-it\tilde\chi_{1\dalpha})-\raiz 
[Q^{+},s_{1\dalpha}]
,\ret
\big[Q^{+}(t),\tilde H_{2\dalpha}\big]&=&
2\raiz i([\tilde\chi_{2\dalpha},\phi]+it\tilde\chi_{2\dalpha})-\raiz 
[Q^{+},s_{2\dalpha}],\ret
\label{final}
\eea
If we now set $t=-m/\raiz$ we see that eqs. (\ref{final}) reduce
precisely to  the $Q^{+}(m)$ transformations \eqs{silla} and
\eqs{papaya}. The transformations (\ref{final}), when applied to the
gauge fermion
\bea
\Psi&=& \fr{1}{ e^2_0}\int_X  d^4 x\,\sqrt{g}\, \tr
\Bigl\{\,-\fr{1}{4}
\tilde \chi^{\dalpha\alpha}\bigl (\,\tilde H'_{\alpha\dalpha}-\raiz 
s_{\alpha\dalpha}\,\bigr )
-\fr{1}{ 4}
\chi^{\alpha\beta}\bigl (\, H'_{\alpha\beta}-s_{\alpha\beta}
\,\bigr )\,\Bigr\}
\ret  
&+&\fr{1}{ e^2_0}\int_X  d^4 x\,\sqrt{g}\,
\tr
\Bigl\{\, \fr{1}{{2\raiz}}\bar\phi\,\bigl
(\,\deriv_{\alpha\dalpha}\psi^{\dot\alpha\alpha}+i\raiz
\,[\tilde\psi_{\alpha\beta},B^{\alpha\beta}] -i\raiz\,[\zeta,C]\,\bigr
)\,\Bigr\}
\ret
&-&\fr{1}{ e^2_0}\int_X  d^4 x\,\sqrt{g}\, \tr
\bigl\{\,\fr{i}{4}\eta[\phi,\bar\phi]\,\bigr\},\ret
\eea
reproduce the original unperturbed action plus the mass
terms
\eqs{sumathi} and \eqs{fujita}. To obtain the remaining mass terms we note
that, as is standard in topological (cohomological) field theories, there
remains the possibility of adding to the action a $Q^{+}(t)$-exact piece
without  -- hopefully -- disturbing the theory. As discussed in \cite{eq},
the requisite  piece can be interpreted as the equivariantly-exact
differential form which is  conventionally added to prove localization in
equivariant integration. It  has the form $\{Q^{+}(t),\omega_{X_{\cal
M}}\}$, where
$\omega_{X_{\cal M}}$  is the differential form given by $\omega_{X_{\cal
M}}(Y)=
\langle X_{\cal M}|Y\rangle$, $Y$ being a vector field on ${\cal M}$. In 
view of the form of the vector field $X_{\cal M}$ \eqs{vectorcillo} and of 
the metric \eqs{cernch}, and keeping in mind that the ghosts 
$(\psi, \tilde\psi^{+},\zeta)$ provide a basis of differential forms on 
${\cal M}$, this form gives a contribution 
\begin{align}
\Biggl\{ Q^{+}(t), -\fr{it}{2}\int_X\tr\Bigr(\tilde\psi^{+}_{22}
(-iB^{+}_{11})+\tilde\psi
^{+}_{11}(iB^{+}_{22})\Bigl)&+\fr{it}{2}\int_X\tr\bigg((-i)(B^{+}_{12}-iC)
\left(\tilde\psi^{+}_{12}+\fr{i}{2}\zeta\right)\ret &+i(B^{+}_{12}+iC)
\left(\tilde\psi^{+}_{12}-\fr{i}{2}\zeta\right)\bigg)\Biggr\}.
\ret
\label{morel}
\end{align}

But these are precisely the terms
\eqs{Rimbaud} and
\eqs{Apollinaire}, which as we have seen give correctly the remaining mass
terms. 

\section{The partition function for $G=SU(N)$}

In this section we will consider the Vafa-Witten theory for gauge 
group $SU(N)$. As all the fields are in the adjoint representation of the
gauge group, it is  possible to consider non-trivial gauge configurations 
in $SU(N)/\IZ_N$ and  compute the partition function for a fixed
value of the 't Hooft flux $v\in  H^{2}(X,\IZ_N)$. 
We know that the twisted $\cn=4$ supersymmetric action breaks up into a
$Q^{+}$-exact piece  plus a topological term proportional to  the
instanton number of the gauge configuration,
\begin{equation} 
{\cal S}_{\mbox{\rm\tiny twisted}}= \{Q^{+},\Psi\}-2\pi ik_v\tau_0,
\label{sunramala}
\end{equation} 
with $k_v$ the instanton number of
a gauge bundle with 't Hooft flux $v$. This is an integer for
$SU(N)$ bundles ($v=0$), but for non-trivial
$SU(N)/\IZ_N$ bundles with $v\not=0$ one has 
\begin{equation}
k_v= -\frac{N-1}{2N}\, v\cdot v\quad\mod~\IZ, 
\label{instnum}
\end{equation}
where $v\cdot v$ stands for $\int_X v\wedge v$ . 
Therefore, as pointed out in 
\cite{vw}, one  would expect the
$SU(N)$ partition function to be invariant to $\tau_0\to\tau_0+1$, 
while the $SU(N)/\IZ_N$ theory should be only invariant under 
$\tau_0\to\tau_0+2N$ on arbitrary four-manifolds, and under
$\tau_0\to\tau_0+N$ on spin four-manifolds (where $v\cdot v$ is even.)
In any case, for odd $N$, we have invariance under
$\tau_0\to\tau_0+N$ on any four-manifold.  

As we have argued above,
one can compute the partition function  in terms of the vacuum degrees
of freedom of the 
${\cal N}=1$ theory which results from giving bare masses to all
the three  chiral multiplets of the ${\cal N}=4$
theory. The partition functions 
on $K3$ for gauge group $SU(N)$ and trivial 't Hooft
fluxes have been computed by Vafa and collaborators 
in \cite{estrings}. We will extend their results to 
arbitrary 't Hooft fluxes and compute the partition function on more
general K\"ahler four-manifolds. These results have
been presented in \cite{faro}\cite{sun}.

\subsection{The partition function on $K3$}

As a first step towards the derivation of the formula for the
partition  function we will consider the theory on $K3$, where
some explicit results are already available. For $X$ a $K3$
surface the canonical divisor is trivial, so there exists a
nowhere vanishing section of the bundle of $(2,0)$ forms.
Therefore, the mass perturbation $\omega$   does not vanish
anywhere and the above analysis of the vacuum  structure of the
$\cn=1$ theory carries over without change. 

 The structure of the partition function for trivial 't Hooft
flux was conjectured in \cite{vw}. This conjecture has been
confirmed in
\cite{estrings} by studying the effective theory on $N$
coincident $M5$-branes wrapping around $K3\times T^2$. The
partition function for zero 't Hooft flux is {\sl almost} a Hecke
transformation  of order $N$ \cite{apostol} of
$G(\tau)=\eta(\tau)^{-24}$, with
$\eta(\tau)$ the Dedekind function  -- see eq. (3.7) in
\cite{estrings}:    

\begin{equation} 
Z_{v=0}\equiv Z_N=\frac{1}{N^2}\sum_{ \genfrac{}{}{0pt}{2}{
0\leq a,b,d\in\IZ}{ad=N,~  b<d}}
d\, G\left(\frac{a\tau_0+b}{d}\right).
\label{suocho}
\end{equation}
Notice that the number of terms in \eqs{suocho} equals the sum of
the positive divisors of $N$ as we mentioned above. When $N$ is
prime the formula is considerably simpler 
\begin{equation}  Z_{v=0}=\frac{1}{N^2}G(N\tau_0) + \frac{1}{N}
\sum_{m=0}^{N-1}\,
 G\left(\frac{\tau_0+m}{N}\right).
\label{suochoo}
\end{equation} There are $N+1$ terms, the first one corresponding
to the irreducible embedding, and the other $N$ to the vacua of
the $\cn=1$
$SU(N)$ SYM theory.

The $SU(N)$ partition function is defined from \eqs{suocho} as
$Z_{SU(N)}=
\frac{1}{N} Z_{v=0}$. From it, the $SU(N)/\IZ_N$ partition
function
$Z_{SU(N)/\IZ_N}= \sum_v Z_v$ can be obtained via a modular
transformation \cite{vw} (see the appendix in sect. \ref{apendice2} for
details)

\begin{equation}
 Z_{SU(N)/\IZ_N}(\tau_0)= N^{\chi/2}\left(\frac{\tau_0}{i}
\right)^{\chi/2}Z_{SU(N)}(-1/\tau_0)=
\frac{1}{N^2}\sum_{ \genfrac{}{}{0pt}{2}{a,b,d}{p=\gcd\,(b,d)}} 
d^{12} p^{11}
G\left(\frac{a\tau_0+b}{d}\right).
\label{sudiez}
\end{equation}
Notice the first equality in \eqs{sudiez},  
which is, up to some correction factors which vanish in flat
space,  the original Montonen-Olive conjecture. 

To generalize \eqs{suochoo} for gauge configurations with
arbitrary 't Hooft flux we proceed as in \cite{vw}. The $N$
contributions coming from the $\cn=1$ pure gauge theory vacua are
related by an anomalous  chiral symmetry which takes
$\tau_0\to\tau_0+1$. The anomaly is 
$2Nk_v-(N^2-1)\left(\frac{\chi+\sigma}{4}\right)=-(N-1)v\cdot
v+\cdots$,  which is half the anomaly in Donaldson-Witten theory.
Hence,  the contributions from each vacuum pick anomalous phases 
$\ex^{-i\pi m k_v}=\ex^{i\pi\frac{N-1}{N}m v^2}$.  As for the
contribution coming from the irreducible embedding, modular
invariance  requires that  it vanishes unless $v=0$. Hence, 

\begin{equation} Z_v = \frac{1}{N^2} G(N\tau_0)\delta_{v,0} +
\frac{1}{N}
\sum_{m=0}^{N-1}\,
\ex^{i\pi\frac{N-1}{N}m v^2}\, G\left(\frac{\tau_0+m}{N}\right). 
\la{sudiezsiete}
\end{equation} The $Z_v$ transform into each other under the
modular group as predicted in 
\cite{vw} 
\bea Z_v(\tau_0+1)\!\!\! &=&\!\!\!\ex^{-i\pi\frac{N-1}{N}v^2} 
Z_v(\tau_0), 
\ret\ret  Z_v(-1/\tau_0)\!\!\! &=&\!\!\! N^{-11}\left(\frac{\tau_0}{i}
\right)^{-12}
\sum_u \ex^{\frac{2i\pi u\cdot v}{N}} Z_u(\tau_0). \ret 
\label{sucuatro2}
\eea 
To evaluate the sum over $u$ we use formulas \eqs{terran3}
and
\eqs{terran4} in the appendix to this chapter\footnote{Note that $K3$ has
$\chi=24$,
$\sigma=-16$, $b_1=0$ and $b_2=22$.}.

By summing over $v$ in \eqs{sudiezsiete} we can check
\eqs{sudiez} 
\bea Z_{SU(N)/\IZ_N}\!\!\!&=&\!\!\! \sum_v Z_v
\ret\!\!\!&=&\!\!\!\frac{1}{N^2} G(N\tau_0)+ N^{21} G(\tau_0/N) +
N^{10}\sum_{m=1}^{N-1}\,
 G\left(\frac{\tau_0+m}{N}\right). \ret
\la{supino}
\eea
The above results only hold for prime $N$. We have not been able
to find the appropriate generalization for arbitrary $N$.

\hyphenation{Mont-o-nen}

\subsection{More general K\"ahler manifolds}

On more general K\"ahler manifolds the spatially dependent mass
term vanishes where $\omega$ does, and  we will assume as in
\cite{vw}\cite{wijmp} that $\omega$  vanishes with multiplicity
one on a union of disjoint,  smooth complex curves $C_j$,
$j=1,\ldots n$ of genus $g_j$ which represent the  canonical
divisor $K$ of $X$. The vanishing of $\omega$ introduces
corrections  involving $K$ and additional modular functions 
whose precise form is not known a priori. In the $G=SU(2)$ case, 
each of the
$\cn=1$ vacua bifurcates along each of the components $C_j$ of
the  canonical divisor into two strongly coupled massive vacua.
This vacuum  degeneracy is believed to stem \cite{vw}\cite{wijmp}
from the spontaneous  breaking of a $\IZ_2$ chiral symmetry which
is unbroken in bulk. This is exactly the same pattern that arises
in all known examples of twisted $\cn=2$ theories with gauge
group $SU(2)$ as the Donaldson-Witten theory and its 
generalizations \cite{polynom}\cite{moorewitten}\cite{wijmp}. 
This in turn
seems to be related to the possibility of rewritting the corrections
near the canonical divisor in terms of the Seiberg-Witten
invariants
\cite{monopole}. In fact, it is known that the Vafa-Witten
partition function for $G=SU(2)$ can be rewritten in terms of the
Seiberg-Witten  invariants \cite{coreatres}.

The form of the corrections for $G=SU(N)$ is more involved. From
related results on Donaldson-Witten theory \cite{mmtwo} we know
that  the higher-rank case presents some new features. We have
not been able to disentangle the structure of the vacua near the
canonical  divisor from first principles. Instead, we will
exploit the expected behaviour of the partition function under
blow-ups of $X$. This, together with the modular invariance of
the partition function will suffice to completely determine the
unknown functions.

\subsubsection{Behaviour under blow-ups}

Blowing up a point on a K\"ahler manifold $X$  replaces it  with
a new K\"ahler manifold
$\hat X$ whose second cohomology lattice is 
$H^{2}({\hat X},\IZ)= H^{2}({X},\IZ)\oplus I^{-}$, where $I^{-}$
is the  one-dimensional lattice spanned by the Poincar\'e dual of
the exceptional divisor 
$B$ created by the blow-up. Any allowed $\IZ_N$ flux $\hat v$ on
$\hat X$  is of the form $\hat v=v\oplus r$, where $v$ is a  flux
in $X$ and $r=\lambda B$, $\lambda=0,1,\ldots N-1$. The main
result concerning the $SU(2)$ partition function in \cite{vw} is that
under blowing up a point on a K\"ahler  four-manifold with
canonical divisor as above, the partition functions for  fixed 't
Hooft fluxes $\hat Z_{\hat X,\hat v}$ factorize as $Z_{X,v}$
times  a level $1$ character of the $SU(2)$ WZW model. It would be
natural to expect that the same factorization holds for $G=SU(N)$, 
but now with the level 1 $SU(N)$ characters. 
In fact, the same behaviour under blow-ups  has been proved by  
Yoshioka \cite{yoshio} for the generating function of Euler
characteristics of $SU(N)$ instanton moduli space on K\"ahler
manifolds. This should not come out as a surprise since it is
known  that, on certain four-manifolds, the partition function
of  Vafa-Witten theory computes Euler characteristics of 
instanton moduli spaces \cite{estrings}\cite{vw}. This can be
confirmed  by realizing the Vafa-Witten theory as the low-energy
theory of
$M5$-branes wrapped on $X\times T^2$ \cite{dijkfvbr}. It seems   
therefore natural to assume that the same factorization holds for
the partition function with $G=SU(N)$. Explicitly, given a 't
Hooft flux 
$\hat v=v\oplus \lambda B$, $\lambda=0,1,\ldots N-1$, on $\hat
X$,  we assume the factorization \cite{yoshio}
\begin{equation} 
Z_{\hat X,\hat v}(\tau_0)=
Z_{X,v}(\tau_0)\,\,\frac{\chi_{\lambda}(\tau_0)}{\eta(\tau_0)},
\label{factoriz}
\end{equation} 
where $\chi_{\lambda}(\tau_0)$ is the appropriate level $1$
character of
$SU(N)$ -- see the appendix to this chapter for details. This assumption
fixes almost completely the form of the partition functions. Some loose
ends can be tied up by demanding modular invariance of the
resulting expression.

\subsection{The formula for the partition function}

Given the assumptions above, and taking into account the
structure of the partition function on $K3$, we are in a position
to write down the  formula for  K\"ahler four-folds $X$ with
$h^{(2,0)}\not=0$. We will first assume that the canonical
divisor $K$ is connected and with genus 
$g-1=2\chi+3\sigma$. The formula is then

\bea
Z_v =&&\!\!\!\!\!\!\!\!\! \left (\sum_{\lambda=0}^{N-1}
\left(\frac{\chi_{\lambda}}{\eta}
\right)^{1-g} \delta_{v,\lambda [K]_N}\right)\left 
(\frac{1}{N^2} G(N\tau_0)\right)^{\nu/2} 
\ret &&\!\!\!\!\!\!\!\!\!\!\!\!\!\!+ N^{1-b_1}
\sum_{m=0}^{N-1}\left( \sum_{\lambda=0}^{N-1} 
\left(\frac{\chi_{m,\lambda}}{\eta}\right)^{1-g} 
\ex^{\frac{2i\pi}{N}\lambda v\cdot[K]_N}\right)
\ex^{i\pi\frac{N-1}{N}m v^2}\left( \frac{1}{N^2}
G\left(\frac{\tau_0+m}{N}\right)\right)^{\nu/2}, \ret
\label{suno}
\eea
where $\nu$=$\frac{\chi+\sigma}{4}$, $G(\tau)$=$\eta(\tau)^{-24}$
(with $\eta$  the Dedekind function) and $[K]_N$
is the reduction modulo $N$ of the Poincar\'e dual of $K$. 
In (\ref{suno})
$\chi_\lambda$ are the
$SU(N)$ characters at level $1$ (see the appendix \ref{apendice2}) 
and
$\chi_{m,\lambda}$  are certain linear combinations thereof
\begin{equation}
\chi_{m,\lambda}(\tau_0)=\frac{1}{N} \sum_{\lambda'=0}^{N-1}
\ex^{-\frac{2 i\pi}{N}
\lambda\lambda'}\ex^{i\pi\frac{N-1}{N}m (\lambda')^2}
\chi_{\lambda'}(\tau_0),\qquad0\leq
m,\lambda\leq N-1.
\label{suonce2}
\end{equation}

The structure of the corrections near the canonical divisor in 
\eqs{suno} suggests that the mechanism at work in this case is
not chiral symmetry  breaking. Indeed, near $K$ there is an
$N$-fold bifurcation of the vacuum, and the functions
$\chi_\lambda$, $\chi_{m,\lambda}$ (with $m$ fixed) are not
related  by a shift in $\tau_0$ as it would be the case were chiral
symmetry breaking responsible for the bifurcation.  A plausible
explanation for this bifurcation could be found in the spontaneous
breaking of the center of the gauge group (which for $G=SU(N)$
is  precisely $\IZ_N$.)  This could come about as follows. Let
us focus on the irreducible embedding.  For trivial canonical
divisor the gauge group is almost but not completely Higgsed in
this vacuum. In fact, since the scalar fields transform in the
adjoint representation of $SU(N)$, the center $\IZ_N\subset
SU(N)$ remains unbroken. The $SU(N)$ gauge threory has 
$\IZ_N$ string-like solitons \cite{gthooft} which carry
non-trivial $\IZ_N$-valued electric and magnetic quantum numbers.
If these solitons condense, the center $\IZ_N$ is completely
broken giving rise to an $N$-fold degeneracy of the vacuum. Each
vacuum is singled out by a different value of the $\IZ_N$-valued
flux. Now for non-trivial canonical divisor
$K$ as above, the irreducible vacuum separates into
$N$ vacua with magnetic fluxes $\lambda[K]_N$! One could be
tempted to speculate further and identify the surface $K$
(or the $C_j$ below) with the world-sheet of the condensed
string soliton.       

As in \cite{vw} we can generalize the above formula for the case
that  the canonical divisor consists of $n$ disjoint smooth
components 
$C_j$, $j=1,\ldots,n$ of genus $g_j$ on which $\omega$ vanishes
with multiplicity one. The resulting expression very similar to
that in 
\cite{vw}

\begin{align}
&Z_v = \left (\sum_{\vec\varepsilon}
\delta_{v,w_N(\vec\varepsilon\,)}
\prod_{j=1}^{n}\prod_{\lambda=0}^{N-1} 
\left(\frac{\chi_{\lambda}}{\eta}
\right)^{(1-g_j)\delta_{\varepsilon_j,\lambda}}\right)
\left (\frac{1}{N^2} G(N\tau_0)\right)^{\nu/2} 
\ret &+ N^{1-b_1}
\sum_{m=0}^{N-1}\left[\prod_{j=1}^{n}
\left( \sum_{\lambda=0}^{N-1} 
\left(\frac{\chi_{m,\lambda}}{\eta}\right)^{1-g_j} 
\ex^{\frac{2i\pi}{N}\lambda v\cdot[C_j]_N}\right)\right]
\ex^{i\pi\frac{N-1}{N}m v^2}
\left( \frac{1}{N^2} G\Big(\frac{\tau_0+m}{N}
\Big)\right)^{\nu/2}, 
\ret 
\label{sudos}
\end{align}
where $[C_j]_N$ is the reduction modulo $N$ of the Poincar\'e dual of
$C_j$,  and 

\begin{equation} 
w_N(\vec\varepsilon)=\sum_{j} \varepsilon_j [C_j]_N,
\label{sutres}
\end{equation}

where $\varepsilon_j=0,1,\ldots N-1$ are chosen independently. Notice that
\eqs{sudos} reduces to \eqs{suno} when $n=1$.

The above formulae for the partition function do not apply directly to 
the $SU(2)$
case. For 
$N=2$ there are some extra relative phases $t_i$ -- see equations 
(5.45) and
(5.46)  in \cite{vw} -- that we have not considered here. 
Modulo these extra  phases, (\ref{suno}) and \eqs{sudos} are a
direct generalization of Vafa and Witten's  results. They reduce
on $K3$ to  the formula of Minahan,  Nemeschansky, Vafa and
Warner \cite{estrings}  and generalize their results to non-zero
't Hooft fluxes.  Moreover, as we will show momentarily, they transform as
expected under duality and factorize appropriately under blow-ups.
Therefore, the above results (\ref{suno}) and (\ref{sudos}) can be seen as
predictions for the Euler numbers of instanton moduli spaces on K\"ahler 
four-manifolds with $b_2^+>1$. 

\vskip.5cm

We can actually be more specific about the possibility of including
relative phases between the contributions of the different vacua along
the cosmic string. In \cite{vw} there is one such phase $t_j$, $j =
1,\ldots,n$  for each component of the canonical divisor. These phases
reflect an anomaly in a $\IZ_2$ symmetry that permutes the bifurcated
vacua along each $C_i$. In the present case there seems to be no such
potentially anomalous symmetry permuting the $N$ vacua along each $C_j$,
so one could think that there is no reason to include any relative phase.
But let us  just ignore this for the time being and try to see whether
eqns. \eqs{suno} and
\eqs{sudos} can be generalized to incorporate such phases while still
preserving the required modular properties. The generalization we seek
amounts to replacing
\eqs{sudos} with:

\begin{align}
&Z_v = \left (\sum_{\vec\varepsilon}
\delta_{v,w_N(\vec\varepsilon\,)}
\prod_{j=1}^{n}\prod_{\lambda=0}^{N-1} 
\,(t_\lambda{}^j)^{\delta_{\varepsilon_j,\lambda}}\,
\left(\frac{\chi_{\lambda}}{\eta}
\right)^{(1-g_j)\delta_{\varepsilon_j,\lambda}}\right)
\left (\frac{1}{N^2} G(N\tau_0)\right)^{\nu/2} 
\ret &+ N^{1-b_1}
\sum_{m=0}^{N-1}\left[\prod_{j=1}^{n}
\left( \sum_{\lambda=0}^{N-1} 
\left(\frac{\chi_{m,\lambda}}{\eta}\right)^{1-g_j}\,t_\lambda{}^j\, 
\ex^{\frac{2i\pi}{N}\lambda v\cdot[C_j]_N}\right)\right]
\ex^{i\pi\frac{N-1}{N}m v^2}\left( \frac{1}{N^2}
G(\frac{\tau_0+m}{N})\right)^{\nu/2}, 
\ret 
\label{sudosphases}
\end{align}
where the $nN$ quantities $t_\lambda{}^j$, with $\lambda = 0,\ldots, N-1$
and $j = 1,\ldots, n$, are a priori arbitrary complex numbers. By
rescaling the partition function we can always set $t_0{}^j = 1$,
$j=1,\ldots,n$.

Let us first consider factorization under blow-ups. Let $B$ be the
exceptional divisor created by the blow-up. The partition function on the
blown-up manifold is exactly the same as above but with an extra term in
the product over the components of the canonical divisor which accounts
for $B$. This new term contains
$N$ complex numbers $t_\lambda^B$, with $t_0^B = 1$ as above. Going
through the details (see section $5.5.3.1$ below) it follows that the
partition function factorizes properly under the blow-up provided
$t_\lambda^B = 1$ for all
$\lambda$. In the $SU(2)$ case a similar result holds which follows from
the constraint $\prod_j t_j =(-1)^\nu$. ($\nu$ doesn't change under
blow-ups, so on the blown-up manifold $(\prod_j t_j) t_B =(-1)^\nu$,
which forces $t_B =1$.) We don't know if a similar constraint holds for
$N>2$, but the above result strongly suggests that it does. 

The behaviour under $\tau_0\to-1/\tau_0$ imposes two strong constraints 
on the
$t_\lambda{}^j$. Under $\tau_0\to-1/\tau_0$ the contribution from the
irreducible embedding is mapped to the term with $m=0$ in the contribution
from the $\cn = 1$ vacua, and this holds true for \eqs{sudosphases}
provided $t_\lambda{}^j = t_{N-\lambda}{}^j$ for all $j$. In addition, the
term with $m=k>0$ is mapped to the term with $m = h$, where $kh =
-1~\mod~N$, \ and this requires $t_\lambda{}^j =
t_{k\lambda}{}^j=t_{h\lambda}{}^j$. By considering all possible values of
$k$ we find that $t_{\lambda}{}^j=t_{k\lambda}{}^j$ for all
$k=1,\ldots,N-1$, which implies $t_1{}^j = t_2{}^j = \cdots =
t_{N-1}{}^j$. Hence, we end up with a single $t^j$ for each component
$C_j$ of the canonical divisor. It is now clear that the above result is
not consistent with having an anomalous symmetry permuting the vacua along
each $C_j$, and not knowing of any other mechanism that could be
responsible for having $t^j \not =1$,  we will assume that the $t^j = 1$
for all $j$.

Finally, it could be
interesting to investigate whether \eqs{suno} and \eqs{sudos} can
be rewritten in terms of the Seiberg-Witten invariants. 
We believe that this is not the case for the following reason. 
Let us suppose that it is actually possible to do so. Then one would
expect, by analogy with the result for $SU(2)$ \cite{coreatres},
that the  Donaldson-Witten partition function for $SU(N)$
\cite{mmtwo} should be recovered from the Vafa-Witten $SU(N)$
partition function in the decoupling limit $q_0\to 0$, $m\to\infty$
with $m^{2N}q_0$ fixed.  In particular, one would expect that the
structure of the corrections involving the canonical divisor should
be preserved in this limit. Now in the DW partition function in
\cite{mmtwo}, these corrections are written
in terms of the Seiberg-Witten classes $x$ 
\cite{monopole}. 
For $G=SU(N)$ these basic classes appear in the generic form
$\sum_{x_1,\ldots,  x_{N-1}}n_{x_1}\cdots n_{x_{N-1}}$ ($n_{x_l}$ are the
Seiberg-Witten invariants \cite{monopole}). Therefore, for $G=SU(N)$ there
are $N-1$ independent basic classes contributing to the above sum. On a
K\"ahler manifold with canonical divisor $K=C_1\cup C_2\cup\cdots\cup
C_n$, with the $C_j$ disjoint and with multiplicity one, each of these
basic classes can be written as
\begin{equation}
x_l=\sum_{\rho_l^j}
\rho_l^j C_j,
\end{equation}
with each $\rho_l^j=\pm 1$ \cite{monopole}, and the sum over
the basic classes can be traded for a sum over the $\rho_l^j$. This is
analogous to the sum over the $\varepsilon_j$ in \eqs{sudos}, and both
sums should contain the same number of terms were it possible to rewrite
\eqs{sudos} in terms of the basic classes. However, while in the sum over
the
$\rho_l^j$ there are 
$2^{n(N-1)}$ terms, the sum over the $\varepsilon_j$ contains $N^n$ terms. 
Notice that this two numbers do coincide when $N=2$, as it should be, but
for
$N\not=2$ this is no longer the case. 

It would certainly be most interesting to extend these results to all 
$N$ (not necessarily prime), and to investigate what the large $N$ limit 
of \eqs{suno} and \eqs{sudos} correspond to on the gravity side in the
light of the AdS/CFT correspondence. This remains as yet an open problem.

\subsubsection{Blow-ups} 

Given \eqs{sudos}, we can see explicitly how the factorization 
property \eqs{factoriz} works. Let $X$ be a K\"ahler four-fold with 
Euler characteristic $\chi=2(1-b_1)+b_2$, signature $\sigma=b_2^{+}-
b_2^{-}$ and canonical divisor $K=\cup_{j=1}^{n}C_j$, and let $\hat X$ be
its  one blow-up at a smooth point. Then $\hat b_1=b_1$, $\hat
b_2=b_2+1$,   
$\hat\chi=\chi+1$, $\hat\sigma=\sigma-1$ and $\hat K=K\cup B$, where $B$ is 
the exceptional divisor, which satisfies $B\cdot C_j=0$ and 
$B^2=-1=g_B-1$. Consider a 't Hooft  flux $\hat v=v\oplus \hat\lambda B$
in $\hat X$, where $v$ is a flux in $X$ and 
$\hat\lambda$ is an integer defined modulo $N$. Now $\hat \nu=\nu$, $\hat v^2=
v^2-{\hat\lambda}^2$, 
$\hat v\cdot C_j=v\cdot C_j$, $\hat v\cdot B=\hat\lambda B^2=-\hat\lambda$ and 
$\hat w_N(\vec\epsilon)=
\sum_{j=1}^{n} \varepsilon_j [C_j]_N+\hat\varepsilon \,B$. 
Thus, the partition function 
\eqs{sudos} takes the form 
\begin{align}
\hat Z_{\hat X,\hat v} &= \left (\sum_{\vec\varepsilon,\hat\varepsilon}
\delta_{v,w_N(\vec\varepsilon\,)}\delta_{\hat\lambda,\hat\varepsilon}
\prod_{j=1}^{n}\prod_{\lambda=0}^{N-1} 
\left(\frac{\chi_{\lambda}}{\eta}
\right)^{(1-g_j)\delta_{\varepsilon_j,\lambda}}
\left(\frac{\chi_{\lambda}}{\eta}
\right)^{(1-g_B)\delta_{\hat\varepsilon,\lambda}}\right)
\left (\frac{1}{N^2} G(N\tau_0)\right)^{\nu/2} 
\ret &+ N^{1-b_1}
\sum_{m=0}^{N-1}\left[\prod_{j=1}^{n}
\left( \sum_{\lambda=0}^{N-1} 
\left(\frac{\chi_{m,\lambda}}{\eta}\right)^{1-g_j} 
\ex^{\frac{2i\pi}{N}\lambda v\cdot[C_j]_N}\right)
\left( \sum_{\lambda=0}^{N-1} 
\left(\frac{\chi_{m,\lambda}}{\eta}\right)^{1-g_B} 
\ex^{-\frac{2i\pi}{N}\lambda\hat\lambda}\right)\right]\ret 
&\ex^{i\pi\frac{N-1}{N}m v^2}\ex^{-i\pi\frac{N-1}{N}m {\hat\lambda}^2}
\left( \frac{1}{N^2}
G\Big(\frac{\tau_0+m}{N}\Big)\right)^{\nu/2}, 
\notag 
\end{align}
and therefore
\begin{align}
\hat Z_{\hat X,\hat v} &=\left(\frac{\chi_{\hat\lambda}}{\eta}
\right) \left (\sum_{\vec\varepsilon}
\delta_{v,w_N(\vec\varepsilon\,)}
\prod_{j=1}^{n}\prod_{\lambda=0}^{N-1} 
\left(\frac{\chi_{\lambda}}{\eta}
\right)^{(1-g_j)\delta_{\varepsilon_j,\lambda}}\right)
\left (\frac{1}{N^2} G(N\tau_0)\right)^{\nu/2} 
\ret +& N^{1-b_1}
\sum_{m=0}^{N-1}
\left( \sum_{\lambda=0}^{N-1} 
\left(\frac{\chi_{m,\lambda}}{\eta}\right) 
\ex^{-\frac{2i\pi}{N}\lambda\hat\lambda}\ex^{-i\pi\frac{N-1}{N}m
{\hat\lambda}^2}\right)
\left[\prod_{j=1}^{n}
\left( \sum_{\lambda=0}^{N-1} 
\left(\frac{\chi_{m,\lambda}}{\eta}\right)^{1-g_j} 
\ex^{\frac{2i\pi}{N}\lambda v\cdot[C_j]_N}\right)
\right]\ret 
&\ex^{i\pi\frac{N-1}{N}m v^2}
\left( \frac{1}{N^2}
G\Big(\frac{\tau_0+m}{N}\Big)\right)^{\nu/2}.\ret 
\label{sudos2}
\end{align}
Now, from \eqs{suonce2} it follows that 
\begin{equation}
 \sum_{\lambda=0}^{N-1} 
\left(\frac{\chi_{m,\lambda}}{\eta}\right) 
\ex^{-\frac{2i\pi}{N}\lambda\hat\lambda}\ex^{-i\pi\frac{N-1}{N}m
{\hat\lambda}^2}=\frac{1}{N}\sum_{\lambda,\lambda'} 
\ex^{-\frac{2 i\pi}{N}
\lambda(\lambda'+\hat\lambda)}
\ex^{i\pi\frac{N-1}{N}m ((\lambda')^2-{\hat\lambda}^2)}
\left(\frac{\chi_{\lambda'}}{\eta}\right).
\la{suonce3}
\end{equation}
Summing over $\lambda$ and using \eqs{prono} we get 
\begin{align}
\frac{1}{N}&\sum_{\lambda,\lambda'} 
\ex^{-\frac{2 i\pi}{N}
\lambda(\lambda'+\hat\lambda)}
\ex^{i\pi\frac{N-1}{N}m ((\lambda')^2-{\hat\lambda}^2)}
\left(\frac{\chi_{\lambda'}}{\eta}\right)\ret&=
\sum_{\lambda'}\delta_{\lambda'+\hat\lambda,0}\, 
\ex^{i\pi\frac{N-1}{N}m ((\lambda')^2-{\hat\lambda}^2)}
\left(\frac{\chi_{\lambda'}}{\eta}\right)\ret
&=\frac{\chi_{-\hat\lambda}}{\eta}=\frac{\chi_{N-\hat\lambda}}{\eta}=
\frac{\chi_{\hat\lambda}}{\eta}.
\la{suonce4}
\end{align}
Hence, 
\begin{align}
\hat Z_{\hat X,\hat v} &=\left(\frac{\chi_{\hat\lambda}}{\eta}
\right) \left (\sum_{\vec\varepsilon}
\delta_{v,w_N(\vec\varepsilon\,)}
\prod_{j=1}^{n}\prod_{\lambda=0}^{N-1} 
\left(\frac{\chi_{\lambda}}{\eta}
\right)^{(1-g_j)\delta_{\varepsilon_j,\lambda}}\right)
\left (\frac{1}{N^2} G(N\tau_0)\right)^{\nu/2} 
\ret &+ N^{1-b_1}
\sum_{m=0}^{N-1}
\left(\frac{\chi_{\hat\lambda}}{\eta}
\right)
\left[\prod_{j=1}^{n}
\left( \sum_{\lambda=0}^{N-1} 
\left(\frac{\chi_{m,\lambda}}{\eta}\right)^{1-g_j} 
\ex^{\frac{2i\pi}{N}\lambda v\cdot[C_j]_N}\right)
\right]\ret 
&\ex^{i\pi\frac{N-1}{N}m v^2}
\left( \frac{1}{N^2} G\Big(\frac{\tau_0+m}{N}\Big)\right)^{\nu/2}
=\left(\frac{\chi_{\hat\lambda}}{\eta}
\right) Z_{X,v},\ret 
\label{sudos3}
\end{align}
as expected. 

\subsubsection{Modular transformations}
We will now study the modular properties of the partition functions 
\eqs{suno} and \eqs{sudos}.  With the formulas in section \ref{apendice2} one
can  check that they have the expected  modular  behaviour\footnote{We
assume as in
\cite{vw} that there is no torsion in $H_2(X,\iz)$. Were this not case, Eqs.
\eqs{sucinco} and \eqs{sudos} above should be modified along the lines 
explained in \cite{wiads2}.}

\begin{align}
Z_v(\tau_0+1)
&=\ex^{\frac{i\pi}{12}N(2\chi+3\sigma)}
\ex^{-i\pi\frac{N-1}{N}v^2} 
Z_v(\tau_0), 
\ret 
Z_v(-1/\tau_0) &= N^{-b_2/2}\left(\frac{\tau_0}{i}
\right)^{-\chi/2}
\sum_u \ex^{\frac{2i\pi u\cdot v}{N}} Z_u(\tau_0), 
\la{sucuatro}\\ 
\intertext{and also, with  $Z_{SU(N)}= N^{b_1-1} Z_{0}$ and 
$Z_{SU(N)/\IZ_N}=\sum\limits_{v} Z_v$,} 
Z_{SU(N)}(\tau_0+1)&=\ex^{\frac{i\pi}{12}N(2\chi+3\sigma)}
Z_{SU(N)}(\tau_0),\ret
Z_{SU(N)/\IZ_N}(\tau_0+N)&=\ex^{\frac{i\pi}{12}N^2(2\chi+3\sigma)}
Z_{SU(N)/\IZ_N}(\tau_0), \la{sucinco1}\\
\intertext{and}
Z_{SU(N)}(-1/\tau_0)&= N^{-\chi/2}\left(\frac{\tau_0}{i}
\right)^{-\chi/2}Z_{SU(N)/\IZ_N}(\tau_0),
\label{sucinco}
\end{align}
which is the Montonen-Olive relation. Notice that since $N$ is
odd,  the
$SU(N)$ (or $SU(N)/\IZ_N$) partition function is modular (up to a
phase) for $\Gamma_0(N)$ on any four-manifold.
On the other hand, for {\sl even} $N$ one would expect on 
general grounds \cite{vw} modularity for 
$\Gamma_0(2N)$, or at most $\Gamma_0(N)$ on spin manifolds.

\subsubsection{The partition function on $T^4$}
We will finish by considering the twisted theory on  
$T^4$, where an unexpected result emerges. As $K3$, $T^4$ is a 
compact hyper-K\"ahler manifold (hence with trivial canonical 
divisor). It has $b_1\,$=$\,4$, $b_2\,$=$\,6$ and $\chi\,$=
$0\,$=$\,\sigma$.  
On $T^4$ the partition function \eqs{suno} reduces to its bare
bones
\begin{equation}
Z_v = \delta_{v,0} + \frac{1}{N^3}
\sum_{m=0}^{N-1}\,
\ex^{i\pi\frac{N-1}{N}m v^2},
\la{sudiezsiete2}
\end{equation}
and does not depend on $\tau_0$! This should be compared with the
formulas in \cite{italia}. The $Z_v$ are self-dual in the
following sense 
\begin{equation}
Z_v =\frac{1}{N^3}
\sum_u \ex^{\frac{2i\pi u\cdot v}{N}} Z_u. 
\la{sumatra}
\end{equation}
Notice that
since $T^4$ is a spin manifold, $v^2\in 2\IZ$, and therefore the sum
over $m$ in \eqs{sudiezsiete2} vanishes unless $v^2=0$ (modulo $N$), 
so $Z_v$ reduces to the rather simple form
\begin{equation}
Z_{T^4,v} = \delta_{v,0} + \frac{1}{N^2}\delta_{v^2,0},
\la{borneo}
\end{equation}
which gives the partition function for the physical $\cn=4$ $SU(N)$
theory in the sector of 't Hooft flux $v$ and in the limit
$\bar\tau_0\to\infty$ (which would explain the discrepancy with the
results in \cite{italia}.)  

\section{Appendix}
\la{apendice2}
Here we collect some useful formulas which should help the 
reader follow the computations in the preceding section.
\subsection{Modular forms}

The function $G$ is defined as 
\begin{equation}
G(\tau)=\frac{1}{\eta(\tau)^{24}},
\la{lage}
\end{equation}
and is a modular form of weight $-12$
\begin{equation}
G(\tau)\,\mapright{\tau\to\tau+1}\, 
G(\tau),\qquad\qquad G(\tau)\,\mapright{\tau\to
-1/\tau}   \tau^{-12}  G(\tau),
\la{supadre}
\end{equation}
From \eqs{supadre} we can determine the modular behaviour of the
different modular forms in the $K3$ partition function 

\begin{align} 
G(N\tau)&\mapright{\tau\to -1/\tau}\tau^{-12} N^{12}
G(\tau/N),\ret \ret
G\left(\frac{\tau+m}{N}\right)&\mapright{\tau\to
-1/\tau}
\tau^{-12}G\left(\frac{\tau+h}{N}\right),\ret &{}
\label{susi}
\end{align}
where $1\leq h \leq N-1$, $mh=-1~\mod~ N$ and $N$ prime. 

For arbitrary $N$ one has to consider the modular forms
$G\left(\frac{a\tau+b}{d}\right)$, where $ad =N$ and $b< d$
\cite{estrings}. These functions transform as follows
\begin{equation}
G\left(\frac{a\tau+b}{d}\right)\mapright{\tau\to
-1/\tau}\tau^{-12}
\left(\frac{a}{p}\right) ^{12}G\left(\frac{p\tau+ab'}{a\tilde
d}\right),
\label{susiete}
\end{equation}
where $p=\gcd~ (b,d)$, $\tilde d=d/p$, $\tilde b=b/p$, $b'\tilde
b=-1~
\mod~ \tilde d$. If $b=0$, then $p=d$ and $b'=0=\tilde b$. Notice
that for prime $N$ \eqs{susiete} reduces to \eqs{susi}.

\subsection{Flux sums}

The basic sums we have to consider are of the form 
\begin{equation} 
I(m,N)=
\sum_{\lambda=0}^{N-1} \ex^{\frac{i\pi m}{N} \lambda (N-\lambda)}=
\sum_{\lambda=0}^{N-1} \ex^{i\pi\frac{N-1}{N}m \lambda^2},
\la{zergs}
\end{equation} 
for $1\leq m \leq N-1$, and discrete Fourier
transformations thereof
\begin{equation}
\sum_{\lambda=0}^{N-1} \ex^{\pm\frac{2 i\pi}{N}
\lambda\lambda'}\ex^{i\pi\frac{N-1}{N}m \lambda^2},
\la{protoss}
\end{equation} from which the sums over fluxes can be easily
computed.  The basic sum \eqs{zergs} is related to a standard
Gauss sum
$G(m,N)=\sum_{r~ \text{\scriptsize mod $N$}} \ex^{2i\pi m r^2/N}$
\cite{numbertheory}. In fact, $I(m,N)=I(m+N,N)$ and, since $N$ is
odd, it suffices to consider the case where $m$ is even. But in
this case 
\begin{equation} I(2a,N)=\sum_{\lambda=0}^{N-1}
\ex^{i\pi\frac{N-1}{N}2a\lambda^2}= 
\sum_{\lambda} \ex^{-2i\pi a\lambda^2/N}=\overline{G(a,N)}.
\la{mutalisk}
\end{equation} 
Now, when $a=1$, 
\begin{equation}
G(1,N)=\frac{\sqrt{N}}{2}(1+i)\left(1+\ex^{-\frac{i\pi
N}{2}}\right),
\la{Gaussum}
\end{equation} (\cite{numbertheory}, p. $165$.) Moreover, for
$a>1$ and $N$ and {\sl odd prime}, 
\begin{equation}
G(a,N)=\left(\frac{a}{N}\right)\,G(1,N),
\la{Gaussum2}
\end{equation}
where $\left(\frac{a}{N}\right)$ is the Legendre symbol
\cite{numbertheory}, which is $+1$ if $a$ is a perfect square
($\mod~N$) and $-1$ otherwise. Hence, taking
\eqs{mutalisk}-\eqs{Gaussum2} into account we have the result
\begin{equation}
\sum_{\lambda=0}^{N-1} \ex^{i\pi\frac{N-1}{N}m \lambda^2}
=\epsilon(m)\sqrt{N} \ex^{-\frac{i\pi}{8}(N-1)^2},
\la{supinador}
\end{equation}
where

\begin{equation}
\epsilon(m)=\begin{cases}
\left(\frac{m/2}{N}\right), & \text{$m$
even},\\ \\
\left(\frac{(m+N)/2}{N}\right), & \text{$m$ odd},
\end{cases}
\la{sudiezseis}
\end{equation}

If $kh=-1~\mod~ N$, $\epsilon(k)=\epsilon(h)$ for $N=5 ~\mod~ 4$,
and 
$\epsilon(k)=-\epsilon(h)$ for $N=3 ~\mod ~4$. This property is
essential in proving the second relation in \eqs{sucuatro}. 

We also have the identity

\begin{equation}
\sum_{\lambda=0}^{N-1} \ex^{\pm \frac{2 i\pi}{N}
\lambda\lambda'}=N\delta_{\lambda',0},
\la{prono}
\end{equation}

and the fundamental result

\begin{equation}
\sum_{\lambda=0}^{N-1} \ex^{\pm\frac{2 i\pi}{N}
\lambda\lambda'}\ex^{i\pi\frac{N-1}{N}m \lambda^2}=
\epsilon(m)\sqrt{N}\ex^{-\frac{i\pi}{8}(N-1)^2}
\ex^{i\pi\frac{N-1}{N}h(\lambda')^2},
\la{pronador}
\end{equation}
with $mh=-1~\mod~N$ and $N$ an {\sl odd prime}. 

Now, given \eqs{supinador}, the basic sum over fluxes 
$\sum_{v} \ex^{i\pi\frac{N-1}{N}m v^2}$ can be computed in terms
of
\eqs{zergs} as follows -- see \cite{vw}, eq. (3.21)-(3.22):
\begin{equation}
\sum_{v\in H^{2}(X,\IZ_N)} \ex^{i\pi\frac{N-1}{N}m v^2} =
I(m,N)^{b_2^{+}} \,\overline {I(m,N)}{\,}^{b_2^{-}},
\la{terran}
\end{equation}
so one has (for {\sl prime} $N$)
\begin{equation}
\sum_{v\in H^{2}(X,\IZ_N)} \ex^{i\pi\frac{N-1}{N}m v^2} =
\big(\epsilon(m)\big)^{b_2}N^{b_2/2}\ex^{-\frac{i\pi}{8}(N-1)^2
\sigma},
\la{terran2}
\end{equation}
and also, from \eqs{prono} and \eqs{pronador}  
\begin{equation}
\sum_{v\in H^{2}(X,\IZ_N)} \ex^{ \frac{2 i\pi}{N} u\cdot
v}=N^{b_2}\delta_{u,0},
\la{terran3}
\end{equation}

\begin{equation}
\sum_{v\in H^{2}(X,\IZ_N)} \ex^{\frac{2 i\pi}{N} u\cdot
v}\ex^{i\pi\frac{N-1}{N}m v^2}=
\big(\epsilon(m)\big)^{b_2}N^{b_2/2}\ex^{-\frac{i\pi}{8}(N-1)^2
\sigma}
\ex^{i\pi\frac{N-1}{N}hu^2},
\la{terran4}
\end{equation}
with $mh=-1~\mod~N$ as above.

\vskip1cm

\subsection{$SU(N)$ characters}
We have seen above that the corrections to the $SU(N)$ partition
function near the canonical  divisor of the four-manifold $X$ are
given in terms  of the level one characters $\chi_\lambda$ of the
$SU(N)$ WZW model. These are defined as \cite{izzy}
\begin{equation}
\chi_\lambda (\tau)=\frac{1}{\eta(\tau)^{N-1}}\sum_{\vec w\in
[\lambda]}
\ex^{i\pi\tau \vec w^2},\qquad \lambda\in\IZ~\mod~N,
\la{chars}
\end{equation}
where $[\lambda]$ is the $\lambda$-th conjugacy class of $SU(N)$,
and the identification $\chi_\lambda (\tau)=\chi_{\lambda+N}
(\tau)$ is understood. Also, from the symmetry properties of the inverse
Cartan matrix \eqs{cartan} it follows that $\chi_\lambda =
\chi_{N-\lambda}$. 
$\lambda\,$=$\,0~\mod~N$  corresponds to $\vec w$ in the root
lattice, while for 
$1\leq\lambda\leq N-1$,    
$[\lambda]=\{\vec w\in \Lambda_{\mbox{\rm\tiny weight}}:\vec
w=\vec\alpha^\lambda+\sum_{n^{\lambda'}\in\IZ}n^{\lambda'}
\vec\alpha_{\lambda'}\}$. $\vec\alpha_{\lambda}$ are the simple
roots and 
$\vec\alpha^\lambda$ the fundamental weights of $SU(N)$,
normalized in such a way that the inverse Cartan matrix
$A^{\lambda\lambda'}$ has the standard form 
\begin{equation} 
A^{\lambda\lambda'}=\vec\alpha^\lambda\cdot\vec\alpha^{\lambda'}=
{\rm Inf}~\{\lambda,\lambda'\}-\frac{\lambda\lambda'}{N},\qquad
1\leq\lambda,\lambda'\leq N-1.
\la{cartan}
\end{equation} 
The characters \eqs{chars} have definite properties under the
modular  group \cite{izzy}
\bea 
\chi_\lambda (\tau+1)&=&
\ex^{-\frac{i\pi}{12}(N-1)}\ex^{i\pi\frac{N-1}{N}\lambda^2}
\chi_\lambda (\tau),\ret
\chi_\lambda (-1/\tau)&=&\frac{1}{\sqrt{N}}\sum_{\lambda'=0}^{N-1} 
\ex^{-\frac{2 i\pi}{N}\lambda\lambda'}\chi_{\lambda'}(\tau).\ret
\la{sudoce}
\eea

From the characters $\chi_\lambda$ we introduce the linear
combinations  ($N>2$ and prime)
\begin{equation}
\chi_{m,\lambda}(\tau)=\frac{1}{N} \sum_{\lambda'=0}^{N-1}
\ex^{-\frac{2 i\pi}{N}
\lambda\lambda'}\ex^{i\pi\frac{N-1}{N}m (\lambda')^2}
\chi_{\lambda'}(\tau),\quad 0\leq m,\lambda\leq N-1, 
\label{suonce}
\end{equation} 
which have the ciclicity property
$\chi_{m+N,\lambda}=\chi_{m,\lambda}=\chi_{m,\lambda+N}$ since
$N$ is odd. Under the modular group one has
\bea 
\chi_{m,\lambda} (\tau+1)\!\!\! &=& \!\!
\ex^{-\frac{i\pi}{12}(N-1)}
\chi_ {m+1,\lambda}(\tau),\ret
\chi_{0,\lambda} (-1/\tau)\!\!\! &=& \!\!
\frac{1}{\sqrt{N}} \chi_{\lambda}(\tau),\ret
\chi_{m,\lambda} (-1/\tau)\!\!\! &=&
\!\!\epsilon(m)\,\ex^{-\frac{i\pi}{8}(N-1)^2}
\ex^{i\pi\frac{N-1}{N}h\lambda^2}\chi_{m,h\lambda} (\tau),\quad
m>0,
\ret
\la{suquince}
\eea
with $mh=-1~\mod~ N$.


\chapter{Adjoint non-Abelian monopoles}
\la{chtwist}
\markboth{\footnotesize\bfseries Adjoint non-Abelian monopoles
}{\footnotesize\bfseries Adjoint non-Abelian monopoles}
\markright{\textsc {Duality in Topological Quantum Field Theories}}

 In this chapter we will study the second possible twist of the $\cn=4$
theory. The twisted theory is well-defined on spin four-manifolds, and can be
deformed  by giving masses to two of the
chiral multiplets. The massive theory is still topological on arbitrary spin
manifolds and is in fact the twisted counterpart of the mass deformed
$\cn=4$ theory. Using the low-energy effective description of the physical
theory we compute, within the $u$-plane approach of Moore and Witten
\cite{moorewitten}, the generating function of topological correlation
functions for gauge group
$SU(2)$ and arbitrary values of the 't Hooft fluxes, and analyze the
duality properties of the resulting formulas.  

\section{The twisted theory}
The theory is defined by the splitting ${\bf 4}\too ({\bf 2},{\bf 1})\oplus 
({\bf 1},{\bf 1})\oplus({\bf 1},{\bf 1})$ of the fundamental representation of 
$SU(4)_I$ in representations of $SU(2)_L\otimes SU(2)_R$. 
As explained in \cite{yamron}, this amounts to breaking  $SU(4)_I$ 
 down to a subgroup $SU(2)_A\otimes SU(2)_F\otimes U(1)$,  and then 
replacing  $SU(2)_L$ with the diagonal sum
$SU(2)'_L$ of
$SU(2)_L$ and $SU(2)_A$. The subgroup
$SU(2)_F\otimes U(1)$ remains in the theory as an internal symmetry group. Hence,
we observe that, as a by-product of the twisting procedure, it remains in the
theory a $U(1)$ symmetry which was not present in the original
$\cn=4$ theory, and which becomes, as we shall see in a moment, the ghost number
symmetry associated to the topological theory. With respect to the new symmetry
group
${\cal H'} =SU(2)'_L\otimes SU(2)_R\otimes SU(2)_F\otimes U(1)$ the supercharges
$Q^v{}_{\!\alpha}$ split up into three supercharges
$Q_{(\beta\alpha)}$, $Q$ and $Q^{i}{}_{\!\alpha}$, where the index $i$
labels the representation ${\bf 2}$ of $SU(2)_F$. In more detail,
\begin{equation} 
Q^{v=1,2,3,4}_\alpha \to\begin{cases}
Q^{v=1,2}_\alpha \to
Q^{\beta}_{\!\alpha}\to\begin{cases} 
Q=-Q^\alpha{}_{\!\alpha}\equiv
-Q^{v=2}{}_{\!2}-Q^{v=1}{}_{\!1},\\
Q_{(\beta\alpha)}=-C_{\gamma(\beta}Q^\beta{}_{\alpha)},
\end{cases}\\
Q^{v=3}_{\!\alpha} \to Q^{i= 1}_\alpha,
\\ Q^{v=4}_{\!\alpha} \to Q^{i= 2}_\alpha.
\end{cases}
\la{nuria}
\end{equation}
The conjugate supercharges 
$\bar Q_{v\dot\alpha}$ split up accordingly into a vector isosinglet and a
right-handed spinor isodoublet supercharge, $\bar Q_{\alpha\dot\alpha}$
and $\bar Q_{i\dot\alpha}$. 

The fields of the $\cn=4$ multiplet give rise, after the twisting, to the 
following topological multiplet (in the notation of reference \cite{yamron}):
\begin{align}
A_{\alpha\dalpha} &\too A^{(0)}_{\!\alpha\dalpha},\ret
\lambda_{v\alpha}&\too
\chi^{(-1)}_{\beta\alpha},~
\eta^{(-1)},~\lambda^{(+1)}_{i\alpha},\ret
\bar\lambda^v{}_{\dot\alpha} &\too
\psi^{(+1)}_{\!\alpha\dot\alpha},~\zeta^{(-1)}_{i\dot\alpha},\ret 
\phi_{uv}&\too
B^{(-2)},~C^{(+2)},~G^{(0)}_{\! i\alpha},
\la{cien}
\end{align} 
where  we have indicated the ghost number  carried by the fields after the
twisting by a superscript. Notice that the twisted theory contains several
spinor fields. This means that  the theory is not well defined on those
manifolds $X$ that do not admit a spin  structure. As explained in
\cite{corea}\cite{nabm}\cite{baryon} -- see \cite{monopole} for a related
discussion --, one  could try to avoid this problem by coupling the spinor
fields to a fixed  (background) Spin$_{c}$ structure. This is essentially a 
fixed Abelian gauge configuration with magnetic fluxes through the two-cycles of
$X$ quantized in $\IZ/2$ instead of in $\IZ$, which would be the standard Dirac 
quantization condition. Quantum mechanics of electric charges is not consistent
in this background, but it turns out that the obstruction to defining consistent
quantum propagation can be fixed to cancel the obstruction to defining spinors
on $X$ \cite{shawking}. We will not follow this path here,
and therefore we  will take
$X$ to be an -- otherwise arbitrary -- spin  four-manifold.  

Some of the definitions in \eqs{cien} need clarification.
Our choices for the anticommuting fields are:
\begin{align}
\lambda_{v\alpha}&= \begin{cases}
\lambda_{(v=1,2)\alpha}\to
\lambda_{\beta\alpha}\to\begin{cases}
\chi_{
\beta\alpha}=\lambda_{(\beta\alpha)},\\
\eta=2\lambda_{[\beta\alpha]},
\end{cases}\\
\lambda_{(v=3,4)\alpha}\to
\lambda_{i\alpha},
\end{cases}\ret
\bar\lambda^v{}_{\dot\alpha} &=
\begin{cases}
\bar\lambda^{v=1,2}{}_{\!\dot\alpha}\to
\psi^{\alpha}{}_{\!\dot\alpha},\\
\bar\lambda^{v=3,4}{}_{\dot\alpha}\to
\zeta^{i}{} _{\!\dot\alpha},
\end{cases}
\la{ccuno}
\end{align}
whereas for the commuting ones we set:
\begin{align}
B&=\phi_{12},& C&=\phi_{34},\ret 
G_{(i=1)1}&=\phi_{13},&  G_{(i=2)1}&=\phi_{14},\ret 
G_{(i=1)2}&=\phi_{23},&
G_{(i=2)2}&=\phi_{24}.
\la{cctres}
\end{align}

In terms of the twisted fields, the action for the theory (on flat $\IR^4$)
takes the form:
\begin{align}
{\cal S}^{(0)}=& \fr{1}{ e^2_0}\int d^4 x\, \tr\,
\bigl\{\,\half\nabla_{\!\alpha\dalpha} B\nabla^{\dalpha\alpha}C
-\fr{1}{4}\nabla_{\!\alpha\dalpha} G_{i\beta}\nabla^{\dalpha\alpha}
G^{i\beta}  -i\psi^{\beta}{}_{\dot\alpha}\nabla^{\dot\alpha\alpha}
\chi_{\alpha\beta}
\ret 
-&\fr{i}{ 2}\psi_{\alpha\dalpha}\nabla^{\dalpha\alpha}
\eta -i\zeta^i{}_{\dot\alpha}\nabla^{\dot\alpha\alpha}\lambda_{i\alpha}-
\fr{1}{4}F_{mn} F^{mn}-
\fr{i}{\raiz}\,\chi^{\alpha\beta}[\chi_{\alpha\beta},C]
\ret
-&\fr{i}{\raiz}\,\lambda^{i\alpha}[\lambda_{i\alpha},B] 
+i\raiz\,\chi^{\alpha\beta}[\lambda_{i\alpha},G^i{}_\beta]
+\fr{i}{\raiz}\,\eta[\lambda_i{}^\alpha,G^i{}_\alpha]
-\fr{i}{{2\raiz}}\,\eta[\eta,C]
\ret
+&\fr{i}{\raiz}\,\psi_{\alpha\dalpha}[\psi^{\dalpha\alpha},B]
+i\raiz\,\psi^\alpha{}_{\dalpha}[\zeta^{i\dalpha},G_{i\alpha}]+
\fr{i}{\raiz}\,\zeta_{i\dot\alpha}[\zeta^{i\dot\alpha},C]-
\half[B,C]^2
\ret 
-&[B,G_{i\alpha}][C,G^{i\alpha}]  +\fr{1}{
4}[G_{i\alpha},G_{j\beta}] [G^{i\alpha},G^{j\beta}]\,\bigr\}
-\fr{i\theta_0}{ 32\pi^2}\int d^4 x\,\tr\,\bigl\{\, *  F_{mn}F^{mn}
\,\bigr\} .
\ret
\la{cccuatro}
\end{align}

To obtain the corresponding topological symmetry we proceed as follows. First of
all, we recall that the $\cn=4$ supersymmetry transformations
\eqs{Vian4} are generated by $\xi_v{}^\alpha Q^v{}_\alpha + 
\bar\xi^v{}_\dalpha \bar Q_v{}^\dalpha$. According to our conventions, to obtain
the $Q$-transformations we must set $\bar\xi^v{}_\dalpha=0$ and make the
replacement:  
\begin{equation}
\xi_{v\alpha}=\begin{cases}
\xi_{(v=1,2)\alpha}\to\epsilon C_{\beta\alpha},\\
\xi_{(v=3,4)\alpha}\to 0.
\end{cases}
\la{nndos}
\end{equation} 
The resulting transformations turn out to be:
\begin{align}
\delta A_{\alpha\dalpha} &= 2i\epsilon\psi_{\alpha\dalpha},&
\delta G_{j\alpha}&=-\raiz\epsilon\lambda_{j\alpha},\ret
\delta\psi_{\alpha\dot\alpha} &=  -i{\raiz}\epsilon\nabla_{\alpha\dot\alpha}C,&
\delta\lambda_{j\alpha}&=-2i\epsilon [G_{j\alpha},C],\ret
\delta C&=0,&
\delta F^{+}_{\alpha\beta}&=2\epsilon\nabla_{(\alpha}{}^{\dot\alpha}
\psi{}_{\beta)\dot\alpha},\ret
\delta\chi_{\alpha\beta} &=  -i\epsilon F^{+}_{\alpha\beta}
-i\epsilon[G_{i\alpha},G^i{}_{\beta}],&
\delta B &=\raiz\epsilon\eta,\ret 
\delta\zeta^j{}_{\dot\alpha}&=-i{\raiz}\epsilon\nabla_{\alpha\dot\alpha}
G^{j\alpha},&
\delta\eta &=2i\epsilon[B,C].\ret
\la{cccinco}
\end{align} 

The generator $Q$ of the transformations \eqs{cccinco} satisfies the
on-shell algebra 
$\{Q,Q\}=\delta_g(C)$ where by $\delta_g(C)$ we mean a non-Abelian gauge
transformation generated by $C$.   It is possible to realize this algebra
off-shell, \ie, without the input of the equations of motion for some of the fields
in the theory. A minimal off-shell formulation can be constructed by introducing in
the theory the auxiliary fields $N_{\alpha\beta}$ (symmetric in its spinor indices)
and
$P^i{}_{\!\alpha}$, both with ghost number $0$.  The off-shell BRST
transformations which correspond to the enlarged topological multiplet can be cast
in the form:
\begin{align}
[Q, A_{\alpha\dalpha}]& = 2i\psi_{\alpha\dalpha},&
\{Q,\psi_{\alpha\dot\alpha}\} &= 
-i{\raiz}\nabla_{\alpha\dot\alpha}C,\ret
[Q,F^{+}_{\alpha\beta}]&=2\nabla_{(\alpha}{}^{\dot\alpha}
\psi{}_{\beta)\dot\alpha},&
[Q, C]&=0,\ret
[Q, G_{j\alpha}]&=-\raiz\lambda_{j\alpha},&
\{Q,\lambda_{j\alpha}\}&=-2i[G_{j\alpha},C],\ret
\{Q,\chi_{\alpha\beta}\}& = N_{\alpha\beta},&  
[Q,N_{\alpha\beta}]&=2\raiz i\,[\chi_{\alpha\beta},C],\ret
\{Q,\zeta^j{}_{\dot\alpha}\}&=P^j{}_{\!\dalpha},&
[Q,P^i{}_{\!\dalpha}]&=2\raiz i\,[\zeta^i{}_{\!\dalpha},C],\ret
[Q, B]&=\raiz\eta,&
\{Q,\eta\,\}&=2i[B,C].\ret
\la{Medeiros4}
\end{align}

After some suitable manipulations, the off-shell action which corresponds
to the topological symmetry \eqs{Medeiros4} is:
\begin{align}
{\cal S}^{(1)}=&\fr{1}{ e^2_0}\int d^4 x\, \tr\,
\Bigl\{\ret&
\half\nabla_{\!\alpha\dalpha} B\nabla^{\dalpha\alpha}C
+\fr{1}{4}P_i{}^{\dalpha}\,\bigl (\, P^i{}_{\!\dalpha}+2\raiz
i\,\nabla_{\!\alpha
\dalpha}G^{i\alpha}\,\bigr )  -i\psi^{\beta}{}_{\dot\alpha}\nabla^{\dot\alpha\alpha}
\chi_{\alpha\beta}
\ret 
-&\fr{i}{ 2}\psi_{\alpha\dot\alpha}\nabla^{\dot\alpha\alpha}
\eta -i\zeta^j{}_{\dot\alpha}\nabla^{\dot\alpha\alpha}\lambda_{j\alpha}+
\fr{1}{4}N_{\alpha\beta}\,\bigl (\, N^{\alpha\beta}+2i F^{+\alpha\beta}
+2i[G_i{}^{\alpha},G^{i\beta}]\,\bigr )
\ret 
-&\fr{i}{\raiz}\,\chi^{\alpha\beta}[\chi_{\alpha\beta},C]
-\fr{i}{\raiz}\,\lambda^{i\alpha}[\lambda_{i\alpha},B]+ 
i\raiz\,\chi^{\alpha\beta}[\lambda_{i\alpha},G^i{}_\beta]+
\fr{i}{\raiz}\,\zeta_{i\dot\alpha}[\zeta^{i\dot\alpha},C]
\ret 
+&\fr{i}{\raiz}\,\eta[\lambda_i{}^\alpha,G^i{}_\alpha]
-\fr{i}{{2\raiz}}\,\eta[\eta,C]
+\fr{i}{\raiz}\,\psi_{\alpha\dot\alpha}[\psi^{\dalpha\alpha},B]+
i\raiz\,\psi^\alpha{}_{\dot\alpha}[\zeta^{i\dot\alpha},G_{i\alpha}]
\ret 
-&\half[B,C]^2  -[B,G_{i\alpha}][C,G^{i\alpha}]\,\Bigr\} 
-2\pi i\tau_0\fr{1}{
32\pi^2}\int d^4 x\,\tr\,\bigl\{\, *  F_{mn}F^{mn}
\,\bigr\}.
\ret
\la{cinza4}
\end{align}
The $\tau_0$-independent part of the  topological action above is, as it 
could be
expected, BRST-exact, that is, it can be written as $\{Q,\Psi\,\}$. The 
appropriate gauge fermion is easily seen to be: 
\bea
\Psi\!\!&=&\!\!\fr{1}{ e^2_0}\int d^4 x\,\tr\,\bigl\{\,\, 
\fr{1}{4}\zeta_i{}^{\!\dot\alpha}\,\bigl (\,P^i{}_{\!\dot\alpha}+ 2\raiz
i\,\nabla_{\!\alpha\dalpha}G^{i\alpha}\,\bigr )
\ret
\!\!&+&\!\!\fr{1}{4}\chi_{\alpha\beta}\,\bigl (\, N^{\alpha\beta}+
2iF^{+\alpha\beta} +2i[G_{i}{}^{\alpha},G^{i\beta}]\,\bigr )\,\bigr \}
\ret 
\!\!&-&\!\!\fr{1}{
e^2_0}\int d^4 x\,\tr\,\bigl\{\,\, \fr{i}{{2\raiz}}\,B\,\bigl (\,
\nabla_{\!\alpha\dalpha}\psi^{\dalpha\alpha}
-\raiz\,[G_{i\alpha},\lambda^{i\alpha}]\,\bigr )\,\bigr \}
\ret 
\!\!&-&\!\!\fr{1}{ e^2_0}\int
d^4 x\,\tr\,\bigl\{\,\, \fr{i}{ 4}\,B[\eta,C]\,\bigr\}.\ret
\la{cerne4}
\eea

The next  step will consist of the coupling the theory to an arbitrary background
metric $g_{\mu\nu}$ of Euclidean signature. To achieve this goal we make use of the
covariantized version of the topological symmetry 
\eqs{Medeiros4} (which is trivial to obtain), and of the gauge fermion $\Psi$,  
and then
define the topological action to be ${\cal
S}^{(1)}_{c}=\{Q,\Psi\,\}_{\mbox{\rm\tiny cov}}-2 \pi i k \tau_0$.  The resulting
action is:
\begin{align}
{\cal S}^{(1)}_{c}&=\fr{1}{ e^2_0} \int_X  d^4 x\,\sqrt{g}\,\tr\,
\Bigl\{\ret&\half\deriv_{\alpha\dalpha} B\deriv^{\dalpha\alpha}C
+\fr{1}{4}P_i{}^{\dalpha}\,\bigl (\, P^i{}_{\!\dalpha}+2\raiz i\,\deriv_
{\alpha\dot\alpha}G^{i\alpha}\,\bigr ) 
-i\psi^{\beta}{}_{\!\dot\alpha}\deriv^{\dot\alpha\alpha}
\chi_{\alpha\beta}
\ret
& -\fr{i}{ 2}\psi_{\alpha\dot\alpha}\deriv^{\dot\alpha\alpha}
\eta -i\zeta^j{}_{\dot\alpha}\deriv^{\dot\alpha\alpha}\lambda_{j\alpha}+
\fr{1}{4}N_{\alpha\beta}\,\bigl (\, N^{\alpha\beta}+2i F^{+\alpha\beta}
+2i[G_i{}^{\!\alpha},G^{i\beta}]\,\bigr )
\ret 
&-\fr{i}{\raiz}\,\chi^{\alpha\beta}[\chi_{\alpha\beta},C]
-\fr{i}{\raiz}\,\lambda^{i\alpha}[\lambda_{i\alpha},B]+ 
i\raiz\,\chi^{\alpha\beta}[\lambda_{i\alpha},G^i{}_\beta]+
\fr{i}{\raiz}\,\zeta_{i\dot\alpha}[\zeta^{i\dot\alpha},C]
\ret 
&+\fr{i}{\raiz}\,\eta[\lambda_i{}^\alpha,G^i{}_\alpha]
-\fr{i}{{2\raiz}}\,\eta[\eta,C]
+\fr{i}{\raiz}\,\psi_{\alpha\dot\alpha}[\psi^{\dalpha\alpha},B]+
i\raiz\,\psi^\alpha{}_{\dot\alpha}[\zeta^{i\dot\alpha},G_{i\alpha}]
\ret 
&-\half[B,C]^2  -[B,G_{i\alpha}][C,G^{i\alpha}]\,\Bigr\} -2\pi
i\tau_0\fr{1}{ 32\pi^2}\int_X  d^4 x\,\tr\,\bigl\{\, *  F_{\mu\nu}
F^{\mu\nu}
\,\bigr\} .
\ret
\la{covcinza4}
\end{align} 

Eqs. \eqs{Medeiros4} and \eqs{covcinza4} summarize what we could say was 
the ``standard"
formulation of the second  twist as discussed by Yamron \cite{yamron}. 
However, we think
there are several subtleties that demand  clarification. Since the twisted theory
contains several spinor fields taking  values in the fundamental representation of the
internal $SU(2)_F$ symmetry group, and these fields are necessarily complex, as they
live in complex representations of the rotation group and of the isospin group, it can
be  seen that the action
\eqs{covcinza4} is not real. Moreover, there are more fields  in the
twisted theory than in the physical theory. To see this, pick for example
the scalar fields
$\phi_{uv}$ in the physical $\cn=4$ theory. They are $6$ real fields that
after the twisting become the   scalar fields $B$ and $C$ (which can be
safely taken to be real, thus making a total of $2$ real fields) and  the
isospin doublet bosonic spinor field 
$G_{i\alpha}$, which is necessarily complex and thus is built out of
$2\times2
\times2=8$ real fields. Thus we see that $6$ real fields in the $\cn=4$
theory  give rise to $10$ real fields in the twisted theory. With the
anticommuting fields this overcounting is even worse.  In what follows we
will break
$SU(2)_F$  explicitly and rearrange the resulting fields wisely so as to
avoid   these problems. The outcome of this reformulation is that we will 
make contact with the non-Abelian monopole theory formulated in 
\cite{nabm}\cite{polynom}\cite{eq} (see \cite{tesis} for a review). As an
aside we would like  to comment on the reality conditions we have been
forced to impose in this and  the previous twist. It is well known that a
similar problem arises in  Donaldson-Witten theory. We are aware that it
would be desirable to have at our  disposal a systematic way of handling
these issues. However, at present there is  no such unified scheme -- but
see \cite{BnTH} for a concrete proposal in this direction. The best way
to proceed in our opinion, is to consider  the theory from the viewpoint
of the Mathai-Quillen approach (whenever this is  possible), where the
geometrical content of the theory as well as the role played  by each
field are explicit.  

We start with the fields $G_{i\alpha}$, which we rearrange in a complex  commuting 
two-component Weyl spinor $M_{\alpha}\equiv G_{2\alpha}$ and its complex  conjugate 
$\overline M^{\alpha}\equiv G_{1}{}^{\alpha}$. The constraint $G_{1}{}^{\alpha}
=(G_{2\alpha})^*$   looks rather natural when considered from the viewpoint of the
physical 
$\cn=4$ theory,  in terms of which -- recall eqn. \eqs{Inclan} and
\eqs{cctres} -- 
\begin{equation} 
G_{1\alpha}=\begin{pmatrix}
\phi_{13}\\
\phi_{23}
\end{pmatrix}=\begin{pmatrix}
B^{\dag 2}\\
-B^{\dag 1}
\end{pmatrix},
\qquad  G_{2\alpha}=\begin{pmatrix}
\phi_{14}\\
\phi_{24}
\end{pmatrix}=\begin{pmatrix}
-B_{1}\\
-B^{2}\end{pmatrix}.
\la{sienna}
\end{equation}
Similarly, for the other isodoublets in the theory we make the
rearrangements:
\begin{align}
\lambda_{1\alpha}&=\bar\mu_\alpha,&
\zeta_{1\dalpha}&=\bar\nu_\dalpha,\ret
\lambda_{2\alpha}&=\mu_\alpha,& 
\zeta_{2\alpha}&=\nu_\dalpha,\ret
P_{1\dalpha}&=\bar h_{\dalpha},&
P_{2\dalpha}&=h_{\dalpha}.
\la{bologna}
\end{align} 
Finally, after redefining $\psi\to -i\psi$, $C\to\phi$, $B\to\lambda$ and 
$N_{\alpha\beta}\to H_{\alpha\beta}$ ($A$ and $\eta$ remain the same), the action
\eqs{covcinza4} becomes:
\begin{align}
{\cal S}^{(1)}_c&=\fr{1}{ e^2_0}\int_X  d^4 x\,\sqrt{g}\,\tr\,
\bigl\{\,\half\deriv_{\alpha\dalpha}
\phi\deriv^{\dalpha\alpha}\lambda +\fr{1}{4}\bar h^{\dalpha}\bigl (
h_{\dalpha}+2\raiz i\,\deriv_{\alpha\dalpha}M^{\alpha}\bigr )
\ret
&-\fr{1}{4}h^{\dalpha}\bigl (\bar h_{\dalpha}+2\raiz i\,\deriv_{\alpha
\dalpha}\overline M^{\alpha}\bigr )- 
\psi^{\beta}{}_{\!\dalpha}\deriv^{\dalpha\alpha}
\chi_{\alpha\beta}  -\fr{1}{ 2}\psi_{\alpha\dalpha}\deriv^{\dalpha\alpha}
\eta
\ret
&-i\nu_{\dalpha}\deriv^{\dalpha\alpha}\bar\mu_{\alpha}+
i\bar\nu_{\dalpha}\deriv^{\dalpha\alpha}\mu_{\alpha}+
\fr{1}{4}H_{\alpha\beta}\bigl ( H^{\alpha\beta}+2i F^{+\alpha\beta}
+4i[\overline M^{(\alpha},M^{\beta)}]\bigr )
\ret
&-\fr{i}{\raiz}\,\chi^{\alpha\beta}[\chi_{\alpha\beta},\phi]
+i\raiz\,\bar\mu^{\alpha}[\mu_{\alpha},\lambda]
\ret
&+i\raiz\,\chi^{\alpha\beta}[\bar\mu_{\alpha},M_\beta]-
 i\raiz\,\chi^{\alpha\beta}[\mu_{\alpha},\overline M_\beta]+
i\raiz\,\bar\nu_{\dalpha}[\nu^{\dalpha},\phi]
\ret 
&-\fr{i}{\raiz}\,\eta[\bar\mu_\alpha,M^\alpha]
+\fr{i}{\raiz}\,\eta[\mu_\alpha,\overline M^\alpha]
-\fr{i}{{2\raiz}}\,\eta[\eta,\phi]
-\fr{i}{\raiz}\,\psi_{\alpha\dalpha}[\psi^{\dalpha\alpha},\lambda]
\ret
&+\raiz\,\psi^\alpha{}_{\!\dalpha}[\nu^{\dalpha},\overline M_{\alpha}]-
\raiz\,\psi^\alpha{}_{\!\dalpha}[\bar\nu^{\dalpha},M_{\alpha}]- 
\half[\phi,\lambda]^2  
\ret
&-[\lambda,\overline M_{\alpha}][\phi,M^{\alpha}]+
[\lambda,M_{\alpha}][\phi,\overline M^{\alpha}]
\,\bigr\} -2\pi i\tau_0\fr{1}{ 32\pi^2}\int_X  d^4
x\,\tr\,\sqrt{g}\bigl\{\, *   F_{\mu\nu}F^{\mu\nu}
\,\bigr\}.
\ret
\la{coventry4}
\end{align}

Let us now focus on the bosonic part of the action not containing the
scalar fields
$\phi$ and $\lambda$. After integrating out the auxiliary fields, this  part reads:
\begin{equation}
\int_X  d^4 x\,\sqrt{g}\,\tr\, \bigl\{\,-\deriv_{\alpha
\dalpha}\overline M^{\alpha}\deriv_{\beta}{}^{\dalpha}M^{\beta} +\fr{1}{4} 
\,(\,F^{+\alpha\beta} +2[\overline M_{(\alpha},M_{\beta)}]\,)^2\,\bigr\}.
\la{Assur}
\end{equation}
Expanding the squares we obtain the contributions:
\begin{align}
\int_X  d^4 x\,\sqrt{g}\,\tr\,&\bigl\{-g^{\mu\nu}\deriv_{\mu}
\overline M^{\alpha}\deriv_{\nu}M_{\alpha} -\fr{1}{4}R\,\overline
M^{\alpha}M_{\alpha}-\fr{1}{2} F^{+}_{\mu\nu}F^{+\mu\nu}
\ret 
&+[\overline M_{(\alpha},M_{\beta)}] [\overline M^{(\alpha},M^{\beta)}]
\bigr\},
\ret
\la{Dorma}
\end{align}
where we discover a non-minimal coupling to the curvature scalar $R$ of
$X$.  In the derivation of \eqs{Dorma} we have used the Weitzenb\"ock
formula,
\begin{equation}
\deriv_{\alpha\dalpha}\deriv^{\dalpha\beta}=\half\delta_\alpha{}^{\!\beta}
\,\deriv_{\gamma\dalpha}\deriv^{\gamma\dalpha}+\fr{1}{4}\delta_\alpha{}
^{\!\beta}\,R +F_{\;\alpha}^{+a\beta}T^a
\la{weitzenbock}
\end{equation} 
being $R$ the scalar curvature and 
$T^a$, $a=1,\dots,$dim$(G)$, the generators of the  gauge group in the
appropriate representation.

The corresponding BRST symmetry is readily obtained from \eqs{Medeiros4}:
\begin{align}
[Q, A_{\alpha\dalpha}] &= 2\psi_{\alpha\dalpha},& 
\{Q,\psi_{\alpha\dot\alpha}\}& =  {\raiz}\deriv_{\alpha\dot\alpha}\phi,\ret
[Q,F^{+}_{\alpha\beta}]&=2\deriv_{(\alpha}{}^{\dot\alpha}
\psi{}_{\beta)\dot\alpha},& 
[Q, \phi]&=0,\ret 
[Q, M_{\alpha}]&=-\raiz\mu_{\alpha}, &
\{Q,\mu_{\alpha}\}&=-2i [M_{\alpha},\phi],\ret
\{Q,\chi_{\alpha\beta}\} &= H_{\alpha\beta},&  
[Q,H_{\alpha\beta}]&=2\raiz i\,[\chi_{\alpha\beta},\phi],\ret
\{Q,\nu_{\dalpha}\}&= h_{\dalpha},&
[Q,h_{\dalpha}]&=2\raiz i\,[\nu_{\dalpha},\phi],\ret 
[Q, \lambda]&=\raiz\eta,& 
\{Q,\eta\,\}&=2i[\lambda,\phi].\ret
\la{covMedeiros4}
\end{align} 

The covariantized gauge fermion \eqs{cerne4} takes now the form:
\begin{align}
\Psi&=\fr{1}{ e^2_0}\int_X  d^4 x\,\sqrt{g}\,\tr\,\bigl\{\,\, 
\fr{1}{4}\bar\nu^{\dalpha}\bigl (h_{\dalpha}+ 2\raiz
i\,\deriv_{\alpha\dalpha}M^{\alpha}\bigr )- \fr{1}{4}\nu^{\dalpha}\bigl
(\bar h_{\dalpha}+ 2\raiz i\,\deriv_{\alpha\dalpha}\overline
M^{\alpha}\bigr )
\ret
&+\fr{1}{4}\chi_{\alpha\beta}\,\bigl (\, H^{\alpha\beta}+
2i(F^{+\alpha\beta} +2[\overline M^{(\alpha},M^{\beta)}])\,\bigr )\,\bigr
\}
\ret 
&-\fr{1}{ e^2_0}\int_X 
d^4 x\,\tr\,\bigl\{\,\, \fr{1}{{2\raiz}}\,\lambda\,\bigl (\,
\deriv_{\alpha\dalpha}\psi^{\dalpha\alpha} +i\raiz\,[\overline
M^{\alpha},\mu_{\alpha}] -i\raiz\,[\bar\mu^{\alpha},M_{\alpha}]
\,\bigr )\,\bigr \}
\ret 
&-\fr{1}{ e^2_0}\int_X  d^4 x\,\tr\,\bigl\{\,\, 
\fr{i}{4}\,\lambda[\eta,\phi]\,\bigr\}.\ret
\la{covcerne4}
\end{align}

The resulting theory is equivalent to the theory of  non-Abelian monopoles
discussed in
\cite{kungfu}\cite{corea}\cite{nabm}\cite{polynom}\cite{eq}\cite{tesis},   but
with the monopole multiplet in the adjoint representation of the gauge  group.
That theory in turn is a generalization of the Abelian monopole  equations
proposed in
\cite{monopole}.  

\subsection{The massive theory}

Let us now deform the  $\cn=4$ theory by turning on equal bare masses for two of
the chiral multiplets. The mass terms break preserve two supersymmetries 
and break the $SU(4)_I$ symmetry group down to $SU(2)_I\otimes U(1)_B$. Of the
four gauginos of the $\cn=4$ theory, two remain massles and transform as a
doublet under $SU(2)_I$ and carry no $U(1)_B$ charge, while the other two 
get a mass and transform as a singlet under $SU(2)_I$ and are charged under  
$U(1)_B$, which is in fact the baryon number symmetry of the $\cn=2$ theory. 
If we now twist $SU(2)_I$ with $SU(2)_L$, say, the resulting theory is a
non-Abelian monopole theory with massive monopoles in the adjoint
representation of the gauge group. This theory has a well-defined scalar
supersymmetry and is therefore topological for any value of the mass on any spin
four-manifold. This was
shown for the $\cn=2$ supersymmetric theory with one hypermultiplet  in the
fundamental representation in 
\cite{kungfu}, and it is straightforward to extend this result to the present
situation. In fact, it was shown in \cite{eq} (again for the $\cn=2$,
$N_f=1$ theory), that the twisted theory is the equivariant extension of
the massless theory with respect to the 
$U(1)_B$ symmetry. Notice that from the viewpoint of the twisting, the $U(1)_B$ 
is the Cartan subgroup of $SU(2)_F$.

\section{The Mathai-Quillen approach}

As we saw before, the model arising from the second twist is  equivalent to the
theory of non-Abelian monopoles with the monopole fields in the adjoint 
representation of the gauge group.    The
relevant basic equations for this  model  involve the self-dual part 
of the gauge
connection $F^{+}$ and a certain complex spinor field $M$  
taking values in the
adjoint representation of some compact finite dimensional  
Lie group $G$:  
\begin{equation}
\begin{cases} 
F^{+}_{\alpha\beta}+[\overline M_{(\alpha},M_{\beta)}]=0,\\
\deriv_{\alpha\dalpha}M^{\alpha}=0,
\end{cases}
\la{marinho}
\end{equation}  
where $\overline M$ is the complex conjugate of $M$. 

\subsection{The topological framework}

The geometrical setting is a certain oriented, closed Riemannian  four-manifold
$X$, that we will also assume to be spin. We will denote  the positive and
negative chirality spin bundles by $S^{+}$ and $S^{-}$  respectively. The   
field space is
${\cal M}={\cal A}\times\Gamma(X,S^{+}\otimes\ad P)$, where ${\cal A}$ is the
space of connections on a principal 
$G$-bundle $P\to X$, and the second factor denotes the space of sections of the
product bundle $S^{+}\otimes \ad P$, that is, positive chirality spinors taking
values in the Lie algebra of  the gauge group.  The group
${\cal G}$ of gauge transformations of the bundle $P$ has an action on the field
space which is given locally by:

\begin{align}
g^{*}(A)&=i(dg)g^{-1}+gAg^{-1},\ret   
g^{*}(M)&=gMg^{-1},
\la{mahler}
\end{align}   
where $M\in \Gamma(X,S^{+}\otimes\ad P)$ and $A$ is the gauge
connection.  In terms of the covariant derivative $d_A =d+i[A,~]$, the
infinitesimal form of the transformations \eqs{mahler}, with $g={\hbox{\rm
exp}}(-i\phi)$ and
$\phi\in \Omega^0(X,\ad P)$, takes the form:
\begin{align}
\delta_g(\phi)A&=d_A \phi,\ret
\delta_g(\phi)M&=i[M,\phi].
\la{mohler}
\end{align}   
The tangent space to the field space at the point $(A,M)$ is
the vector space
$T_{(A,M)}{\cal M}=\Omega^1(X,\ad P)\oplus\Gamma(X,S^{+}\otimes\ad P)$. On
$T_{(A,M)}{\cal M}$ we can define a gauge-invariant Riemannian metric given by:
\begin{equation}
\langle (\psi,\mu),(\theta,\omega)\rangle =\int_X
\tr\,(\psi\wedge *\theta)+\half\int_X \tr\,(\bar\mu^\alpha\omega_\alpha+\bar
\omega^\alpha\mu_\alpha),
\la{mmetrica}
\end{equation}   
where $\psi,\theta\in\Omega^1(X,\ad P)$ and
$\mu,\omega\in \Gamma(X,S^{+}\otimes\ad P)$. 

The basic equations \eqs{marinho} are introduced in this framework  
as sections of the
trivial vector bundle 
${\cal V}=
\mani\times{\cal F}$, where the fibre is in this case 
 ${\cal F}= \Omega^{2,+}(X,\ad P)\oplus\Gamma (X,S^{-}\otimes\ad P)$.  
Taking into
account the form of the basic  equations, the section reads, up to some 
harmless
normalization factors that we  introduce for reasons that will become 
apparent soon:
\begin{equation}   
s(A,M) =\bigl ( -2(F^{+}_{\alpha\beta}+[\overline M_{(\alpha},M_ {\beta)}]),\,
 \raiz\deriv_{\alpha\dalpha}M^{\alpha}\,
\bigr ).
\la{secante}
\end{equation}   
The section \eqs{secante} can be alternatively seen as a gauge
ad-equivariant  map from the principal ${\cal G}$-bundle ${\cal M}\to {\cal
M}/{\cal G}$ to   the vector space ${\cal F}$, and in this way it descends
naturally to a  section 
${\tilde s}$ of the associated vector bundle ${\cal M}\times_{\cal G} {\cal F}$,
whose zero locus  gives precisely the  moduli space of the  topological theory. 
It
would be desirable to compute  the dimension of this moduli space. The relevant
deformation complex (which allows one to compute, in a  general situation, the
virtual dimension of the moduli space) is the following:
\bea
0&&\!\!\!\!\!\!\!\!\!\too\Omega^0(X,\ad P)\mapright{{\cal C}}\Omega^1(X,\ad
P)\oplus\Gamma(X,S^{+}\otimes\ad P)\ret 
&&\mapright{ds}\Omega^{2,+}(X,\ad
P)\oplus\Gamma(X,S^{-}\otimes\ad P).\ret
\la{complex}
\eea   
The map ${\cal C}:\Omega^0(X,\ad P)\too T{\cal M}$ is given by:
\begin{equation}   
{\cal C}(\phi)=(d_A \phi,i[M,\phi]),\quad \phi
\in\Omega^0(X,\ad P),
\la{barbarella}
\end{equation}  
while the map $ds:T_{(A,M)}{\cal M}\too {\cal F}$ is 
provided by the
linearization of the basic equations 
\eqs{marinho}:
\bea
ds(\psi,\mu)=\bigl
(-4\sigma^{\mu\nu}_{\!\alpha\beta}\deriv_{\mu}\psi_{\nu} -2[\bar\mu_{(\alpha},
M_{\beta)}]-2[\overline M_{(\alpha},\mu_{\beta)}],\ret
\qquad\raiz\deriv_{\alpha\dalpha}\mu^{\alpha}+\raiz [\psi_{\alpha\dalpha},
M^{\alpha}]\bigr ).
\ret
\la{mozarella}
\eea
 Under suitable conditions, the index of the complex \eqs{complex} computes
de  dimension of ${\hbox{\rm Ker}}(ds)/{\hbox{\rm Im}}({\cal C})$.
 To calculate the index, the complex \eqs{complex} can be split up into the
ASD-instanton deformation complex:
\begin{align}   
&(1)~0\too\Omega^0(X,\ad P)\mapright{d_{\!A}}\Omega^1(X,\ad P)
\mapright{p^{+}\!d_{\!A}}\Omega^{2,+}(X,\ad P)\too 0,
\la{hitchin}\\
\intertext{whose index is $p_1(\ad P)+{\hbox{\rm dim}}(G)(\chi+\sigma)/2$, 
being
$p_1(\ad P)$ the first Pontryagin class of the adjoint bundle 
$\ad P$, and the complex associated to the twisted Dirac operator}  
&(2)~D:\Gamma(X,S^{+}\otimes\ad P)\too
\Gamma(X,S^{-}\otimes\ad P),
\la{francis}
\end{align}   
whose index is $p_1(\ad P)/2 - {\hbox{\rm dim}}(G)\sigma/8$.
Thus, the index of the total complex  (which gives minus the virtual dimension of
the moduli space) is:

\begin{equation}
-{\hbox{\rm dim}}({\cal M})=\ind(1)-2\times \ind(2)= {\hbox{\rm
dim}}(G)\fr{(2\chi+3\sigma)}{4}
\la{indices4}
\end{equation}  
where $\chi$ is the Euler characteristic of the $4$-manifold $X$ and 
$\sigma$ is its signature. The factor of two appears in \eqs{indices4} 
since we want  to
compute the real dimension of the moduli space.

\vskip 1cm
\subsection{The topological action}

We now proceed as in the previous case. To build a topological theory out of the
moduli problem defined by  the equations \eqs{marinho} we need the following  
multiplet of fields. For the field space ${\cal M}= {\cal
A}\times\Gamma(X,S^{+}\otimes\ad P)$ we introduce commuting fields $(A,M)$,  both
with ghost number $0$, and their  corresponding superpartners,  the anticommuting
fields
$\psi$ and $\mu$, both with ghost number
$1$. For the fibre ${\cal F}= \Omega^{2,+}(X,\ad P)\oplus\Gamma (X,S^{-}\otimes\ad
P)$ we introduce anticommuting fields 
$\chi^{+}$ and  $\nu$ respectively,  both with ghost number $-1$, and their
superpartners, a commuting self-dual two-form
$H^{+}$ and  a commuting negative chirality spinor 
$h$, both with ghost number
$0$ and which appear as auxiliary fields in the associated  field theory. And
finally, associated to the gauge symmetry, we have  a commuting  scalar field
$\phi\in\Omega^{0}(X,\ad P)$ with ghost number
$+2$, and a multiplet of scalar fields $\lambda$ (commuting and with ghost number
$-2$) and $\eta$ (anticommuting and with ghost number
$-1$), both also in $\Omega^{0}(X,\ad P)$ and which enforce the horizontal
projection ${\cal M}\to {\cal M}/{\cal G}$. The BRST symmetry of the model 
is given
by:
\begin{align}
[Q, A_\mu] &=\psi_\mu,&  
\{Q,\psi_{\mu}\} &= \deriv_{\mu}\phi, \ret
[Q,M_{\alpha}]&=\mu_{\alpha},&
\{Q,\mu_{\alpha}\,\} &=i\,[M_{\alpha},\phi],\ret
[Q,\phi]&=0,&  &{}\ret
\{Q,\chi^{+}_{\alpha\beta}\}&=H^{+}_{\alpha\beta},&
 [Q,H^{+}_{\alpha\beta}]&=i\,[\chi^{+}_{\alpha\beta},\phi],\ret
\{Q,\nu_{\dalpha}\} &= h_{\dalpha},&
[Q,h_{\dalpha}]&= i\,[\nu_{\dalpha},\phi],\ret
[Q,\lambda]&=\eta,&
\{Q,\eta\,\} &=i\,[\lambda,\phi].\ret &{} & &{}
\la{darek}
\end{align}   
This BRST algebra closes up to a gauge transformation
generated  by
$\phi$. 

We have to give now the expressions for the different pieces of the  
gauge fermion.
For the localization gauge fermion we have:
\begin{align}
\Psi_{\text{ loc}}&=i\langle(\chi^{+},\nu),s(A,M)
\rangle -\langle (\chi^{+},\nu),( H^{+},h)\rangle= 
\ret 
&\int_X
\sqrt{g}\,\tr\,\bigl\{\,\fr{1}{4}\chi^{+}_{\alpha\beta}\bigl 
(\, H^{+\alpha\beta}
+2i( F^{+\alpha\beta}+[\overline M^{(\alpha},M^{\beta)}])\,\bigr )
\ret 
&+\half\bar\nu^{\dalpha}\bigl (\,h_{\dalpha}-i\raiz
\deriv_{\alpha\dalpha}M^
{\alpha}\bigr ) -\half\nu^{\dalpha}\bigl (\,\bar
h_{\dalpha}-i\raiz\deriv_{\alpha\dalpha}
\overline M^{\alpha}\bigr )
\,\bigr \},\ret  
\la{local}
\end{align}   
and for the projection gauge fermion, which enforces the
horizontal projection,
\begin{align}
\Psi_{\text{proj}}&=\langle\lambda,{\cal C}^{\dag}(\psi,
\mu)\rangle_{\hbox{\bf g}}\ret &=
\int_X \sqrt{g}\,\tr\,\bigl\{\, \lambda\bigl (\,-\deriv_\mu\psi^\mu
+\fr{i}{2}[\bar\mu^{\alpha}, M_{\alpha}]- \fr{i}{2}[\overline M^{\alpha},
\mu_{\alpha}]
\,\bigr )\,\bigr\}.\ret
\la{projecting}
\end{align}

As in the previous case, it is necessary to add an extra 
 piece to the gauge fermion to make full contact with the 
corresponding twisted
supersymmetric theory. In  this case, this extra term is:
\begin{equation}
\Psi_{\text{extra}}=-\int_X \sqrt{g}\,\tr\,\bigl\{\,
\fr{i}{2}\lambda[\eta,\phi]\,\bigr\}.
\la{stranger}
\end{equation}

It is now straightforward to see that, after making the 
following  redefinitions,
\begin{align}
A'&=A,& M'&=M,& 
H^{'+}&=H^{+},\ret  
\psi'&=\fr{1}{2}\psi,& \overline M'&=\half \overline M,& 
\nu'&=-2\nu,\ret
\phi'&=\fr{1}{{2\raiz}}\phi,& \mu'&=-\fr{1}{\raiz}\mu,& 
\bar\nu'&=-\bar\nu,\ret
\lambda'&=-2\raiz \lambda,& 
\bar\mu'&=-\fr{1}{{2\raiz}}\bar\mu,& h'&=2h,\ret 
\eta'&=-2\eta,&
\chi^{'+}&= \chi^{+},&  
\bar h'&=-\bar h,\ret &{} & &{} & &{}
\la{dolphin}
\end{align} 
one recovers, in terms of the primed fields,  the twisted model 
summarized in
\eqs{coventry4} and \eqs{covMedeiros4}.

\newpage

\section{The  observables for $G=SU(2)$}

Let us now focus on the twisted theory for gauge group $SU(2)$. As discussed
above, the twisted
$\cn=4$  action breaks up into a
$Q$-exact piece  plus a topological term proportional to 
the instanton number of the gauge configuration,
\begin{equation}
{\cal S}_{\mbox{\rm\tiny twisted}}= \{Q,\Psi\}-2\pi ik\tau_{0},
\label{ramala}
\end{equation}
with $k=-\xi^2/4 ~\mod~\IZ$  the instanton number of the gauge
configuration with 't Hooft flux $\xi\in H^2(X,\IZ_2)$ (in this chapter we
change our notation to make contact with the results in \cite{htwist};
thus, $\xi$ will denote an
$SO(3)$ 't Hooft flux). This is an  integer for
$SU(2)$ bundles (which have $\xi=0$), but a half-integer for non-trivial
$SO(3)$  bundles on spin four-manifolds. Therefore, as pointed out in \cite{vw},
one  would expect the $SU(2)$ theory to be invariant under $\tau_0\to\tau_0+1$, 
while the $SO(3)$ theory should be only invariant under $\tau_0\to\tau_0+2$ 
on spin manifolds. Notice that, owing to (\ref{ramala}), the partition function 
depends on 
the microscopic couplings $e_{0}$ 
and $\theta_{0}$ only through the combination $2\pi ik\tau_{0}$, 
and in particular this dependence is a priori holomorphic (as usual, if we
reversed the orientation of the manifold $X$, the partition function would
depend anti-holomorphically on $\tau_0$). 

 In the twisting procedure, one couples the twisted action (\ref{ramala}) to 
 arbitrary gravitational backgrounds, so as to deal with its formulation for 
a wide variety of manifolds. In general, the procedure involves the 
covariantization of the flat-space action, as well as the addition of 
curvature terms to render the new action as a $Q$-exact piece plus a 
topological term as in (\ref{ramala}). Actually, on curved space one might think 
of additional topological terms -- such as 
$\int R\wedge R$ or $\int R\wedge*R$, with $R$ the curvature two-form of 
the manifold -- besides the one already present in (\ref{ramala}). Thus, the action 
which comes out of the twisting procedure is not unique (even modulo $Q$-exact 
terms),  since it is always possible to add $c$-number terms, which vanish on 
flat space but are nevertheless topological. In a topological field theory in 
four dimensions, those terms are proportional to the Euler number $\chi$ and 
the signature $\sigma$ of the manifold $X$. In order to keep the holomorphicity 
in $\tau_0$, the proportionality constants must be functions of $\tau_0$. 
At this stage one does not know which particular functions to take, but clearly 
good transformation properties under duality could be spoiled if one does not 
make the right choice. It seems therefore that there exists a preferred choice 
of those terms, which is compatible with duality. This issue was treated in 
detail in \cite{vw},  where it was shown that a $c$-number of the form 
$-i\pi \tau_0 \chi/6$ was needed in the topological action in order to have 
a theory with good transformation properties under duality. In the case at
hand, this $c$-number has the form
$-i\pi\tau_0(\chi+\sigma)/2$ if $2\chi+3
\sigma=0~\mod~32$, 
and $i\pi\tau_0(2\chi+3\sigma)/8$ otherwise. This will be shown in subsection
\ref{halconviajes} below. 

Topological invariants are obtained by considering the vacuum 
expectation value of arbitrary products of observables, which 
are operators that are $Q$-invariant but not $Q$-exact. As discussed in 
\cite{ene4}\cite{polynom}, the relevant observables for this theory  
and gauge group $SU(2)$ or $SO(3)$, are precisely the same as in the 
Donaldson-Witten theory \cite{tqft}:
\begin{align}
W_0 &= \fr{1}{ 8\pi^2}\tr(\phi^2), &
W_1 &= \fr{1}{4\pi^2}\tr(\phi\psi), \ret
W_2 &= \fr{1}{ 8\pi^2}\tr(2\phi F+\psi \wedge \psi), &
W_3 &= \fr{1}{4\pi^2}\tr(\psi \wedge  F). \ret
\label{guta}
\end{align}
The operators $W_i$ have positive ghost numbers given by $4-i$ 
and satisfy the descent equations 
\begin{equation}[Q, W_i\} = d W_{i-1},
\label{descent}
\end{equation}which imply that
\begin{equation}{\cal O}^{(\gamma_j)} = \int_{\gamma_j} W_j,
\label{noguta}
\end{equation}
$\gamma_j$ being homology cycles of $X$, are observables. 

The vacuum expectation value of an arbitrary product of observables has 
the general form (modulo a term which involves the exponential of a linear 
combination of $\chi$ and $\sigma$),
\begin{equation}
\left\langle \prod_{\gamma_j} {\cal O}^{(\gamma_j)}\right\rangle
= \sum_{k} 
\left\langle \prod_{\gamma_j} {\cal O}^{(\gamma_j)}\right\rangle_k
\ex^{-2\pi i k \tau_0},
\label{pumba}
\end{equation}
where $k$ is the instanton number and 
$\langle \prod_{\gamma_j} {\cal O}^{(\gamma_j)}\rangle_k$
is the vacuum expectation value computed 
at a fixed value of $k$. These quantities are independent of the 
coupling constant $e_0$. When
analysed in the weak coupling limit, the contributions to the functional
integral come from field configurations which are solutions to eqs.  
(\ref{marinho}). All the dependence of the observables on $\tau_0$ is
contained in the phases $\exp(-2\pi i k \tau_0)$ in (\ref{pumba}). 
The question 
therefore arises as to whether
the vacuum expectation values of these observables have good modular
properties under $SL(2,{\IZ})$ transformations acting on $\tau_0$. 
Below it will be shown that this is indeed the case,  
at least for  spin four-manifolds of simple type 
(although one could easily extend the arguments presented here to 
all simply-connected spin manifolds with $b^+_2>1$).

The ghost-number anomaly of the theory restricts the possible non-trivial 
topological invariants to be those for which the overall ghost number of 
the operator insertions matches the anomaly $-(3/4)(2\chi+3\sigma)$. Notice 
that since any arbitrary product of observables has necessarily positive 
ghost number, there will be no non-trivial topological invariant for those 
manifolds for which $2\chi+3\sigma$ is strictly positive. On the other hand, 
if $2\chi+3\sigma<0$, there is only a finite number -- if any -- of non-trivial 
topological invariants. Finally, when $2\chi+3\sigma=0$, as is the case for 
$K3$, for example, the only non-trivial topological invariant is the partition 
function. Moreover, as the physical and twisted theories are actually the same 
on hyper-K\"ahler manifolds as $K3$, this partition function should coincide 
with the one computed by Vafa and Witten for another twist of the $\cn=4$ 
supersymmetric theory in \cite{vw}. Below we will show that this is indeed 
the case. Notice that this assertion does not apply to the twist first 
considered by Marcus, as it actually involves two independent twists, one on 
each of the $SU(2)$ 
factors of the holonomy group of the manifold \cite{ene4}\cite{yamron}.   
The selection rule on the topological invariants that we have just 
discussed does not apply of course to the massive theory, as the 
mass terms explicitly break the ghost number symmetry. However, when the mass is
 turned on the partition
function remaims invariant (one would expect on general grounds a 
dependence of the
form $Z\sim m^{s}$, where $s$ is some number, but since the partition 
function is
dimensionless, and there is no other dimensionful parameter in the theory, 
the only
possibility is $s=0$), but the observables do get an explicit dependence on $m$
which can be fixed on dimensional grounds. Indeed, the operators $W_0$ and 
$W_2$ in \eqs{guta}, which will be the only ones that we will consider,  
have mass dimensions  $2$ and $1$ respectively (notice that these coincide with 
half the ghost number of each operator), so for a vev   
$\langle \C{O}^{(1)}\cdots \C{O}^{(r)}\rangle$ involving these operators we 
expect a mass dependence of the form   
\begin{equation}
\langle \C{O}^{(1)}\cdots \C{O}^{(r)}\rangle\propto m^{\sum^r_{n=1}g_n/2}.
\la{marflores1}
\end{equation}

\section{Integrating over the $u$-plane}

The functional-integral (or microscopic) approach to twisted
supersymmetric  quantum field theories gives great insight into their
geometric structure,  but it is not useful to make explicit
calculations. Once  the relevant set of field configurations (moduli
space) on which  the functional-integral is supported has been
identified, the computation of  the partition function or, more
generally, of the topological correlation  functions, is reduced to a
finite-dimensional integration over the quantum  fluctuations (zero
modes) tangent to the moduli space. For this to produce   sensible
topological information, it is necessary to give a suitable 
prescription for the integration, and a convenient compactification of
the  moduli space is usually needed as well. This requires an extra
input of  information which, in most of the cases is at the heart of
the subtle  topological information expected to capture with the
invariants  themselves. 

The strategy to circumvent these problems and extract concrete
predictions  rests in taking advantage of the crucial fact that, by
construction, the  generating functional for topological correlation
functions in a topological  quantum field theory is independent of the
metric on the manifold, at least for $b_2^{+}>1$. This implies that, in 
principle, these
correlation functions can be computed  in either the ultraviolet
(short-distance) or infrared (long-distance)  limits. The naive
functional-integral approach focuses on the short-distance  regime,
while long-distance computations require a precise knowledge of  the
vacuum structure and low-energy dynamics of the physical theory.    

Following this line of reasoning, it was proposed by Witten in
\cite{monopole}  that the explicit solution for the low-energy
effective descriptions for a  family of four-dimensional $\cn=2$
supersymmetric field theories presented  in \cite{swi}\cite{swii}
could be used to perform an alternative long-distance  computation of
the topological correlators which relies completely on the  properties
of the physical theory. This idea is at the heart of the  successful
reformulation of the Donaldson invariants, for a certain subset  of
four-manifolds, in terms of the by now well-known Seiberg-Witten
invariants,  which are essentially the partition functions of the
twisted effective Abelian  theories at the singular points on the
moduli space of vacua of the physical, 
$\cn=2$ supersymmetric theories. The same idea has been subsequently
applied  to some other theories
\cite{kanoyang}\cite{polynom}\cite{lns}\cite{mmone}\cite{moorewitten}, 
thereby providing a whole  set of predictions which should be tested
against explicit mathematical results.  The moral of these computations is
that the duality structure of extended  supersymmetric theories
automatically incorporates, in an as yet   not fully understood way, a
consistent compactification scheme for the  moduli space of their
twisted counterparts.    

The standard computation of this sort involves an integration 
over the moduli space of vacua (the $u$-plane) of the physical theory. 
The reason why one has to integrate over all the vacua of the physical
theory rather  than just consider the contribution from a given
vacuum has to do with the compacteness of the spacetime manifold 
\cite{dijklh}. Indeed, on a non-compact manifold the constant
values $\phi_0$ of the scalar fields which define the vacuum state of
the theory are not normalizable, so one does not integrate over them
in the path integral. But if we consider the same theory on a compact
manifold, these constant modes become normalizable (that is, they are
true zero modes) and one has to integrate over them in the path
integral. Alternatively, on a compact manifold there is tunneling
between different vacua, so there is really no point in considering
the contribution from a single vacuum. We thus have to consider the
theory for all vacua. At  a generic vacuum, the only contribution comes
from a twisted
$\cn=2$ Abelian  vector multiplet. The effect of the massive modes is
contained in appropriate  measure factors, which also incorporate the
coupling  to gravity -- these measure factors were derived in
\cite{sdual} by demanding  that they reproduce the gravitational
anomalies of the massive fields --,  and in contact terms among the
observables -- the contact term for the  two-observable for the
$SU(2)$ theory was derived in \cite{moorewitten} and was  subsequently
extended to more general observables and other gauge groups  in
\cite{lns}\cite{mmone}\cite{mmtwo}. 

The total contribution to the generating function thus consists of an 
integration  over the moduli space with the singularities removed --
which is  non-vanishing for $b^{+}_2(X)=1$ \cite{moorewitten} only --
plus a discrete sum  over the contributions of the twisted effective
theories at each of the  singularities. The effective theory at a
given singularity should contain,  together with the appropriate dual
photon multiplet, several charged  hypermultiplets, which correspond
to the states becoming massless at  the singularity. The complete
effective action for these  massless states contains as well certain
measure factors and  contact terms among the observables, which
reproduce the  effect of the massive states which have  been
integrated out. However, it is not possible to fix these a priori 
unknown functions by anomaly considerations only. As first proposed in 
\cite{moorewitten} -- see \cite{mmone}\cite{mmtwo}\cite{mm3} for more 
details and further  extensions --, the alternative strategy takes
advantage of the  {\sl wall-crossing} properties of the $u$-plane integral.
It is a well-known fact that Donaldson theory fails to produce topological 
invariants on four-manifolds with $b_2^{+}=1$. The field theory explanation of 
this failure is as follows. On manifolds $X$ with $b_2^{+}=1$ the Donaldson 
invariants are given by the sum of the contributions from the singularities at 
$u=\pm \Lambda^2$ of the $\cn=2$ SYM theory plus the $u$-plane integral.  
Consider a one-parameter family of metrics $g_\theta=\ex^\theta g$, where $g$ 
is a reference metric on $X$, and let $\omega_\theta$ be the corresponding 
period point (which gives a basis for $H^{2,+}(X)$) normalized as 
$\omega_\theta^2=1$.  
It was shown in \cite{moorewitten} that at those points on the $u$-plane 
where the (imaginary part of the) effective coupling  diverges, the integral 
has a discontinuous variation at those values of $\theta$ where, for a fixed 
gauge configuration $\lambda\in H^{2}(X;\IZ)$, the period 
$\omega_\theta\cdot\lambda$ changes sign. This is commonly referred to as 
``wall crossing" (with $\lambda$ defining a wall on the K\"ahler cone.) 
This discontinuous variation of the integral as we change $\theta$ is precisely 
the proper physical explanation of the lack of topological invariance of the 
correlation functions of the twisted theories on manifolds with $b_2^{+}=1$. 
The points where wall crossing can take place are
the  singularities of the moduli space due to charged matter
multiplets becoming  massless -- the appropriate local effective 
coupling $\tau$ diverges there -- and, in the case of the 
asymptotically free theories, the point at infinity, $u\to\infty$, 
where also the effective electric coupling diverges owing to
asymptotic freedom.

On the other hand, the final expression for the invariants can 
exhibit a wall-crossing behaviour at most at $u\to\infty$, so the 
contribution to  wall crossing from the integral at the  singularities
at finite values of $u$  must cancel against the contributions coming
from the effective theories  there, which also display wall-crossing
discontinuities.   As shown in \cite{moorewitten}, this cancellation
fixes almost completely the  unknown functions in the contributions to
the topological correlation  functions from the singularities.

\vskip2cm


\subsection{The integral for $\cn=4$ supersymmetry}

The complete expression for the $u$-plane integral for the gauge group 
$SU(2)$ and $N_f\leq 4$ was worked out in \cite{moorewitten}. The appropriate 
general formulas for the contact terms can be found in 
\cite{lns}\cite{mmone}\cite{mmtwo}. 
These formulas follow the conventions in 
\cite{swii}, 
according to which, for $N_f=0$, the $u$-plane is the modular curve of 
$\Gamma^0(4)$. In this formalism, the monodromy associated to a 
single matter multiplet 
becoming massless is conjugated to $T$. As for the $\cn=4$ supersymmetric 
theory, it is 
more convenient to use instead a formalism related to $\Gamma(2)$, in 
which the basic monodromies are conjugated to $T^2$. Our formulas follow 
straightforwardly from those in \cite{mmone}\cite{moorewitten}, 
with some  minor changes to conform to our conventions.     

The integral in the $\cn=4$ supersymmetric case, for gauge groups $SU(2)$ 
or $SO(3)$ and on 
simply-connected four-manifolds, and for 't Hooft flux $\xi$, is given by the
formula: 

\begin{equation}
\left\langle \ex^{p{\cal O}+ I(S)}\right\rangle_{\xi}\bigg\vert_{\hbox
{\tiny {\it u}-\rm plane}}= 
Z_u(p,S,m,\tau_0) = \fr{2}{ i}\int_{\cc}
\fr{dz  d\bar z }{  {y}^{1/2}}
\mu(\tau) \ex^{2 p z + S^2 \hat T(z)} \Psi ,
\label{integral}
\end{equation}
where $y=\im\,\tau$. The expression (\ref{integral}) gives the generating function 
for the vacuum expectation values of two of the observables in (\ref{noguta}): 
\bea
{\cal O}&=& \fr{1}{ 8\pi^2}\tr(\phi^2),\ret 
I(S)&=& \int_{S}\fr{1}{ 8\pi^2}\tr(2\phi F+\psi \wedge \psi).\ret
\label{belen}
\eea
Here, $S$ is a two-cycle on $X$ given by the formal sum $S=\sum_{a}
\alpha_{a}S_{a}$, where $\{S_{a}\}$, $a=1,\ldots,b_2(X)$ are two-cycles 
representing a basis of $H_2(X)$, and $S^2\equiv \sum_{a,b}\alpha_{a}
\alpha_{b}\sharp(S_{a}\cap S_{b})$, where $\sharp(S_{a}\cap S_{b})$ 
is the intersection number of $S_{a}$ and $S_{b}$. Notice that since we 
are restricting ourselves to simply-connected four-manifolds, there is no 
non-trivial contribution from the one- and three-observables $W_{1}$ and $W_3$ 
in (\ref{noguta}). 
The generalization to the non-simply-connected case was outlined in
\cite{moorewitten},  and it has been recently put on a more rigorous basis in
\cite{mm3}.  

The measure factor $\mu(\tau)$ is given by the expression:    
\begin{equation}\mu(\tau)= f(m,\chi,\sigma,\tau_0) \fr{ d \bar \tau }{  d
\bar z}
\left(\fr{d a }{  dz } \right)^{1- \half \chi  } \Delta^{\sigma/8},
\label{measure}
\end{equation}where $\Delta$ is the square root of the discriminant of the Seiberg-Witten 
curve (\ref{elliptic}):
\bea
\Delta &=& \eta(\tau_0)^{12}(z-z_1)(z-z_2)(z-z_3) = \fr{\eta(\tau)^{12}}{ 
2^3 (d a/d z)^6}
\ret
&=& -\fr{\eta(\tau_D)^{12}}{ 2^3 (d a_D/d z)^6} = 
\fr{\eta(\tau_d)^{12}}{ 2^3 (d (a_D-a)/d z)^6},
\ret
\label{discriminante}
\eea
where $\eta(\tau)$ is the Dedekind function  
and $f(m,\chi,\sigma,\tau_0)$ is a universal normalization factor which 
cannot be fixed a priori. It can be fixed in the $N_{f}=0$ case by comparing 
with  known 
results for the Donaldson invariants \cite{moorewitten}, but a
first-principles derivation from the microscopic theory in the general 
case is still lacking -- see however \cite{geog}\cite{sdual}, where some steps in
this  direction have already been taken. 

In eq. (\ref{integral}) $\hat T(z)$ is the monodromy-invariant combination:  
\begin{equation}{\hat T}(z)= T(z) + \fr{ (dz/da)^2 }{ 4 \pi {\rm Im} \,
\tau },
\label{karembeu}
\end{equation}where the {\sl contact term} $T(z)$ is given by the general formula 
\cite{mmone}\cite{mmtwo}: 
\begin{equation}T(z)= \fr{4 }{ \pi i} \fr{\partial ^2 {\cal F} }{\partial
\tau_0^2}.
\label{contacto}
\end{equation}
Here ${\cal F}$ is the prepotential governing the low-energy dynamics of 
the theory. For the asymptotically free theories, $\tau_0$ is defined in 
terms of the dynamically generated scale $\Lambda_{N_f}$ of the theory by 
\cite{mmone}: 
$(\Lambda_{N_f})^{4-N_f}=\ex^{i\pi\tau_0}$, while for the finite theories 
$N_f=4$ and $\cn=4$ it corresponds to the microscopic coupling. For the 
$\cn=4$ theory one gets from (\ref{contacto}) \cite{mmone} 
-- see also \cite{zamora} for further details and extensions:  

\begin{equation}
T(z)= -\fr{1 }{ 12} E_2 (\tau) \left( \fr{ dz }{ da }
\right)^2 + E_2 (\tau_0) \fr{z }{ 6} + \fr{m^2 }{ 72} E_4 (\tau_0),
\label{contato}
\end{equation}
where $E_2$ and $E_4$ are the Eisenstein functions of weight $2$ and 
$4$, respectively -- see the appendix to this chapter for further
details. 
 
Under a monodromy transformation acting on $\tau$ (holding $\tau_0$ fixed),
 $\tau\to (a\tau+b)/(c\tau+d)$,  
the contact term (\ref{contacto}) transforms into itself plus a shift:
$T(z)\to T(z)+ \fr{i}{ 2\pi}\fr{c}{ c\tau+d}(dz/da)^2$. Under a
microscopic  duality transformation $\tau_0\to (a\tau_0+b)/(c\tau_0+d)$,
the situation is  slightly more involved. As these duality transformations
interchange the  singularities, they induce a non-trivial monodromy
transformation 
$\tau\to ({\hat a}\tau+{\hat b})/({\hat c}\tau+{\hat d})$ on the effective 
low-energy theory \cite{ferrari}. Under these combined duality transformations 
one has, for example, 
$z\to (c\tau_0+d)^2z$, $(dz/da)\to \fr{(c\tau_0+d)^2}{{\hat c}\tau+
{\hat d} }(dz/da)$, so that \cite{zamora}:   
\begin{equation}
T(z)\to (c\tau_0+d)^4 \left(T(z) + \fr{i}{ 2\pi}\fr{{\hat
c}}{  {\hat c}\tau+{\hat d}}(dz/da)^2\right) -
\fr{i}{\pi}(c\tau_0+d)^3cz
\label{hideous}
\end{equation}
The factor $\Psi$ in (\ref{integral}) is essentially the photon partition
function,  but it contains, apart from the sum over the Abelian line 
bundles of the effective low-energy theory, certain additional terms which 
carry information about the $2$-observable insertions. In the electric frame 
it takes the form:
\bea
\Psi= \exp \biggl (-\fr{1 }{ 4 \pi y} \left(\fr{dz }{ da}\right)^2 
S_-^2 \biggr)& &\!\!\!\!\!\!\!\sum_{\lambda \in \Gamma} 
\,\biggl[ \lambda\cdot\omega + \fr{i }{ 4 \pi y} \fr{dz }{ da}
 S\cdot\omega \biggr]
\ret
& &\!\!\!\!\!\!\!
\exp \biggl[ -2i \pi {\overline \tau} (\lambda_+)^2 - 2i\pi\tau (\lambda_-)^2
-2i \fr{dz }{ da} S\cdot\lambda_- \biggr], \ret
\label{lattice}
\eea
where the lattice $\Gamma$ is $H^2(X,\IZ)$ shifted by a half-integral class 
$\half\xi=\half w_2(E)$ representing a 't Hooft flux for the $SO(3)$ theory, 
that is, $\lambda\in H^2(X,\IZ)+\half w_2(E)$. 
As explained in detail in \cite{sdual}, this shift takes into account the fact 
that in the $SO(3)$ theory, while the rank-$3$ $SO(3)$ bundle $E$ (which 
at a generic vacuum is broken down to $E=(L\oplus L^{-1})^{\otimes 2}=
L^2\oplus{\cal O}\oplus L^{-2}$, ${\cal O}$ being a trivial bundle) is 
always globally defined -- and therefore $L^2$ is represented by an integral 
class $c_1(L^2)=2\lambda\in 2H^2(X,\IZ)+w_2(E)$ --, it is not necessarily 
true that the corresponding $SU(2)$ bundle $F$ (which we can somewhat 
loosely represent at 
low energies by $F=L\oplus L^{-1}$) exists, the obstruction being precisely 
$w_2(E)$: the line bundle $L$ is represented by a class $c_1(L)=\lambda\in 
H^2(X,\IZ)+\half w_2(E)$, which is not integral unless $w_2(E)=0~(\mod~2)$. 
If $w_2(E)=0~\mod~2$, the $SO(3)$ bundle can be lifted to an $SU(2)$ bundle 
and one has $F\otimes F = E\oplus{\cal O}$, where now $F$ is a globally 
defined rank-$2$ $SU(2)$ bundle. 

For a given metric in $X$, $\omega$ in (\ref{lattice}) is the 
unique -- up to sign -- self-dual two-form satisfying -- see for example 
\cite{mmone}\cite{moorewitten}: $\omega\cdot\omega =1$ (recall that, as
explained in 
\cite{moorewitten}, the integral vanishes unless $b^+_2=1$ due to fermion 
zero modes 
{\sl and} topological invariance). Here $\cdot$ denotes the intersection 
pairing on $X$, $\omega\cdot\lambda = \int_X \omega\wedge\lambda$ . Thanks 
to its properties,  $\omega$ acts as a projector onto the self-dual and 
antiself-dual subspaces of the two-dimensional cohomology of $X$: 
$\lambda_+=(\lambda\cdot\omega)\omega$, $\lambda_-=\lambda-\lambda_+$. 
     
From the above formulas it can be readily checked, along the lines explained 
in detail in \cite{mmone}\cite{mmtwo}\cite{moorewitten}, that the integral
(\ref{integral}) is  well defined and, in particular, is invariant under
the monodromy group of  the low-energy theory (for example, this can be
seen almost immediately for  the semiclassical 
monodromy, which at large $z\simeq u$ takes $z\to \ex^{2\pi i}z$ and 
$a\to -a$, $a_D\to -a_D$, while leaving $\tau\simeq\tau_0$ unchanged.  

\subsection{Wall crossing at the singular points}

At each of the three singularities, the corresponding local effective 
coupling diverges: $y_{j} = {\rm Im}\,\  \tau_{j}\rightarrow +\infty$, 
$q_{j} \rightarrow 0$. The first step to analyse the behaviour of the 
integral around the singular points is to make a duality transformation 
(in $\tau$) to rewrite the integrand in terms of the appropriate 
variables: $\tau\to -1/\tau$ near the monopole point, etc. 
Due to the divergence of ${\rm Im}\, \tau_{j}$, one finds 
a discontinuity in $Z_u$ when $\lambda\cdot\omega$ changes sign. 
We begin by considering the 
behaviour near the semiclassical singularity at $z_1$. As 
the BPS state responsible for the singularity is electrically charged, it is 
not necessary to perform a duality transformation in this case: the theory is 
weakly 
coupled in terms of the original effective coupling $\tau$. Let us consider
the integral (\ref{integral}). Fix $\lambda$ and define $\ell(q)$ as follows:
\begin{equation}
\ell(q)=f(m,\chi,\sigma,\tau_0) \fr{ d z }{  d \tau}
\left(\fr{d a }{  dz } \right)^{1- \half \chi  } \Delta^{\sigma/8}
\ex^{2pz+S^2 T(z)-2i(dz/da)S\cdot\lambda}=\sum_{r}c(r)q^{r}.
\label{lara}
\end{equation}
Pick the $n^{th}$ term in the above expansion. The piece of the integral 
relevant to wall crossing is \cite{moorewitten}: 
\begin{equation}
\int^{\infty}_{y_{\hbox{\rm\tiny min}}}\fr{dy}{ y^{1/2}}
\int^{+\half}_{-\half} dx c(n) \ex^{2 \pi i x n - 2 \pi yn}
\ex^{-2\pi i x (\lambda_+^2 + \lambda_-^2)}
\ex^{- 2\pi y (\lambda_+^2 - \lambda_-^2)}
\lambda_{+}. 
\label{guolcrosin} 
\end{equation}
The integration over $x\equiv\re\,\tau$ imposes $n=\lambda^2$; the 
remaining $y$ integral can be easily evaluated with the result: 

\begin{equation}
\int^{\infty}_{0}\fr{dy}{ y^{1/2}}c(\lambda^2) 
\ex^{-4\pi y\lambda_+^2}\lambda_+=\fr{\vert\lambda_+
\vert}{\lambda_+} \fr{c(\lambda^2)}{2} 
\label{masguolcrosin}
\end{equation}(we have set $y_{\hbox{\rm\tiny min}}=0$, as the discontinuity   
comes from the $y\to\infty$ part of the integral). The result of the integral 
(\ref{masguolcrosin}) is discontinuous as $\lambda_{+}=\omega
\cdot\lambda\to 0$:  

\begin{equation}
Z_u\big\vert_{\lambda_+\rightarrow 0^{+}} - Z_u\big\vert_{\lambda_+
\rightarrow 0^{-}} = c(\lambda^2) = \bigl[ q^{-\lambda^2} 
\ell(q) \bigr]_{q^0}=
{\hbox{\rm Res}}_{\,q=0}\bigl[ q^{-\lambda^2-1}\ell(q) \bigr].
\label{reguolcrosin}
\end{equation}Therefore, the wall-crossing discontinuity of $Z_u$ at 
$z_1$ is:
\bea
& &\Delta Z_u\big\vert_{z=z_1}= f(m,\chi,\sigma,\tau_0)\left[ q^{-\lambda^2} 
\fr{d z }{  d \tau}\left(\fr{d a }{  dz } \right)^{1-\half\chi} 
\Delta^{\sigma/8}\ex^{2pz+S^2 T(z)-2i(dz/da)S\cdot\lambda} 
\right]_{q^0}=\ret 
& & {\hbox{\rm Res}}_{\,q=0}
f(m,\chi,\sigma,\tau_0)\left[ \fr{dq}{ q} q^{-\lambda^2} 
\fr{d z }{  d \tau}\left(\fr{d a }{  dz } \right)^{1-\half\chi} 
\Delta^{\sigma/8}\ex^{2pz+S^2 T(z)-2i(dz/da)S\cdot\lambda} 
\right]\ret 
\label{finguolcrosin}
\eea

We have now to evaluate the wall-crossing discontinuities at 
the other two singularities. At the monopole point ($z=z_2$), 
we have to perform a duality transformation to express the 
integral in terms of $\tau_{D}=-1/\tau$, which is the 
appropriate variable there. This duality transformation involves 
a Poisson resummation in (\ref{lattice}), which exchanges the electric 
class $\lambda\in H^{2}(X,\IZ)+\half w_{2}(E)$ with the magnetic 
class $\lambda^{*}\in H^{2}(X,\IZ)$\footnote{Notice that, as the manifold 
$X$ is spin, the magnetic class is an integral class, not a 
Spin$_c$ structure as in \cite{monopole}.}, and inverts the coupling 
constant $\tau$. The details are not terribly important, so we 
just give the final result for the integral:
\begin{align}
Z_u &= f(m,\chi,\sigma,\tau_0)2^{-\frac{b_2}{2}}
\int\fr{dx_{D}dy_{D}}{{y_D}^{1/2}}\fr{ d z}{  d \tau_{D}}
\left(\fr{d a_{D} }{  dz } \right)^{1- \half \chi  } 
\Delta^{\sigma/8}\ex^{2 p z + S^2 \hat T_{D}-\fr{1}{4\pi y_{D}} 
\left(\fr{dz }{ da_{D}}\right)^2 S_-^2}\ret &\sum_{\lambda^{*}} 
\left( \fr{\lambda^{*}\cdot\omega}{2} + \fr{i }{ 4 \pi y_{D}} 
\fr{dz }{ da_{D}} S\cdot\omega \right)(-1)^{\lambda^{*}\cdot\xi}
\ex^{-\half i\pi{\overline
\tau_{D}} (\lambda^{*}_+)^2 - \half i\pi\tau_{D}
(\lambda^{*}_-)^2-i \fr{dz}{ da_{D}} S\cdot\lambda^{*}_-},\ret
\label{dualintegral}
\end{align}
where now   
\begin{equation}
{\hat T}_{D}(z)=-\fr{1 }{ 12} E_2 (\tau_{D}) \left( 
\fr{ dz}{ da_{D} } 
\right)^2 + E_2 (\tau_0) \fr{z }{ 6} + \fr{m^2 }{ 72} E_4 (\tau_0)+
 \fr{ (dz/da_{D})^2 }{ 4 \pi {\rm Im} \, \tau_{D}}.
\label{seedorf}
\end{equation}The functions $\Delta$ and $z$ are exactly the same as before, but expressed 
in terms of $\tau_D$. The crucial point here is that the modular weight 
of the lattice sum cancels against that of the measure.  

From (\ref{dualintegral}) we can easily derive the wall-crossing discontinuity 
at $z_2$ along the lines explained above -- see eqs. 
(\ref{lara})--(\ref{finguolcrosin}). The final result differs from (\ref{finguolcrosin}) 
in several extra numerical factors:
\bea
\Delta Z_u\big\vert_{z=z_2}&=& f(m,\chi,\sigma,\tau_0) 2^{-b_2/2}
(-1)^{\lambda^{*}\cdot\xi}\ret 
{\hbox{\rm Res}}_{\,q_{D}=0}&&\!\!\!\!\!\!\!\!\!\!\!
\left( \fr{dq_{D}}{ q_{D}}q_{D}^{-(\lambda^{*})^2/4} 
\fr{d z }{  d \tau_{D}}\left(\fr{d a_{D}}{  dz } \right)^{1-\fr{\chi}{2}} 
\Delta^{\sigma/8}\ex^{2pz+S^2 T_{D}-i\fr{dz}{ da_{D}}S\cdot\lambda^{*}} 
\right).\ret
\label{finlandia}
\eea

The corresponding expression at the dyon point $z_3$, is exactly 
the same as (\ref{finlandia}) (with $q_{d}$ instead of $q_{D}$) but with 
an extra relative phase $i^{-\xi^2}$ 
\cite{moorewitten}\cite{wijmp}\cite{monopole}, which 
follows from doing the duality transformation 
$\tau\to\tau_{d}=-1/(\tau-1)$ in the lattice sum (\ref{lattice}).

      Ä


\subsection{Contributions from the singularities}

At each of the singularities, the complete effective theory 
contains a dual Abelian vector multiplet\footnote{This is so for the monopole and 
dyon singularities; at the semiclassical singularity, the distinguished 
vector multiplet is the original electric one.} (weakly) coupled to a 
massless charged hypermultiplet representing the BPS configuration 
responsible for the singularity. This effective theory can be twisted in the 
standard way, and the resulting topological theory is the celebrated 
Witten's Abelian monopole theory. Its moduli space is defined by the 
Abelianized version of eqs. (\ref{marinho}). On spin four-manifolds, 
and for a given gauge configuration $\tilde\lambda\in H^{2}(X,\IZ)$, 
the virtual dimension of the moduli space can be seen to be
\begin{equation}
{\hbox{\rm dim}}_{\tilde\lambda}= -\fr{(2\chi+3\sigma)}{4}+
(\tilde\lambda)^{2}.
\label{indiecita}
\end{equation}
A class $\tilde\lambda$ for which ${\hbox{\rm dim}}_{\tilde\lambda}=0$ 
is called a {\sl basic} class. If we define $x=-2\tilde\lambda$, we see 
from (\ref{indiecita}) that for a basic class $x\cdot x=2\chi+3\sigma$. As 
${\hbox{\rm dim}}_{x}=0$, 
the moduli space consists (generically) of a (finite) collection of 
isolated points. The partition function of the theory 
evaluated at each basic class gives the Seiberg-Witten invariant 
$n_{x}$. The complete partition function will therefore be a (finite) sum 
over the different basic classes: $Z_{{\hbox{\rm\tiny singularity}}}\sim 
\sum_{x}n_{x}$. If, on the other hand, the dimension of the moduli space 
of Abelian monopoles is strictly positive, one has to insert 
observables to obtain a non-trivial result. This leads to the definition  
of the generalized Seiberg-Witten invariants \cite{moorewitten}\cite{taubes}: if 
${\hbox{\rm dim}}_{\tilde\lambda}= 2n$ (otherwise the invariant is 
automatically set to zero), 
\begin{equation}SW_n(\tilde\lambda)=\left\langle(\tilde\phi)^n\right\rangle_
{\tilde\lambda},
\label{general}
\end{equation}where $\tilde\phi$ is the (twisted) scalar field in the Abelian $\cn=2$ 
vector multiplet. For a four-manifold $X$ with $b^{+}_{2}>1$, the $u$-plane 
integral vanishes and the only contributions to the topological correlation 
functions come from the effective theories at the singularities. Those  
manifolds with $b^{+}_{2}>1$ for which the only non-trivial contributions 
come from the zero-dimensional Abelian monopole moduli spaces are called of 
{\sl simple type}. No four-manifold with $b^{+}_{2}>1$ is known which is not 
of simple type. We will restrict ourselves to manifolds of simple type. The 
generalization to positive-dimensional monopole moduli spaces should be 
straightforward from the explicit formulas in \cite{moorewitten} and our own results. 
 
The general form of the contribution to the generating function 
$\left\langle\ex^{p{\cal O}+I(S)}\right\rangle_{\xi}$ from  
the twisted Abelian monopole theory at a given singularity was presented in 
\cite{moorewitten}. It contains certain effective gravitational couplings as well as 
contact terms among the observables. We just adapt here eq. (7.12) of 
\cite{moorewitten}:   

\bea
&&\left\langle \ex^{p{\cal O}+I(S)}\right\rangle_{\tilde\lambda_j, z_j,\xi} 
= SW_n(\tilde\lambda)\, {\hbox{\rm Res}}_{\,a_j=0}\Bigg\{\fr{da_{j}}{ 
(a_{j})^{1+\tilde
\lambda_{j}^2/2-(2\chi+3\sigma)/8}}(-1)^{\tilde\lambda_{j}\cdot\xi}\ret 
&&\ex^{2pz - i \fr{d z }{ da_{j}} \tilde\lambda_{j}\cdot S 
+ S^2 T_{j}(z)} C_j(z)^{-\tilde\lambda_j^2} P_j(z)^{\sigma/8}
L_j(z)^{\chi/4}\Bigg\}.
\ret
\label{sw}
\eea
In (\ref{sw}), $a_{j}$ is the distinguished (dual) coordinate at the singularity:
$a-a(z_1)\simeq a-m/\raiz$ at the semiclassical singularity, 
$a_{D}-a_{D}(z_2)$ at the monopole point, and 
$(a-a_{D})-(a-a_{D})(z_3)$ at the dyon point. $C_j$, $P_j$, $L_j$ are a priori 
unknown functions, which can be determined by wall crossing as follows 
\cite{moorewitten}. For $b^{+}_2=1$ and  fixed $\tilde\lambda_{j}$, (\ref{sw}) 
exhibits a 
wall-crossing behaviour when $\omega\cdot\tilde\lambda_j$ changes sign. 
At such points, 
the only discontinuity comes from $SW_n(\tilde\lambda)$, which jumps 
by $\pm 1$. 
Therefore, the discontinuity in (\ref{sw}) is: 
\bea
& &\Delta\left\langle \ex^{p{\cal O}+I(S)}\right\rangle_{\tilde\lambda_j, 
z_j,\xi} = \pm {\hbox{\rm Res}}_{\,a_j=0}\Bigg\{\fr{da_{j}}{ (a_{j})^
{\tilde\lambda_{j}^2/2-\sigma/8}}(-1)^{\tilde\lambda_{j}\cdot\xi}
\ret & &\ex^{2pz - i \fr{d z }{ da_{j}} \tilde\lambda_{j}\cdot S 
+ S^2 T_{j}(z)} C_j(z)^{-\tilde\lambda_j^2} P_j(z)^{\sigma/8}
L_j(z)^{1-\sigma/4}\Bigg\}.
\ret
\label{swdiscont}
\eea
(We have set $\chi+\sigma=4$, which is equivalent to $b^{+}_{2}=1$ 
for $b_{0}(X)=1, b_{1}(X)=0$.) The crucial point now is that the 
complete expression for the generating function cannot have 
wall-crossing discontinuities at finite values of $z$. This is not 
difficult to understand if one realizes that nothing physically (or 
mathematically) special occurs at the singular points: when expressed 
in terms of the appropriate variables, and once all the relevant degrees 
of freedom are taken into account, the low-energy effective description 
is perfectly smooth there. The conclusion is therefore that the 
discontinuity in the $u$-plane integral has to cancel against the 
discontinuity in the contribution from the effective theory at the 
singularity. As shown in \cite{moorewitten}, this suffices to fix the unknown 
functions $C_j$, $P_j$, $L_j$ in (\ref{sw}).

At a generic vacuum, the $SU(2)$ -- or, more generally, the $SO(3)$ --  
rank-$3$  bundle $E$ is broken down to $E=L^2\otimes{\cal O}
\otimes L^{-2}$ by the 
Higgs mechanism, where ${\cal O}$ is the trivial line bundle 
(where the neutral $\cn=4$ multiplet lives), 
while $L^{\pm 2}$ are globally defined line bundles where the charged 
massive $W^{\pm}$ $\cn=4$ multiplets live. With our conventions, 
$c_1(L^2)=2c_1(L)=2\lambda\in 2H^{2}(X,\IZ)+w_2(E)$, which is indeed an 
integral class. The ``monopole" becoming massless at the semiclassical 
singularity is just one of the original electrically charged (massive) 
quarks, which sits in an $\cn=4$ Abelian multiplet together with the 
$\cn=2$ vector multiplet of one of the massive $W$ bosons. The corresponding 
basic classes are therefore of the form: 
\begin{equation}
x=-2\tilde\lambda_{1}=-2c_1(L^2)=-4\lambda\in 4H^{2}(X,\IZ)+2w_2(E),
\label{alice}
\end{equation}
which are even classes since the manifold $X$ is spin\footnote{If the
manifold  is not spin, the basic classes are shifted from being even
classes by the  second Stiefel-Whitney class of the manifold, $w_2(X)$
\cite{monopole}.}.  Notice that, because 
of (\ref{alice}), not all the basic classes of $X$ will contribute to the 
computation at $z_1$. Rather, only those basic classes $x$ satisfying 
\begin{equation}
\fr{x}{2}+w_2(E)= 0~(\mod~2)\Leftrightarrow \left[\fr{x}{2}\right]=w_2(E), 
\label{constraint}
\end{equation}with $\left[\fr{x}{2}\right]$ the mod $2$ reduction of
$\fr{x}{2}$, can give  a non-zero contribution.  

Taking this into account, we can rewrite (\ref{swdiscont}) at $z_1$ as follows:
\bea
&&\Delta\left\langle \ex^{p{\cal O}+I(S)}\right\rangle_{2\lambda,z_1,\xi} 
= \pm \fr{1}{\pi i}{\hbox{\rm Res}}_{\,q=0}\Bigg\{\fr{dq }{ q} (a_{1})^
{-2\lambda^2+\sigma/8}\fr{da}{ dz}\fr{dz}{ d\tau}
\ret &&\ex^{2pz - 2i \fr{d z }{ da} \lambda\cdot S 
+ S^2 T_{1}(z)} C_1(z)^{-4\lambda^2} P_1(z)^{\sigma/8}
L_1(z)^{1-\sigma/4}\Bigg\} 
\ret
\label{swdiscuno}
\eea
(notice that the phase $(-1)^{\tilde\lambda\cdot\xi}$ does not appear 
here), where we have used  
\begin{equation} 
{\hbox{\rm Res}}_{\,a=0}\left[da
F(a)\right]= 2\,{\hbox{\rm Res}}_{\,q=0}\left[dq (da/dq) F(a)\right],
\end{equation} 
and we
have  taken into account that, near $z=z_1$, $a_{1}=a-a(z_1)=a_0 q^{1/2}+\cdots$.
By comparing (\ref{swdiscuno}) with the wall-crossing formula for the integral 
at $z_1$, (\ref{finguolcrosin}), we can determine the unknown functions 
in (\ref{sw}). 
We find, for example, 
\begin{align}
T_1 &= T,& L_1 =&\left(\fr{dz }{  da}\right)^2,\ret
(C_1)^4 &= \fr{{\displaystyle a}_1{}^2}{\displaystyle q},
& P_1 =& \fr{\Delta}{ a_1}. \ret
{}&{}&{}&
\label{alessia}
\end{align}

Putting everything together, the final form for the contribution to 
the generating function at $z_1$ is given by the following formula:
\bea
\left\langle \ex^{p{\cal O}+I(S)}\right\rangle_{\lambda,z_1,\xi}&=& 
SW_n(\tilde\lambda)\, 2\pi if(m,\chi,\sigma,\tau_0)\ret 
{\hbox{\rm Res}}_{\,q=0}&&\!\!\!\!\!\!\!\!\!
\left[ dq q^{-\lambda^2} 
\fr{d z }{  d q}\left(\fr{d a }{  dz } \right)^{1-\half\chi} 
a_{1}{}^{\fr{\chi+\sigma}{4}-1}\Delta^{\sigma/8}\ex^{2pz+S^2 
T(z)-2i(dz/da)S\cdot\lambda} 
\right].\ret 
\label{seibergwitten}
\eea
We can now specialize to the simple-type case, for which $4\lambda^2 = 
(2\chi+3\sigma)/4$. We use the following series expansions around $z_1$:
\bea
z&=&z_1+\kappa_1 q^{\half}+\cdots,\ret 
a_{1}&=& (da/dz)_1(z-z_1)+\cdots = 
(da/dz)_1 \kappa_1 q^{\half}+\cdots,\ret
da/dz &=&(da/dz)_1+\cdots,\ret
\Delta &=&\fr{\eta(\tau)^{12}}{ 2^3 (da/dz)^6}= 
2^{-3} (dz/da)_1{}^6 q^{\half}+\cdots.\ret
\label{serexp}
\eea
The final formula is the following:
\bea
\left\langle \ex^{p{\cal O}+I(S)}\right\rangle_{x,z_1,\xi}&=& 
2^{-\fr{3\sigma}{8}}\pi i
f(m,\chi,\sigma,\tau_0)\, \ret (\kappa_1)^{\nu} 
&&\!\!\!\!\!\!\!\!\!\!\left(\fr{d a }{  dz } \right)_1^{-(\nu+\sigma/4)} 
\ex^{2pz_1+S^2 T(z_1)}\,
\delta_{\left[\fr{x}{2}\right],\xi}\,n_{x}\,
\ex^{\half i(dz/da)_1S\cdot x},  
\ret 
\label{wimoore}
\eea
where $\nu=(\chi+\sigma)/4$. The delta function 
$\delta_{\left[\fr{x}{2}\right],\xi}$ in (\ref{wimoore})\ enforces the 
constraint (\ref{constraint}), and $T(z_1)$ is given by:
\begin{equation}T(z_1)= -\fr{1 }{ 12} ( dz/da)_1{}^2 +
E_2 (\tau_0) \fr{z_1 }{ 6} + \fr{m^2 }{ 72} E_4 (\tau_0).
\label{contralto}
\end{equation}
The corresponding expressions at the monopole and dyon 
singularities can be determined along the same lines. One 
has to take into account the relative factors in each case, 
and the fact that, at these singularities, the basic classes 
are given by $x=-2\tilde\lambda=-2\lambda^{*}$, where 
$\lambda^{*}$ is the appropriate dual class. One finds in this 
way, for the monopole singularity at $z_2$, the following expression: 
\bea
\left\langle \ex^{p{\cal O}+I(S)}\right\rangle_{x,z_2,\xi}&=& 
2^{-\fr{3\sigma}{8}-\fr{b_2}{ 2}}\pi i
f(m,\chi,\sigma,\tau_0)\, \ret (\kappa_2)^{\nu} 
&&\!\!\!\!\!\!\!\!\!\!\left(\fr{d a_{D} }{  dz }
\right)_2^{-(\nu+\sigma/4)} 
\ex^{2pz_2+S^2 T(z_2)}
(-1)^{ \fr{x}{2}\cdot \xi}\,n_{x}\,\ex^{\half i(dz/da_{D})_2 S\cdot x},  
\ret 
\label{monopolo}
\eea
while for the dyon singularity at $z_3$ one finds:
\bea
\left\langle \ex^{p{\cal O}+I(S)}\right\rangle_{x,z_3,\xi}&=&
2^{-\fr{3\sigma}{8}-\fr{b_2}{ 2}}\pi i
f(m,\chi,\sigma,\tau_0)\, \ret i^{-\xi^2}(\kappa_3)^{\nu} 
&&\!\!\!\!\!\!\!\!\!\!\left(\fr{d (a_{D}-a)}{  dz }
\right)_3^{-(\nu+\sigma/4)} 
\ex^{2pz_3+S^2 T(z_3)}
(-1)^{\fr{x}{2}\cdot\xi}\, n_{x}\,\ex^{\half i(dz/d(a_{D}-a))_3 S\cdot
x},  
\ret 
\label{dyon}
\eea
where $T(z_2)$ and $T(z_3)$ are given by expressions analogous to 
(\ref{contralto}):
\bea
T(z_2)&=& -\fr{1 }{ 12} ( dz/da_{D})_2{}^2 +
E_2 (\tau_0) \fr{z_2 }{ 6} + \fr{m^2 }{ 72} E_4 (\tau_0),\ret
T(z_3)&=& -\fr{1 }{ 12} ( dz/d(a_{D}-a))_3{}^2 +
E_2 (\tau_0) \fr{z_3 }{ 6} + \fr{m^2 }{ 72} E_4 (\tau_0). 
\ret
\label{mezzosoprano}
\eea

In \eqs{monopolo} there is an extra phase $(-1)^{\sigma/8}$ that we have 
supressed. The reason for this is that, as stated in a theorem by Rohlin 
\cite{donaldkr}, on smooth, compact, spin four-manifolds, $\sigma\in 16\IZ$. 
This is in contrast with the weaker arithmetic 
result $\sigma\in 8\IZ$ which follows from the evenness of the intersection 
form. We will restrict ourselves to such four-manifolds and hence supress this  
phase in what follows.



\subsection{The formula for the generating function}
\la{halconviajes}
The complete formula for the generating function of the  
theory on simply-connected spin four-manifolds of simple type is given 
by the combination of (\ref{wimoore}), (\ref{monopolo}) and (\ref{dyon}), summed
over  the basic classes (we do not sum over 't Hooft fluxes, though). The 
contribution from the $u$-plane integral is absent, since it vanishes for 
manifolds with $b^{+}_2>1$. As for the as yet unknown function 
$f(m,\chi,\sigma,\tau_0)$, it is not possible to determine it completely
in the context of the $u$-plane approach. 
However, we will propose an ansatz for this function, which is motivated 
by a series of natural conditions that it has to satisfy. 
As we will show, with our ansatz  the partition function is dimensionless
and   displays two properties of the partition function of the twisted
$\cn=4$  supersymmetric theory
considered by Vafa and Witten \cite{vw}: it is a modular form of weight 
$-\chi/2$ and contains the Donaldson invariants in the form shown 
in \cite{coreatres}. In addition, its final expression reduces to 
the Vafa-Witten 
partition function on $K3$. 

Our ansatz for $f(m,\chi,\sigma,\tau_0)$, which turns out to satisfy the  
properties
stated above, is:
\begin{equation}
f(m,\chi,\sigma,\tau_0)= -\fr{i}{\pi}\,2^{(3\chi+7\sigma)/8}\,
m^{-(3\nu+7\sigma)/8} 
\eta(\tau_0)^{-12\nu}.
\label{ansatz}
\end{equation}
In \cite{htwist} a different choice for $f$ was made 
\begin{equation}
f(m,\chi,\sigma,\tau_0)= -\fr{i}{\pi}\,2^{(3\chi+7\sigma)/8}\,
m^{\sigma/8} 
\eta(\tau_0)^{-12\nu},
\label{ansatzold}
\end{equation}
the only difference being that the generating function therein contained  
an overall factor $m^{3(2\chi+3\sigma)/8}$. If we insist that the generating
function be adimensional -- see \cite{geog} -- the correct choice is of course 
\eqs{ansatz}. 

Taking (\ref{wimoore}), (\ref{monopolo}) and (\ref{dyon}), the formula
that one obtains for  the generating function of all the topological correlation
functions for  simply-connected spin manifolds is the following:
\bea
\left\langle\ex^{p{\cal O}+I(S)}\right\rangle_{\xi}= 2^{\nu/2}
2^{(2\chi+3\sigma)/8}m^{-(3\nu+7\sigma)/8}
(\eta(\tau_0))^{-12\nu}&&\!\!\!\!\!\!\Bigg\{
\ret 
(\kappa_1)^{\nu} \left (\fr{da}{dz}\right)^
{-(\nu+\fr{\sigma}{4})}_1\ex^{2pz_1 + S^2 T_1} &&\!\!\!\!\!\!\sum_{x} 
\delta_{\left[\fr{x}{2}\right],\xi} \,n_x \,\ex^{\fr{i}{2} \left (dz
/da\right)_1 x\cdot S}\ret +
2^{-\fr{b_2}{2}}(\kappa_2)^{\nu} \left 
(\fr{{da_D}}{{dz}}\right)^
{-(\nu+\fr{\sigma}{4})}_2\,\ex^{2pz_2 + S^2 T_2} &&\!\!\!\!\!\!\sum_{x} 
(-1)^{\xi \cdot \fr{x}{2}}\,n_x\, \ex^{\fr{i}{2} \left (dz/
da_D\right)_2 x\cdot S}\ret +
2^{-\fr{b_2}{2}} i^{-\xi^2}(\kappa_3)^{\nu} \left 
(\fr{{d(a_D-a)}}{{dz}}\right)^
{-(\nu+\fr{\sigma}{4})}_3\ex^{2pz_3 + S^2 T_3} &&\!\!\!\!\!\!\sum_{x} 
(-1)^{\xi \cdot \fr{x}{2}}\,n_x\, \ex^{\fr{i}{2} \left (dz/
d(a_D-a)\right)_3 x\cdot S}\Bigg\},\ret 
\label{formulauno}
\eea
where the sum $\sum_{x}$ is over {\sl all} the Seiberg-Witten basic 
classes. This formula can be written in terms of modular 
forms by substituting the explicit expressions (\ref{yespadas}) for 
$\kappa_{j}$ and (\ref{zhivago}) for the periods. Notice that there is no need 
to resolve the square roots in (\ref{zhivago}). Indeed, the periods in 
(\ref{formulauno}) are raised to the power $-(\nu+\sigma/4)$. Since the 
manifold $X$ is spin, $\sigma=0~\mod~16$, so $\sigma/4$ is even. As 
for $\nu=(\chi+\sigma)/4$, it is only guaranteed\footnote{
For $x=-2\tilde\lambda=-2c_1(\tilde L)$ a basic class, $\nu$ is the index 
of the corresponding Dirac operator\hfill\break 
$D:\Gamma(X,S^{+}\otimes\tilde L)\to 
\Gamma(X,S^{-}\otimes\tilde L)$, which is always an integer \cite{monopole}.}
that $\nu\in\IZ$. 
Nevertheless, as explained in sect. 11.5 of \cite{moorewitten}, one can 
take advantage of the following property of the Seiberg-Witten invariants 
$n_{x}$ -- see \cite{monopole} for a quick proof:
\begin{equation}
n_{-x}=(-1)^{\nu}n_{x}. 
\label{sherezade}    
\end{equation}
Upon summing over $x$ and $-x$ using (\ref{sherezade}), the factors 
$\ex^{\fr{i}{2}
(dz/da)_{j}x\cdot S}$ average to a cosine 
when $\nu$ is even, and to 
a sine when $\nu$ is odd. Therefore, no odd powers of $(dz/da)_j$ appear 
in (\ref{formulauno}). Note that as $x$ is an even class on 
spin manifolds, $\xi \cdot x/2$ is always an integer, and therefore 
the phase $(-1)^{\xi \cdot \fr{x}{2}}$ does not spoil the argument. 
Likewise, if $x+2\xi=0~\mod~4$, it is also true that 
$-x+2\xi=0~\mod~4$ ($x$ is an even class), so if a given basic class 
$x$ contributes to the first term in (\ref{formulauno}) , so does $-x$.

We will now work out a more explicit formula for the partition function
(setting $p=0$ and $S=0$ in (\ref{formulauno})). Notice that, since the partition 
function does not care 
about whether the manifold is simply-connected or not, at least in the 
simple-type case (in any case, we are not computing correlation 
functions of observables), we can easily extend 
our result to the non simply-connected case. As in \cite{vw} this extension 
involves the introduction of a factor $2^{-b_1}$ in two of the three 
contributions. Notice also that, because 
of (\ref{sherezade}), the partition function is zero when $\nu$ is odd, 
since $n_{x}+n_{-x}=n_{x}(1+(-1)^{\nu})$. Therefore, in the following 
formula for the partition function we assume that $\nu$ is even. Upon 
substituting eqs. (\ref{yespadas}) and (\ref{zhivago}) in (\ref{formulauno}), 
and taking 
into account the identities  \eqs{valiente}, the partition function 
for a fixed 't Hooft flux $\xi$ is given by:
\bea
Z_\xi &=& \Bigg\{
\left(\fr{G(q_0{}^2)}{4}\right)^{\nu/2}
\left(\vartheta_3\vartheta_4\right)^{(2\chi +3\sigma)/2}
\sum_{x}\delta_{\xi,\left[\fr{x}{2}\right]}n_x
\ret
&&+ 2^{1-b_1+\fr{1}{4}(7\chi +11\sigma)} 
\left(G(q_0{}^{1/2})\right)^{\nu/2}
\left(\fr{\vartheta_2 \vartheta_3}{2}\right)^
{(2\chi +3\sigma)/2}\sum_{x}(-1)^{\fr{x}{2}\cdot \xi}\, n_x 
\ret &&
+ 2^{1-b_1+\fr{1}{4}(7\chi +11\sigma)} 
  i^{-\xi^2}
\left(G(-q_0{}^{1/2})\right)^{\nu/2}
\left(\fr{\vartheta_2 \vartheta_4}{2}\right)^
{(2\chi +3\sigma)/2}
\sum_{x} (-1)^{\fr{x}{2}\cdot \xi}\,n_x\Bigg\}.
\ret
\label{formulados}
\eea 

The partition function for gauge groups $SU(2)$ and $SO(3)$ is easily 
obtained from this expression. One finds:
\begin{align}
Z_{SU(2)} &= Z_{\xi=0}/2^{1-b_1} \ret
   &= \Bigg\{2^{b_1-1}
\left(\fr{G(q_0{}^2)}{4}\right)^{\nu/2}
\left(\vartheta_3\vartheta_4\right)^{(2\chi +3\sigma)/2}
\sum_{x=4y}n_x
\ret
&+ 2^{\fr{1}{4}(7\chi +11\sigma)} 
\left(G(q_0{}^{1/2})\right)^{\nu/2}
\left(\fr{\vartheta_2 \vartheta_3}{2}\right)^
{(2\chi +3\sigma)/2}\sum_{x}n_x 
\ret &
+ 2^{\fr{1}{4}(7\chi +11\sigma)} 
 \left(G(-q_0{}^{1/2})\right)^{\nu/2}
\left(\fr{\vartheta_2 \vartheta_4}{2}\right)^
{(2\chi +3\sigma)/2}
\sum_{x}n_x\Bigg\}.
\ret
\label{formulacinco}
\end{align} 
The constraint $x=4y$ in the first term in (\ref{formulacinco}) means that 
one has to consider only those basic classes $x\in 4H^2(X,\IZ)$. 
Notice that this constraint implies that the corresponding 
contribution vanishes unless $x^2 = 2\chi+3\sigma = 8\nu +\sigma = 
16 y^2 = 0~\mod~32$. Notice that as $\nu$ is even and $\sigma\in 16\IZ$, it 
is only guaranteed that $2\chi+3\sigma=0~\mod~16$. When $2\chi+3\sigma=
0~\mod~32$, the first 
singularity contributes to the $SU(2)$ partition function, and the 
leading behaviour of the partition function is 
\begin{equation}
Z_0\sim q_0{}^{-\nu}+{\cal O}(q_0{}^{-\nu+1}).
\label{leading}
\end{equation}
As in \cite{vw}, the leading contribution could be interpreted as 
the contribution of the trivial connection, shifted from $(q_0)^{0}$ to 
$q_0{}^{-\nu}$ by the $c$-number we referred to above. But notice 
that the next power in the series expansion is $q_0{}^{-\nu+1}$. The gap 
between the trivial connection contribution and the first non-trivial 
instanton contribution which was noted in \cite{vw} for the Vafa-Witten 
partition function is not present here: all instanton configurations 
contribute to $Z_{SU(2)}$.  
 
On the other hand, when $2\chi+3\sigma= 16(2j + 1),~j\in\IZ$, the first 
singularity does not contribute to the partition function and the leading 
behaviour comes from the strong coupling singularities. Then $Z_0$ has an 
expansion:    
\begin{equation}Z_0\sim q_0{}^{\fr{2\chi+3\sigma}{16}}+{\cal O}
(q_0{}^{\fr{2\chi+3\sigma}{16}+1}), 
\label{masleading}
\end{equation}again with no gap between the contribution of the trivial connection 
(shifted from $(q_0)^{0}$ by the $c$-number
$q_0{}^{\fr{2\chi+3\sigma}{16}}$)  and higher-order instanton
contributions.

As for the $SO(3)$ partition function, one has to sum (\ref{formulados}) 
over all allowed bundles, which means summing over all allowed 
't Hooft fluxes. One finds in this way:
\begin{align}
Z_{SO(3)} &= \sum_{\xi}Z_{\xi}
\ret&=\Bigg\{
\left(\fr{G(q_0{}^2)}{4}\right)^{\nu/2}
\left(\vartheta_3\vartheta_4\right)^{(2\chi +3\sigma)/2}
\sum_{x}n_x
\ret
&+ 2^{1-b_1+b_2+\fr{1}{4}(7\chi +11\sigma)} 
\left(G(q_0{}^{1/2})\right)^{\nu/2}
\left(\fr{\vartheta_2 \vartheta_3}{2}\right)^
{(2\chi +3\sigma)/2}\sum_{x=4y}n_x 
\ret &
+ 2^{1-b_1+b_2/2+\fr{1}{4}(7\chi +11\sigma)} 
\left(G(-q_0{}^{1/2})\right)^{\nu/2}
\left(\fr{\vartheta_2 \vartheta_4}{2}\right)^
{(2\chi +3\sigma)/2}
\sum_{x} n_x\Bigg\}.
\ret
\label{formulario}
\end{align}

To perform the summation over fluxes in (\ref{formulario}), one uses the 
following identities \cite{vw}:
\bea
\sum_{\xi}\sum_{x}n_{x}\,
\delta_{\left[\fr{x}{2}\right],\xi}&=&\sum_{x}\,n_{x},\ret
\sum_{\xi}\sum_{x}
(-1)^{\fr{x}{2}\cdot\xi}\,n_{x}&=&
2^{b_2}\sum_{x=4y}n_{x},\ret
\sum_{\xi}\,i^{-\xi^2}\sum_{x}
(-1)^{\fr{x}{2}\cdot\xi}\,n_{x}&=&2^{b_2/2}
\sum_{x}\,n_{x},\ret
\label{profident}
\eea
where in the last identity we have supressed again the phase $(-1)^{\sigma/8}$.

\section{Duality transformations of the generating function}

In this section we will study the properties under duality transformations of 
the generating function (\ref{formulauno}). We will start by checking the modular 
properties of $Z_{\xi}(\tau_0)$ as given in (\ref{formulados}). 
As explained in \cite{vw}, one should expect the following behaviour 
under the modular group. Under a $T$ transformation, taking 
$\tau_0\to\tau_0+1$, the partition function for fixed $\xi$ must transform 
into itself with a possible anomalous $\xi$-dependent phase. Indeed, 
(\ref{formulados}) behaves under $T$ in the expected fashion:
\begin{equation}Z_{\xi}(\tau_0+1)=i^{-\xi^2}Z_{\xi}(\tau_0).  
\label{tduality}
\end{equation}Checking (\ref{tduality}) involves some tricky steps that we now explain. 
Let us rewrite (\ref{formulados}) as:   
\begin{equation}Z_\xi
=A_1(\tau_0)\sum_{x}\delta_{\xi,\left[\fr{x}{2}\right]}n_x + 
\left [ A_2(\tau_0)+
i^{-\xi^2}A_3(\tau_0)\right]\sum_{x} (-1)^{\fr{x}{2}\cdot \xi}n_x.
\label{fmdos}
\end{equation}where we have set 
\bea
A_1(\tau_0)&=&\left(G(q_0{}^2)/4
\right)^{\nu/2}
\left(\vartheta_3\vartheta_4\right)^{(2\chi +3\sigma)/2},\ret 
A_2(\tau_0)&=&2^{1-b_1+\fr{1}{4}(7\chi +11\sigma)} 
\left(G(q_0{}^{1/2})\right)^{\nu/2}
\left(\vartheta_2 \vartheta_3/2\right)^
{(2\chi +3\sigma)/2},\ret
\label{mierda} 
\eea
and so on. Under $\tau_0\to\tau_0+1$ we have: 
\begin{equation}A_1\to A_1,\quad A_2\to
\ex^{\fr{i\pi}{8}(2\chi+3\sigma)}A_3= A_3,\quad \quad
A_3\to \ex^{\fr{i\pi}{8}(2\chi+3\sigma)} A_2= A_2, 
\label{procyon}
\end{equation}
and we have taken into account that $\nu\in 2\IZ$ throughout. 
In view of (\ref{procyon}), the partition function in (\ref{fmdos}) transforms 
as
follows:
\begin{equation}
Z_\xi\too \tilde Z_\xi=
A_1(\tau_0)\sum_{x}\delta_{\xi,\left[\fr{x}{2}\right]} n_x + 
\left [ A_3(\tau_0)+
i^{-\xi^2}A_2(\tau_0)\right]\sum_{x} (-1)^{\fr{x}{2}\cdot \xi}n_x, 
\label{alcyone}
\end{equation}
that is 
\begin{equation}\tilde Z_\xi=
i^{-\xi^2}\left(i^{\xi^2}  
A_1\sum_{x}\delta_{\xi,\left[\fr{x}{2}\right]} n_x + \left [ A_2+
i^{-\xi^2}A_3\right]\sum_{x} (-1)^{\fr{x}{2}\cdot \xi}n_x\right). 
\label{deneb}
\end{equation}
The phase $i^{\xi^2}$ in front of $A_1$ seems to spoil 
the invariance of $Z_\xi$ under $T$. That this is not actually so is 
because of the constraint $\xi+x/2=0~\mod~2$. Indeed, 
when $\xi+x/2=0~\mod~2$ we have $\xi^2=x^2/4 ~\mod~4$, and therefore 
\begin{equation}i^{\xi^2}=i^{x^2/4}=i^{(2\chi+3\sigma)/4}=i^{2\nu+\sigma/4}
=1, 
\label{vega}
\end{equation}
which guarantees that $Z_\xi$ is invariant (up to a phase) under $\tau_0
\to\tau_0+1$.  

Likewise, under an $S$ transformation, taking $\tau_0\to -1/\tau_0$,  
one should expect the following behaviour:
\begin{equation}
Z_{\xi}(-1/\tau_0)\propto\sum_{\zeta} (-1)^{\zeta\cdot\xi}
Z_{\zeta}(\tau_0).  
\label{thooftrule}
\end{equation}
It turns out that the partition function (\ref{formulados}) indeed  
satisfies (\ref{thooftrule}). In fact, 
\begin{equation}
Z_{\xi}(-1/\tau_0) = 2^{-b_2/2}
\left(\fr{\tau_0}{ i}\right)^{-\chi/2}
\sum_{\zeta}(-1)^{\zeta\cdot\xi}Z_{\zeta}(\tau_0).
\label{thooft}
\end{equation}
In order to check (\ref{thooft}), one has to use the modular properties 
of the different functions in (\ref{formulados}), which are compiled in the 
appendix in sect. \ref{apendice3}, and handle with care the summation over 
fluxes in (\ref{thooft}). 
The following identities are useful:
\bea
\sum_{\zeta}(-1)^{\zeta\cdot\xi}\sum_{x}n_{x}\,
\delta_{\left[\fr{x}{2}\right],\zeta}&=&\sum_{x}
(-1)^{\fr{x}{2}\cdot\xi}\,n_{x},\ret
\sum_{\zeta}(-1)^{\zeta\cdot\xi}\sum_{x}
(-1)^{\fr{x}{2}\cdot\zeta}\,n_{x}&=&
2^{b_2}\sum_{x}n_{x}\,\delta_{\left[\fr{x}{2}\right],\xi},\ret
\sum_{\zeta}(-1)^{\zeta\cdot\xi}\,i^{-\zeta^2}\sum_{x}
(-1)^{\fr{x}{2}\cdot\zeta}\,n_{x}&=&2^{b_2/2}i^{-\xi^2}
\sum_{x}(-1)^{\fr{x}{2}\cdot\xi}
\,n_{x}.\ret
\label{profiden}
\eea
To prove these identities, we borrow from eq. (5.40) of   
\cite{vw} the following formulas:  
\bea
\sum_{z}(-1)^{z\cdot y} \delta_{z,z'}
&=&(-1)^{y\cdot z'},\ret
\sum_{z}(-1)^{z\cdot y}
&=& 2^{b_2}\delta_{y,0},\ret
\sum_{z}(-1)^{z\cdot y} i^{-z^2}
&=& 2^{b_2/2}i^{-\sigma/2 + y^2}.
\ret
\label{sumident}
\eea
In addition to this, we have to take into account that, since the 
manifold is spin, 
$\sigma=0~\mod~16$, $x\in 2H^{2}(X,\IZ)$, and $x\cdot\xi,\xi^2\in 2\IZ$. 
Similarly, one should not forget to impose the constraint 
$\nu\in 2\IZ$. 

Using (\ref{formulacinco}) and (\ref{formulario}), one finds the following 
duality 
transformation properties for the $SU(2)$ and $SO(3)$ partition 
functions:

\bea
Z_{SU(2)}(\tau_0+1)&=&Z_{SU(2)}(\tau_0),\ret\ret
Z_{SO(3)}(\tau_0+2)&=& Z_{SO(3)}(\tau_0),\ret\ret
Z_{SU(2)}(-1/\tau_0)&=&2^{-\chi/2}\tau_{0}{}^{-\chi/2}
Z_{SO(3)}(\tau_0).\ret
\label{tdual}
\eea
As expected, the partition function for $SO(3)$ does not transform properly 
under $\tau_0\rightarrow \tau_0+1$, but transforms into itself under 
$\tau_0\to\tau_0+2$. Hence, the $SU(2)$ or $SO(3)$ partition funcions are 
invariant under the subgroup $\Gamma^0(2)\subset SL(2,\IZ)$.

We will now analyse the behaviour of the topological correlation functions 
under modular transformations. First, it is useful to work out how the 
different terms entering (\ref{formulauno}) transform. Under $\tau_0\to-1/\tau_0$ we 
have, 
\bea
&z_1\too \tau_0{}^2 z_2,\qquad\qquad &T_1\too \left(\fr{\tau_0}{ i}
\right)^4\left( T_2-\fr{i}{\pi\tau_0}z_2\right),\ret 
&z_2\too \tau_0{}^2 z_1,\qquad\qquad  &T_2\too \left(\fr{\tau_0}{ i}
\right)^4\left( T_1-\fr{i}{\pi\tau_0}z_1\right),\ret 
&z_3\too \tau_0{}^2 z_3, \qquad\qquad &T_3\too  \left(\fr{\tau_0}{
i}\right)^4\left(  T_3-\fr{i}{\pi\tau_0}z_3\right).
\ret 
\label{testadicazzo}
\eea
These formulas entail for the topological correlation functions the 
following behaviour under an $S$ transformation:
\bea
\left\langle\fr{1}{8\pi^2}\tr\,\phi^2\right\rangle
^{SU(2)}_{\tau_0}&=&\left\langle {\cal O}
\right\rangle^{SU(2)}_{\tau_0}=\fr{1}{\tau_0{}^2}\left\langle 
{\cal O}\right\rangle^{SO(3)}_{-1/\tau_0},\ret\ret
\left\langle\fr{1}{8\pi^2}\int_S\tr\,\left(2\phi F + \psi\wedge\psi\right)
\right\rangle^{SU(2)}_{\tau_0}&=&
\left\langle I(S)\right\rangle^{SU(2)}_{\tau_0}= 
\fr{1}{\tau_0{}^2}\left\langle I(S)\right\rangle^{SO(3)}_{-1/\tau_0},
\ret\ret
\left\langle I(S)I(S)\right\rangle^{SU(2)}_{\tau_0}= 
\left(\fr{\tau_0}{ i}\right)^{-4}&&\!\!\!\!\!\!\!\!\!\!\!\!\!\!\left\langle
I(S)I(S)\right\rangle^{SO(3)}_ {-1/\tau_0}+\fr{i}{2\pi}\fr{1}{\tau_0{}^3}
\left\langle {\cal O}\right\rangle^{SO(3)}_{-1/\tau_0}\sharp 
(S\cap S).\ret
\label{bizarre}
\eea

At first sight, the behaviour of $\left\langle I(S)\right\rangle$ under 
$\tau_0\to-1/\tau_0$ seems rather unnatural. Since $I(S)$ is essentially 
the magnetic flux operator of the theory, one would expect that it should 
transform under $S$ into the corresponding electric flux operator 
$J(S)\sim \int_S \tr\,(\phi *F)$ of the dual theory. However, this operator 
(or any appropriate generalization thereof) does not give rise to 
topological invariants, so one could question whether it should play any 
role at all. Likewise, one would like to understand the origin of the 
 shift $\left\langle {\cal O}\right\rangle\sharp 
(S\cap S)$ in the transformation of $\left\langle I(S)I(S)\right\rangle$.   

These a priori puzzling behaviours are quite  
natural when analysed from the viewpoint of Abelian electric-magnetic 
duality\footnote{We thank J.L.F. Barb\'on for useful conversations on this
point.}. In fact,  there exists a simple Abelian topological model whose
correlation  functions mimic the behaviour in (\ref{bizarre}) under
electric-magnetic duality. 
  
This model contains an Abelian gauge field $A$, whose field 
strength is defined as $F=dA$, two neutral scalar fields $\phi$, $\lambda$, 
a Grassmann-odd neutral one-form $\psi$ and a Grassmann-odd neutral two-form 
$\chi$. The Lagrangian is simply the topological density 
\begin{equation}
\fr{i}{4\pi}\tau_{0} F\wedge F =\fr{i\tau_{0}}{
16\pi}\epsilon_{\alpha\beta
\mu\nu} 
F^{\alpha\beta}F^{\mu\nu},
\label{chern}
\end{equation}plus conventional kinetic terms for the rest of the fields:
\begin{equation}{\cal L}_0 = \fr{i}{4\pi}\tau_{0} F\wedge F + \im\,\tau_0\,
d\phi\wedge *  d\lambda + \im\,\tau_0\, \chi\wedge * d\psi.
\label{alvarito} 
\end{equation}This Lagrangian possesses the following BRST symmetry:
\begin{equation}[Q,A]=\psi,\quad \{Q, \psi\}=d\phi,\quad [Q,\phi]=[Q,\lambda]=\{Q,\chi\}=0.
\label{nanette}
\end{equation}Notice that the non-holomorphic metric-dependent dependence on $\tau_0$ in 
(\ref{alvarito}) is BRST-exact:
\begin{equation}\im\,\tau_0 \left ( d\phi\wedge * 
d\lambda + \chi\wedge * d\psi\right)= \im\,\tau_0 \{Q,\psi\wedge * d\lambda 
-\chi\wedge * F\}.
\label{zagier}
\end{equation}Therefore, the partition function $\int{\cal D}[A,\phi,\lambda,\phi,\chi,]
\ex^{\int {\cal L}_0}$ is metric-independent and 
a priori holomorphic in $\tau_0$. 

The presence of magnetic sources in the theory is mimicked by imposing the 
conditions: 
\begin{equation}\int_{S_a}F = 2\pi n^{a},\qquad n^{a}\in\IZ, 
\label{unochern}
\end{equation}where the $\{S_a\}_{a=1,\cdots b_2(X)}$ are two-cycles representing a basis 
of $H_2(X,\IR)$. Notice that indeed $\int_{S}F$ gives the magnetic flux of 
$F$ through $S$. Owing to (\ref{unochern}), $F$ can be decomposed as $F=da+ 
2\pi\sum_a  n^{a}[S_a]$, where $a$ is a one-form in $X$ and $[S_a]$ are 
closed two-forms representing a basis of $H^2(X,\IR)$ dual to $\{S_a\}$. 
With these conventions, the piece in $\int {\cal L}_{0}$ containing the 
field strength is simply 
\begin{equation}i\pi\tau_0\sum_{a,b}n^a Q_{ab}n^b,
\label{accion}
\end{equation}with $Q_{ab}=\int_X [S_a]\wedge[S_b]=\sharp(S_a\cap S_b)$ the intersection 
form of the manifold. The functional integral $\int{\cal D}A\,
\ex^{{\cal L}_0}$ therefore involves a continuous integration over 
$a$ plus a discrete summation over the magnetic fluxes $n^a$. 

We wish to calculate the correlation functions $\langle\phi^2\rangle$ and 
$\langle \int_S (2\phi F +\psi\wedge\psi)\rangle$, and analyse their 
behaviour under 
duality transformations. For this we consider the generating functional: 
\begin{equation}
\int {\cal D}A{\cal D}\phi{\cal D}\lambda{\cal D}\psi{\cal D}\chi\, 
\ex^{\int_X {\cal L}(p,S)},
\label{odette}
\end{equation}where
\bea
{\cal L}(p,S)=\fr{i}{4\pi}\tau_0 F\wedge F \,+&&\!\!\!\!\!\!\!\!\!
\im\,\tau_0\, d\phi\wedge *  d\lambda + \im\,\tau_0\, \chi\wedge * d\psi
\ret +&&\!\!\!\!\!\!\!\!\!\fr{1}{8\pi^2}\left(2\phi F  
+\psi\wedge\psi\right)\wedge [S]+ \fr{p}{8\pi^2}\phi^2.\ret
\label{castor}
\eea

Notice that the operators $\phi^2$ and $\int_S (2\phi F +\psi\wedge\psi)$ are 
BRST-invariant. This guarantees, in the usual fashion, the topological invariance of 
the generating function (\ref{odette}). 

It is possible to rewrite (\ref{odette}) in terms of an equivalent, 
dual theory, which is also a topological Abelian model of the same sort, but 
with an inverted coupling constant $\tau^{D}_{0}=-1/\tau_{0}$. The details 
are similar to those in conventional 
electric-magnetic duality for Abelian
gauge theories \cite{newresults}\cite{verlinde}\cite{sdual}: one introduces  a
Lagrange multiplier $A_{D}$ to enforce the constraint $F=dA$ in the functional 
integral. This $A_{D}$ can be thought of as a connection on a dual magnetic line
bundle.  Its field strength $F_{D}= dA_{D}$ is taken to have quantized fluxes
through the cycles
$S_a$:
$F_{D}= da_{D}+2\pi\sum_a m^a [S_a]$. To deal with the other fields, we augment the quintet
$\{F,\psi,\phi,\lambda,\chi\}$  with a dual Abelian field strength $F_D$,  a dual neutral
Grassmann-odd one-form $\psi_{D}$, dual  neutral scalars $\phi_{D}$, $\lambda_D$, a dual neutral
Grassmann-odd  two-form $\chi_D$, bosonic four-form multipliers $b$, $\tilde b$, a Grassmann-odd
three-form multiplier $\rho$ and a Grassmann-odd two-form multiplier 
$\omega$, and consider the extended Lagrangian:
\bea
\tilde{\cal L}(p,S)&=&\fr{i}{4\pi}\tau_0 F\wedge F +
\fr{\im\,\tau_0}{2\tau_0} d\phi_D\wedge * d\lambda + 
\fr{\im\,\tau_0}{2\bar\tau_0} d\phi\wedge * d\lambda_D\ret &&+ 
\fr{\im\,\tau_0}{2\tau_0}\chi\wedge * d\psi_D  + 
\fr{\im\,\tau_0}{2\bar\tau_0}\chi_D\wedge * d\psi \ret 
&&+ \fr{1}{8\pi^2}\left(
2\phi F +\fr{1}{\tau_0}\psi\wedge\psi_{D}\right)\wedge [S]+
\fr{p}{8\pi^2\tau_0}\phi\phi_{D}\ret &&-\fr{i}{2\pi}F\wedge F_{D}
+ b(\phi_{D}-\tau_0\phi) + \tilde b(\lambda_{D}-\bar\tau_0\lambda)
\ret&&+\rho\wedge(\psi_{D}-\tau_0\psi) 
+\omega\wedge(\chi_{D}-\bar\tau_0
\chi).\ret
\label{bellatrix}
\eea
By integrating out the dual fields $F_{D}$, $\phi_D$, $\psi_{D}$, $\chi_D$ 
and $\lambda_D$, and the multipliers $b$, $\tilde b$, $\rho$, $\omega$, it 
is straightforward to verify that the generating functional:
\begin{equation}\int {\cal D}[F,\phi,\psi,\lambda,\chi,F_{D},\phi_{D},\psi_{D},\lambda_D,
\chi_D,b,\tilde b,\rho,\omega]\, 
\ex^{\int_X \tilde{\cal L}(p,S)},
\label{capella}
\end{equation}where now the integration over $F$ is unrestricted, represents the same 
theory as (\ref{odette}). The dual formulation can be 
achieved by integrating out instead the original fields $F$, $\phi$, 
$\lambda$, $\chi$ and $\psi$, together with the multipliers $b$, $\tilde b$, $\rho$ 
and $\omega$. One obtains in this way 
the dual Lagrangian:
\bea
{\cal L}_{D}(p,S)&=&-\fr{i}{4\pi\tau_0} F_D\wedge F_D +
\im\,\tau^D_0 d\phi_D\wedge * d\lambda_D + \im\,\tau^D_0 \chi_D\wedge * 
d\psi_D \ret
+\fr{1}{\tau_0{}^2}&&\!\!\!\!\!\!\!\!\!\!\!\!\!\fr{1}{8\pi^2}\big(2\phi_{D}
F_{D}+
\psi_{D}\wedge\psi_{D}\big)\wedge [S]+
\fr{p}{8\pi^2\tau_0{}^2}(\phi_{D})^2+\fr{i}{2\pi\tau_0{}^3}
\fr{(\phi_{D})^2}{8\pi^2} [S]\wedge [S].
\ret
\label{pollux}
\eea
Notice that this dual Lagrangian is invariant under the appropriate 
dualized version of (\ref{nanette}):
\begin{equation}[Q,A_D]=\psi_D,\quad \{Q, \psi_D\}=d\phi_D,\quad [Q,\phi_D]=[Q,\lambda_D]=
\{Q,\chi_D\}=0.
\label{dualnanette}
\end{equation}From (\ref{pollux}) we can immediately read off the behaviour of the correlation 
functions under $\tau_0\to-1/\tau_0$:
\begin{align}
\left\langle\fr{1}{8\pi^2}\phi^2\right\rangle_{\tau_0}&=
\left\langle {\cal O}\right\rangle_{\tau_0}=
\fr{1}{\tau_0{}^2}\left\langle\fr{1}{8\pi^2}(\phi_{D})^2
\right\rangle_{-1/\tau_0}=
\fr{1}{\tau_0{}^2}\left\langle {\cal O}\right\rangle^{D}_{-1/\tau_0},
\ret\ret
\left\langle\fr{1}{8\pi^2}\int_S\left(2\phi F + \psi\wedge\psi\right)
\right\rangle_{\tau_0}&=
\left\langle I(S)\right\rangle_{\tau_0}=\fr{1}{\tau_0{}^2} 
\left\langle\fr{1}{8\pi^2}\int_S\left(2\phi_{D} F_{D} + \psi_{D}
\wedge\psi_{D}\right)\right\rangle_{-1/\tau_0}\ret
&=\fr{1}{\tau_0{}^2}\left\langle I(S)\right\rangle^{D}_{-1/\tau_0},
\ret\ret
\left\langle I(S)I(S)\right\rangle_{\tau_0}&=
\fr{1}{\tau_0{}^4}\left\langle I(S)I(S)\right\rangle^{D}_{-1/\tau_0}+
\fr{i}{2\pi\tau_0{}^3}
\left\langle {\cal O}\right\rangle^{D}_{-1/\tau_0}\sharp 
(S\cap S),\ret
\label{antares}
\end{align}
which, as promised, faithfully reproduces the modular properties (\ref{bizarre}) of 
the corresponding topological correlation functions of the full, non-Abelian 
theory.

\section{More properties of the generating function}

In this section we will analyse more properties of the generating function
(\ref{formulauno}).  First, we will study its behaviour in the massless limit 
$m\to 0$, then we will show that the Donaldson invariants are contained in the
partition function  by studying the $\cn=2$ limit ($m\to\infty$) and, finally, 
we
will show that on $K3$ it reduces to the one obtained by Vafa and Witten  for
another twist of the $\cn=4$ supersymmetric theory.

\subsection{Massless limit}

We wish to analyse the behaviour of (\ref{formulauno}) and (\ref{formulados}) in 
the limit $m\to 0$ holding $\tau_0$ fixed. Since we have factorized out the overall
dependence on $m$, the analysis here is much  simpler than that in 
\cite{htwist} but leads to a somewhat puzzling result. Notice that all the
dependence on
$m$ in the generating function is contained in the observables. Consider a 
certain
correlation function 
$\langle \C{O}^{(1)}\cdots \C{O}^{(r)}\rangle$. The observable insertions carry an
explicit mass dependence dictated by their ghost number \eqs{marflores1}, 
so we have 
\begin{equation}
\langle \C{O}^{(1)}\cdots \C{O}^{(r)}\rangle\propto m^{\sum^r_{n=1}g_n/2},
\la{marflores}
\end{equation}
where $g_n$ is the ghost number of the observable $\C{O}^{(n)}$. As the
exponent in \eqs{marflores} is always positive, all the  correlation
functions vanish in the limit $m\to 0$ whatever the values of $\chi$ and
$\sigma$ be. 
Notice that one would generically expect
that the correlation functions should be non-zero unless 
$2\chi +3\sigma\geq0$. However, on manifolds with $b_2^{+}>1$, the topological 
correlation functions are given by the contributions  
from the singularities in the moduli space of the physical theory. In the 
massless limit, the three singularities of the
low-energy description of the physical $\cn=4$ theory collapse to a
unique singularity at $u=0$. This is a superconformal
point where we generically expect all the correlation functions other than the 
partition function (which is itself the vacuum expectation value of the identity 
operator) to vanish. Thus, the vanishing of the topological correlation
functions could be thought of as an effect of the superconformal
invariance of the undelying $\cn=4$ theory. Anyhow, one should not take these 
results too seriously. The generating function was derived for generic values 
of the mass $m$ (and in particular for non-zero values of the mass), so 
it is not clear that setting $m=0$ is meaningful at all.

\subsection{The $\cn=2$ limit and the Donaldson-Witten invariants}

We would like to analyse the fate of our formulas for the generating  function
under the decoupling limit $m\to\infty$, 
$q_0\to 0$, holding $\Lambda_{0}$, the scale of the $N_f=0$ theory, fixed:
$4m^4q_0=\Lambda_{0}{}^4$. In this limit, the singularities  at strong coupling
evolve to the singularities of the $\cn=2$, $N_f=0$ $SU(2)$  theory, while
the semiclassical singularity goes to infinity and  disappears. While this
limit is perfectly well-defined for the  Seiberg-Witten curve, it is not
clear whether the corresponding  explicit expressions for the
prepotentials and the periods should  remain non-singular as well.  The
question therefore arises as to whether taking this naive limit in the
twisted theory could give a non-singular result, that is whether, 
starting from (\ref{formulauno}) or (\ref{formulados}), one could recover
the  corresponding expressions for the twisted (pure) $SU(2)$ $\cn=2$ 
supersymmetric theory. This limit has been  systematically studied for the
twisted $\cn=2$ gauge theory with
$N_f$ hypermultiplets in the fundamental representation (and $N_f<4$)  in
\cite{geog}. Dijkgraaf et al. \cite{coreatres} considered the same limit for the
Vafa-Witten partition function, and they were able to single out  a piece which
corresponds to the partition function of the twisted $\cn=2$  supersymmetric theory
as first computed by Witten
\cite{monopole}. We will go  a step further and  recover, in
the same limit, the full generating function for the Donaldson-Witten invariants.

We will focus on the generating function (\ref{formulauno}). We will keep the 
leading terms in the series expansion of the different modular functions 
in powers of $q_{0}$. We will use the explicit formulas:
\begin{align}
G(q_0{}^2)&=1/q_0{}^2+\cdots,& \vartheta_3(q_0)\vartheta_4(q_0)&=1+\cdots,\ret
G(q_0{}^{1/2})&=1/q_0{}^{1/2}+\cdots,& 
\vartheta_2(q_0)\vartheta_3(q_0)/2&=q_0{}^{1/8}+\cdots,   \ret
G(-q_0{}^{1/2})&=-1/q_0{}^{1/2} +\cdots,& 
\vartheta_2(q_0)\vartheta_4(q_0)/2&=q_0{}^{1/8}+\cdots.\ret
\label{buchenwald}
\end{align} 
As for the modular functions entering the observables, we have the 
following behaviour:
\begin{align}
z_1&=\fr{1}{4}m^2 e_1(\tau_0)= \fr{1}{6}m^2 + O(\Lambda_0{}^4/m^2),\ret
z_2&=\fr{1}{4}m^2 e_2(\tau_0)= -\fr{1}{12}m^2 -4\Lambda_0{}^2+
O(\Lambda_0{}^4/m^2),\ret z_3&=\fr{1}{4}m^2 e_3(\tau_0)= -\fr{1}{12}m^2
+4\Lambda_0{}^2+ O(\Lambda_0{}^4/m^2),\ret 
\label{lore}
\end{align}
and
\begin{align}
(dz/da)_1{}^2&=\half m^2(\vartheta_3\vartheta_4)^4= 
\half m^2+ O(\Lambda_0{}^4/m^2),\ret
(dz/da_D)_2{}^2&=\half m^2(\vartheta_2\vartheta_3)^4= 
16 \Lambda_0{}^2+ O(\Lambda_0{}^4/m^2),\ret
(dz/da_d)_3{}^2&=-\half m^2(\vartheta_2\vartheta_4)^4= 
-16 \Lambda_0{}^2+ O(\Lambda_0{}^4/m^2),\ret
\label{decameron}
\end{align}
and, for the contact terms $T_i$ (\ref{contralto}), (\ref{mezzosoprano}):
\begin{equation}T_1= O(\Lambda_0{}^4/m^2),\quad 
T_2= -2\Lambda_0{}^2 + O(\Lambda_0{}^4/m^2),\quad 
T_3= 2\Lambda_0{}^2 + O(\Lambda_0{}^4/m^2).
\label{hades}
\end{equation}
While the contribution from the semiclassical singularity behaves  
as 
\begin{equation}
2^{-\nu}q_0{}^{-\nu}\ex^{p m^2/3}\ldots,
\label{quark}
\end{equation}
the contributions from the strong coupling singularities give 
the following result:
\bea
&&\fr{1}{2^{b_1}}  q_0{}^{-\nu} 
q_0{}^{3\nu/4}
q_0{}^{(2\chi+3\sigma)/16}\ex^{- p m^2/6}\ret&&\left\{
2^{1+ \fr{1}{4}(7\chi +11\sigma)}\left(
\ex^{2p+\fr{S^2}{2}}
\sum_{x}(-1)^{\fr{x}{2}\cdot\xi}\ex^{S\cdot x}\,n_x 
+i^{\nu-\xi^2} \ex^{-2p-\fr{S^2}{2}}
\sum_{x}(-1)^{\fr{x}{2}\cdot\xi}\ex^{-iS\cdot x}\,n_x 
\right)\right\},\ret
\label{donaldwitten}
\eea
(we have set $\Lambda_0{}^2=-1/4$), 
where the quantity in brackets is precisely Witten's generating  
function for the twisted $\cn=2$ $SO(3)$ gauge theory!

\subsection{The partition function on $K3$}
We now specialize to $K3$. This is a compact hyper-K\"ahler manifold, 
and as such one would expect \cite{wijmp} the physical and the twisted 
theories to coincide. Therefore, our formulas are to be considered 
as true predictions for the partition function and a selected set of 
correlation functions of 
the physical $\cn=4$ $SO(3)$ gauge theory on 
$K3$. 

Only the zero class $x=0$ contributes on $K3$, and $n_{x=0}=1$. 
Moreover, $\chi=24$ 
and $\sigma=-16$, so $\nu=2$ and $2\chi+3\sigma=0$.  The answer for $K3$ is
therefore:  
\begin{equation}
Z_\xi^{K3} = \fr{1}{4}G(q_0{}^2)\delta_{\xi,0}
+ \fr{1}{2}G(q_0{}^{1/2})
+ i^{-\xi^2}\fr{1}{2}G(-q_0{}^{1/2}),
\label{formol}
\end{equation}
which happily coincides with the formula given by Vafa and Witten 
\cite{vw}. We can go even further and present the full generating 
function on $K3$:
\begin{align}
\left\langle\ex^{p{\cal O}+I(S)}\right\rangle^{K3}_{\xi}&=\ret   
\fr{1}{4}G(q_0{}^2)&\ex^{2pz_1 + S^2
T_1}\,\delta_{\xi,0} +
\fr{1}{2}G(q_0{}^{1/2})\ex^{2pz_2 + S^2 T_2} +
i^{-\xi^2}\fr{1}{2}G(-q_0{}^{1/2})\ex^{2pz_3 + S^2 T_3}.
\ret
\label{formaldehido}
\end{align}

Notice that the correlation functions, which follow from (\ref{formaldehido}), 
are proportional to the mass $m$, and therefore all vanish (except for 
the partition function) when $m\to 0$, as expected.

The generating function for $SU(2)$ is obtained from 
(\ref{formaldehido}) by simply setting $\xi=0$ and dividing by $2$:
\bea
\left\langle\ex^{p{\cal O}+I(S)}\right\rangle^{K3}_{SU(2)}&=&\ret   
\fr{1}{8}G(q_0{}^2)\,\ex^{2pz_1 + S^2 T_1}&&\!\!\!\!\!\!\!\!\!\!\!\!
+ \fr{1}{4}G(q_0{}^{1/2})\,\ex^{2pz_2 + S^2 T_2}
+ \fr{1}{4}G(-q_0{}^{1/2})\,\ex^{2pz_3 + S^2 T_3}.
\ret
\label{aldehido}
\eea

The corresponding expression for $SO(3)$ bundles is given by the 
sum of (\ref{formaldehido}) over all 't Hooft fluxes. As explained in 
\cite{vw}, the allowed 't Hooft fluxes on $K3$ can be grouped into 
different diffeomorphism classes, which are classified by the value 
of $\xi^2$ modulo $4$ ($K3$ is spin, so $\xi^2$ is always even). 
There are three different possibilities and, correspondingly, three  
different generating functions to be computed: $\xi=0$, with 
multiplicity $n_0=1$, gives just the $SU(2)$ partition function; 
$\xi\not=0, \xi^2\in 4\IZ$, with multiplicity $n_{\rm even}=
(2^{22}+2^{11})/2 -1$; and $\xi^2=2~\mod~4$, with multiplicity 
$n_{\rm odd}=(2^{22}-2^{11})/2$. The $SO(3)$ answer is the sum 
of the three generating functions (with the appropriate multiplicities):
\bea
\left\langle\ex^{p{\cal O}+I(S)}\right\rangle^{K3}_{SO(3)}&&\!\!\!\!\!\!\!=
\left\langle\ex^{p{\cal O}+I(S)}\right\rangle^{K3}_{\xi=0}+
n_{\rm even}
\left\langle\ex^{p{\cal O}+I(S)}\right\rangle^{K3}_{\rm even}+
n_{\rm odd}
\left\langle\ex^{p{\cal O}+I(S)}\right\rangle^{K3}_{\rm odd}\ret =  
\fr{1}{4}G(q_0{}^2)&&\!\!\!\!\!\!\!\ex^{2pz_1 + S^2 T_1}
+ 2^{21}G(q_0{}^{1/2})\ex^{2pz_2 + S^2 T_2}
+ 2^{10}G(-q_0{}^{1/2})\ex^{2pz_3 + S^2 T_3}.
\ret
\label{smirnoff}
\eea

All these generating functions behave under duality as dictated by 
(\ref{tduality}), (\ref{thooftrule}) and (\ref{bizarre}). In particular, the 
$S$ transformation 
exchanges the $SU(2)$ and $SO(3)$ generating functions according to 
(\ref{tdual}) and (\ref{bizarre}). 

\section{Appendix}
\la{apendice3}
Here we collect some useful formulas which should help the reader 
follow the computations in this chapter. A more detailed account can be
found in appendices A and B of \cite{moorewitten}. A very useful  
review containing definitions and properties of many modular forms 
can be found in appendices A and F of \cite{kiritsis}.

\subsection{Modular forms}

Our conventions for the Jacobi theta functions are:
\begin{align}
\vartheta_2 (\tau ) &= \sum _{n=-\infty}^{\infty}  q^{(n+1/2)^2/2}=
2q^{1/8}(1+ q + q^3 + \cdots),\ret 
\vartheta_3 (\tau ) &=\sum _{n=-\infty}^{\infty} q^{n^2/2} = 1 + 2
q^{1/2} + 2 q^2 + \cdots,\ret
\vartheta_4 (\tau ) &=\sum _{n=-\infty}^{\infty} (-1)^n q^{n^2/2}
= 1 - 2 q^{1/2} + 2 q^2 +\cdots,\ret
\label{jacobi}
\end{align} 
where $q=\ex^{2\pi i\tau}$.  They satisfy the identity:
\begin{equation}
\vartheta_3 (\tau )^4=\vartheta_2 (\tau )^4
+\vartheta_4 (\tau )^4,
\label{cris}
\end{equation}
and they have the following properties under modular transformations:
\bea 
&\vartheta_2 (-1/\tau ) = \sqrt{\fr{\tau}{i}}\vartheta_4
(\tau ), \qquad\qquad 
&\vartheta_2 (\tau+1 ) = \ex^{i\pi/4}\vartheta_2 (\tau ),\ret 
&\vartheta_3 (-1/\tau ) = 
\sqrt{\fr{\tau}{i}}\vartheta_3 (\tau ),\qquad\qquad  
&\vartheta_3 (\tau+1 ) = \vartheta_4 (\tau),\ret 
&\vartheta_4 (-1/\tau ) = \sqrt{\fr{\tau}{i}}\vartheta_2
(\tau ),\qquad\qquad 
&\vartheta_4 (\tau+1 ) = \vartheta_3 (\tau ).
\ret
\label{jacobino}
\eea  
From these, the modular properties of the functions 
$e_j$ (\ref{spin}) follow straightforwardly:
\bea 
&e_1(-1/\tau_0)= \tau_0{}^2 e_2(\tau_0),\qquad\qquad 
&e_1(\tau_0+1)= e_1(\tau_0),\ret &e_2(-1/\tau_0)= \tau_0{}^2
e_1(\tau_0),\qquad\qquad &e_2(\tau_0+1)= e_3(\tau_0),\ret
&e_3(-1/\tau_0)= \tau_0{}^2 e_3(\tau_0),\qquad\qquad &e_3(\tau_0+1)=
e_2(\tau_0).\ret
\label{irredento}
\eea
 Notice that, from their definition, $e_1+e_2+e_3=0$.  Likewise,
we can determine explicitly the behaviour of the  functions
$\kappa_{j}$ (\ref{yespadas}) and of the periods  (\ref{zhivago})
under modular transformations:
\begin{align}
\kappa_{1}(-1/\tau_0)&=\tau_0{}^2\kappa_{2}(\tau_0),  
&\kappa_{1}(\tau_0+1)&=-\kappa_{1}(\tau_0),\ret
\kappa_{2}(-1/\tau_0)&=\tau_0{}^2 \kappa_{1}(\tau_0),
&\kappa_{2}(\tau_0+1)&=\kappa_{3}(\tau_0),\ret
\kappa_{3}(-1/\tau_0)&=-\tau_0{}^2
\kappa_{3}(\tau_0), 
&\kappa_{3}(\tau_0+1)&=\kappa_{2}(\tau_0),\ret 
&{}&{}&
\label{calatayud}
\end{align}
and
\begin{align}
(da/dz)_1{}^2\vert_{-\frac{1}{\tau_0}}& =
\tau_0{}^{-4} (da_{D}/dz)_2{}^2\vert_{\tau_0},
&(da/dz)_1{}^2\vert_{\tau_0+1}&= (da/dz)_1{}^2\vert_{\tau_0},\ret
(da_{D}/dz)_2{}^2\vert_{-\frac{1}{\tau_0}}&= 
\tau_0{}^{-4} (da/dz)_1{}^2\vert_{\tau_0},&
(da_{D}/dz)_2{}^2\vert_{\tau_0+1}&= (da_{d}/dz)_3{}^2
\vert_{\tau_0},\ret 
(da_{d}/dz)_3{}^2\vert_{-\frac{1}{\tau_0}}&= 
\tau_0{}^{-4}(da_{d}/dz)_3{}^2\vert_{\tau_0},
 &(da_{d}/dz)_3{}^2\vert_{\tau_0+1}&= (da_{D}/dz)_2{}^2
\vert_{\tau_0},\ret
\label{calahorra}
\end{align} 
where we have set $a_{d}\equiv a_{D}-a$.

The Dedekind eta function is defined as follows:
\begin{equation}
\eta(\tau)=q^{1/24}\prod^{\infty}_{n=1}(1-q^n)=\sum_{-\infty}
^{\infty}(-1)^{n}q^{\frac{3}{2}(n-1/6)^2}= q^{1/24}(1 - q - q^2 + 
\cdots),
\label{dedekind}
\end{equation}
with $q=\exp(2i\pi\tau)$. Under the modular group:
\begin{equation}
\eta(-1/\tau)=\sqrt{\frac{\tau}{i}}\eta(\tau),\qquad 
\eta(\tau+1)=\ex^{i\pi/12}\eta(\tau).
\label{borrell}
\end{equation}
The following identities are useful:
\begin{align}
\eta(\tau)^3&=\half\vartheta_2 (\tau )\vartheta_3 (\tau )\vartheta_4
(\tau ) ,&
\vartheta_2 (\tau )&= 2 \frac{\eta(2\tau)^2}{\eta(\tau)},\ret
\vartheta_3 (\tau )&=
\frac{\eta(\tau)^5}{\eta(\tau/2)^2\eta(2\tau)^2},
&
\vartheta_4 (\tau )&= \frac{\eta(\tau/2)^2}{\eta(\tau)}.
\ret
&{}&{}&
\label{almunia}
\end{align} 
With these formulas we can rewrite the functions $G(q)$ 
featuring in  the Vafa-Witten formula in terms of standard modular
forms:
\begin{align} 
G(q)&=\frac{1}{\eta(q)^{24}},&  G(q^2)&=
\frac{1}{\eta(2\tau)^{24}}=
\left(\frac{2}{\eta(\tau)\vartheta_2(\tau)}\right)^{12},\ret
 G(q^{1/2})&= \frac{1}{\eta(\frac{\tau}{2})^{24}}=
\frac{1}{\bigl(\eta(\tau)\vartheta_4(\tau)\bigr) ^{12}},&
G(-q^{1/2})&= \frac{1}{\eta(\fr{\tau+1}{2})^{24}}=
-\frac{1}{\bigl(\eta(\tau)\vartheta_3(\tau)\bigr)^{12}}.
\ret 
\label{valiente} 
\end{align} 
These functions have the following modular properties \cite{vw}:
\begin{align} 
G(q^{2})&{\buildrel\tau\to\tau+1\over\too}   
G(q^{2}),& G(q^{2})&{\buildrel\tau\to -1/\tau\over\too}   
2^{12}\tau^{-12}  G(q^{1/2}),\ret
G(q^{1/2})&{\buildrel\tau\to\tau+1\over\too}  G(-q^{1/2}),&
G(q^{1/2})&{\buildrel\tau\to -1/\tau\over\too}  2^{-12}\tau^{-12} 
G(q^{2}),\ret 
G(-q^{1/2})&{\buildrel\tau\to\tau+1\over\too}
G(q^{1/2}),& G(-q^{1/2})&{\buildrel\tau\to
-1/\tau\over\too} \tau^{-12}  G(-q^{1/2}).
\ret
\label{conhazo}
\end{align}

The Eisenstein series of weights 2 and 4 are:
\bea E_2  &=& \fr{12}{ i\pi}\partial_{\tau}{\rm log}\,\eta = 
1-24\sum_{n=1}^{\infty} \frac{n q^n}{1-q^n}=1 - 24 q + \cdots
,\ret E_4 & =&\half\left
(\vartheta_2^8+\vartheta_3^8+\vartheta_4^8\right)
=1+240\sum_{n=1}^{\infty} \frac{n^3 q^n}{1-q^n}.\ret
\label{cazzo}
\eea While $E_4$ is a modular form of weight $4$ for $Sl(2,\IZ)$,
$E_2$ is  not quite a modular form: under $\tau\to
(a\tau+b)/(c\tau+d)$  we have
\begin{equation}E_2\left(\frac{a\tau+b}
{c\tau+d}\right)=(c\tau+d)^2 E_2(\tau)-
\frac{6ic}{\pi}(c\tau+d). 
\label{uno} 
\end{equation}The  non-holomorphic combination $\hat E_2 = E_2 -
3/(\pi\im\,\tau)$ is a  modular form of weight $2$, which enters in
the definition of the  contact term $\hat T$ in (\ref{karembeu}). 
 
\subsection{Lattice sums}  
Here we quote some of the results in appendix
B of \cite{moorewitten}  to which we refer the reader for more
details. These  formulas are quite useful when performing the duality 
transformations among the different frames on the $u$-plane. 

We introduce the following theta function:
\begin{align}
\Theta_{\Gamma} (\tau, \alpha,\beta; P, \gamma)
&\equiv
 \exp\left[\frac{\pi}{2 y} ( \gamma_+^2 - \gamma_-^2) \right]
\ret
\sum_{\lambda\in\Gamma}
\exp\biggl\{ i \pi \tau (\lambda+ \beta)_+^2 + i \pi \bar \tau
(\lambda+ \beta)_-^2 & + 2 \pi i
(\lambda+\beta)\cdot\gamma - 2 \pi i
\left(\lambda+\half \beta\right)\cdot\alpha \biggr\} \ret &=
\ex^{i \pi \beta\cdot\alpha}\,
 \exp\left[\frac{\pi}{2 y} ( \gamma_+^2 - \gamma_-^2)\right]
\ret
\sum_{\lambda\in\Gamma}
\exp\biggl\{ i \pi \tau (\lambda+ \beta)_+^2 + i \pi \bar \tau
(\lambda+ \beta)_-^2 & + 2 \pi i
(\lambda+\beta)\cdot\gamma - 2 \pi i (\lambda+  \beta)\cdot\alpha
\biggr\}, \ret
\label{caspa}
\end{align} 
where $\Gamma$ is a lattice of signature $(b_+,b_-)$ on which 
an orthogonal projection $P_\pm(\lambda)= \lambda_\pm$ and  a pairing
$(\alpha,\beta)\to \alpha\cdot\beta\in\IR$ are defined.  

The main transformation law is:
\begin{equation}
\Theta_{\Gamma }\left(-1/\tau, \alpha,\beta; P, \frac{\gamma_+}
{\tau} + \frac{\gamma_-}{\bar \tau}\right) = 
\sqrt{\frac{\vert \Gamma \vert}{\vert \Gamma^{*} \vert}} (-i
\tau)^{b_+/2} (i \bar
\tau)^{b_-/2}
\Theta_{\Gamma^{*}} ( \tau, \beta,-\alpha ; P, \gamma),
\label{lolita}
\end{equation}
where $\Gamma^{*}$ is the dual lattice. Given a characteristic
vector $w_2\in\Gamma$, such that
\begin{equation}
\lambda\cdot\lambda = \lambda\cdot w_2 ~ {\rm mod} ~2
\label{vangaal}
\end{equation}
for all $\lambda\in\Gamma$, then we have:
\begin{equation}
\Theta_{\Gamma} (\tau+1, \alpha,\beta; P, \gamma) = \ex^{-i
\pi\beta\cdot w_2/2}
\Theta_{\Gamma }\left( \tau, \alpha-\beta-\half w_2,\beta ; P,
\gamma\right).
\label{cruyff}
\end{equation}

\vfill
\newpage


\chapter{The amphicheiral theory}
\la{chmarcus}
\markboth{\footnotesize\bfseries The amphicheiral theory
}{\footnotesize\bfseries The amphicheiral theory}
\markright{\textsc {Duality in Topological Quantum Field Theories}}

\section{The twisted theory}
The last theory we will consider was briefly introduced at the end of
reference
\cite{yamron}, and afterwards it was  considered in detail in
\cite{blauthomp}\cite{ene4}\cite{marcus}. It corresponds to the
decomposition (see chapter \ref{ctwists})
${\bf 4}\too ({\bf 2},{\bf 1})\oplus({\bf 1},{\bf 2})$, but it is
easier (and equivalent) to start from the second twisted theory  and
replace
$SU(2)_R$ with the diagonal sum $SU(2)'_R$ of $SU(2)_R$ itself and 
the remaining isospin group $SU(2)_F$ (this is very much alike to a
conventional $\cn=2$ twisting). This introduces in the theory a second
BRST-like symmetry, which comes from the $\cn=4$ spinor supercharges
$\bar Q_{v\dalpha}$. As we pointed out at before,
there are several unusual features to this theory that we think 
deserve a detailed analysis. We begin by recalling the fundamentals of
the second twist. The symmetry group
${\cal H}=SU(2)_L\otimes SU(2)_R\otimes SU(4)_I$ of the original
$\cn=4$ supersymmetric gauge theory is twisted to give the symmetry 
group 
${\cal H'}=SU(2)'_L\otimes SU(2)_R\otimes SU(2)_F\otimes U(1)$  (we
will refer to this as the $L$ twist) of the half-twisted theory. The
supersymmetry charges
$Q^v_{\!\alpha}$ and $\bar Q_{v\dalpha}$ decompose under
${\cal H'}$ as:
\begin{equation} Q^v{}_\alpha \oplus \bar Q_{v\dot\alpha}\too   
Q^{(+1)}\oplus Q^{(+1)}_{(\alpha\beta)}\oplus Q^{(-1)}_{i\alpha} \oplus
\bar Q^{(-1)}_{\alpha\dot\alpha}\oplus \bar Q^{(+1)}_{i\dalpha}.
\la{Reeve}
\end{equation} But one can also twist with $SU(2)_R$ thus obtaining its
corresponding ``mirror" $\tilde T$ twist with symmetry group  
${\tilde{\cal H'}}=SU(2)_L\otimes SU(2)'_R\otimes SU(2)_F\otimes
U(1)$  ($R$ twist). Both formulations are related \eqs{titin4} through
an  orientation reversal and a change of sign in $\theta_0$. Now we can
twist
$SU(2)_F$ away in four different ways. Two of these ($LL$ and $RR$)
take us back to the Vafa-Witten twists $T$ and $\tilde T$. The other
two ($LR$ and 
$RL$) should lead us to the twist considered in
\cite{blauthomp}\cite{marcus} and its corresponding $\tilde T$ twist.
The non-trivial result is that either of these two different choices
leads to the  same topological theory. This can be seen as follows.
Pick one of the possibilities,  say, $LR$. After the first twist we
have the half-twisted  theory with symmetry group ${\cal H'}$ and
supersymmetry charges
\eqs{Reeve}. If we now twist $SU(2)_F$ with $SU(2)_R$ we obtain, from
the last charge in
\eqs{Reeve}, a second scalar charge  $\tilde Q$ given by:
\begin{equation}
\bar Q_{i\dalpha}\to \bar Q_{{\dot\beta}\dalpha}\to \tilde Q=
C^{{\dot\beta}\dalpha}\bar Q_{{\dot\beta}\dalpha}.
\la{Sontag}
\end{equation}

Notice that both the anticommuting symmetries, $Q$ and $\tilde Q$,
have the same ghost number, so they are both to be considered either
as BRST or anti-BRST operators. This is in contrast with the situation
we  found in the Vafa-Wittten twist where, after explicitly breaking
the isospin group
$SU(2)_F$ down to its
$T_3$ subgroup, we ended up with two scalar charges $Q^{(+)}$ and
$Q^{(-)}$ with opposite ghost numbers, which were then interpreted as a
BRST-antiBRST system.   

The fields of the new theory can be obtained from those in the
half-twisted theory as follows:
\begin{equation}
\begin{matrix} 
A_{\alpha\dalpha}\\ \lambda_{v\alpha}\\    
\bar\lambda^v{}_{\!\dot\alpha}\\
\phi_{uv}
\end{matrix}
\quad
\begin{matrix}
\too\\\too\\\too\\\too
\end{matrix}
\quad
\begin{matrix}
&A^{(0)}_{\alpha\dalpha}\\
\chi^{(-1)}_{\beta\alpha},&\!
\eta^{(-1)},&\!\lambda^{(+1)}_{i\alpha}\\
\psi^{(+1)}_{\alpha\dalpha},&\!\zeta^{(-1)}_{i\dalpha}\\
 B^{(-2)},&\!C^{(+2)},&\!G^{(0)}_{i\alpha}
\end{matrix}
\quad
\begin{matrix}
\too\\\too\\\too\\\too
\end{matrix}
\quad
\begin{matrix}
&A^{(0)}_{\alpha\dalpha}\\
\chi^{(-1)}_{\beta\alpha},&
\!\eta^{(-1)},&\!\tilde\psi^{(+1)}_{\alpha\dot\alpha}\\
\psi^{(+1)}_{\alpha\dalpha},&\!\tilde\eta^{(-1)},&\!\tilde\chi^{(-1)}
_{\dalpha{\dot\beta}}\\
B^{(-2)},&\!C^{(+2)},&\!V^{(0)}_{\alpha\dalpha}
\end{matrix}  
\la{pcien}
\end{equation}
where we have included also the corresponding fields of
the $\cn=4$ theory and the ghost number carried by the twisted fields.
The notation is similar to that in ref.
\cite{marcus}. Notice that if we exchange $SU(2)_L$ by $SU(2)_R$ the
field content in \eqs{pcien} does not change. This in turn implies
that the $LR$ and $RL$ twists are in fact the same,
\begin{equation}  
{\cal S}^{LR}_{X}={\cal S}^{RL}_{X},
\la{anfiuno}
\end{equation}  or, in other words, the third twist leads to an
amphicheiral topological quantum field theory.
Since it is known that the two twists are related by
$ {\cal S}^{LR}_{X}={\cal S}^{RL}_{\tilde X}\big|_{\tau_0\rightarrow
-\bar\tau_0}$ ($\tilde X$ denotes the manifold $X$ with the opposite
orientation), it follows that by reversing the sign of the
$\theta$-angle one can jump from 
$X$ to $\tilde X$:
\begin{equation}  
{\cal S}_{X}={\cal S}_{\tilde
X}\big|_{\tau_0\rightarrow - \bar\tau_0}.
\la{anfidos}
\end{equation}  We will see in a moment that this information is
encoded in the discrete  conjugation symmetry introduced in
\cite{marcus}.

The definitions in \eqs{pcien} are almost self-evident. The only ones
which need clarification are those corresponding to $\tilde \eta$ and
$\tilde\chi_{\dalpha{\dot\beta}}$. Our conventions are:
\begin{equation}
\zeta^{i}{}_{\!\dot\alpha}\to \zeta^{\dot\beta}{}_{\!\dot\alpha}\to
\begin{cases}
\tilde\eta =-\zeta^{\dot\alpha}{}_{\dot\alpha},\\
\tilde\chi_{\dot\alpha\dot\beta}=-C_{\dot\gamma(\dot\beta}\zeta^{\dot
\gamma}{}_{\dot\alpha)}.
\end{cases}
\la{Moon4}
\end{equation}

In terms of the fields in \eqs{pcien}, the on-shell action
\eqs{cccuatro}  takes the form:
\begin{align} 
{\cal S}^{(0)}&=\frac{1}{ e^2_0}\int d^4 x\,
\tr\,\Big\{\,\half\nabla_{\!\alpha\dalpha}
 B\nabla^{\dalpha\alpha}C -\frac{1}{4}\nabla_{\!\beta{\dot\beta}}
V_{\alpha\dalpha}\nabla^{{\dot\beta}\beta}V^{\dalpha\alpha} 
-i\psi^{\beta}{}_{\dalpha}\nabla^{\dalpha\alpha}
\chi_{\alpha\beta}
\ret  &-\frac{i}{ 2}\psi_{\alpha\dalpha}\nabla^{\dalpha\alpha}
\eta +\frac{i}{
2}\tilde\eta\nabla^{\dalpha\alpha}\tilde\psi_{\alpha\dalpha}+
i\tilde\chi_{\dalpha{\dot\beta}}\nabla^{\dalpha\alpha}\tilde\psi
^{\dot\beta}{}_{\alpha}-\frac{1}{ 4}F_{mn} F^{mn}
\ret  &-\frac{i}{\raiz}\,\chi^{\alpha\beta}[\chi_{\alpha\beta},C]
-\frac{i}{\raiz}\,\tilde\psi^{\dalpha\alpha}[\tilde\psi_{\alpha\dalpha},B]+ 
i\raiz\,\chi^{\alpha\beta}[\tilde\psi_{\alpha\dalpha},V_\beta{}^{\dalpha}]
+\frac{i}{\raiz}\,\eta[\tilde\psi^\alpha{}_{\dalpha},V_\alpha{}^{\dalpha}]
\ret &-\frac{i}{{2\raiz}}\,\eta[\eta,C]
+\frac{i}{\raiz}\,\psi_{\alpha\dalpha}[\psi^{\dalpha\alpha},B]-
\frac{i}{\raiz}\,\tilde\eta[\psi^\alpha{}_{\dalpha},V_\alpha{}^\dalpha]-
i\raiz\,\tilde\chi^{\dalpha{\dot\beta}}[\psi_{\alpha\dalpha},V^\alpha{}_{\dot\beta}]
\ret
&+\frac{i}{\raiz}\,\tilde\chi_{\dalpha{\dot\beta}}[\tilde\chi^{\dalpha{\dot\beta}},C]
+\frac{i}{{2\raiz}}\,\tilde\eta[\tilde\eta,C]-\half[B,C]^2 
-[B,V_{\alpha\dalpha}][C,V^{\dalpha\alpha}]
\ret &+\frac{1}{4}[V_{\alpha\dalpha},V_{\beta{\dot\beta}}]
[V^{\dalpha\alpha},V^{{\dot\beta}\beta}]\,\Big\} -\frac{i\theta_0}{
32\pi^2}\int d^4 x\,\tr\,\bigl\{\, *  F_{mn}F^{mn}
\,\bigr\}.
\la{Light4}
\end{align}   
The next thing to do is to obtain the symmetry
transformations which correspond to the new model. Recall that we have
now two fermionic charges 
$Q$ and $\tilde Q$. The transformations generated by $Q$ are easily
obtained from those in the previous twist \eqs{cccinco}. To obtain
the transformations generated by $\tilde Q$ we must return to the
$\cn=4$ theory. Let us  recall that the $\cn=4$ supersymmetry
transformations are generated by $\xi_v{}^\alpha Q^v{}_\alpha + 
\bar\xi^v{}_\dalpha \bar Q_v{}^\dalpha$. The transformations
corresponding to
$\tilde Q$ are readily extracted by setting 
$\bar\xi^1=\bar\xi^2=0$ and making the replacement

\begin{equation}
\bar\xi^{\,3,4}{}_\dalpha\to \bar\xi^{\,i}{}_{\!\dalpha}\to
\bar\xi^{\,{\dot\beta}}{}_{\!\dalpha}\to 
\tilde\epsilon\,\delta^{\dot\beta}{}_{\!\dalpha}.
\la{pera4}
\end{equation} In this way one gets the following set of
transformations:

\begin{align}
\delta A_{\alpha\dalpha} &= 2i\epsilon\psi_{\alpha\dalpha},&
\tilde\delta A_{\alpha\dalpha}
&=-2i\tilde\epsilon\tilde\psi_{\alpha\dalpha},\ret   
\delta F^{+}_{\alpha\beta}&=2\epsilon\nabla_{(\alpha}{}^{\dot\alpha}
\psi{}_{\beta)\dot\alpha},& 
\tilde\delta F^{+}_{\alpha\beta}&=-2\tilde\epsilon\nabla_{(\alpha}{}^
{\dot\alpha}\tilde\psi{}_{\beta)\dot\alpha},\ret
\delta\psi_{\alpha\dot\alpha} &=
-i{\raiz}\epsilon\nabla_{\alpha\dot\alpha}C,&
\tilde\delta\psi_{\alpha\dot\alpha} &=
-2i\tilde\epsilon[V_{\alpha\dot\alpha},C]\ret
\delta\tilde\eta&=i{\raiz}\epsilon\nabla_{\alpha\dot\alpha}
V^{\dalpha\alpha},&
\tilde\delta\tilde\eta&=2i\tilde\epsilon[B,C],\ret
\delta\tilde\chi_{\dot\alpha\dot\beta}&=-i{\raiz}\epsilon\nabla_{\alpha
(\dot\alpha}V^{\alpha}{}_{\dot\beta)},& 
\tilde\delta\tilde\chi_{\dot\alpha\dot\beta}&=i\tilde\epsilon F^{-}_
{\dot\alpha\dot\beta}
-i\tilde\epsilon[V_{\alpha\dot\alpha},V^{\alpha}{}_{\dot\beta}],
\ret 
\delta\chi_{\alpha\beta} &=  -i\epsilon F^{+}_{\alpha\beta}
-i\epsilon[V_{\alpha\dot\alpha},V_{\beta}{}^{\dot\alpha}],&
\tilde\delta\chi_{\alpha\beta}&=i{\raiz}\,\tilde\epsilon\,\nabla^{\dot
\alpha}{}_{(\alpha}V_{\beta)\dot\alpha},
\ret
\delta\eta&=2i\epsilon[B,C],& 
\tilde\delta\eta & =i{\raiz}\,\tilde\epsilon\,\nabla_{\alpha\dot\alpha}
V^{\dalpha\alpha},
\ret
\delta\tilde\psi_{\alpha\dot\alpha}&=-2i\epsilon
[V_{\alpha\dot\alpha},C],&
\tilde\delta\tilde\psi_{\alpha\dot\alpha}&=-i\raiz\tilde\epsilon\nabla
_{\alpha\dot\alpha}C ,
\ret
\delta B&=\raiz\epsilon\eta,&
\tilde\delta B&=-\raiz\tilde\epsilon\tilde\eta,
\ret
\delta C&=0,& \tilde\delta C&=0,
\ret 
\delta
V_{\alpha\dot\alpha}&=-\raiz\epsilon\tilde\psi_{\alpha\dot\alpha}, &
\tilde\delta V_{\alpha\dot\alpha}&=\raiz\tilde\epsilon\psi_
{\alpha\dot\alpha}.
\ret
\la{melon4}
\end{align}

Since there are no half-integer spin fields in the theory it is
preferable to convert the spinor indices into vector indices. To do
this we make the  following definitions:
\begin{equation}
\begin{pmatrix} V\\ \psi \\ \tilde\psi 
\end{pmatrix}_{\!\alpha\dalpha}\equiv
\sigma^m{}_{\!\alpha\dalpha}\left(
\begin{matrix} V\\ \psi \\ \tilde\psi 
\end{matrix}
\right)_{\!m},\qquad
\chi_{\alpha\beta}=\sigma^{mn}{}_{\!\alpha\beta}\chi^{+}_{mn},\qquad
\tilde\chi_{\dalpha\dot\beta}=\bar\sigma^{mn}{}_{\!\dalpha\dot\beta}\chi^{-}
_{mn}
\la{uva4}
\end{equation}
where $\chi^{\pm}{}_{\!mn}=(1/2)\{\chi_{mn}\pm(1/2)\epsilon_{mn
pq}\chi^{pq}\}$. In order to extract a manifestly real action we also
make the replacements $\psi\to -i\psi$, $\chi^{+}\to i\chi^{+}$,
$\tilde\eta\to  i\tilde\eta$ and $\tilde Q\to i\tilde Q$.   The
resulting action is:
\begin{align} 
{\cal S}^{(0)}&=\frac{1}{ e^2_0}\int d^4 x\, \tr\,
\bigl\{\,-\nabla_m B\nabla^m C -\nabla_m V_n\nabla^m V^n 
+4\psi^m\nabla^n\chi^{+}_{mn}  +\psi^m\nabla_m\eta
\ret &+\tilde \psi^m\nabla_m\tilde\eta+
4\tilde\psi^m\nabla^n\chi^{-}_{mn}-\frac{1}{ 4}F_{mn} F^{mn}-
i\raiz\,\chi^{+mn}[\chi^{+}_{mn},C]
\ret  &+i\raiz\,\tilde\psi^m[\tilde\psi_m,B]- 
4i\raiz\,\chi^{+}_{mn}[\tilde\psi^m,V^n]
+i\raiz\,\eta[\tilde\psi_m,V^m] -\frac{i}{{2\raiz}}\,\eta[\eta,C]
\ret  &+i\raiz\,\psi_m[\psi^m,B]- i\raiz\,\tilde\eta[\psi_m,V^m]+
4\raiz\,i\chi^{-}_{mn}[\psi^m,V^n]-i\raiz\,\chi^{-}_{mn}[\chi^{-mn},C]
\ret &-\frac{i}{{2\raiz}}\,\tilde\eta[\tilde\eta,C]-\half[B,C]^2 
+2[B,V_m][C,V^m]+  [V_m,V_n] [V^m,V^n]\,\bigr\}
\ret  &-\frac{i\theta_0}{ 32\pi^2}\int d^4 x\,\tr\,\bigl\{\,* 
F_{mn}F^{mn}
\,\bigr\},
\la{pLight}
\end{align}   and the corresponding transformations become:
\begin{align}
\delta A_m &= 2\epsilon\psi_m,   &\tilde\delta A_m &=
-2\tilde\epsilon\tilde\psi_m,\ret 
\delta\psi_m &= {\raiz}\epsilon\nabla_m C,  &\tilde\delta\psi_m&= 
-2i\tilde\epsilon[V_m,C]\ret
\delta\tilde\eta&=-2{\raiz}\epsilon\nabla_m V^m,
&\tilde\delta\tilde\eta&=-2i\tilde\epsilon[B,C],\ret
\delta\chi^{-}_{mn}&=2{\raiz}\epsilon (\nabla_{[m}V_{n]})^{-},
&\tilde\delta\chi^{-}_{mn}&=\tilde\epsilon F^{-}_ {mn}
-2i\tilde\epsilon ([V_m,V_n])^{-},\ret  
\delta\chi^{+}_{mn} &=-\epsilon F^{+}_{mn} +2i\epsilon
([V_m,V_n])^{+},  &\tilde\delta\chi^{+}_{mn} &=
2{\raiz}\tilde\epsilon(\nabla_{[m}V_{n]})^{+},\ret
\delta\eta  &=2i\epsilon[B,C],
&\tilde\delta\eta&=-2{\raiz}\tilde\epsilon\nabla_m V^m,\ret
\delta\tilde\psi_m&=-2i\epsilon [V_m,C],
&\tilde\delta\tilde\psi_m&=-\raiz\tilde\epsilon\nabla_m C ,\ret 
\delta B&=\raiz\epsilon\eta,  &\tilde\delta
B&=-\raiz\tilde\epsilon\tilde\eta,\ret
\delta C&=0,  &\tilde\delta C&=0,\ret
\delta V_m&=-\raiz\epsilon\tilde\psi_m, &\tilde\delta V_m
&=-\raiz\tilde\epsilon\psi_ m,
\la{melones4}
\end{align}  
where $(X_{[mn]})^{\pm}\equiv \half(X_{[mn]}\pm
*X_{[mn]})$, and
$X_{[mn]}\equiv \half(X_{mn}-X_{nm})$. The generators $Q$ and $\tilde
Q$ satisfy the on-shell algebra:
\bea &&\{\,Q,Q\,\}=\delta_g (C),\ret &&\{\,\tilde Q,\tilde
Q\,\}=\delta_g (C),\ret &&\{\,Q,\tilde Q\,\}=0.
\la{aceituna4}
\eea

Now consider the following discrete transformations acting on the
fields of the theory:
\begin{align} B&\too B,& C&\too C,\ret  A&\too A,& V&\too -V,\ret 
\eta&\too -\tilde\eta,& \psi&\too-\tilde\psi,\ret 
\tilde\eta&\too -\eta,&\tilde\psi&\too-\psi,\ret  
\chi^{+}&\longleftrightarrow -\chi^{-} \Rightarrow \begin{cases}
\chi\to-\chi,\\  *\chi\to *\chi,
\end{cases} & F^{+}&\longleftrightarrow F^{-}\Rightarrow
\begin{cases}  F\to F,\\ *F\to -*F.\end{cases}\ret
\la{melocoton4}
\end{align}  Notice that these transformations involve the simultaneous
replacement
$\epsilon_{mnpq}\to -\epsilon_{mnpq}\,$, which is equivalent to a
reversal of the orientation of the four-manifold $X$. Because of this
orientation reversal, the sign of the theta-term  in \eqs{pLight}
is also reversed. Thus the ${\IZ}_2$-like transformations
\eqs{melocoton4} map the action on $X$ to the same action on
$\tilde X$ but with  $-\theta_0$. This is precisely the information
encoded in
\eqs{anfidos}.   

It is also noteworthy that the transformations 
\eqs{melocoton4} exchange the BRST generators $Q$ and $\tilde Q$ -- 
one can see this by looking at \eqs{melones4}. Indeed, had we not
known about the existence of one of the topological symmetries, say
$\tilde Q$, we would have discovered it immediately with the aid of
the symmetry \eqs{melocoton4}. In addition to this, one can readily
see that the replacements dictated by \eqs{melocoton4} preserve the
ghost number assignments of the fields. In what follows, we will
usually refer to the transformations 
\eqs{melocoton4}  by ${\IZ}_2$, but the reader must be aware of this
abuse of notation.

Several things remain to be done. It would be desirable to obtain an
off-shell formulation of the model. Besides, it would be interesting
to find out whether the off-shell action (provided that it exists) can
be written as a $Q$- (or $\tilde Q$, or both) commutator, and write
down the explicit expression for the corresponding gauge fermion. And
finally, the theory should be generalized to any arbitrary
four-manifold of euclidean signature. 

We have found a complete off-shell  formulation involving both  BRST
symmetries simultaneously such that  the action \eqs{pLight} is (up
to  appropriate  theta-terms) $Q$ and $\tilde Q$-exact. 

Let us examine these results  in more detail. The on-shell algebra
\eqs{aceituna4}  can be extended off-shell by introducing the
auxiliary fields 
$N^{+}_{mn}$, $N^{-}_{mn}$ and $P$, which have zero ghost number and 
are taken to transform under ${\IZ}_2$ as $N^{+}\leftrightarrow 
-N^{-}$, $P\to -P$. In terms of these fields, the transformations
\eqs{melones4}  are modified as follows:
\begin{align} 
\delta\tilde\eta&=-2{\raiz}\epsilon\nabla_m V^m +\epsilon P,\ret 
\delta P&=-4\epsilon\nabla_m\tilde\psi^m +4\raiz i\epsilon
[\psi^m,V_m]+ 2\raiz i\epsilon[\tilde \eta,C],
\ret 
\delta\chi^{-}_{mn}&=2{\raiz}\epsilon (\nabla_{[m}V_{n]})^{-}
+\epsilon N^{-}_{mn},
\displaybreak\ret 
\delta N^{-}_{mn}&=4\epsilon\nabla_{[m}\tilde\psi^{-}_{n]}- 4\raiz
i\epsilon [\psi_{[m},V_{n]}]^{-} +2\raiz
i\epsilon[\chi^{-}_{mn},C],\ret 
\delta\chi^{+}_{mn}&=  -\epsilon F^{+}_{mn}+ 2i\epsilon
([V_m,V_n])^{+}+\epsilon N^{+}_{mn},\ret  
\delta N^{+}_{mn}&=4\epsilon\nabla_{[m}\psi^{+}_{n]} +4\raiz i\epsilon
[\tilde\psi_{[m},V_{n]}]^{+}+ 2\raiz i\epsilon[\chi^{+}_{mn},C].
\ret
\la{Basler}
\end{align}  
The other transformations in \eqs{melones4} remain the
same.  Equivalent formulas hold for $\tilde Q$ and are related to
those in 
\eqs{Basler} through the ${\IZ}_2$ transformation. In this off-shell 
realization the auxiliary fields appear in the action only
quadratically, that  is, 
\begin{equation} {\cal S}^{(1)}={\cal S}^{(0)}+\int
\tr\left\{\,\half(N^{+})^2+
\half(N^{-})^2+\frac{1}{ 8}P^2\right\}.
\la{meca}
\end{equation} The action ${\cal S}^{(1)}$ can be written either as a
$Q$ commutator or as a
$\tilde Q$ commutator and is invariant under both, 
$Q$ and $\tilde Q$, that is,
\begin{equation}  {\cal S}^{(1)}=\{Q,\hat \Psi^{+}\}-2\pi ik\tau_0
=\{\tilde Q,\hat \Psi^{-}\} -2\pi ik\bar\tau_0\,;\quad  [Q, {\cal
S}^{(1)}]=0=[\tilde Q,{\cal S}^{(1)}],
\la{boomerang4}
\end{equation}  where the gauge fermions $\hat\Psi^{\pm}$ are not
equal but are formally interchanged by the ${\IZ}_2$ transformation
and $k$ is the instanton number
\eqs{knumber}. It is possible to redefine the auxiliary fields to cast
either  the $Q$ or the $\tilde Q$ transformations  (but not both
simultaneously) in the standard form,
\begin{align}
\{Q,\textsc {antighost}
\}&=\textsc{auxiliary field}, \ret  [Q,\textsc{auxiliary field}]&=
\delta^{\text {\tiny gauge}}\textsc{antighost},\ret
\end{align}  which is essential to make contact with the Mathai-Quillen
interpretation.  Performing the shifts,
\begin{align} P&\too P+2{\raiz}\nabla_m V^m,\ret  N^{-}_{mn}&\too
N^{-}_{mn} -2{\raiz} (\nabla_{[m}V_{n]})^{-},\ret  N^{+}_{mn}&\too
N^{+}_{mn}+ F^{+}_{mn} -2i([V_m,V_n])^{+},\ret
\la{Allou}
\end{align}  which can be guessed from \eqs{Basler}, the $Q$
transformations take the  simple form:
\begin{align}
\delta A_m &= 2\epsilon\psi_m,& 
\delta\psi_m&=\raiz\,\epsilon\nabla_m C,\ret
\delta V_m&=-\raiz\epsilon\tilde\psi_m, 
&\delta\tilde\psi_m&=-2i\epsilon [V_m,C],\ret
\delta C&=0,  &\delta\eta &=2i\epsilon [B,C],\ret
\delta B&=\raiz\epsilon\eta,  &\delta P&=2\raiz i\epsilon
[\tilde\eta,C],\ret
\delta\tilde\eta&=\epsilon P,& &{}\ret
\delta\chi^{+}_{mn}&=\epsilon N^{+}_{mn}, 
 &\delta N^{+}_{mn}&=2\raiz i\epsilon[\chi^{+}_{mn},C],\ret
\delta\chi^{-}_{mn}&=\epsilon N^{-}_{mn},  &\delta N^{-}_{mn}&=2\raiz
i\epsilon [\chi^{-}_{mn},C].
\ret
\la{butterfly4}
\end{align}  The point is that if instead of \eqs{Allou}  we make the
${\IZ}_2$ conjugate  shifts, 
\begin{align} P&\too P+2{\raiz}\nabla_m V^m,\ret  N^{+}_{mn}&\too
N^{+}_{mn} -2{\raiz} (\nabla_{[m}V_{n]})^{+},\ret  N^{-}_{mn}&\too
N^{-}_{mn}- F^{-}_{mn}+2i([V_m,V_n])^{-},
\ret
\la{Proust4}
\end{align}
 it is $\tilde\delta\equiv\tilde\epsilon\tilde Q$ the one which can be
cast in the  simple form:
\begin{align}
\tilde\delta A_m &= -2\tilde\epsilon\tilde\psi_m,
&\tilde\delta\tilde\psi_m&=-\raiz\,\tilde\epsilon\nabla_m C,\ret
\tilde\delta V_m&=-\raiz\tilde\epsilon\psi_m,
&\tilde\delta\psi_m&=-2i\tilde\epsilon [V_m,C],\ret
\tilde\delta C&=0, &\tilde\delta\tilde\eta &=-2i\tilde\epsilon
[B,C],\ret
\tilde\delta B&=-\raiz\tilde\epsilon\tilde\eta,& &{}\ret
\tilde\delta\eta&=\tilde\epsilon P, &\tilde\delta P&=2\raiz
i\tilde\epsilon [\eta,C],\ret
\tilde\delta\chi^{+}_{mn}&=\tilde\epsilon N^{+}_{mn}, &\tilde\delta
N^{+}_{mn}&=2\raiz i\tilde\epsilon [\chi^{+}_{mn},C],\ret
\tilde\delta\chi^{-}_{mn}&=\tilde\epsilon N^{-}_{mn}, &\tilde\delta
N^{-}_{mn}&=2\raiz i\tilde\epsilon [\chi^{-}_{mn},C].
\ret
\la{boheme}
\end{align}  Notice that since the appropriate shifts are in each case
different, the one which simplifies the $Q$ transformations makes the
corresponding
$\tilde Q$ transformations (not shown)  much more complicated and
conversely, the shift which simplifies the $\tilde Q$ transformations
makes the corresponding $Q$ transformations (not shown) much more
complicated.

Keeping these  results in mind, from now on we will focus on the $Q$
formulation, that is, on the off-shell formulation in which the $Q$
transformations take  the form
\eqs{butterfly4}.  The off-shell action which corresponds to this
formulation is:
\begin{align} 
{\cal S}^{(2)}&= \frac{1}{ e^2_0}\int d^4 x\, \tr\,
\bigl\{\,-\nabla_m B\nabla^m C +\half N^{+}_{mn}\bigl
(\,N^{+mn}+2F^{+mn}-4i[V^m,V^n]^{+}\,
\bigr )
\ret    &+\half N^{-}_{mn}\bigl (\,N^{-mn}-4\raiz
(\nabla^{[m}V^{n]})^{-}\,\bigr )  +\frac{1}{8}P\bigl (\,P+4\raiz
\,\nabla_m V^m\,\bigr)
\ret &+4\psi^m\nabla^n\chi^{+}_{mn} +\psi^m\nabla_m\eta +\tilde
\psi^m\nabla_m\tilde\eta+ 4\tilde\psi^m\nabla^n\chi^{-}_{mn}
\ret  &-i\raiz\,\chi^{+mn}[\chi^{+}_{mn},C]+
i\raiz\,\tilde\psi^m[\tilde\psi_m,B]-
4\raiz\,i\chi^{+}_{mn}[\tilde\psi^m,V^n]
\ret &+i\raiz\,\eta[\tilde\psi_m,V^m]
-\frac{i}{{2\raiz}}\,\eta[\eta,C]+ i\raiz\,\psi_m[\psi^m,B]-
i\raiz\,\tilde\eta[\psi_m,V^m]
\ret &+4\raiz\,i\chi^{-}_{mn}[\psi^m,V^n]-
i\raiz\,\chi^{-}_{mn}[\chi^{-mn},C]
-\frac{i}{{2\raiz}}\,\tilde\eta[\tilde\eta,C]-\half[B,C]^2
\ret &+2[B,V_m][C,V^m]\,\bigr\} -2\pi i\tau_0\frac{1}{ 32\pi^2}\int d^4
x\,
\tr\,\bigl\{\,*F_{mn}F^{mn}\,\bigr\},
\ret
\la{isinha4}
\end{align}  
and reverts to \eqs{pLight}  after integrating  out the
auxiliary fields. The
$\tau_0$-independent part of the action \eqs{isinha4} is $Q$-exact,
that is,   it can be written as a $Q$-commutator. The appropriate gauge
fermion  is:
\begin{align}
\Psi^{+}=&\frac{1}{ e^2_0}\int d^4
x\,\tr\,\bigl\{\,\,\half\chi^{+}_{mn}\,\bigl 
(\,N^{+mn}+2F^{+mn}-4i[V^m,V^n]^{+}\,\bigr )
\ret  &+\frac{1}{2} \chi^{-}_{mn}\bigl (\,N^{-mn}-4\raiz
(\nabla^{[m}V^{n]})^{-}\,
\bigr) +\frac{1}{8}\tilde\eta\,\bigl (\,P+4\raiz \,\nabla_m V^m\,\bigr
)\,\bigr \}
\ret   &+\frac{1}{ e^2_0}\int d^4
x\,\tr\,\bigl\{\,\frac{1}{\raiz}B\,\bigl
 (\,\nabla_m \psi^m  +i\raiz [\tilde\psi_m,V^m]\,\bigr )\,\bigr\}
\ret  &+\frac{1}{ e^2_0}\int d^4 x\,\tr\,\bigr\{\,\frac{i}{4}\eta
[B,C]\,\bigr\}.
\ret
\la{sesamo4}
\end{align}  
notice that $\Psi^-$ would correspond to the ${\IZ}_2$
transformation  of
$\Psi^+$. The gauge fermions $\hat\Psi^+$ and $\hat\Psi^-$ in 
\eqs{boomerang4} are easily obtained after undoing the shifts
\eqs{Allou} and \eqs{Proust4},
 respectively.

Now we switch on an arbitrary background metric $g_{\mu\nu}$ of
euclidean  signature. This is straightforward once we have expressed
the model in the  form of eqs. \eqs{butterfly4} and \eqs{sesamo4}.
The covariantized transformations  are  the following: 
\begin{align}
\delta A_\mu &= 2\epsilon\psi_\mu, 
&\delta\psi_\mu&=\raiz\,\epsilon\deriv_\mu C,\ret
\delta V_\mu&=-\raiz\epsilon\tilde\psi_\mu,
&\delta\tilde\psi_\mu&=-2i\epsilon [V_\mu,C],\ret
\delta C&=0,& &{}\ret 
\delta B&=\raiz\epsilon\eta,
 &\delta\eta &=2i\epsilon [B,C],\ret
\delta\tilde\eta&=\epsilon P, &\delta P&=2\raiz i\epsilon
[\tilde\eta,C],\ret
\delta\chi^{+}_{\mu\nu}&=\epsilon N^{+}_{\mu\nu}, &\delta
N^{+}_{\mu\nu}&=2\raiz i\epsilon [\chi^{+}_{\mu\nu},C],\ret
\delta\chi^{-}_{\mu\nu}&=\epsilon N^{-}_{\mu\nu}, &\delta
N^{-}_{\mu\nu}&=2\raiz i\epsilon [\chi^{-}_{\mu\nu},C],
\ret
\la{butter4}
\end{align}  and the action for the model is defined to be ${\cal
S}^{(2)}_c=\{Q,\Psi^{+}_c\}- 2\pi i k \tau_0$, with the gauge fermion 
(appropriately covariantized):
\begin{align}
\Psi^{+}_c=&\frac{1}{ e^2_0}\int_X d^4
x\,\sqrt{g}\,\tr\,\bigl\{\,\,\half\chi^{+}_ {\mu\nu}
\,\bigl (\,N^{+\mu\nu}+2F^{+\mu\nu}-4i[V^\mu,V^\nu]^{+}\,
\bigr )
\ret &+\frac{1}{2}\chi^{-}_{\mu\nu}\bigl (\,N^{-\mu\nu}-4\raiz
(\deriv^{[\mu}V^{\nu]}) ^{-}\,
\bigr ) +\frac{1}{8}\tilde\eta\,\bigl (\,P+4\raiz \,\deriv_\mu
V^\mu\,\bigr )\,
\bigr \}
\ret   &+\frac{1}{ e^2_0}\int_X d^4
x\,\sqrt{g}\,\tr\,\bigl\{\,\frac{1}{\raiz}B\,\bigl  (\,\deriv_\mu
\psi^\mu +i\raiz [\tilde\psi_\mu,V^\mu]\,\bigr )\,\bigr\}
\ret &+\frac{1}{ e^2_0}\int_X d^4 x\,\sqrt{g}\,
\tr\,\bigr\{\,\frac{i}{4}\eta [B,C]\,\bigr\}.
\ret
\la{epi}
\end{align}   
The resulting action reads:
\begin{align}  
{\cal S}^{(2)}_c&= \frac{1}{ e^2_0}\int_X d^4 x\,\sqrt{g}\,
\tr\, 
\bigl\{\,-\deriv_\mu B\deriv^\mu C +\half N^{+}_{\mu\nu}\bigl
(\,N^{+\mu\nu}+2F^{+\mu\nu}- 4i[V^\mu,V^\nu]^{+}\,
\bigr )
\ret  &+\half N^{-}_{\mu\nu}\bigl (\,N^{-\mu\nu}-4\raiz
(\deriv^{[\mu}V^{\nu]}) ^{-}\,\bigr ) +\frac{1}{8}P\bigl (\,P+4\raiz
\,\deriv_\mu V^\mu\,\bigr )
\ret &+4\psi^\mu\deriv^\nu\chi^{+}_{\mu\nu} +\psi^\mu\deriv_\mu\eta
+\tilde
\psi^\mu\deriv_\mu\tilde\eta+
4\tilde\psi^\mu\deriv^\nu\chi^{-}_{\mu\nu}
\ret &-i\raiz\,\chi^{+\mu\nu}[\chi^{+}_{\mu\nu},C]+
i\raiz\,\tilde\psi^\mu[\tilde\psi_\mu,B]-
4\raiz\,i\chi^{+}_{\mu\nu}[\tilde\psi^\mu,V^\nu]
+i\raiz\,\eta[\tilde\psi_\mu,V^\mu]
\ret & -\frac{i}{{2\raiz}}\,\eta[\eta,C]+ i\raiz\,\psi_\mu[\psi^\mu,B]-
i\raiz
\,\tilde\eta[\psi_\mu,V^\mu]+
4\raiz\,i\chi^{-}_{\mu\nu}[\psi^\mu,V^\nu]
\ret &-i\raiz\,\chi^{-}_{\mu\nu}[\chi^{-\mu\nu},C]
-\frac{i}{{2\raiz}}\,\tilde\eta[\tilde\eta,C]-\half[B,C]^2 
+2[B,V_\mu][C,V^\mu]\,\bigr\}
\ret  & -2\pi i\tau_0\frac{1}{ 32\pi^2}\int_X d^4
x\,\sqrt{g}\tr\,\bigl\{\, *  F_{\mu\nu} F^{\mu\nu}\,\bigr\}.
\ret
\la{mitocondria4}
\end{align}  If we integrate out the auxiliary fields in
\eqs{mitocondria4} we recover the action
\eqs{pLight}.  
To close the analysis in this section we  shall
state the relation between our model and those presented in 
\cite{blauthomp}\cite{marcus}. After integrating out the auxiliary
fields $N^{\pm}$,  our action (with  $\theta_0$  set to zero)
and transformations match those by Marcus \cite{marcus} in the $\alpha
=1$ gauge,  after the following  redefinitions (Marcus'  fields and
charges are denoted by a subscript $M$):
\begin{align} Q_M &= \half (Q-i\tilde Q),&
\tilde Q_M &= \half (Q+i\tilde Q),\ret A_M &=-A,&
 V_M &=\raiz V,\ret
\psi_M &= i\psi -\tilde\psi,&
\tilde\psi_M &= i\psi +\tilde\psi,\ret
\eta_M &=-\half(\eta+i\tilde\eta),&
\tilde\eta_M &=-\half(\eta-i\tilde\eta),\ret
\chi_M &= 2(\chi^{+}-i\chi^{-}),& P_M &= -\half P,\ret C_M
&=-\frac{i}{\raiz}B,& B_M &=i\raiz C.
\ret
\la{SPICE}
\end{align}  According to Marcus' conventions, one has to replace
simultaneously our  hermitean group generators $T$ with the
antihermitean generators $T_M =-iT$.

Similarly, the relation to the theory proposed by Blau and
Thompson
\cite{blauthomp}  can be stated by  means of the dictionary, 
\begin{align} 
Q_{BT} &= \frac{\omega}{4} (Q+\tilde Q),&
\bar Q_{BT} &= \frac{\omega}{4} (Q-\tilde Q),\ret  
A_{BT} &=-A, & V_{BT}&=\raiz V,\displaybreak\ret
\psi_{BT} &= \frac{\omega}{2}(\tilde\psi -\psi),&
\bar\psi_{BT} &= -\frac{\omega}{2}(\psi +\tilde\psi),\ret
\eta_{BT} &=-\frac{\omega}{2}(\eta-\tilde\eta),&
\bar\eta_{BT} &=-\frac{\omega}{2}(\eta+\tilde\eta),\ret
\chi_{BT} &= -2\omega(\chi^{+}+\chi^{-}),&
 u_{BT} &= -\frac{1}{4} P,\ret
\bar\phi_{BT} &=-{\raiz}B,&
\phi_{BT} &=\frac{i}{\raiz} C .
\ret
\la{GIRLS}
\end{align}  
where $\omega=1-i$. Again,  $T_{BT} =-iT$.

\section{The theory in the Mathai-Quillen approach}  

The relevant basic equations for this model can be identified from the
fixed points of the $\delta$ symmetry transformations \eqs{melones4}. 
They involve the self-dual part of the gauge connection
$F^{+}$ and a  real vector field
$V_\mu$   taking values in the adjoint  representation of some finite
dimensional  compact  Lie group
$G$:
\begin{equation}
\begin{cases}  F^{+}_{\mu\nu}-i[V_{\mu},V_{\nu}]^{+}=0,\\
\bigl (\,\deriv_{[\mu}V_{\nu]}\,\bigr )^{-}=0,\\
\deriv_\mu V^\mu=0.
\end{cases}
\la{morequations4}
\end{equation}

\subsection{The topological framework}

The geometrical setting is a certain compact, oriented Riemannian 
four-manifold
$X$, and the  field space is ${\cal M}={\cal A}\times\Omega^1(X,\ad
P)$,  where
${\cal A}$ is the space of connections on a principal 
$G$-bundle $P\to X$, and the second factor denotes, as we have 
already seen before, 
$1$-forms on $X$ taking  values in the Lie algebra of $G$.   The group
${\cal G}$ of gauge transformations  of the bundle $P$ has an action
on the field space which is given locally by:
\begin{align} g^{*}(A)&=i(dg)g^{-1}+gAg^{-1},\ret  
g^{*}(V)&=gVg^{-1},\ret &{}
\la{mjauje}
\end{align}   where $V\in \Omega^1(X,\ad P)$ and $A$ is the gauge
connection. In terms of the covariant derivative $d_A =d+i[A,~]$, the
infinitesimal  form of the transformations
\eqs{mjauje}, with $g={\hbox{\rm exp}}(-iC)$ and
$C\in \Omega^0(X,\ad P)$, takes the form:
\begin{align}
\delta_g(C)A&=d_A C,\ret
\delta_g(C)V&=i[V,C].\ret &{}
\la{mjauja}
\end{align}   The tangent space to the field space at the point
$(A,V)$ is the vector space
$T_{(A,V)}{\cal M}=\Omega^1_{(A)}(X,\ad P)\oplus\Omega^1_{(V)}(X,\ad
P)$, where
$\Omega^1_{(A)}(X,\ad P)$ denotes the tangent space to ${\cal A}$ at
$A$, and 
$\Omega^1_{(V)}(X,\ad P)$ denotes the tangent space to 
$\Omega^1(X,\ad P)$ at $V$. On
$T_{(A,V)}{\cal M}$, the  gauge-invariant Riemannian metric (inherited
from that on
$X$) is defined as:
\begin{equation}
\langle (\psi,\tilde\psi),(\theta,\tilde\omega)\rangle =\int_X
\tr(\psi\wedge *\theta)+\int_X \tr (\tilde\psi\wedge *\tilde\omega)
\la{mmetrica4}
\end{equation}   where $\psi,\theta\in\Omega^1_{(A)}(X,\ad P)$ and
$\tilde\psi,\tilde\omega\in \Omega^{1}_{(V)} (X,\ad P)$. 

The basic equations \eqs{morequations4} are introduced in this
framework   as sections of the trivial vector bundle 
${\cal V}=
\mani\times{\cal F}$, where the fibre is in this case 
 ${\cal F}= \Omega^{2,+}(X,\ad P)\oplus\Omega^{2,-} (X,\ad P)\oplus
\Omega^0(X,\ad P)$. Taking into account the form of the basic 
equations, the section reads:
\begin{equation}    s(A,V) =\bigl (
-2(F^{+}_{\mu\nu}-i[V_{\mu},V_{\nu}]^{+}),~
 4(\,\deriv_{[\mu}V_{\nu]}\,)^{-},
\raiz\deriv_\mu V^\mu
\bigr ).
\la{mseccion}
\end{equation}    The section \eqs{mseccion}, being gauge
ad-equivariant, descends to a  section
${\tilde s}$ of the associated vector bundle ${\cal M}\times_{\cal G}
{\cal F}$ whose zero locus  gives precisely the  moduli space of the
topological theory. It would be desirable to compute  the dimension of
this moduli space. The relevant deformation complex   is the following:
\bea
 0&&\!\!\!\!\!\!\!\!\too\Omega^0(X,\ad P)\mapright{{\cal C}}
\Omega^1_{(A)}(X,\ad P)\oplus\Omega^1_{(V)}(X,\ad P)\ret
&&\mapright{ds}\Omega^{2,+}(X,\ad P)\oplus\Omega^{2,-}(X,\ad P)\oplus
\Omega^0(X,\ad P)\too 0.\ret
\la{complexo}
\eea The map ${\cal C}:\Omega^0(X,\ad P)\too T{\cal M}$ is given by:
\begin{equation}    {\cal C}(C)=(d_A C,i[V,C]),\quad
C\in\Omega^0(X,\ad P),
\la{barbacoa}
\end{equation}   while the map $ds:T_{(A,V)}{\cal M}\too {\cal F}$ is 
given by the linearization of the basic equations 
\eqs{morequations4}:
\bea ds(\psi,\tilde\psi)&&\!\!\!\!\!\!\!\!=\bigl
(-4(\deriv_{[\mu}\psi_{\nu]})^{+} +4i[\tilde\psi_{[\mu},
V_{\nu]}]^{+},\ret &&4(\deriv_{[\mu}\tilde\psi_{\nu]})^{-}
+4i[\psi_{[\mu}, V_{\nu]}]^{-},\raiz\deriv_\mu\tilde\psi^\mu +\raiz
i[\psi_\mu ,V^\mu]\bigr ).
\ret
\la{casper}
\eea    Under suitable conditions, the index of the complex
\eqs{complexo} computes the  dimension of ${\hbox{\rm
Ker}}(ds)/{\hbox{\rm Im}}({\cal C})$. To calculate its index, the
complex  can be split up into the ASD-instanton deformation complex:
\begin{align}    &(1)~0\too\Omega^0(X,\ad
P)\mapright{d_{\!A}}\Omega^1(X,\ad P)
\mapright{p^{+}\!d_{\!A}}\Omega^{2,+}(X,\ad P)\too 0,
\la{athisng}\\
\intertext{and the complex associated to the operator} 
&(2)~D=p^{-}\!d_A+d^{*}_{\!A}:\Omega^1(X,\ad P)\too
\Omega^0(X,\ad P)\oplus\Omega^{2,-} (X,\ad P),
\la{kouba}
\end{align}    which is easily seen to correspond to the instanton
deformation complex  for self-dual (SD) connections. Thus, the index
of the total complex  (which gives minus the virtual dimension of the
moduli space) is:

\begin{align} -{\hbox{\rm dim}}({\cal
M})&=\ind(1)-\ind(2)=\ind({\hbox{\rm ASD}})+ 
\ind({\hbox{\rm SD}})\ret   &= p_1(\ad P)+\half{\hbox{\rm
dim}}(G)(\chi +\sigma) - p_1(\ad P)+\half{\hbox{\rm
dim}}(G)(\chi-\sigma)\ret  &= {\hbox{\rm dim}}(G)\chi,
\ret &{}
\la{indice}
\end{align}   where $p_1(\ad P)$ is the first Pontryagin class of the
adjoint bundle 
$\ad P$, $\chi$ is the Euler characteristic of the $4$-manifold $X$
and 
$\sigma$ is its signature. 

\vskip 1cm
\subsection{The topological action}

We now proceed as in the previous cases. To build a topological theory
out of the moduli problem defined by  the equations
\eqs{morequations4}, we need the following   multiplet of fields. For
the field space ${\cal M}= {\cal A}\times\Omega^1(X,\ad P)$ we
introduce the gauge connection $A_\mu$ and the one-form
$V_{\mu}$, both commuting and with ghost number $0$. For the
(co)tangent  space
$T_{(A,V)}{\cal M}=\Omega^1_{(A)}(X,\ad P)\oplus\Omega^1_{(V)}(X,\ad
P)$ we introduce the anticommuting fields
$\psi_\mu$ and $\tilde\psi_{\mu}$, both with ghost number
$1$ and which can  be seen as differential forms on the moduli space.
For the fibre
${\cal F}= \Omega^{2,+}(X,\ad P)\oplus\Omega^{2,-} (X,\ad
P)\oplus\Omega^0(X,\ad P)$ we have anticommuting fields with the
quantum numbers  of the equations, namely a  self-dual two-form
$\chi^{+}_{\mu\nu}$,  an  anti-self-dual two-form $\chi^{-}_{\mu\nu}$
and a
$0$-form $\tilde
\eta$, all with ghost number $-1$, and their superpartners, a
commuting self-dual two-form
$N^{+}_{\mu\nu}$,  a commuting anti-self-dual two-form
$N^{-}_{\mu\nu}$ and a commuting $0$-form $P$, all with ghost number
$0$ and which appear as auxiliary fields in the associated  field
theory. And finally, associated to the gauge symmetry, we introduce  a
commuting scalar field
$C\in\Omega^{0}(X,\ad P)$ with ghost number
$+2$, and a multiplet of scalar fields $B$ (commuting and with ghost
number $-2$) and $\eta$ (anticommuting and with ghost number
$-1$), both also in $\Omega^{0}(X,\ad P)$ and which enforce the
horizontal projection ${\cal M}\to {\cal M}/{\cal G}$ \cite{moore}.
The BRST symmetry of the model is given by:
\begin{align} [Q, A_\mu] &=\psi_\mu,&
\{Q,\psi_{\mu}\} &= \deriv_{\mu}C, \ret [Q,V_\mu]&=\tilde\psi_\mu,&
\{Q,\tilde\psi_\mu\,\} &=i\,[V_\mu,C],\ret [Q,C]&=0,& &{}\ret
\{Q,\chi^{+}_{\mu\nu}\}&=N^{+}_{\mu\nu},& 
 [Q,N^{+}_{\mu\nu}]&=i\,[\chi^{+}_{\mu\nu},C],\ret
\{Q,\chi^{-}_{\mu\nu}\} &= N^{-}_{\mu\nu},&
[Q,N^{-}_{\mu\nu}]&=i\,[\chi^{-}_{\mu\nu},C],\ret  
\{Q,\tilde\eta\} &= P,& [Q, P]&=i\,[\tilde\eta,C],\ret [Q,B]&=\eta,&
\{Q,\eta\,\} &=i\,[B,C].\ret &{} & &{}
\la{marek}
\end{align}   This BRST algebra closes up to a gauge transformation
generated  by
$C$. 

We have to give now the expressions for the different pieces of  the 
gauge fermion. For the localization gauge fermion we have:
\begin{align}
\Psi_{\text{ loc}}&=\langle(\chi^{+},\chi^{-},\tilde\eta),s(A,V)
\rangle +\langle (\chi^{+},\chi^{-},\tilde\eta),(
N^{+},N^{-},P)\rangle=
\ret  &\int_X
\sqrt{g}\,\tr\,\bigl\{\,\,\half\chi^{+}_{\mu\nu}\bigl (\,N^{+\mu\nu} -2
F^{+\mu\nu}+2i[V^{\mu},V^{\nu}]^{+}\,\bigr )
\ret & +\half\chi^{-}_{\mu\nu}\bigl
(\,N^{-\mu\nu}+4(\deriv^{[\mu}V^{\nu]}) ^{-}\,\bigr )+\tilde\eta\bigl 
(\,P+\raiz\deriv_\mu V^\mu\,\bigr )
\,\bigr \},\ret
\la{localizac}
\end{align}  while for the projection gauge fermion, which enforces the
horizontal projection, we have:
\begin{equation}
\Psi_{\text{proj}}=\langle B,{\cal
C}^{\dag}(\psi,\tilde\psi)\rangle_{\hbox{\bf g}}=
\int_X \sqrt{g}\,\tr\,\bigl\{\, B\bigl (\,-\deriv_\mu\psi^\mu +
i[\tilde\psi_{\mu}, V^{\mu}]\,\bigr )\,\bigr\}.
\la{projectione}
\end{equation}

As in the other cases we have studied, it is necessary to add an
extra   piece to the gauge fermion to make full contact with the
corresponding twisted supersymmetric theory. In  this case, this extra
term is:
\begin{equation}
\Psi_{\text{extra}}=\int_X \sqrt{g}\,\tr\,\bigl\{\,\,
\frac{i}{2}\eta[B,C]\,\bigr\}.
\la{extrah}
\end{equation}

It is now straightforward to see that, with the redefinitions
\begin{align} A'&=A,& V'&=-\frac{1}{\raiz}V,&
\tilde\eta'&=-2\raiz\tilde\eta,\ret
\psi'&=\frac{1}{2}\psi,&  \tilde\psi'&=\half\tilde\psi,& P'&=-2\raiz
P,\ret C'&=\frac{1}{{2\raiz}}C,& \chi^{'+}&= -\chi^{+}, & 
\chi^{'-}&=\chi^{-},\ret
 B'&=-2\raiz B,& N^{'+}&=-N^{+},& 
  N^{'-}&=N^{-},\ret
\eta'&=-2\eta,& &{} & &{}\ret
\la{redefin}
\end{align}   one recovers, in terms of the primed fields,  the
twisted model summarized in
\eqs{butter4} and  \eqs{epi},  which corresponds to the topological
symmetry $Q$.

It is worth to remark that one could  also consider the ``dual"
problem built out of the  basic equations:
\begin{equation}
\begin{cases}  F^{-}_{\mu\nu}-i[V_{\mu},V_{\nu}]^{-}=0,\\
\bigl (\,\deriv_{[\mu}V_{\nu]}\,\bigr )^{+}=0,\\
\deriv_\mu V^\mu=0.
\end{cases}
\la{reversequations}
\end{equation}   The resulting theory corresponds precisely to the
second type of theory obtained in the previous section in our
discussion of the third twist. The corresponding action has the form
$\{\tilde Q, \Psi^- \}$ where $\tilde Q$ is given in \eqs{boheme}  
and $\Psi^-$ is the result of performing a   
${\IZ}_2$-transformation  (see \eqs{melocoton4}) on the gauge fermion
$\Psi^+$ in
\eqs{sesamo4}.

\section{Observables}

In this section we will analyze the structure of the observables of
the theory. Observables are operators which are
$Q$-invariant but are not
$Q$-exact. A quick look at the $Q$-transformations \eqs{melones4} 
shows that the observables are basically the same as in ordinary
Donaldson-Witten theory. However, as there are two anticommuting BRST
charges with the same ghost number, there exist correspondingly,
two possible sets of operators: 
\begin{align} W_0 =& \tr(C^2), & 
\tilde W_0 =& \tr(C^2), \ret W_1 =& \raiz\tr(C\wedge\psi), & 
\tilde W_1 =& -\raiz\tr(C\wedge\tilde \psi), \ret W_2=& \tr(\half
\psi \wedge \psi + \frac{1}{\raiz}C\wedge F), & 
\tilde W_2 =& \tr(\half \tilde\psi \wedge \tilde\psi + \frac{1}{\raiz}
C\wedge F), \ret W_3 =&\frac{1}{2}\tr(\psi \wedge  F), &
\tilde W_3 =& -\frac{1}{2}\tr(\tilde\psi \wedge  F). \ret
\la{Chicholina}
\end{align}  
They satisfy the descent equations $[Q,W_i\}=dW_{i-1}$,
$[\tilde Q,\tilde W_i\}=d\tilde W_{i-1}$, which as we know imply that 
\begin{equation}
{\cal O}^{(\gamma_i)}=\int_{\gamma_i}W_i,\qquad \text{and}\qquad
\tilde{\cal O}^{(\gamma_j)}=\int_{\gamma_i}\tilde W_i, 
\end{equation}
for given
homology cyles $\gamma_j$ of $X$, are observables. It is 
interesting to note that
$W_{0,1}$ and 
$\tilde W_{0,1}$ give observables which are invariant under both,
$Q$ and 
$\tilde Q$. What is more, the operator $W_1$ is $\tilde Q$-exact, 
$W_1=-\frac{1}{\raiz}[\tilde Q,\tr(C\wedge V)]$, so it does not
contribute  to vacuum expectation values containing only $\tilde
Q$-closed observables like 
${\cal O}^{(\gamma_0)}$, ${\cal O}^{(\gamma_1)}$ itself and, as we
shall see  in a moment, ${\cal O}^{(\gamma_2)}$ (besides the
$\tilde{\cal O}^{(\gamma_i)}$).  Indeed, it is possible to modify
$W_2$  adding an irrelevant $Q$-exact piece so that  the resulting
observable is invariant under both,  $Q$ and $\tilde Q$.  The
perturbed two-form, 
\begin{align}  
W'_2 &= \tr\bigl\{\half \psi \wedge \psi
+\half\tilde\psi\wedge\tilde\psi+  \frac{1}{\raiz}C\wedge (F+2iV\wedge
V)\bigr\}\ret&=W_2+\frac{1}{{2\raiz}}\bigl\{\,Q,
\tr(\tilde\psi\wedge V)\,\bigr\}\ret
\la{scarey}
\end{align}  is both $Q$ and $\tilde Q$-closed up to exact forms
\begin{equation}  [Q,W'_2]=[Q,W_2]=dW_1,\qquad\qquad [\tilde
Q,W'_2]=d\tilde W_1.
\la{geri4}
\end{equation}  The two-form $W'_2$ descends to a three-form in the
cohomology  of $Q$, 
\begin{equation}  
dW'_2 =\{Q,W'_3\}, \quad W'_3 =
\half\tr\bigl(\psi\wedge F+\frac{1}{\raiz} d(\tilde\psi\wedge V)\bigr
),
\la{melanie}
\end{equation}  which is  again the original three-form $W_3$ plus a
(irrelevant) perturbation. However, this form is not $\tilde
Q$-closed. We have not been able to find an  appropriate additional
irrelevant perturbation which would render $W_3$ $\tilde  Q$-closed.

Completely analogous arguments hold for the operators 
$\tilde W_i$. Thus, the even operators $W_0$, $W_2'$, $\tilde W_0$ and
$\tilde W_2'$ lead to observables corresponding to the even homology
classes of the  four-manifold $X$ which are invariant under both $Q$
and $\tilde Q$.

Topological invariants are obtained by considering the vacuum
expectation  value of arbitrary products of observables:
\begin{equation}
\left\langle \prod_{\gamma_j} {\cal O}^{(\gamma_j)} \right\rangle,
\la{leoondos4}
\end{equation}  
where it should be understood
that 
$\prod_{\gamma_j} {\cal O}^{(\gamma_j)}$ denotes products of operators
${\cal O}^{(\gamma_i)}$ and $\tilde{\cal O}^{(\gamma_j)}$. The general
form of this vacuum expectation value is,
\begin{equation}
\left\langle \prod_{\gamma_j} {\cal O}^{(\gamma_j)} \right\rangle =
\sum_{k} 
\left\langle \prod_{\gamma_j} {\cal O}^{(\gamma_j)}\right \rangle_k
\ex^{-2\pi i k \tau_0},
\la{pumba4}
\end{equation}  where $k$ is the instanton number and 
$\langle \prod_{\gamma_j} {\cal O}^{(\gamma_j)} \rangle_k$ is the
vacuum expectation value computed  at a fixed value of $k$ with an
action which is
$Q$-exact,
\begin{equation}
\left\langle \prod_{\gamma_j} {\cal O}^{(\gamma_j)} \right\rangle_k
=\int [df]_k
\ex^{\{Q,\Psi\}} 
\prod_{\gamma_j} {\cal O}^{(\gamma_j)}.
\la{timon4}
\end{equation}  In this equation $[df]_k$ denotes collectively the
measure indicating that only gauge configurations of instanton number
$k$  enter in the functional integral. These quantities are
independent of the coupling constant $e_0$. When analyzed in the weak
coupling limit the contributions to the functional integral come from
field configurations which are solutions to the equations
\eqs{morequations4}. All the dependence of the observables on
$\tau_0$ is contained in the sum \eqs{pumba4}.

The $Q$-symmetry of the theory imposses a selection rule for the
products entering
\eqs{leoondos4} which could lead to a possibly non-vanishing result:
the ghost number of  \eqs{leoondos4} must be equal to the virtual
dimension of the corresponding moduli space. In this case the virtual
dimension is not zero but it is independent of the instanton number
$k$, so one could
obtain contributions from many values of $k$. Possibly non-trivial
topological invariants for these cases correspond to products of
operators
\eqs{leoondos4} such that their ghost number matches the  virtual
dimension dim$(G)\chi$. One important
question is again whether  the vacuum expectation values of these
observables have good modular properties under $SL(2,{\IZ})$ 
transformations. We will now show that in fact these vacuum expectation
values are actually independent of $\tau_0$. Thus, in some sense the
invariance under $SL(2,{\IZ})$ is trivially realized in this case.

To see how this comes about, let us consider the action
\eqs{boomerang4} (in its covariantized form) in which the auxiliary
fields appear quadratically. The bosonic part of this action involving
only the field strength $F_{\mu\nu}$ and the vector field
$V_\mu$ can be written in three equivalent forms. The form of the
action
${\cal S}= \{Q,\hat\Psi^+\} - 2 \pi i k \tau_0$, leads to

\begin{align} 
-\int_X d^4 x\,\sqrt{g}\, \tr\, &\Big\{\,\,
 \frac{1}{2e^2_0}\bigl (\,F^{+\mu\nu}-2i[V^\mu,V^\nu]^{+}\,
\bigr )^2
 +\frac{4}{ e^2_0}\bigl (\,(\deriv^{[\mu}V^{\nu]}) ^{-}\,\bigr )^2\ret
+\frac{1}{ e^2_0}
\bigl (\,\deriv_\mu V^\mu\,\bigr )^2\,&\Big\} -2\pi i\tau_0\frac{1}{
32\pi^2}\int_X d^4 x\,\sqrt{g}\tr\,\bigl\{\, *  F_{\mu\nu}
F^{\mu\nu}\,\bigr\},\ret
\la{golgi4}
\end{align}  
the form ${\cal S}= \{\tilde Q,\hat\Psi^-\} - 2
\pi i k \bar\tau_0$, to
\begin{align} 
-\int_X d^4 x\,\sqrt{g}\, \tr\, &\Big\{\,\,
 \frac{1}{2e^2_0}\bigl (\,F^{-\mu\nu}-2i[V^\mu,V^\nu]^{-}\,
\bigr )^2
 +\frac{4}{ e^2_0}\bigl (\,(\deriv^{[\mu}V^{\nu]}) ^{+}\,\bigr )^2\ret
+\frac{1}{ e^2_0}
\bigl (\,\deriv_\mu V^\mu\,\bigr )^2\,&\Big\} -2\pi
i\bar\tau_0\frac{1}{ 32\pi^2}\int_X d^4 x\,\sqrt{g}\tr\,\bigl\{\, * 
F_{\mu\nu} F^{\mu\nu}\,\bigr\},
\ret
\la{otrogolgi4}
\end{align}  
and, finally, the form
${\cal S}= \half \{Q,\hat\Psi^+\} + \half \{\tilde Q,\hat\Psi^-\}  - 2
\pi i k {\hbox{\rm Re}}(\tau_0) $, to
\begin{align} 
-\frac{1}{e^2_0}\int_X\tr\, \Big\{\,\,
 &\frac{1}{4}\bigl (\,F^{\mu\nu}-2i[V^\mu,V^\nu]\,
\bigr )^2
 +2(\deriv^{[\mu}V^{\nu]})^2+
\bigl (\,\deriv_\mu V^\mu\,\bigr )^2\,\Bigr\}
\ret &-2\pi i\,{\hbox{\rm Re}}(\tau_0) \frac{1}{ 32\pi^2}\int_X d^4
x\,\sqrt{g}\tr\,\bigl\{\, *  F_{\mu\nu} F^{\mu\nu}\,\bigr\}.\ret
\la{ateate}
\end{align}   
Now in the short-distance limit the path integral is dominated by 
the quadratic bosonic terms in the action. In the first
case, the contributions come from the moduli space defined by Eqs.
\eqs{morequations4}.  Notice that the normalization factor for
$V_\mu$ in
\eqs{redefin} has to be taken into account since \eqs{golgi4} 
corresponds to the action resulting after the twisting. Similarly, in
the second case the contributions come from 
the moduli space defined by Eqs.
\eqs{reversequations}. As for the third case, however, the
contributions come from configurations which solve the following set of
equations:
\begin{equation}
\begin{cases}  F_{\mu\nu}-2i[V_{\mu},V_{\nu}]=0,\\
\deriv_{[\mu}V_{\nu]}=0,\\
\deriv_\mu V^\mu=0,
\end{cases}
\la{jontas4}
\end{equation} 
 which define a moduli space which is the intersection of the other
two. This is the moduli space which appears in the formulation of the
third twist presented in
\cite{blauthomp}\cite{marcus}. 
 
There are two different ways to understand that the theory indeed
localizes on the moduli  space defined by \eqs{jontas4}. On  the one
hand, from the identity ${\cal S}=
\half \{Q,\hat\Psi^+\} +  \half \{\tilde Q,\hat\Psi^-\} - 2 \pi i k
{\hbox{\rm Re}}(\tau_0) $, one can see that  vacuum expectation values
$\langle
\prod_{j} {\cal O}^{(j)} \rangle$ are independent of the coupling
constant $e_0$ as long as the observables are  simultaneously invariant
under both, $Q$ and
$\tilde Q$.  In the case at hand this is  true at least for  vacuum
expectation values involving products of operators ${\cal
O}^{(\gamma_{0})}$,
${\cal O}^{(\gamma_{2})}$, and $\tilde{\cal O}^{(\gamma_{0})}$,
$\tilde{\cal O}^{(\gamma_{2})}$.   On the other hand, one could extend
Witten's arguments leading to his fixed-point  theorem
\cite{yau}, to argue that the theory should localize onto the fixed
points  of both the BRST symmetries $Q$ and
$\tilde Q$ simultaneously, that is to say on the intersection of the
zero loci
\eqs{morequations4} and \eqs{reversequations}. This could be  thought
of as some sort of generalized localization principle for topological
field  theories with more than one BRST symmetry. 
 
Notice that the three points of view lead to three different types of
dependence on
$\tau_0$. The first one implies that vacuum expectation values are
holomorphic in
$\tau_0$, the second that they are antiholomorphic, and the third that
they depend only on the real part of
$\tau_0$. We will solve this puzzle showing that actually the vacuum
expectation values are just real numbers and not functions of 
$\tau_0$.

We first prove that any solution of \eqs{jontas4} must involve a gauge
connection whose instanton number is zero. Indeed, from the identity,
\begin{align}
\int_X d^4 x\,\sqrt{g}\, \tr\, &\bigl\{\,
  *  F_{\mu\nu}\bigl (\,F^{\mu\nu}-2i[V^\mu,V^\nu]\,
\bigr )-4 * \deriv^{[\mu}V^{\nu]}\deriv_{[\mu}V_{\nu]}\,\bigr\}
\ret &=
\int_X d^4 x\,\sqrt{g}\,\tr\,\bigl\{\, * 
F_{\mu\nu}F^{\mu\nu}\,\bigr\},\ret
\la{gadget4}
\end{align}  follows that any solution of $\eqs{jontas4}$ must have
$k=0$. This implies that only configurations with vanishing instanton
number contribute and therefore:
\begin{equation}
\left\langle \prod_{\gamma_j} {\cal O}^{(\gamma_j)} \right\rangle =  
\left\langle \prod_{\gamma_j} {\cal O}^{(\gamma_j)}
\right\rangle_{k=0},
\la{mana4}
\end{equation}  which is clearly independent of $\tau_0$. From
\eqs{golgi4} and \eqs{otrogolgi4} follows that for $k=0$ a solution to
the equations of the first moduli space
\eqs{morequations4} is also a solution to the ones of the second
\eqs{reversequations} and therefore also to the ones of the third
\eqs{jontas4}. For
$k\neq 0$, however, one can have solutions to the equations of the
first moduli space which are not solutions to the equations of the
second and therefore neither to the ones of the third. For $k\neq 0$
the quantities
$\langle \prod_{\gamma_j} {\cal O}^{(\gamma_j)} \rangle_{k}$ are
different in each point of view.  They clearly vanish in the third
case. On the other hand, there is no reason why they should also
vanish in the other two cases. Our results, however, suggest that they
do vanish. We have shown it at least for vacuum expectation values
involving products of operators ${\cal O}^{(\gamma_{0})}$, 
${\cal O}^{(\gamma_{2})}$, and $\tilde{\cal O}^{(\gamma_{0})}$,
$\tilde{\cal O}^{(\gamma_{2})}$.

We will end this section by recalling a vanishing theorem which has
already been  discussed in \cite{marcus} and which tells us when the
third moduli space  \eqs{jontas4} reduces to the moduli space of flat
connections. The equations \eqs{jontas4} have the immediate solution
$V=0$, $F=0$, that is, the moduli space of flat connections is
contained in  the moduli space defined by the equations \eqs{jontas4}.
We will show that under certain conditions both moduli spaces are in
fact the same.  To see this note that since,  
\begin{align} &\int_X d^4 x\,\sqrt{g}\, \tr\, \bigl\{\,
 \frac{1}{4}\bigl (\,F^{\mu\nu}-2i[V^\mu,V^\nu]\,
\bigr )^2+2\bigl(\deriv^{[\mu}V^{\nu]}\bigr )^2+
\bigl (\,\deriv_\mu V^\mu\,\bigr )^2\,\bigr\}
\ret &=
\int_X d^4 x\,\sqrt{g}\,\tr\,\bigl\{\,\deriv_\mu V_\nu \deriv^\mu 
V^\nu +R_{\mu\nu}V^\mu V^\nu+\frac{1}{4}F_{\mu\nu}F^{\mu\nu}-
([V_\mu,V_\nu])^2\,\},\ret &{}
\la{axon}
\end{align}  
it follows that if the Ricci tensor is such that 
\begin{equation} 
R_{\mu\nu}V^\mu V^\nu>0\quad {\hbox{\rm for}}~
V\not=0,
\la{vanishth}
\end{equation}   
the solutions to the equations \eqs{jontas4} are
necessarily of the form
$V=0$, 
$F=0$, and thus the moduli space is the space of flat gauge
connections on 
$X$. 

\newpage


\chapter*{Appendix}
\markboth{\footnotesize\bfseries Appendix
}{\footnotesize\bfseries Appendix}
\markright{\textsc {Duality in Topological Quantum Field Theories}}
\addcontentsline{toc}{chapter}{\numberline{}Appendix}
\def\theequation{A.\arabic{equation}}
\def\thesection{A.\arabic{section}}

\setcounter{equation}{0}
\vskip1.5cm

We will now summarize the conventions used in this work. 
Basically we will describe the elements of the positive and
negative chirality spin bundles $S^+$ and $S^-$ on a
four-dimensional spin manifold $X$ endowed with a vierbein
$e^{m\mu}$ and a spin connection
$\omega_{\mu}^{m n}$. The spaces of sections of the spin
bundles $S^+$  and $S^-$ correspond, from the field-theory
point of view, to the  set of two-component Weyl spinors
defined on the manifold $X$. These are  the simplest
irreducible representations of the holonomy group $SO(4)$. 
We will denote positive-chirality (or negative-chirality)
spinors by indices 
$\alpha,\beta,\ldots=1,2$ (or
$\d{\alpha},\d{\beta},\ldots=1,2$). Spinor indices  are
raised and lowered with the $SU(2)$ invariant tensor
$C_{\alpha\beta}$  (or
$C_{\d{\alpha}\d{\beta}}$) and its inverse
$C^{\alpha\beta}$  (or
$C^{\d{\alpha}\d{\beta}}$), with the conventions
$C_{21}=C^{12}=+1$,  so that, 
\bea
&C_{\alpha\beta}C^{\beta\gamma}=\delta_{\alpha}{}
^{\gamma},
\qquad\qquad
&C_{\alpha\beta}C^{\gamma\delta}=\delta_\alpha{}
^\delta
\delta_\beta{}^\gamma-\delta_\alpha{}^\gamma
\delta_\beta{}^\delta,\ret
&C_{\d{\alpha}\d{\beta}}C^{\d{\beta}\d{\gamma}}=\delta_{
\d{\alpha}}{}
^{\d{\gamma}},
\qquad\qquad
&C_{\d{\alpha}\d{\beta}}C^{\d{\gamma}\d{\delta}}=
\delta_\d{\alpha}{}
^\d{\delta}
\delta_\d{\beta}{}^\d{\gamma}-\delta_\d{\alpha}{}^\d{\gamma}
\delta_\d{\beta}{}^\d{\delta}.\ret
\label{setas}
\eea 

The  spinor representations and the vector representation
associated to 
$S^+\!\otimes S^-$ are related by the Clebsch-Gordan 
$\sigma^m{}_{\!\alpha\d{\alpha}}=(i{\bf 1}, \vec\tau)$ and 
$\bar\sigma^{m\dalpha\alpha}= (i{\bf 1}, -\vec\tau)$, where
${\bf 1}$ is the $2\times2$ unit matrix and 
$\vec\tau=(\tau^1,\tau^2,\tau^3)$ are the Pauli matrices,  
\begin{equation}\tau^1 = \left(
\begin{matrix} 
0&{1}\\
{1}&0
\end{matrix}
\right),\,\,\,\,\,\,\,\,\,\,\,
\tau^2 = \left(
\begin{matrix}
0&{-i}\\
{i}&0
\end{matrix}\right),\,\,\,\,\,\,\,\,\,\,\,
\tau^3 = \left(
\begin{matrix}
1&{0}\\
{0}&-1
\end{matrix}\right).
\label{lados}
\end{equation}
The Pauli matrices satisfy: 
\begin{equation}\tau_a\tau_b = 
i\epsilon_{abc}\tau_c+\delta_{ab}{\bf 1},
\label{latres}
\end{equation}where $\epsilon_{abc}$ is the totally 
antisymmetric
tensor with
$\epsilon_{123}=1$.

Under an infinitesimal $SO(4)$ rotation a Weyl spinor
$M_\alpha$,
$\alpha=1,2$, associated to $S^+$, transforms as:
\begin{equation}\delta M_\alpha = \frac{i}{ 2} \epsilon_{mn}
(\sigma^{mn})_\alpha{}^\beta M_\beta,
\label{laocho}
\end{equation}where $\epsilon_{mn}=-\epsilon_{nm}$ are the
infinitesimal parameters of the transformation. On the
other hand, a Weyl spinor $N^{\dot\alpha}$,
$\dot\alpha=1,2$, associated to $S^-$, transforms as,
\begin{equation}\delta N^{\dot\alpha} = \frac{i}{ 2}
\epsilon_{mn} (\bar\sigma_{mn})^{\dot\alpha}{}_{\dot\beta}
N^{\dot\beta}.
\label{lanueve}
\end{equation}The matrices $\sigma^{mn}$ and $\bar\sigma^{mn}$ are
antisymmetric in $m$  and $n$ and are defined as follows:
\begin{equation}\sigma^{mn}{}_{\!\alpha}{}^{\!\beta}=\frac{i}{
2}\sigma^{[m}{}_{\!\alpha\dalpha}\bar\sigma
^{n]\dalpha\beta},\qquad\qquad
\bar\sigma^{mn\dalpha}{}_{\!\d{\beta}}=\frac{i}{
2}\bar\sigma^{[m\dalpha\alpha}\sigma^{n]}{}_{\!\alpha
\d{\beta}}.
\label{ap1Pollux}
\end{equation}They satisfy the self-duality properties,
\begin{equation}\sigma^{mn} =\half
\epsilon^{mnpq}\sigma_{pq},
\qquad\qquad 
\bar\sigma^{mn} = -\half
\epsilon^{mnpq}
\bar\sigma_{pq},
\label{Io}
\end{equation}and the $SO(4)$ algebra,
\begin{equation}[\sigma_{mn},\sigma_{pq}]= i(\delta_{mp}\sigma_{nq}-
\delta_{mq}\sigma_{np}-
\delta_{np}\sigma_{mq}+
\delta_{nq}\sigma_{mp}).
\label{Ramses}
\end{equation}The same algebra is fulfilled by $\bar\sigma^{mn}$.

Let us consider the covariant derivative $ D_\mu$ on the
manifold
$X$. Acting on an element of $\Gamma(X,S^+)$ it has the
form:
\begin{equation}D_\mu M_\alpha = \partial_\mu M_\alpha +
\frac{i}{ 2}
\omega_{\mu}^{m n} (\sigma_{mn})_\alpha{}^\beta M_\beta,
\label{carve}
\end{equation}where $\omega_{\mu}^{m n}$ is the spin connection.
Defining 
$ D_{\alpha\dot\alpha}$ as,
\begin{equation}D_{\alpha\dot\alpha} = (\sigma_n)_{\alpha\dot\alpha}
e^{n\mu}  D_\mu,
\label{cave}
\end{equation}where $e^{n\mu}$ is the vierbein on $X$, the Dirac
equation for 
$M\in \Gamma(X,S^+)$ and $N\in \Gamma(X,S^-)$  can be
simply written as,
\begin{equation}D_{\alpha\dot\alpha} M^\alpha=0,
\,\,\,\,\,\,\,\,\,\,\,\,\, D_{\alpha\dot\alpha}
N^{\dot\alpha}=0.
\label{dirac}
\end{equation}Let us now introduce a principal 
$G$-bundle $P\to X$ with its associated connection one-form
$A$,  and let us consider that the Weyl spinors $M_\alpha$
realize locally an element of 
$\Gamma(S^+\otimes \ad P)$, \ie, they transform under a $G$
gauge transformation in the adjoint representation
-- indeed,  $\ad P$ is the vector bundle  associated  to $P$
through the adjoint representation of the gauge group on 
its Lie algebra: 

\begin{equation}\delta M_\alpha^a =
i[M_\alpha,\phi]^a=-i(T^c)^{ab}M_\alpha^b \phi^c,
\label{gaugetransf}
\end{equation}where $(T^a)^{bc}=-if^{abc}$, $a=1,\cdots,{\rm
dim}(G)$,  are the generators of $G$ in the adjoint
representation, which are traceless and 
hermitian. In (\ref{gaugetransf}) 
$\phi^a$, $a=1,\cdots,
{\rm dim}(G)$, denote the infinitesimal parameters of the
gauge transformation.

In terms of the gauge connection $A$, the covariant
derivative 
(\ref{carve}) can be promoted to a full covariant derivative
acting on   sections in $\Gamma(X, S^+\otimes\ad P)$,
\begin{equation}{\cal D}_\mu M_\alpha = \partial_\mu M_\alpha +
\frac{i}{ 2}
\omega_{\mu}^{m n} (\sigma_{mn})_\alpha{}^\beta M_\beta
+i[A_\mu,M_\alpha],
\label{carvemore}
\end{equation}and its analogue in (\ref{cave}):
\begin{equation}{\cal D}_{\alpha\dalpha} = (\sigma_n)_{\alpha\dalpha}
e^{n\mu}
 {\cal D}_\mu.
\label{cavemore}
\end{equation}In terms of the full covariant derivative the Dirac
equations (\ref{dirac})  become:
\begin{equation}{\cal D}_{\alpha\dalpha} M^{\alpha}=0,
\,\,\,\,\,\,\,\,\,\,\,\,\, {\cal D}_{\alpha\dalpha}
N^{\dalpha}=0.
\label{diraci}
\end{equation}Given an element of $\Gamma(X,S^+\otimes \ad P)$, $M_\alpha
= (a,b)$ we define $\overline M^{\alpha} = (a^{*},b^{*})$.
In this way,  given $M,N \in \Gamma(X, S^+\otimes \ad P)$,
the gauge-invariant quantity entering the metric 
\begin{equation}\half\tr\,\bigl (\,\overline M^{\alpha} N_{\alpha} +
\overline N^{\alpha} M_{\alpha}\,\bigr ),
\label{mmmetric}
\end{equation}is positive definite. With similar arguments the
corresponding gauge  invariant metric in the fibre 
$\Gamma(X,S^-\otimes \ad P)$, which  we define as 
\begin{equation}\half\tr\,\bigl (\,\overline M_{\dalpha} N^{\dalpha} +
\overline N_{\dalpha} M^{\dalpha}\,\bigr ),
\label{mmmetrica}
\end{equation}for $M,N\in\Gamma(X,S^-\otimes \ad P)$, can be seen to
be  positive definite, too. For self-dual two-forms $Y,Z\in
\Gamma(X,\Lambda^{2,+}T^{*}X\otimes\ad P)\equiv
\Omega^{2,+}(X,\ad P)$ our definition of the metric is the
following:
\begin{equation}\langle Y,Z\rangle =\int_X\tr\,\bigl (\,Y\wedge
*Z\,\bigr)=\half 
\int_X\tr\,\bigl (\,Y_{\mu\nu}Z^{\mu\nu}\,\bigr
)=-\frac{1}{4}
\int_X\tr\,\bigl (\,Y_{\alpha\beta}Z^{\alpha\beta}\,\bigr ),
\label{gangan}
\end{equation}where
$Y_{\alpha\beta}=\sigma^{\mu\nu}_{\alpha\beta}
Y_{\mu\nu}$ (and similarly for $Z$),
and  we have used the identity 
\begin{equation}
Y_{\alpha\beta}Z^{\alpha\beta
}=-2Y^+_{\mu\nu}Z^{+\mu\nu}.
\label{ap1caca}
\end{equation}
Acting on an element of $\Gamma(X,S^+\otimes\ad P)$ the
covariant derivatives satisfy:
\begin{equation}[{\cal D}_\mu,{\cal D}_\nu]M_\alpha = i[F_{\mu\nu},
M_\alpha] + \frac{i}{2}
R_{\mu\nu}{}^{mn}(\sigma_{mn})_\alpha{}^\beta M_\beta,
\label{commutador}
\end{equation}where $F_{\mu\nu}$ are the components of the two-form
field strength:
\begin{equation}F = d A + iA\wedge A,
\label{fuerzacampo}
\end{equation}and $R_{\mu\nu}{}^{mn}$ the components of the curvature
two-form,
\begin{equation}
R^{mn} = d\omega^{mn}+\omega^{mp}\wedge \omega^{pn},
\label{curvatura}
\end{equation}being $\omega^{mn}$ the spin connection one-form. The
scalar curvature is defined as:
\begin{equation}
R=e^{\mu}_m e^{\nu}_n R_{\mu\nu}{}^{mn},
\label{curvescalar}
\end{equation}
and the Ricci tensor as:
\begin{equation}R_{\kappa\lambda}=e^{\mu}{}{\!_m} 
e_{n\lambda}
R_{\mu\kappa}{}^{mn}.
\label{Ricciten}
\end{equation}
The components of the curvature two-form (\ref{curvatura}) 
are
related to the  components of the Riemann tensor as follows:
\begin{equation}R_{\mu\nu\kappa\lambda}=e_{\kappa m}e_{\lambda
n}R_{\mu\nu}{}^{\!mn}.
\label{riemann}
\end{equation}
The Riemann tensor satisfies the following algebraic
properties:
\vskip.5cm
(a) Symmetry:
\begin{equation}
R_{\lambda\mu\nu\kappa}=R_{\nu\kappa\lambda\mu},
\label{weinbuno}
\end{equation}

(b) Antisymmetry:
\begin{equation}
R_{\lambda\mu\nu\kappa}=-R_{\mu\lambda\nu\kappa}=
-R_{\lambda\mu\kappa\nu}=+R_{\mu\lambda\kappa\nu},
\label{weinbdos}
\end{equation}

(c) Cyclicity:
\begin{equation}
R_{\lambda\mu\nu\kappa}+R_{\lambda\kappa\mu\nu}+
R_{\lambda\nu\kappa\mu}=0.
\label{weinbtres}
\end{equation}
Notice that (\ref{weinbtres}) implies that 
\begin{equation}
\epsilon^{\mu\nu\kappa\sigma}R_{\lambda\mu\nu\kappa}=0.
\label{weinbcuatro}
\end{equation}

\newpage


\vfil

\newpage

\end{document}